\documentclass[a4paper,12pt,notitlepage]{report}

\usepackage{graphics}
\usepackage{latexsym}
\title{\textbf{\huge{A~Review~on~Tachyon~Condensation \vspace{0.5cm}\\
in~Open~String~Field~Theories}} \\ \vspace{1.5cm}}
\author{Kazuki Ohmori\thanks{E-mail: 
ohmori@hep-th.phys.s.u-tokyo.ac.jp}
\vspace{1cm} \\
\small{\textit{Department of Physics, Faculty of Science, University of 
Tokyo}} \\
\small{\textit{Hongo 7-3-1, Bunkyo-ku, Tokyo 113-0033, Japan}}}
\date{}
\newcommand{\ap}{\alpha^{\prime}}
\newcommand{\cA}{\mathcal{A}}
\newcommand{\cB}{\mathcal{B}}
\newcommand{\cC}{\mathcal{C}}
\newcommand{\cD}{\mathcal{D}}
\newcommand{\cG}{\mathcal{G}}
\newcommand{\cH}{\mathcal{H}}
\newcommand{\cI}{\mathcal{I}}
\newcommand{\cK}{\mathcal{K}}
\newcommand{\cL}{\mathcal{L}}
\newcommand{\cM}{\mathcal{M}}
\newcommand{\cN}{\mathcal{N}}
\newcommand{\cO}{\mathcal{O}}
\newcommand{\cP}{\mathcal{P}}
\newcommand{\cQ}{\mathcal{Q}}
\newcommand{\cS}{\mathcal{S}}
\newcommand{\cT}{\mathcal{T}}
\newcommand{\cV}{\mathcal{V}}
\newcommand{\cW}{\mathcal{W}}
\newcommand{\cX}{\mathcal{X}}
\newcommand{\zetto}{\mathbf{Z}}
\newcommand{\aaru}{\mathbf{R}}
\newcommand{\shii}{\mathbf{C}}
\newcommand{\gh}{\#_{\mathrm{gh}}}
\newcommand{\pic}{\#_{\mathrm{pic}}}
\newcommand{\ca}{{}^{\circ}\!\!\!\>\!{}_{\circ}}
\newcommand{\llk}{\langle\!\langle}
\newcommand{\rrk}{\rangle\!\rangle}
\newcommand{\bllk}{\biggl\langle\!\!\!\biggl\langle}
\newcommand{\brrk}{\biggr\rangle\!\!\!\biggr\rangle}
\newcommand{\sech}{\mathop{\mathrm{sech}}\nolimits}

\begin{document}
\maketitle
\baselineskip 6mm
\begin{abstract}
We review the recent studies of tachyon condensation in
string field theory. After introducing the open string field 
theory both for bosonic string and for superstring, we use them 
to examine the conjecture 
that the unstable configurations of the D-brane will decay 
into the `closed string vacuum' through the tachyon condensation.
And we describe the attemps to construct 
a lower dimensional bosonic D-brane as an unstable lump 
solution of the string field equation. 
We obtain exact results from another formulation,
background independent open string field theory.
We also discuss some other topics which are related to
tachyon condensation in string theory,
such as the construction of a D-brane as a noncommutative 
soliton and some field theory models. 
This paper is based on my master's thesis submitted to 
Department of Physics, Faculty of Science, University of Tokyo on January 
2001. 
\end{abstract}
\tableofcontents
%\listoffigures
%\listoftables
\baselineskip 6mm

\chapter{Introduction~and Conventions}
\section{Introduction}
The spectrum of bosonic open strings living on a bosonic D-brane contains a 
tachyonic mode\footnote{For earlier works on tachyon condensation, 
see \cite{Halp}.}, which indicates that the bosonic D-brane is unstable. It has 
been conjectured that the potential for the tachyon field has a non-trivial 
minimum where the sum of the D-brane tension and the negative energy density 
from the tachyon potential vanishes so that the minimum represents the usual 
vacuum of closed string theory without any D-brane or open 
string~\cite{Descent}. Moreover, it has also been conjectured that, instead 
of spatially homogeneous tachyon condensation, solitonic lump configurations 
where the tachyon field asymptotically approaches the vacuum value represent 
lower dimensional D-branes~\cite{Descent,Reck}.

Though there are no tachyonic modes on a BPS D-brane of Type II superstring 
theory, by considering the unstable systems, such as a non-BPS D-brane or 
a coincident D-brane anti-D-brane pair, tachyonic modes appear. In these 
systems, similar conjectures have been made: At the minimum of the tachyon 
potential the energy density of the system vanishes and the D-brane 
disappears. And the tachyonic kink solution which interpolates between two 
inequivalent (but degenerate) minima represents a lower dimensional 
D-brane~\cite{cycle,Sen,DK,Hora}.
\medskip

In the framework of the conventional string theory which is first-quantized 
and is formulated only on-shell, various arguments supporting the above 
conjectures have already been given. But they can only provide indirect 
evidences because the concept itself of the potential for the zero-momentum 
(\textit{i.e.} spacetime independent) tachyon is highly off-shell. So we 
need an off-shell formulation of string theory to obtain direct evidence for 
the conjectures. As such, \textit{open string field theory} has recently 
been studied in this context. In this paper, we review the various results 
which have been obtained about the tachyon physics as well as the 
formulations of open string field theories. 
\medskip

This paper is organized as follows. In the remainder of this chapter, we 
collect the formulae we may use in later chapters without explanation. 
We follow the convention that $\hbar=c=1$, but explicitly keep the Regge 
slope parameter $\ap$ almost everywhere (except for part of chapter 3). 
In chapter 2, we introduce the Witten's formulation of open string field 
theory. After writing down the form of the cubic action, we define the 
3-string interaction vertex in terms of the two dimensional conformal field 
theory (CFT) correlators and show examples of calculations. Using the 
\textit{level truncation} method, we will find the `nonperturbative vacuum' 
which minimizes the potential for the tachyonic string field and obtain 
the numerical evidence for the D-brane annihilation conjecture. Further, 
we explore the nature of the new vacuum, including the open string spectrum 
of the fluctuations around it. In chapter 3, we calculate the tachyon 
potential in the level truncation scheme in superstring theory. For that 
purpose, we introduce three candidates for superstring field theory: 
Witten's cubic (Chern-Simons--like) open superstring field theory, modified 
(0-picture) cubic superstring field theory, and Berkovits' 
Wess-Zumino-Witten--like superstring field theory. In chapter 4, we construct 
a tachyonic lump solution on a D-brane and compare its tension with the 
tension of the expected lower dimensional D-brane in bosonic string field 
theory (except for section~\ref{sec:kink}). The modified level expansion 
scheme gives us very accurate results that are regarded as evidence for the 
conjecture that the lump solution is identified with a D-brane of lower 
dimension. In chapter 5, we obtain exact tachyon potential and lump-like 
tachyon condensate whose tension exactly agrees with the expected one from 
another formalism of string field theory, called background independent 
open string field theory or boundary string field theory, both in 
bosonic string theory and in superstring theory. In chapter 6, we take up 
some topics which are useful in understanding the tachyon condensation in 
string field theory. In section 6.1 we see how the noncommutativity enables 
us to construct exact soliton (lump) solutions in the effective 
tachyon(-gauge) field theory. In section 6.2 we consider tachyonic scalar 
field theory models whose lump (or kink) solutions have several interesting 
features. 

\section{Conventions and Useful Formulae}\label{sec:Conv}
We enumerate here the notations and conventions used throughout the paper. 
We follow mostly the conventions of the text by Polchinski~\cite{Pol}.

\vspace{1cm}

\begin{itemize}
\item Spacetime metric (flat)
\begin{equation}
\eta_{\mu\nu}=\mathrm{diag}(-+\ldots +).
\end{equation}

\item Free bosonic string world-sheet action (matter)
\begin{eqnarray}
\cS &=& \frac{1}{4\pi\ap}\int d^2\sigma \sqrt{g} g^{\alpha\beta}
\partial_{\alpha}
X^{\mu}\partial_{\beta}X_{\mu} \nonumber \\ &=&\frac{1}{2\pi\ap}\int d^2z 
\partial X^{\mu}\bar{\partial}X_{\mu}. \label{eq:wsS}
\end{eqnarray}

\item (ghost)
\begin{equation}
\cS_{\mathrm{g}}=\frac{1}{2\pi}\int d^2z \  (b\bar{\partial}c+
\tilde{b}\partial \tilde{c}). \label{eq:wsSg}
\end{equation}
where $z=e^{-iw}$ for closed string, $z=-e^{-iw}$ for open string; and 
$w=\sigma^1+i\sigma^2$. \\

\item $\mathcal{N}=(1,1)$ superstring world-sheet action (matter)
\begin{equation}
\cS=\frac{1}{2\pi\ap}\int d^2z\left(\partial X^{\mu}\bar{\partial}X_{\mu}
+\frac{\ap}{2}\left(\psi^{\mu}\bar{\partial}\psi_{\mu}+\tilde{\psi}^{\mu}
\partial\tilde{\psi}_{\mu}\right)\right). \label{eq:wsSs} 
\end{equation}

\item (ghost)
\begin{equation}
\cS_{\mathrm{g}}=\frac{1}{2\pi}\int d^2z\left(b\bar{\partial}c+\beta
\bar{\partial}\gamma+\tilde{b}\partial\tilde{c}+\tilde{\beta}\partial
\tilde{\gamma}\right). \label{eq:wsSsg}
\end{equation}

\item Operator product expansion (OPE) for closed string (sphere) 
or disk interior
\begin{eqnarray}
X^{\mu}(z_1,\bar{z_1})X^{\nu}(z_2,\bar{z_2}) &\sim& -\frac{\ap}{2}
\eta^{\mu\nu}\ln |z_{12}|^2, \label{eq:XX} \\
\partial X^{\mu}(z_1)\partial X^{\nu}(z_2) &\sim& -\frac{\ap}{2}\frac{
\eta^{\mu\nu}}{(z_1-z_2)^2}, \label{eq:dXdX} \\
\psi^{\mu}(z)\psi^{\nu}(0)&\sim& \frac{\eta^{\mu\nu}}{z}, \label{eq:psipsi} \\
b(z_1)c(z_2)&\sim&\frac{1}{z_1-z_2}, \label{eq:bcOPE} \\
\beta (z_1)\gamma (z_2)&\sim&-\frac{1}{z_1-z_2}. \label{eq:betagamma}
\end{eqnarray}

\item On the disk boundary, $XX$ OPE is modified as
\begin{equation}
X^{\mu}(x_1)X^{\nu}(x_2) \sim -\ap\eta^{\mu\nu}\ln (x_1-x_2)^2. \label{eq:XXd}
\end{equation}

\item Energy-momentum tensor (matter)
\begin{eqnarray}
T(z)&=&-\frac{1}{\ap}:\partial X^{\mu}\partial X_{\mu}:, \label{eq:emT} \\
T(z)T(0)&\sim&\frac{c}{2z^4}+\frac{2}{z^2}T(0)+\frac{1}{z}\partial T(0),
\label{eq:TTOPE} \\
\left(\frac{\partial z'}{\partial z}\right)^2T'(z')&=&T(z)-\frac{c}{12}
\{z',z\}; \label{eq:Ttransf} \\ 
\{z',z\}&=&\frac{2\partial_z^3z'\partial_zz'-3\partial_z^2z'
\partial_z^2z'}{2\partial_zz'\partial_zz'} \quad \mbox{(Schwarzian)}.
\label{eq:Schwarz} 
\end{eqnarray}

\item (ghost)
\begin{equation}
T_g(z)=:(\partial b)c:-2\partial :bc:=-:(\partial b)c:-2:b\partial c:.
\label{eq:Tgh}
\end{equation}

\item Supercurrent, energy-momentum tensor in superstring case (matter)
\begin{eqnarray}
G^{\mathrm{m}}(z)&=&i\sqrt{\frac{2}{\ap}}\psi^{\mu}\partial X_{\mu}(z) \quad
\mbox{weight } h=\frac{3}{2}, \label{eq:Gmat} \\
T^{\mathrm{m}}(z)&=&-\frac{1}{\ap}\partial X^{\mu}\partial X_{\mu}-\frac{1}{2}
\psi^{\mu}\partial\psi_{\mu} \quad h=2. \label{eq:emTs}
\end{eqnarray}

\item (ghost)
\begin{eqnarray}
G^{\mathrm{g}}(z)&=&-\frac{1}{2}(\partial\beta)c+\frac{3}{2}\partial(\beta c)
-2b\gamma, \label{eq:Ggh} \\
T^{\mathrm{g}}(z)&=&(\partial b)c-2\partial(bc)+(\partial\beta)\gamma-\frac{3}
{2}\partial(\beta\gamma), \label{eq:Tsgh}
\end{eqnarray}

\item Ghost number current
\begin{eqnarray}
j&=&-:bc:, \label{eq:jgh} \\
\frac{\partial z'}{\partial z}j'(z')&=&j(z)+\frac{3}{2}\frac{\partial_z^2z'}
{\partial_zz'}, \label{eq:jtransf}
\end{eqnarray}

\item Assignment of ghost number ($\gh$ is an operator that counts the 
ghost number of its argument, defined in the text.)
\begin{eqnarray*}
\gh(b)&=&-1 \> , \quad \gh(c)=1, \\
\gh(\beta)&=&-1 \> , \quad \gh(\gamma)=1. 
\end{eqnarray*}

\item Virasoro algebra
\begin{equation}
[L_m,L_n]=(m-n)L_{m+n}+\frac{c}{12}(m^3-m)\delta_{m+n,0}. \label{eq:Vir}
\end{equation}
\begin{eqnarray}
[L_m,\cO_n]
&=& [(h-1)m-n] \cO{}_{m+n} \label{eq:LOcom} \\ & & \cO : 
\mbox{a holomorphic tensor of weight} (h,0). \nonumber 
\end{eqnarray}
\begin{equation}
\{G_r,G_s\}=2L_{r+s}+\frac{c}{12}(4r^2-1)\delta_{r+s,0}, \label{eq:VirGG}
\end{equation}
\begin{equation}
[L_m,G_r]=\frac{m-2r}{2}G_{m+r}. \label{eq:VirLG}
\end{equation}

\item Mode expansion
\begin{eqnarray}
\partial X^{\mu}&=&-i\sqrt{\frac{\ap}{2}}\sum_{m=-\infty}^{\infty}
\frac{\alpha^{\mu}_m}{z^{m+1}}, \label{eq:delX} \\
X^{\mu}(z,\bar{z})&=&x^{\mu}-i\frac{\ap}{2}p^{\mu}\ln |z|^2+i\sqrt{\frac{\ap}
{2}}\sum_{m=-\infty \atop m\neq 0}^{\infty}\frac{1}{m}\left(\frac{
\alpha^{\mu}_m}{z^m}+\frac{\tilde{\alpha}^{\mu}_m}{\bar{z}^m}\right); 
\nonumber \\ \mathrm{with} & & p^{\mu}=\sqrt{\frac{2}{\ap}}
\alpha_0^{\mu}=\sqrt{\frac{2}{\ap}}\tilde{\alpha}_0^{\mu} \quad 
\mbox{for closed string}, \label{eq:Xcl} \\
X^{\mu}(z,\bar{z})&=&x^{\mu}-i\ap p^{\mu}\ln |z|^2+i\sqrt{\frac{\ap}
{2}}\sum_{m=-\infty \atop m\neq 0}^{\infty}\frac{\alpha_m^{\mu}}{m}
\left(z^{-m}+\bar{z}^{-m}\right); 
\nonumber \\ \mathrm{with} & & p^{\mu}=\frac{1}{\sqrt{2\ap}}
\alpha_0^{\mu} \quad 
\mbox{for open string}. \label{eq:Xop}
\end{eqnarray}

\item Virasoro generators (bosonic string)
\begin{eqnarray}
L_0^{\mathrm{m}}&=&\frac{\ap p^2}{4}+\sum_{n=1}^{\infty}\alpha^{\mu}_{-n}
\alpha_{\mu n} \quad \mbox{for closed string}, \label{eq:L0mc} \\
L_0^{\mathrm{m}}&=&\ap p^2+\sum_{n=1}^{\infty}\alpha^{\mu}_{-n}
\alpha_{\mu n} \quad \mbox{for open string}, \label{eq:L0mo} \\
L_m^{\mathrm{m}}&=&\frac{1}{2}\sum_{n=-\infty}^{\infty}
\ca \alpha^{\mu}_{m-n}\alpha_{\mu n} \ca , \label{eq:Lmm} \\
L_0^{\mathrm{g}}&=&\sum_{n=-\infty}^{\infty}n\ \ca c_{-n}b_n\ca -1 ,
\label{eq:L0g} \\
L_m^{\mathrm{g}}&=&\sum_{n=-\infty}^{\infty}(2m-n)\ca b_nc_{m-n}\ca \quad 
\mathrm{for} \, m\neq 0, \label{eq:Lmg}
\end{eqnarray}
where $\ca \cdots \ca$ represents the oscillator normal ordering. \\

\item Virasoro generators (superstring)
\begin{eqnarray}
L_m^{\mathrm{m}}&=&\frac{1}{2}\sum_{n\in \zetto}\ca\alpha_{m-n}^{\mu}
\alpha_{\mu n}\ca +\frac{1}{4}\sum_{r\in \zetto +\nu}(2r-m)\ca\psi_{m-r}^{\mu}
\psi_{\mu r}\ca +a^{\mathrm{m}}\delta_{m,0}, \nonumber \\
& &\mathrm{with}\quad a^{\mathrm{m}}=\left\{
	\begin{array}{lc}
	D/16 & \mathrm{R} \\
	0 & \mathrm{NS}
	\end{array}
\right. , \label{eq:Lmms} \\
& &\mbox{($D$ is the spacetime dimensionality.)} \nonumber \\
G_r^{\mathrm{m}}&=&\sum_{n\in\zetto}\alpha_n^{\mu}\psi_{\mu,r-n},
\label{eq:Grm} \\
L_m^{\mathrm{g}}&=&\sum_{n\in\zetto}(m+n)\ca b_{m-n}c_n\ca +\sum_{r\in\zetto
+\nu}\frac{m+2r}{2}\ca \beta_{m-r}\gamma_r\ca +a^{\mathrm{g}}\delta_{m,0},
\nonumber \\ & &\mathrm{with}\quad a^{\mathrm{g}}=\left\{
	\begin{array}{lc}
	-5/8 & \mathrm{R} \\
	-1/2 & \mathrm{NS}
	\end{array}
\right. , \label{eq:Lmgs} \\
G_r^{\mathrm{g}}&=&-\sum_{n\in\zetto}\left[\frac{2r+n}{2}\beta_{r-n}c_n
+2b_n\gamma_{r-n}\right]. \label{eq:Grg}
\end{eqnarray}

\item Bosonization
\begin{equation}
\beta(z)\cong e^{-\phi}\partial\xi , \quad \gamma(z)\cong \eta e^{\phi} , 
\quad \delta(\gamma)\cong e^{-\phi} , \quad 
\delta(\beta)\cong e^{\phi}.\label{eq:boso}
\end{equation}
\begin{equation}
T^{\phi}=-\frac{1}{2}\partial\phi\partial\phi-\partial^2\phi \, , \quad 
T^{\eta\xi}=-\eta\partial\xi. \label{eq:Tbos}
\end{equation}
\begin{equation}
\xi(z)\eta(w)\sim\frac{1}{z-w} \, , \quad \partial\phi(z)\partial\phi(w)\sim
-\frac{1}{(z-w)^2}. \label{eq:xietaOPE}
\end{equation}

\item Commutation relations
\begin{eqnarray}
[\alpha_m^{\mu},\alpha_n^{\nu}]&=&m\delta_{m+n,0}\eta^{\mu\nu}\> ,\quad
[ x^{\mu},p^{\nu}]=i\eta^{\mu\nu}, \\
\{\psi_r^{\mu},\psi_s^{\nu}\}&=&\eta^{\mu\nu}\delta_{r+s,0}, \\
\{b_m,c_n\}&=&\delta_{m+n,0} \> , \quad [\gamma_r,\beta_s]=\delta_{r+s,0}.
\end{eqnarray}

\item State-Operator correspondences
\begin{eqnarray}
\alpha_{-m}^{\mu}|k\rangle &\cong&i\sqrt{\frac{2}{\ap}}\oint \frac{dz}{2\pi i}
z^{-m}\partial X^{\mu}(z)\cdot e^{ik\cdot X(0)} \nonumber \\
&=&\sqrt{\frac{2}{\ap}}\frac{i}{(m-1)!}
:\partial^mX^{\mu}e^{ik\cdot X}(0):, \label{eq:alphaX} \\
b_{-m}&\cong& \frac{1}{(m-2)!}\partial^{m-2}b(0), \label{eq:bdb} \\
c_{-m}&\cong& \frac{1}{(m+1)!}\partial^{m+1}c(0). \label{eq:cdc}
\end{eqnarray}

\item BRST charge (bosonic string)
\begin{eqnarray}
j_B&=&:cT^{\mathrm{m}}:+\frac{1}{2}:cT^{\mathrm{g}}:+\frac{3}{2}\partial^2c \\
&=&:cT^{\mathrm{m}}:+:bc\partial c:+\frac{3}{2}\partial^2c, \label{eq:BRSTjB}
\end{eqnarray}
where the third term was added to make $j_B$ a true $(1,0)$ tensor. As it is a
 total divergence, it does not contribute to the BRST charge.
\begin{eqnarray}
\{Q_B,b_m\}&=&L_m^{\mathrm{m}}+L_m^{\mathrm{g}}=L_m^{\mathrm{tot}}, 
\label{eq:QbL} \\
Q_B^{\mathrm{holom.}}&=&\sum_{n=-\infty}^{\infty}c_nL_{-n}^{\mathrm{m}} 
\label{eq:Qbos} \\ &+&
\sum_{m,n=-\infty}^{\infty}\frac{m-n}{2}\ca c_mc_nb_{-m-n}\ca -c_0.
\nonumber
\end{eqnarray}

\item BRST charge (superstring)
\begin{eqnarray}
j_B&=&cT^{\mathrm{m}}+\gamma G^{\mathrm{m}}+\frac{1}{2}(cT^{\mathrm{g}}+\gamma
G^{\mathrm{g}}) \label{eq:BRSTjBsup} \\
&=& cT^{\mathrm{m}}+\gamma G^{\mathrm{m}}+bc\partial c+\frac{3}{4}(\partial c)
\beta\gamma+\frac{1}{4}c(\partial\beta)\gamma-\frac{3}{4}c\beta\partial\gamma
-b\gamma^2. \nonumber \\
\{Q_B,b_n\}&=&L_n^{\mathrm{tot}} \>, \quad [Q_B,\beta_r]=G_r^{\mathrm{tot}}.
\label{eq:QbeG} \\
Q_B^{\mathrm{holom.}}&=&\sum_mc_{-m}L_m^{\mathrm{m}}+\sum_r\gamma_{-r}
G_r^{\mathrm{m}}-\sum_{m,n}\frac{n-m}{2}\ca b_{-m-n}c_mc_n\ca \label{eq:Qsup} 
\\& &+\sum_{m,r}\left[\frac{2r-m}{2}\ca\beta_{-m-r}c_m\gamma_r\ca -\ca b_{-m}
\gamma_{m-r}\gamma_r\ca \right]+a^{\mathrm{g}}c_0. \nonumber
\end{eqnarray}

\item Picture-changing operator
\begin{eqnarray}
\cX (z)&=&\{Q_B,\xi(z)\}=G(z)\delta(\beta(z))-\partial b(z)\delta^{\prime}
(\beta(z)),\label{eq:PCO} \\
Y(z)&=&c\partial\xi e^{-2\phi}(z). \label{eq:PCOinv}
\end{eqnarray}

\item Path integrals on the disk boundary
\begin{eqnarray}
\left\langle\prod_{i=1}^ne^{ik_iX(y_i)}\prod_{j=1}^p\partial X^{\mu_j}(
y_j^{\prime})\right\rangle_{\mathrm{matter}}
 &=& (2\pi)^d\delta^d\left(\sum_{i=1}^nk_i\right)
\prod^n_{i,j=1 \atop i<j}|y_i-y_j|^{2\ap k_i\cdot k_j} \nonumber \\
& &\times \bigg\langle\prod_{j=1}^p[v^{\mu_j}(y_j^{\prime})+q^{\mu_j}
(y_j^{\prime})]\bigg\rangle_{\mathrm{norm.}} \label{eq:diskamp} \\
v^{\mu}(y)=-2i\ap\sum_{i=1}^n\frac{k_i^{\mu}}{y-y_i}  &,& \, 
\langle q^{\mu}(y)q^{\nu}(y^{\prime})\rangle_{\mathrm{norm.}}=-2\ap
\frac{\eta^{\mu\nu}}{(y-y^{\prime})^2}, \nonumber \\
\langle c(z_1)c(z_2)c(z_3)\rangle =|z_{12}z_{23}z_{13}| \quad 
&\leftrightarrow&
\quad \langle 0|c_{-1}c_0c_1|0\rangle =1. \label{eq:diskgh}
\end{eqnarray}

\item D-brane tension
\begin{eqnarray}
\tau_p&=&\frac{1}{(2\pi)^p\ap{}^{\frac{p+1}{2}}g_c} , \label{eq:Tp} \\
\frac{\tau_p}{\tau_{p-1}}&=&\frac{1}{2\pi\sqrt{\ap}}, \\
g_c&=&\frac{\tau_{\mathrm{F1}}}{\tau_{\mathrm{D1}}}=e^{\Phi}, 
\label{eq:gdil}\\
\tau_{\mathrm{NS5}}&=&\frac{1}{(2\pi)^5\ap{}^3g_c^2},\label{eq:NSfiveten}
\end{eqnarray}
\item Yang-Mills coupling on a D$p$-brane
\begin{equation}
g_{\mathrm{D}p}^2=\frac{1}{(2\pi\ap)^2\tau_p}=(2\pi)^{p-2}g_c\ap{}^{
\frac{p-3}{2}}. \label{eq:YMcoup}
\end{equation}

\end{itemize}

\chapter{String Field Theory}\label{ch:sft}
In bosonic open string theory, it is known that the physical 
spectrum contains a tachyonic mode. In terms of D-brane, the existence of the 
tachyonic mode signals that the bosonic D-branes are unstable, and it was 
conjectured that at the minimum of the tachyon potential the D-brane decays 
into the `closed string vacuum' without any D-brane. In this chapter we 
introduce bosonic open string field theory as an off-shell formalism of 
string theory, and then calculate the tachyon potential to examine the brane 
annihilation conjecture. 

\section{String Field}
To begin with, let us recall the Hilbert space $\cH$ of the first-quantized 
string theory (for more detail see~\cite{Pol}). In the Fock space 
representation, any state in $\cH$ is constructed by acting with the 
negatively moded oscillators $\alpha_{-n}^{\mu},b_{-m},c_{-\ell}$ on the 
oscillator vacuum $|\Omega\rangle$ which is defined by the properties 
\begin{equation}
\left.
	\begin{array}{lc}
	\alpha_n^{\mu}|\Omega\rangle=0 & n>0 \\
	b_n|\Omega\rangle=0 & n\ge 0 \\
	c_n|\Omega\rangle=0 & n>0 \\
	p^{\mu}|\Omega\rangle\propto\alpha_0^{\mu}|\Omega\rangle=0. & 
	\end{array}
\right.
\end{equation}
The relation between $|\Omega\rangle$ and the `$SL(2,\aaru)$ invariant 
vacuum' $|0\rangle$ is given by $|\Omega\rangle=c_1|0\rangle$. Under the 
state-operator isomorphism, these vacua are mapped as
\[ |0\rangle\sim \mathbf{1}\ \mbox{ (unit operator) }, \quad 
|\Omega\rangle\sim c(0). \]
A basis for $\cH$ is provided by the collection of states of the form
\[\alpha^{\mu_1}_{-n_1}\cdots \alpha^{\mu_i}_{-n_i}b_{-m_1}\cdots b_{-m_j}
c_{-\ell_1}\cdots c_{-\ell_k} |\Omega\rangle, \]
where $n>0,m>0,\ell\ge 0$, and $i,j,k$ are arbitrary positive integers. Then 
any state $|\Phi\rangle \in \cH$ can be expanded as
\begin{equation}
|\Phi\rangle =\left(\phi(x)+A_{\mu}(x)\alpha^{\mu}_{-1}+B_{\mu\nu}(x)
\alpha^{\mu}_{-1}\alpha^{\nu}_{-1}+\cdots\right)c_1|0\rangle\equiv \Phi(z=0)
|0\rangle, \label{eq:C}
\end{equation}
where the coefficients in front of basis states have the dependence on 
the center-of-mass coordinate $x$ of the string. As we think of the 
coefficient functions as (infinitely many) spacetime particle fields, we call 
$|\Phi\rangle$ a `\textit{string field}'. 
The vertex operator $\Phi(z)$ defined above 
is also called a string field. Of course, 
if we equip the open string with the Chan-Paton degrees of freedom, $\Phi$ and
 coefficient functions become matrix-valued.
\medskip

Next we construct the physical Hilbert space $\cH_{\mathrm{phys}}$ by 
imposing the physical conditions on the full space $\cH$. 
In the \textit{old covariant quantization (OCQ)} approach, 
we ignore the ghost sector and impose 
on the states $|\psi\rangle \in\cH$ the following conditions
\begin{equation}
\left.
	\begin{array}{ccc}
	(L_0^{\mathrm{m}}-1)|\psi\rangle =0, & & \\
	L_n^{\mathrm{m}}|\psi\rangle =0 & \mbox{for} & n>0,
	\end{array}
\right. \label{eq:D}
\end{equation}
where $L_0^{\mathrm{m}},L_n^{\mathrm{m}}$ are the matter Virasoro 
generators~(\ref{eq:L0mo}), (\ref{eq:Lmm}). $-1$ in the first line can be 
considered as the ghost $(c_1)$ contribution. A state satisfying 
conditions~(\ref{eq:D}) is called \textit{physical}. Let $\bar{\cH}$ denote 
the Hilbert space restricted to the `physical' states in the above sense. 
If a state $|\chi\rangle$ has the form
\begin{equation}
|\chi\rangle=\sum_{n=1}^{\infty}L_{-n}^{\mathrm{m}}|\chi_n\rangle,\label{eq:E}
\end{equation}
its inner product with any physical state $|\psi\rangle$ vanishes because
\begin{equation}
\langle\psi |\chi\rangle=\sum_{n=1}^{\infty}\langle\psi |L_{-n}^{\mathrm{m}}
|\chi\rangle=\sum_{n=1}^{\infty}(L_n^{\mathrm{m}}|\psi\rangle)^{\dagger}|\chi
\rangle=0.\label{eq:F}
\end{equation}
A state of the form~(\ref{eq:E}) is called \textit{spurious}, and if a 
spurious state is also physical, we refer to it as a \textit{null} state.
Eq.~(\ref{eq:F}) means that we should identify
\[ |\psi\rangle\cong|\psi\rangle +|\chi\rangle \]
for a physical state $|\psi\rangle$ and any null state $|\chi\rangle$. So the 
real physical Hilbert space is the set of equivalence classes,
\[ \cH_{\mathrm{phys}}=\bar{\cH}/\cH_{\mathrm{null}}. \]

Now we introduce the \textit{BRST quantization} approach, which is equivalent 
to the OCQ method. The BRST charge $Q_B$~(\ref{eq:Qbos}) is nilpotent in the 
critical dimension $d$ ($d=26$ in bosonic string theory): This property gives 
rise to important consequences. The physical condition in this approach is 
expressed as 
\begin{equation}
Q_B|\psi\rangle =0. \label{eq:G}
\end{equation}
In cohomology theory, such a state is called \textit{closed}. A null state in 
OCQ corresponds to an \textit{exact} state of the form
\begin{equation}
Q_B|\chi\rangle . \label{eq:H}
\end{equation}
And we require the physical states to satisfy one more condition that they 
should have ghost number $+1$: In OCQ, ghost part of the physical state is 
always $c(z)$ in the vertex operator representation, which has ghost number 
$+1$. Thus, in order for the BRST physical Hilbert space to agree with 
the one in the OCQ approach, we need to properly restrict the ghost sector 
in the above mentioned way. Hereafter we denote by $\cH^1_{\cdots}$ 
the restriction of $\cH_{\cdots}$ to the ghost number $+1$ states. 
Then the real physical Hilbert space is given by 
\[ \cH_{\mathrm{phys}}=\cH^1_{\mathrm{closed}}/\cH^1_{\mathrm{exact}}, \]
namely, the cohomology of $Q_B$ with ghost number 1.

\medskip

Let's rewrite the physical conditions in the string field language. For that 
purpose, we consider the following spacetime action $S_0$ in the full space 
$\cH^1$:
\begin{equation}
S_0=\langle\Phi |Q_B|\Phi\rangle . \label{eq:I}
\end{equation}
Since the string field $|\Phi\rangle$ obeys the reality condition\footnote{
We will mention the reality condition at the end of 
section \ref{sec:sample}.} 
essentially written as~\cite{WiSFT1}
\[ \Phi[X^{\mu}(\pi-\sigma)]=\Phi^*[X^{\mu}(\sigma)], \]
the equation of motion (derived by requiring that $S_0$ be stationary with 
respect to the variation of $\Phi$) is
\[ Q_B|\Phi\rangle =0, \]
which is the same as the physical condition~(\ref{eq:G}). Furthermore, the 
action~(\ref{eq:I}) is invariant under the \textit{gauge transformation} 
\begin{equation}
\delta |\Phi\rangle =Q_B|\chi\rangle \label{eq:J}
\end{equation}
due to the nilpotence of $Q_B$. It is nothing but the exact state 
of~(\ref{eq:H}). Note that ghost number matching of both sides of~(\ref{eq:J})
 demands that the gauge parameter $|\chi\rangle$ should have the ghost number 
0. We then conclude that an \textit{on-shell} state, that is, a solution to 
the equation of motion derived from the action~(\ref{eq:I}), corresponds to 
a physical state in the first-quantized string theory. So by extending the 
spacetime action to include higher order terms in $\Phi$, we can hope to 
obtain the interacting string field theory. We describe in the next section 
the Witten's work which contains the cubic interaction term.

\section{Cubic String Field Theory Action}\label{sec:cubic}
Witten has proposed one way to formulate the field theory of open 
string~\cite{WiSFT1}. Its 
starting point is quite axiomatic: An associative noncommutative algebra $B$ 
with a $\mathbf{Z}_2$ grading, and some operations on $B$. The elements of 
$B$ will be regarded as the string fields later. The multiplication law $*$ 
satisfies the property that the $\zetto_2$ degree 
of the product $a*b$ of two elements 
$a,b\in B$ is $(-1)^a\cdot (-1)^b$, where $(-1)^a$ is the $\zetto_2$ degree 
of $a$. And there exists an odd `derivation' $Q$ acting in $B$ as
\[ Q(a*b)=Q(a)*b+(-1)^aa*Q(b).\] $Q$ is also required to be nilpotent: 
$Q^2=0$. These properties remind us of the BRST operator $Q_B$.

The final ingredient is the `integration', which maps $a\in B$ to a complex 
number $\int a\in \shii$. This operation is linear, $\displaystyle \int (a+b)
=\int a+\int b$, and satisfies $\displaystyle \int (a*b)=(-1)^{ab} \int 
(b*a)$ where $(-1)^{ab}$ is defined to be $-1$ only if both $a$ and $b$ are 
odd elements of $B$. Also, $\int Q(a)=0$ for any $a$.
\smallskip

Looking at the above axioms, one may notice that each element or operation 
has its counterpart in the theory of differential forms on a manifold. The 
correspondence is shown in Table~\ref{tab:A}. 
\begin{table}[htbp]
	\begin{center}
	 \begin{tabular}{|c|c|c|}
	 \hline
	  & algebra $B$ & space of differential forms \\
	 \hline\hline
	 element & string field & differential $k$-form \\
	 \hline
	 degree & $(-1)^a$ & $(-1)^k$ \\
	 \hline
	 multiplication & $*$-product & $\wedge$ (wedge product) \\
	 \hline
	 derivation & $Q$ & exterior derivative $d$ \\
	 \hline
	 integration & $\int$ & $\int$ on a $k$-dimensional manifold \\
	 \hline
	 \end{tabular}
	\end{center}
	\caption{Comparison between the abstract algebra $B$ and the space of 
	differential forms.}
	\label{tab:A}
\end{table}
The formal correspondence ceases to be valid in that $\psi \wedge \omega =\pm 
\omega \wedge \psi$ holds in the case of differential forms even without 
integration, whereas $a*b$ and $b*a$ have no simple relation in $B$.

\begin{figure}[htbp]
	\hspace{-1cm}
	\includegraphics{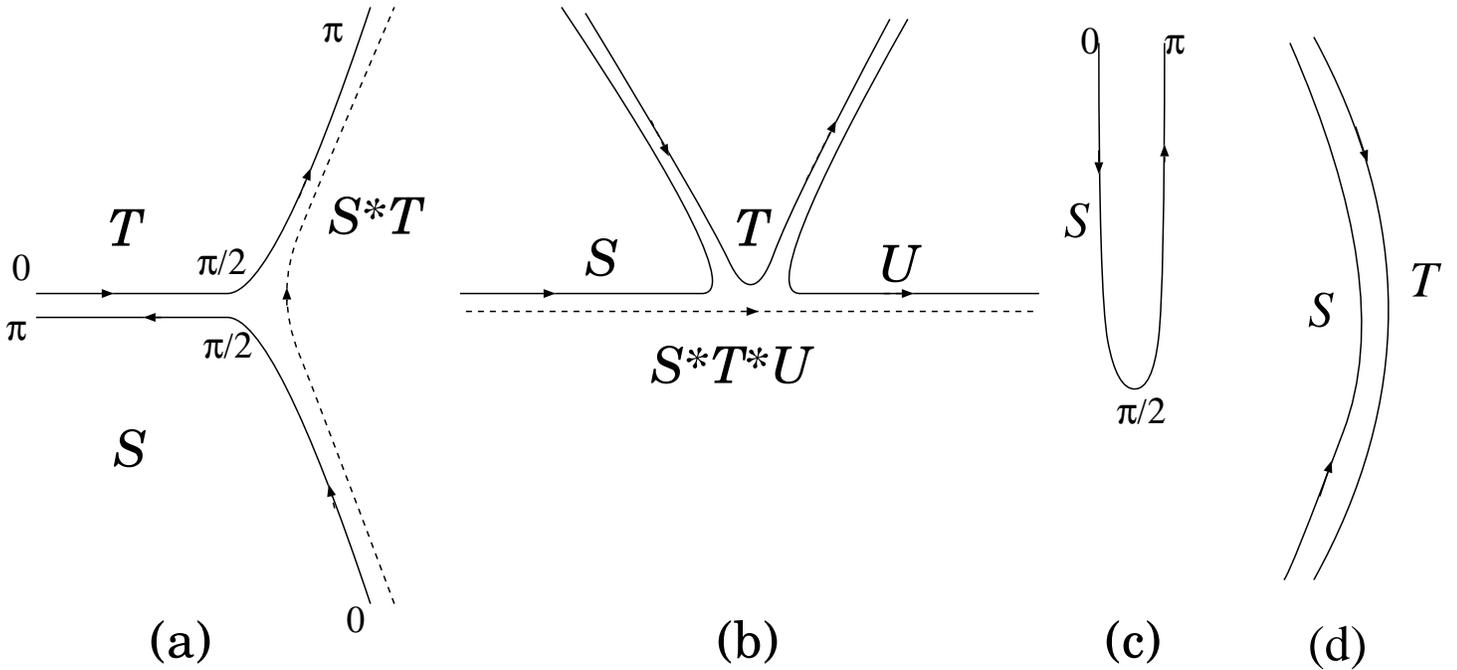}
	\caption{(a) Gluing of two strings $S,T$. 
	         (b) Gluing of three strings $S,T,U$ makes the associativity 
	         clear.
	         (c) Integration operation. 
	         (d) Multiplication followed by an integration, $\int (a*b)$.}
	\label{fig:A}
\end{figure}

Let's take a close look at the multiplication $*$. As discussed in detail 
in~\cite{WiSFT1}, in order for the multiplication to be associative, 
\textit{i.e.} $(a*b)*c=a*(b*c)$, we must interpret $*$-operation as gluing 
two half-strings together. In more detail, take two strings $S,T$, whose 
excitations are described by the string fields $a$ and $b$, respectively. 
Each string is labeled by a coordinate $\sigma \ (0\le \sigma \le \pi)$ with 
the midpoint $\sigma=\pi/2$. Then the gluing procedure is as follows: 
The right hand piece $(\pi/2\le\sigma\le\pi)$ of the string $S$ and the left 
hand piece $(0\le\sigma\le\pi/2)$ of the string $T$ are glued together, and 
what is left behind is the string-like object, consisting of the left half 
of $S$ and the right half of $T$. This is the product $S*T$ in the gluing 
prescription, and the resulting string state on $S*T$ is the string field 
$a*b$, as is illustrated in Figure~\ref{fig:A}(a). From Figure~\ref{fig:A}(b),
the $*$-operation is manifestly associative, at least na\"{\i}vely.
\smallskip 

Next we give the integration operation a precise definition. The axioms 
involve the statement about integration that $\int (a*b)=\pm \int (b*a)$ 
($\pm$ is correctly $(-1)^{ab}$; we do not care that point below).  Since 
$a*b$ and $b*a$ are in general thought of as representing completely 
different elements, their agreement under the integration suggests that the 
integration procedure still glues the remaining sides of $S$ and $T$. 
If we restate it for a 
single string $S$, the left hand piece is sewn to the right hand piece under 
the integration, as in Figure~\ref{fig:A}(c).

\begin{figure}[htbp]
\begin{center}
	\includegraphics{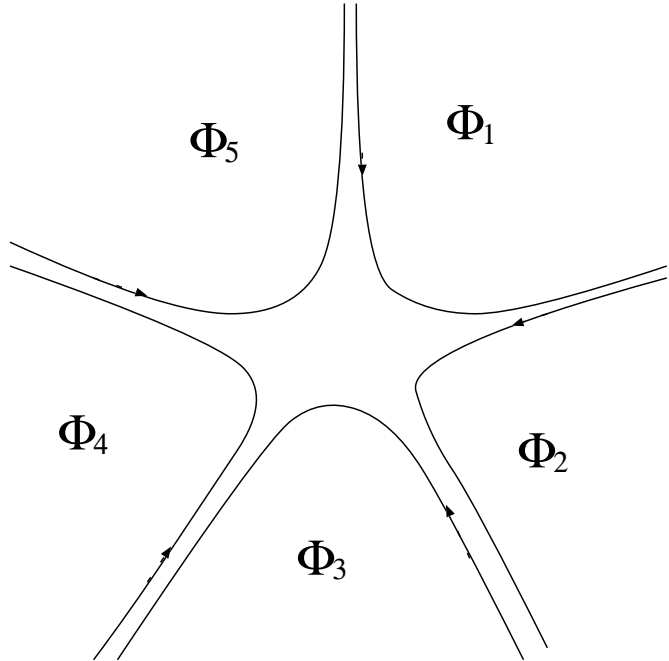}
	\caption{5-string vertex.}
	\label{fig:B}
\end{center}
\end{figure}

Using the above definition of $*$ and $\int$, we can write the $n$-string 
interaction vertex as $\displaystyle \int \Phi_1*\cdots *\Phi_n$, where 
$\Phi_i$ denotes a string field on the $i$-th string Hilbert space. It is 
indicated in Figure~\ref{fig:B} for the case of $n=5$. Such an interaction, 
where each string is divided at the midpoint into the left- and right-piece 
and then glued together, is termed `Witten vertex'.
\bigskip

At last, we have reached a stage to give the string field theory action. 
Using the above definitions of $*$ and $\int$, the quadratic 
action $S_0$ is also written 
as
\begin{equation}
S_0=\langle \Phi |Q_B|\Phi\rangle =\int\Phi *Q_B\Phi. \label{eq:K}
\end{equation}
Of course, BRST charge $Q_B$ is qualified as a derivation operator $Q$ in the 
axioms. And we consider the above mentioned $n$-string vertex
\begin{equation}
S_n=\int \underbrace{\Phi *\cdots *\Phi}_n. \label{eq:L}
\end{equation}
Now let us equip the algebra $B$ with a $\zetto$ grading by the ghost number. 
If we define $\#_{\mathrm{gh}}$ as an operator that counts the ghost 
number of its argument, then
\begin{equation}
\gh(\Phi)=1 \, , \> \gh(Q_B)=1 \, , \> \gh(*)=0 \, .\label{eq:N}
\end{equation}
These assignments are different from those in~\cite{WiSFT1}. The reason is 
traced to the fact that in~\cite{WiSFT1} 
the ghost number refers to a state ($-\frac{1}{2}$ for 
a physical state), whereas we are counting the ghost number of vertex 
operators here. 
The discrepancy comes from the fact that the ghost number current $j=-bc$ is 
not a tensor field on the world-sheet, but the difference doesn't matter. We 
note that the integrands of $S_0$ and $S_n$ have the ghost number 3 and $n$ 
respectively. In the CFT prescription mentioned later, the actions $S_0$ and 
$S_n$ are calculated as 2- or $n$-point correlation functions on the disk. 
According to the Riemann-Roch theorem, the correlation functions vanish 
unless the equation 
\[(\mbox{the number of $c$ ghosts inserted in the correlator})-
(\mbox{that of $b$ ghosts})=3\chi\]
holds for a Riemann surface with Euler characteristics $\chi$. Since the left 
hand side is simply the total ghost number, 
only $S_0$ and $S_3$ can be nonzero in the case of the disk ($\chi=1$). 
Thus we have determined the possible 
candidates for the spacetime action, but we can still fix the relative 
normalization between $S_0$ and $S_3$ by requiring the gauge invariance.
\medskip

We will impose on the action the gauge invariance under the following 
infinitesimal gauge transformation 
\begin{equation}
\delta \Phi=Q_B\Lambda+g_o(\Phi *\Lambda -\Lambda *\Phi), \label{eq:M}
\end{equation}
where $\Lambda$ is a gauge parameter with $\gh(\Lambda)=0$, and $g_o$ is the 
open string coupling constant. The form of gauge transformation is chosen to 
generalize that of the non-abelian gauge theory: If the string fields 
contained $n\times n$ Chan-Paton matrices and the $*$-product were the 
ordinary product of matrices, variation~(\ref{eq:M}) would reduce to the 
non-abelian gauge transformation for the component $A_{\mu}$ 
in~(\ref{eq:C}). Note that the gauge parameters form a subalgebra of $B$ 
as the ghost number of the product of two gauge parameters under the 
$*$-multiplication remains zero : $
\gh(\Lambda *\Lambda')=\gh(\Lambda )+\gh(*)+\gh(\Lambda')=0$. It is desirable 
because in the ordinary gauge theory we think of the gauge parameters as 
simple `functions' (\textit{i.e.} 0-forms) which are closed under the 
$\wedge$-product, so this serves as a nontrivial check of the 
assignments~(\ref{eq:N}). Now we show that the integral of the Chern-Simons 
three-form (with an extra parameter $a$) 
\begin{equation}
S=\int\left(\Phi *Q_B\Phi +\frac{2a}{3}g_o\Phi *\Phi* \Phi \right)\label{eq:O}
\end{equation}
is invariant under the 
infinitesimal gauge transformation~(\ref{eq:M}) if and only if $a=1$. 
The variation of the first term becomes 
\begin{eqnarray*}
\delta S_1 &=& \int (\delta\Phi *Q_B\Phi+\Phi *Q_B\delta\Phi) \\
&=&\int \left(\delta\Phi *Q_B\Phi-Q_B(\Phi *\delta\Phi)+Q_B\Phi *\delta \Phi
\right) \\ &=&2\int \delta\Phi *Q_B\Phi.
\end{eqnarray*}
In the second line we used the fact that $Q_B$ is an odd derivation. In the 
third line, $\displaystyle \int Q_B(\ldots)=0$ and 
$\displaystyle \int Q_B \Phi *\delta
\Phi =\int\delta\Phi *Q_B\Phi$ hold for the ghost number 2 element 
$Q_B \Phi$. The second term of~(\ref{eq:O}) gives rise to 
\begin{eqnarray*}
\delta S_2&=& \frac{2a}{3}g_o\int\left(\delta\Phi *\Phi *\Phi+\Phi *(\delta 
\Phi *\Phi)+(\Phi *\Phi)*\delta\Phi\right) \\ &=& 2ag_o\int\delta\Phi *
\Phi *\Phi
\end{eqnarray*}
because the parenthesized terms have ghost number 2. 
Substituting~(\ref{eq:M}), the total variation is
\begin{eqnarray*}
\delta S&=&\int\biggl(2Q_B\Lambda *Q_B \Phi+2g_o\Phi *(\Lambda *Q_B\Phi)-
2g_o\Lambda *\Phi *Q_B\Phi \\ & & +2ag_oQ_B\Lambda *\Phi *\Phi +2ag_o^2
\Phi *(\Lambda *\Phi *\Phi)-2ag_o^2\Lambda *\Phi *\Phi *\Phi\biggr).
\end{eqnarray*}
The first term vanishes because $\int Q_B(\Lambda *Q_B\Phi)=0$ and $Q_B^2=0$. 
The 5-th and 6-th terms cancel each other. The remaining three terms can be 
arranged as
\begin{eqnarray*}
\delta S&=& 2g_o\int\left(\Lambda *Q_B\Phi *\Phi-\Lambda *\Phi *Q_B\Phi+
(1+(a-1))Q_B\Lambda *\Phi *\Phi\right) \\ &=& 2g_o\int Q_B(\Lambda *\Phi 
*\Phi)+2g_o(a-1)\int Q_B \Lambda *\Phi *\Phi.
\end{eqnarray*}
The first term vanishes. To guarantee the last term to vanish for arbitrary 
$\Lambda$, we must choose $a=1$. Therefore we have proved that the gauge 
invariant action must be of the form~(\ref{eq:O}) with $a=1$. 
Note that a small number of axioms we saw at the beginning of this section 
warrant the gauge invariance of the action~(\ref{eq:O}) under (\ref{eq:M}). 
In particular, the properties of the operator $Q$ and the associativity of 
the $*$-product play crucial roles in proving it. 
Though the form~(\ref{eq:O}) of the action 
is sufficient, for future use we rewrite it slightly. First we rescale 
the string field as $\Phi\to (\ap/g_o)\Phi$ and change the overall 
normalization (the latter can be absorbed into the definition of $g_o$) 
such that
\begin{equation}
S=-\frac{1}{g_o^2}\left(\frac{1}{2\ap}\int\Phi *Q_B\Phi+\frac{1}{3}\int \Phi 
*\Phi *\Phi\right). \label{eq:CSFT}
\end{equation}
Note that $g_o$ is a dimensionful parameter in general. We will see it 
later more explicitly.

\section{Evaluation of the Action}\label{sec:eva}
Since the action~(\ref{eq:CSFT}) is derived quite formally, it is not 
suitable for concrete calculations. In particular, $*$ and $\int$ 
have been defined 
only geometrically as the gluing procedure. Though the quadratic part is 
equivalently represented as a Fock space inner product $\langle\Phi |Q_B|\Phi
\rangle$, we have no such simple translation as to the cubic term $\int\Phi *
\Phi *\Phi$. So in this section we will argue the methods of calculation. 
\medskip

The first approach is the operator formulation opened up in~\cite{GJ,CST,
Sam86}. In this method the 3-string interaction is represented using the 
3-point vertex $\langle V_3|$ as
\begin{equation}
\int \Phi *\Phi *\Phi =\langle V_3|\Phi\rangle_1\otimes|\Phi\rangle_2\otimes
|\Phi\rangle_3, \label{eq:P}
\end{equation}
where the subscript 1,2,3 label three strings which interact. $\langle V_3|$ 
is explicitly written in terms of the oscillators as
\begin{eqnarray}
\langle V_3| &=&\langle 0|_1c_{-1}^{(1)}c_0^{(1)}\otimes\langle 0|_2
c_{-1}^{(2)}c_0^{(2)}\otimes\langle 0|_3c_{-1}^{(3)}c_0^{(3)}\int dp_1dp_2
dp_3(2\pi)^d\delta^d(p_1+p_2+p_3) \nonumber \\ & &\times \exp\left(\frac{1}{2}
\sum_{r,s=1}^3\sum_{n,m=0}^{\infty}\alpha_n^{(r)\mu}N_{nm}^{rs}
\alpha_m^{(s)\nu}\eta_{\mu\nu}
+\sum_{r,s=1}^3\sum_{n\ge 2 \atop m\ge -1}c_n^{(r)}
X_{nm}^{rs}b_m^{(s)}\right). \label{eq:Vthree}
\end{eqnarray}
where $r,s$ stand for strings, and $n,m$ are mode numbers. The 
\textit{Neumann coefficients} $N_{nm}^{rs},X_{nm}^{rs}$ represent 
the effect of conformal 
transformations $h_r$ of the upper half-disks of three open strings. 
For example, $N_{nm}^{rs}$ is given by 
\[N_{nm}^{rs}=\frac{1}{nm}\oint\frac{dz}{2\pi i}z^{-n}h_r^{\prime}(z)
\oint\frac{dw}{2\pi i}w^{-m}h_s^{\prime}(w)\frac{1}{(h_r(z)-h_s(w))^2}.\]
Once all the Neumann coefficients are given, the 3-point 
interaction~(\ref{eq:P}) involves purely algebraic manipulations only, so 
this method is well automated. However, it seems not suited for by-hand 
calculations: we must expand the exponential and pick out all terms which 
do not commute with the oscillators in $|\Phi\rangle_r$, and exploit the 
commutation relations. For this reason, we avoid using the operator method 
in this paper, and instead mainly rely on another, CFT, method. For more 
details about the operator formulation, see~\cite{GJ,CST,Sam86,LPP,Tay}.
\medskip

\begin{figure}[htbp]
	\includegraphics{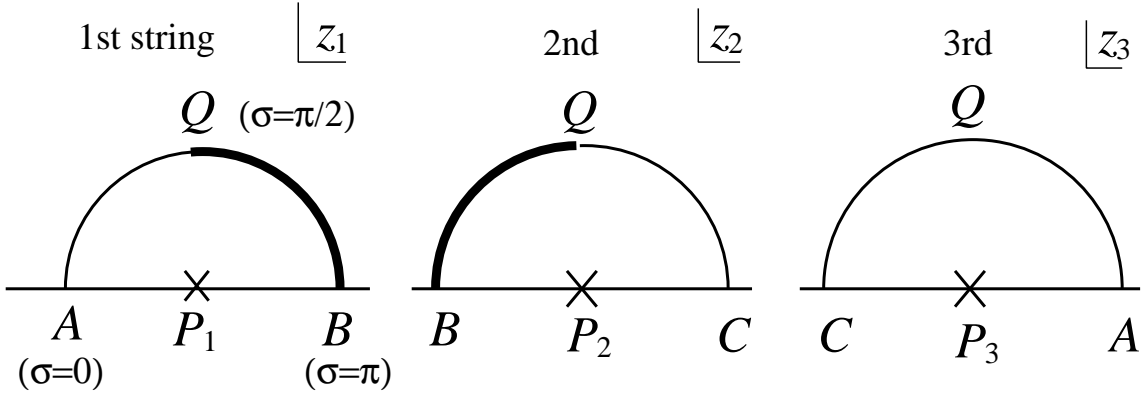}
	\caption{Three strings which will interact.}
	\label{fig:D}
\end{figure}

\begin{figure}[htbp]
	\includegraphics{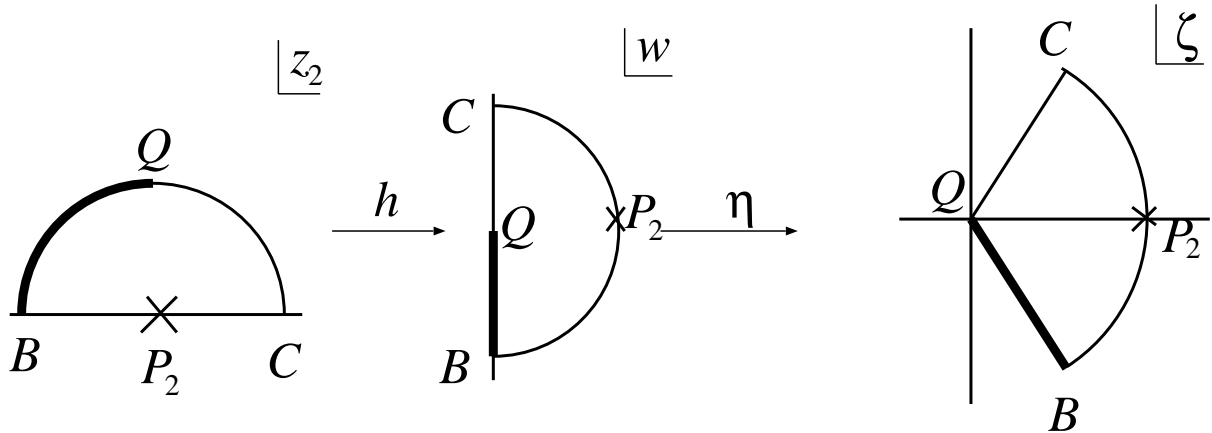}
	\caption{From an upper half disk to one third of a unit disk.}
	\label{fig:E}
\end{figure}

\begin{figure}[htbp]
	\hspace{-0.5cm}
	\includegraphics{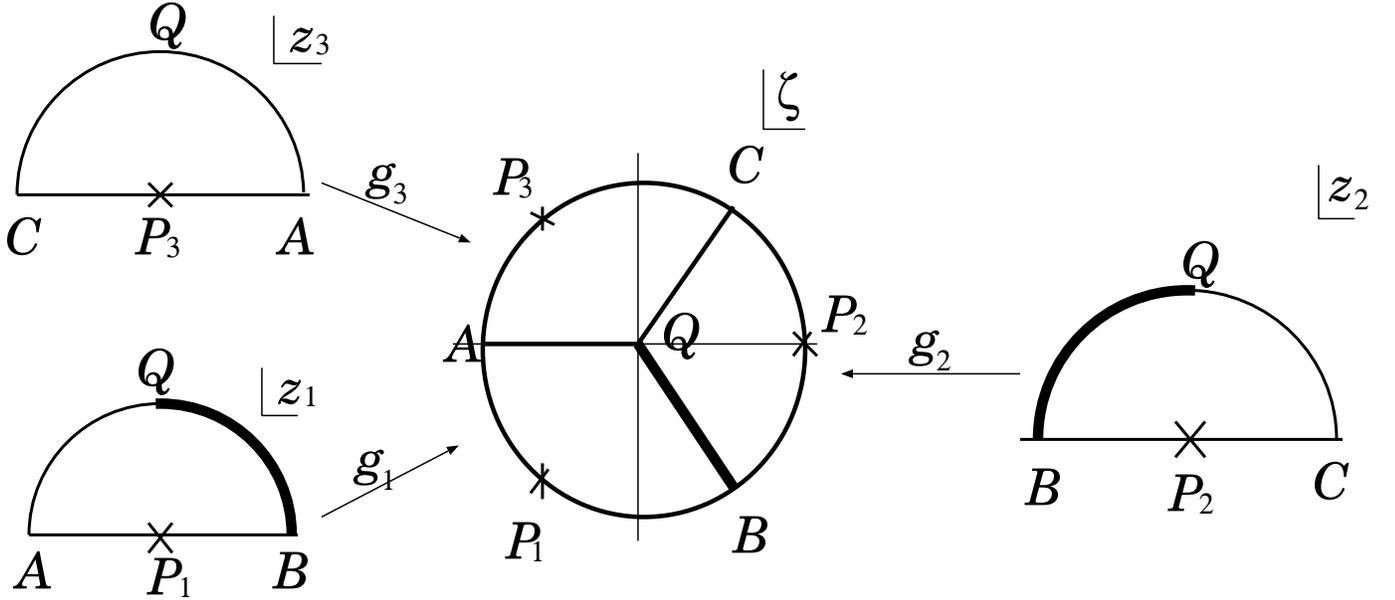}
	\caption{3-string vertex.}
	\label{fig:F}
\end{figure}

The second approach involves conformal mappings and calculation of the 
correlation functions on the disk~\cite{LPP,Univ,RasZw}. 
Let us first consider the case of 3-string vertex. The idea is 
to map the three upper half-disks, each of which represents the propagation 
of one of three open strings, to one disk on a 
conformal plane realizing the Witten vertex. We will describe it in 
more detail below. 
In Figure~\ref{fig:D} three open string 
world-sheets are indicated as upper half-disks. 
Along the time-evolution, at $t=-\infty$, 
which corresponds to $z_i=0 \ (P_i)$, each string appeared (in the 
CFT language, corresponding vertex operator was inserted) and then propagated 
radially, and now $(t=0)$ it has reached the interaction point 
$|z_i|=1$. We want to map three half disks parametrized by their own local 
coordinates $z_i$'s to the interior of a unit disk 
with global coordinate $\zeta$. 
To do so, we first carry out the transformation satisfying following 
properties:
\begin{itemize}
\item It maps the common interaction point $Q \ (z_j=i , j=1,2,3)$ to the 
center $\zeta=0$ of the unit disk.
\item The open string boundaries, which are represented as line segments on 
the real axes in Figure~\ref{fig:D}, are mapped to the boundary of the unit 
disk. 
\end{itemize}
For definiteness, consider the second string. Then the transformation 
\begin{equation}
z_2 \quad \mapsto \quad w=h(z_2)=\frac{1+iz_2}{1-iz_2} \label{eq:Q}
\end{equation}
turns out to satisfy these two properties. But since the angle $\angle BQC$ 
is $180^{\circ}$, three half disks cannot be put side by side to form a unit 
disk if they are left as they are. Hence we perform the second transformation
\begin{equation}
w\quad \mapsto \quad \zeta=\eta(w)=w^{2/3}, \label{eq:R}
\end{equation}
which maps the right half-disk to a wedge with an angle of $120^{\circ}$. 
A series of transformations is illustrated in Figure~\ref{fig:E}. 
Once the 
mapping of the second world-sheet is constructed, that of the first and third 
is easily found. All we have to do is to rotate the $120^{\circ}$ wedge by an 
angle $\mp 120^{\circ}$ respectively, being careful to sew the right hand 
piece of the first string with the left hand piece of the second string, and 
the same is repeated cyclically\footnote{`Left' and `right' are reversed 
between in \cite{RasZw} and in ours. In our paper, 
we follow the convention that 
coordinate $z$ and $(\sigma_1,\sigma_2)$ are related by $z=-\exp(-i\sigma_1
+\sigma_2)$. So for fixed $\sigma_2$ (time), $z$ goes around clockwise as 
$\sigma_1$ increases.} in accordance with the gluing procedure described in 
section~\ref{sec:cubic}. This will be achieved by 
\begin{eqnarray}
g_1(z_1) &=& e^{-\frac{2\pi i}{3}}\left(\frac{1+iz_1}{1-iz_1}\right)^{
\frac{2}{3}}, \nonumber \\
\eta\circ h(z_2)=g_2(z_2) &=& \left(\frac{1+iz_2}{1-iz_2}\right)^{\frac{2}
{3}}, \label{eq:S} \\
g_3(z_3) &=& e^{\frac{2\pi i}{3}}\left(\frac{1+iz_3}{1-iz_3}\right)^{
\frac{2}{3}}. \nonumber
\end{eqnarray}
The above mappings are shown in Figure~\ref{fig:F}. 
Using these mappings, we 
can give the 3-string vertex $\int \Phi *\Phi *\Phi$ the CFT representation as
a 3-point correlation function. That is 
\begin{equation}
\int\Phi *\Phi *\Phi=\langle g_1\circ\Phi(0)g_2 \circ\Phi(0)g_3\circ\Phi(0)
\rangle , \label{eq:T}
\end{equation}
where $\langle \ldots \rangle$ is the correlator on the global disk 
constructed 
above, evaluated in the combined matter and ghost CFT. Its normalization will 
be determined later. $g_i\circ\Phi(0)$ means the conformal transform of 
$\Phi(0)$ by $g_i$. If $\Phi$ is a primary field of conformal weight $h$, 
then $g_i\circ\Phi(0)$ is given by 
\begin{equation}
g_i\circ\Phi(0)=\bigl(g_i^{\prime}(0)\bigr)^h\Phi(g_i(0)). \label{eq:U}
\end{equation}
At this point, one may think that more general conformal transformations can 
be chosen if we wish only to reproduce the Witten vertex. For instance, the 
angles of wedges are not necessarily $120^{\circ}$. But when we demand the 
cyclicity of the 3-point vertex, $\int\Phi_1*\Phi_2*\Phi_3=\int\Phi_2
*\Phi_3*\Phi_1$, three transformations $g_1,g_2,g_3$ are constrained to 
satisfy
\[ g_3=g \, , \quad g_2=T\circ g \, , \quad g_1=T^2\circ g, \]
where $T\in SL(2,\shii)$ obeys $T^3=1$. This condition singles out the 
transformation~(\ref{eq:S}) almost uniquely. But notice that we have so far 
considered only the unit disk representation of a Riemann surface with a 
boundary. The open string world-sheet is also represented as an 
upper half plane, which is mapped bijectively to the unit disk by an 
$SL(2,\shii)$ transformation. In fact, the $SL(2,\shii)$ invariance of the CFT
correlators guarantees that these two representations give the same results.

\begin{figure}[htbp]
	\begin{center}
	\includegraphics{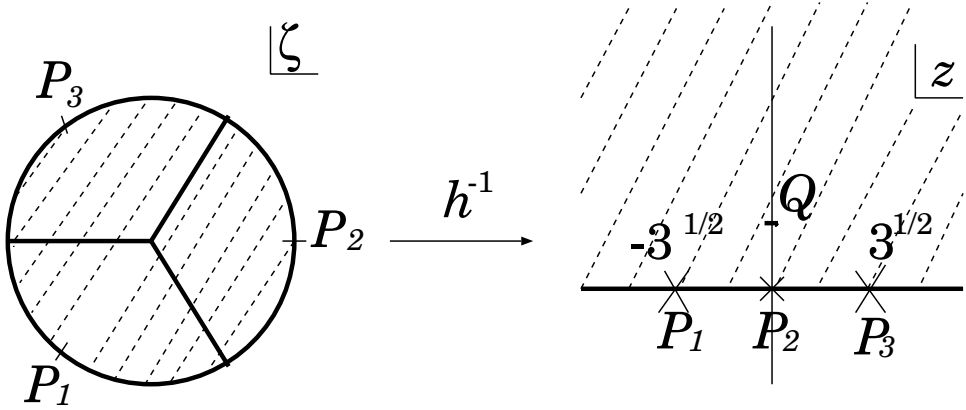}
	\end{center}
	\caption{From a unit disk to an upper half plane.}
	\label{fig:G}
\end{figure}

Now let's construct the transformation that maps the unit disk to the upper 
half plane. Such a role is well played by
\begin{equation}
z=h^{-1}(\zeta)=-i\frac{\zeta-1}{\zeta+1}, \label{eq:V}
\end{equation}
where $h$ is an $SL(2,\shii)$ transformation that has already appeared 
in~(\ref{eq:Q}), and $h^{-1}$ is its inverse function. It is shown in 
Figure~\ref{fig:G}.
Our final expression for the 3-point vertex is 
\begin{eqnarray}
\int\Phi*\Phi*\Phi &=& \langle f_1\circ\Phi(0)f_2\circ\Phi(0)f_3\circ\Phi(0)
\rangle , \label{eq:W} \\
f_i(z_i)&=&h^{-1}\circ g_i(z_i), \nonumber
\end{eqnarray}
and $g_i$'s are given in~(\ref{eq:S}).

\medskip

Now that we have obtained the 3-point vertex, we consider generalizing it to 
arbitrary $n$-point vertices. This can be done quite straightforwardly. Define
\begin{eqnarray}
g_k(z_k)&=& e^{\frac{2\pi i}{n}(k-1)}\left(\frac{1+iz_k}{1-iz_k}\right)^{
\frac{2}{n}} \, , \quad 1\le k \le n \label{eq:X} \\
f_k(z_k)&=& h^{-1}\circ g_k(z_k). \nonumber
\end{eqnarray}
Each $g_k$ maps an upper half disk to a $(360/n)^{\circ}$ wedge, and $n$ such 
wedges gather to make a unit disk. Then $n$-point vertex is evaluated as
\[\int\Phi* \cdots *\Phi=\langle f_1\circ\Phi(0)\cdots f_n\circ\Phi(0)
\rangle .\]

\begin{figure}[htbp]
	\includegraphics{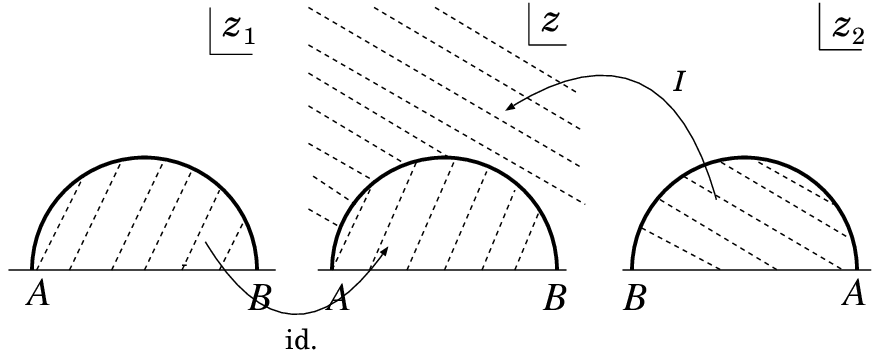}
	\caption{2-string vertex.}
	\label{fig:H}
\end{figure}

Finally, we construct the CFT expression for quadratic term $\displaystyle 
\int\Phi *Q_B\Phi$. For that purpose, it suffices to consider the $n=2$ case 
in~(\ref{eq:X}). We explicitly write down the functions $f_1,f_2$ as
\begin{eqnarray}
f_1(z_1)&=& h^{-1}\left(\frac{1+iz_1}{1-iz_1}\right) =z_1=\mathrm{id}(z_1),
\label{eq:Ya} \\
f_2(z_2)&=& h^{-1}\left(-\frac{1+iz_2}{1-iz_2}\right)=-\frac{1}{z_2}\equiv
\cI (z_2). \label{eq:Yb}
\end{eqnarray}
These mappings are shown in Figure~\ref{fig:H}.
The 2-point vertex is then written as
\begin{equation}
\int\Phi *Q_B\Phi =\langle \cI\circ\Phi(0)Q_B\Phi(0)\rangle .\label{eq:Z}
\end{equation}

By now, we have finished rewriting the string field theory action 
in terms of CFT correlators as
\begin{equation}
S=-\frac{1}{g_o^2}\left(\frac{1}{2\ap}\langle\cI\circ\Phi(0)Q_B\Phi(0)\rangle
+\frac{1}{3}\langle f_1\circ\Phi(0)f_2\circ\Phi(0)f_3\circ\Phi(0)\rangle
\right). \label{eq:CFTSFT}
\end{equation}
The CFT correlators are normalized such that
\[\langle e^{ikX}\rangle_{\mathrm{matter}}=(2\pi)^d\delta^d(k) \> , \quad 
\left\langle \frac{1}{2}\partial^2c\partial c\,c\right\rangle_{\mathrm{ghost}}
=1,\]
when we are considering open strings on D($d-1$)-branes. Other 
choice of normalization convention is also possible, but once we have 
fixed it we must not change it. Though the definition of $g_o$ 
has some ambiguity in this stage, we will 
relate it to the mass of the D-brane on which the open string endpoints live.

Now we have learned how to evaluate any interaction vertex using CFT 
correlator. Since we also know another method of dealing with the interaction 
vertex, namely operator formulation (also called Neumann function method), 
one would naturally ask whether these two prescriptions are equivalent. 
Though we do not present the details here, it is argued in~\cite{LPP} that 
the answer is certainly Yes. 

\section{Sample calculations}\label{sec:sample}

In the preceding section, we gave a scheme of computation using CFT 
correlators. In this section, we will show how to apply it to concrete 
calculations.
\medskip

In the beginning, recall that the string field theory action~(\ref{eq:CSFT}) 
possesses the gauge invariance~(\ref{eq:M}). We will carry out gauge-fixing 
by choosing so-called Feynman-Siegel gauge
\begin{equation}
b_0|\Phi\rangle =0 \label{eq:AA}
\end{equation}
in the Fock space representation. Here we discuss the validity of this gauge 
choice according to~\cite{SenZw}.

We first show that 
\begin{itemize}
\item the Feynman-Siegel gauge can always be chosen, at least at the 
linearized level.
\end{itemize}
Now the proof. Let's consider a state $|\Psi\rangle$ with 
$L_0^{\mathrm{tot}}=L_0^{\mathrm{matter}}+L_0^{\mathrm{ghost}}$ eigenvalue 
$h$, not obeying~(\ref{eq:AA}). Define $|\Lambda\rangle =b_0|\Psi\rangle$ 
and gauge-transform $|\Psi\rangle$ to a new state $|\tilde{\Psi}\rangle$ as 
\begin{equation}
|\tilde{\Psi}\rangle =|\Psi\rangle -\frac{1}{h}Q_B|\Lambda\rangle . 
\label{eq:AB}
\end{equation}
Since
\begin{eqnarray*}
b_0|\tilde{\Psi}\rangle &=& b_0|\Psi\rangle -\frac{1}{h}b_0Q_Bb_0|\Psi\rangle 
\\ &=&b_0|\Psi\rangle -\frac{1}{h}b_0\{Q_B,b_0\}|\Psi\rangle \\ &=& b_0|\Psi
\rangle-\frac{1}{h}b_0L_0^{\mathrm{tot}}|\Psi\rangle \\ &=& b_0|\Psi\rangle
-b_0|\Psi\rangle =0,
\end{eqnarray*}
the new state $|\tilde{\Psi}\rangle$ satisfies the gauge 
condition~(\ref{eq:AA}). The linearized gauge 
transformation~(\ref{eq:AB}) is always 
possible if $h\neq 0$. So the proposition was shown to be true for a state 
with $h\neq 0$. Then we also want to show that in the case of $h\neq 0$
\begin{itemize}
\item there are no residual gauge transformations preserving the 
gauge~(\ref{eq:AA}).
\end{itemize}
Suppose that both $b_0|\Psi\rangle =0$ and $b_0(|\Psi\rangle +Q_B|\xi\rangle)
=0$ can hold, that is, $|\eta\rangle\equiv Q_B|\xi\rangle$ is a 
residual gauge degree of freedom. Then
\[ h|\eta\rangle =L_0^{\mathrm{tot}}|\eta\rangle =\{Q_B,b_0\}Q_B|\xi\rangle =
Q_Bb_0(Q_B|\xi\rangle)+b_0(Q_B)^2|\xi\rangle =0. \]
Since $h\neq 0$, $|\eta\rangle$ must vanish, which completes the proof.
Thus, we have seen that the Feynman-Siegel gauge is a good choice at the 
linearized level, in other words, near $\Phi=0$. But this does not ensure 
that the same conclusions hold even nonperturbatively. Though the arguments 
about the validity of the Feynman-Siegel gauge have been given 
in~\cite{Hata,0101014}, we postpone them until section~\ref{sec:Siegel} 
because the arguments there are based on the specific solutions to the 
equations of motion which will be found in later sections or chapters. 
Here we simply assume the validity of this gauge condition.

\bigskip

We then explain the notion of level truncation. As in eq.(\ref{eq:C}), 
we expand 
the ghost number 1 string field using the Fock space basis as
\begin{eqnarray}
|\Phi\rangle &=& \int d^dk
\Biggl(\phi +A_{\mu}\alpha_{-1}^{\mu}+i\alpha b_{-1}c_0+
\frac{i}{\sqrt{2}}B_{\mu}\alpha_{-2}^{\mu}+\frac{1}{\sqrt{2}}B_{\mu\nu}
\alpha_{-1}^{\mu}\alpha_{-1}^{\nu} \nonumber \\ & & {}+\beta_0b_{-2}c_0
+\beta_1b_{-1}c_{-1}+i\kappa_{\mu}\alpha_{-1}^{\mu}b_{-1}c_0+\cdots 
\Biggr)c_1|k\rangle . \label{eq:AC}
\end{eqnarray}
Since $L_0^{\mathrm{tot}}$ is given by
\begin{equation}
L_0^{\mathrm{tot}}=\ap p^2+\sum_{n=1}^{\infty}\alpha_{-n}^{\mu}\alpha_{\mu n}
+\sum_{n=-\infty}^{\infty}n\ \ca c_{-n}b_n\ca -1, \label{eq:AD}
\end{equation}
where $\ca \cdots \ca $ denotes the usual oscillator--normal ordering, 
each term 
in $|\Phi\rangle$ is an $L_0^{\mathrm{tot}}$ eigenstate. In general, 
\textit{level} of a ($L_0^{\mathrm{tot}}$ eigen-)state is defined to be the 
sum of the level numbers $n$ of the creation operators acting on $c_1|k
\rangle$, \textit{i.e.} sum of the second and third 
term of~(\ref{eq:AD}). This definition is adjusted so that the zero momentum 
tachyon $c_1|0\rangle$ should be at level 0. 
And the level of a component field ($\phi,A_{\mu},\cdots$) is defined to be 
the level of the state associated with it. In some cases, 
this definition is modified to 
include the contribution from the momentum-dependent term, as will 
be explained in chapter~\ref{ch:lump}.

Now that we have defined the level number for the expansion of the string 
field, level of each term in the action is also defined to be the sum of the 
levels of the fields involved. For example, if states $|\Phi_1\rangle ,
|\Phi_2\rangle ,|\Phi_3\rangle$ have level $n_1,n_2,n_3$ respectively, 
we assign level $n_1+n_2+n_3$ to the interaction term 
$\langle \Phi_1,\Phi_2,\Phi_3\rangle$.

Then truncation to level $N$ means that we keep only those terms with level 
equal to or less than $N$. When we say `level $(M,N)$ truncation', it
means that the string field includes the terms with level $\le M$ while the 
action includes ones with level $\le N$.

The level truncation is a means of approximation which is needed simply 
because we cannot deal with an infinite number of terms. But it is not clear 
whether this is a good approximation scheme. Concerning this point, 
some arguments supporting 
the validity of the approximation are given: In~\cite{KS}, it is said 
that the level $n$ terms in the action contain the factor of 
$(4/3\sqrt{3})^n\simeq (0.77)^n$, so that they decrease exponentially as $n$ 
increases. In~\cite{Tay}, effective field theories of tachyon and of gauge 
field are studied numerically. Up to level 20, successive 
approximations seem to be well convergent, obeying neither exponential nor 
power-law fall off. From the point of view of the world-sheet 
renormalization group, higher level states 
correspond to the irrelevant operators in the infrared regime, 
so for the static problems such as tachyon 
condensation it is natural that the higher level terms are quite suppressed. 
However, since there is no very small parameter which validates the 
perturbation expansion ($4/3\sqrt{3}$ seems not small enough to account for 
the rapid convergence exhibited later), it is very interesting if we can 
fully understand the convergence property in the purely theoretical, not 
numerical, way.
\smallskip

In terms of component fields, gauge transformation~(\ref{eq:M}) involves 
transformations of an infinite number of particle fields, and the 
action~(\ref{eq:CSFT}) is invariant under this full gauge transformation. 
Hence the procedures of level truncation, where the fields of levels higher 
than some chosen value are always set to zero, break the gauge invariance. 
As a result, the potential does not have flat directions corresponding to 
degrees of freedom of gauge transformation, even if we do not explicitly 
gauge-fix. But since the lifting of the potential is not under control, we use 
the level truncation after gauge-fixing in the Feynman-Siegel gauge.

\bigskip

Here we show the calculations in detail. We will work in the 
Feynman-Siegel gauge. In this gauge, we simply drop the terms containing 
$c_0$ because the $SL(2,\aaru)$ invariant vacuum satisfies $b_0|0\rangle =0$. 
If we truncate the string field up to level 2, it becomes
\begin{eqnarray}
\hspace{-3cm}
|\Phi\rangle &=& \int d^dk\biggl(\phi(k)+A_{\mu}(k)\alpha_{-1}^{\mu}+\frac{i}{
\sqrt{2}}B_{\mu}(k)\alpha_{-2}^{\mu}+\frac{1}{\sqrt{2}}B_{\mu\nu}(k)
\alpha_{-1}^{\mu}\alpha_{-1}^{\nu}+\beta_1(k)b_{-1}c_{-1}\biggr)c_1|k\rangle
\nonumber \\ &=& \int d^dk \biggl(\phi(k)c(0)+\frac{i}{\sqrt{2\ap}}
A_{\mu}(k)c\partial X^{\mu}(0)-\frac{1}{2\sqrt{\ap}}B_{\mu}(k)c\partial^2
X^{\mu}(0) \label{eq:AE} \\ & & \quad -\frac{1}{2\sqrt{2}\ap}B_{\mu\nu}(k)
c\partial X^{\mu}\partial X^{\nu}(0)-\frac{1}{2}\beta_1(k)\partial^2c(0)
\biggr)|k\rangle, \nonumber
\end{eqnarray}
where $|k\rangle =e^{ik\cdot X(0)}|0\rangle$. In the second line, we 
reexpressed the string field in terms of the vertex operators. 
The CFT method becomes very 
complicated because the level 2 vertex operators associated with the fields 
$B_{\mu},B_{\mu\nu},\beta_1$ are not primary fields. Here we only calculate 
the action~(\ref{eq:CSFT}) to level (1,3) to avoid such complications. 
Conformal transforms of the vertex operators by $\cI(z)=-1/z$ are given by 
\begin{eqnarray}
\cI\circ (ce^{ikX}(\epsilon)) &=& \left(\frac{1}{\epsilon^2}\right)^{-1+
\ap k^2}ce^{ikX}\left(-\frac{1}{\epsilon}\right), \nonumber \\
\cI\circ (c\partial X^{\mu}e^{ikX}(\epsilon))&=& \left(\frac{1}{\epsilon^2}
\right)^{\ap k^2}c\partial X^{\mu}e^{ikX}\left(-\frac{1}{\epsilon}\right),
\label{eq:AF}
\end{eqnarray}
with $\epsilon$ eventually taken to zero. Then the quadratic term becomes
\begin{eqnarray*}
\langle \cI\circ\Phi(0)Q_B\Phi(0)\rangle &=& \int d^dk\ d^dq \, \Biggl\langle
\Biggl[\left(\frac{1}{\epsilon^2}\right)^{-1+\ap k^2}\phi(k)ce^{ikX}\left(
-\frac{1}{\epsilon}\right) \\ & &+ \frac{i}{\sqrt{2\ap}}\left(\frac{1}{
\epsilon^2}\right)^{\ap k^2}A_{\mu}(k)c\partial X^{\mu}e^{ikX}\left(-\frac{1}{
\epsilon}\right)\Biggr]\oint\frac{dz}{2\pi i}(cT^{\mathrm{m}}(z)+bc\partial c
(z)) \\ & &\times \left[\phi(q)ce^{iqX}(\epsilon)+\frac{i}{\sqrt{2\ap}}
A_{\nu}(q)c\partial X^{\nu}e^{iqX}(\epsilon)\right]\Biggr\rangle_{\epsilon\to 
0} \\ &=&\int d^dk\ d^dq\left\langle c\left(-\frac{1}{\epsilon}\right)
\partial c\ c(\epsilon)\right\rangle_{\mathrm{ghost}} \\ & &\times
\Biggl(\left(\frac{1}{\epsilon^2}
\right)^{-1+\ap k^2}(\ap q^2-1)\phi(k)\phi(q)\langle e^{ikX(-1/\epsilon)}
e^{iqX(\epsilon)}\rangle_{\mathrm{matter}} \\ & &- \frac{1}{2\ap}\ap q^2
\left(\frac{1}{\epsilon^2}\right)^{\ap k^2}A_{\mu}(k)A_{\nu}(q)\langle 
\partial X^{\mu}e^{ikX}(-1/\epsilon)\partial X^{\nu}e^{iqX}(\epsilon)
\rangle_{\mathrm{matter}}\Biggr)_{\epsilon\to 0} \\ 
&=& (2\pi)^d\ap \int d^dk\left\{\left(k^2-\frac{1}{\ap}\right)\phi(-k)\phi(k)
+k^2A_{\mu}(-k)A^{\mu}(k)\right\},
\end{eqnarray*}
where we used
\begin{eqnarray*}
\left\langle e^{ikX(-1/\epsilon)}e^{iqX(\epsilon)}\right\rangle_{\mathrm{
matter}}&=&(2\pi)^d\delta^d(k+q)\bigg|-\frac{1}{\epsilon}-\epsilon\bigg|^{
2\ap k\cdot q}, \\ \left\langle\frac{1}{2}\partial^2c\partial c c
\right\rangle_{\mathrm{ghost}}&=&1, \\
\partial X^{\mu}(-1/\epsilon)\partial X^{\nu}(\epsilon)&\sim& -2\ap
\eta^{\mu\nu}\frac{1}{(-\epsilon-(1/\epsilon))^2} \quad 
\mbox{on the boundary}.
\end{eqnarray*}
Notice that the $\epsilon$-dependence has totally cancelled each other, so 
$\epsilon\to 0$ limit can be taken trivially. Fourier-transforming to the 
position space as
\[ \phi(k)=\int\frac{d^dx}{(2\pi)^d}\phi(x)e^{-ikx} \> , \quad A_{\mu}(k)
=\int\frac{d^dx}{(2\pi)^d}A_{\mu}(x)e^{-ikx} \> ,\]
the quadratic part of the action becomes
\begin{eqnarray}
S_{\mathrm{quad}}&=&-\frac{1}{2\ap g_o^2}\langle\cI\circ\Phi(0)Q_B\Phi(0)
\rangle \nonumber \\ &=&\frac{1}{g_o^2}\int d^dx\left(-\frac{1}{2}
\partial_{\mu}\phi\partial^{\mu}\phi+\frac{1}{2\ap}\phi^2-\frac{1}{2}
\partial_{\mu}A_{\nu}\partial^{\mu}A^{\nu}\right). \label{eq:AG}
\end{eqnarray}
This action describes a real\footnote{Reality of the component fields are 
mentioned at the last of this section.} scalar field $\phi$ with mass${}^2
=-1/\ap$ and a massless gauge field $A_{\mu}$. The reason why the gauge 
field does not have the standard kinetic term $-\frac{1}{4}F_{\mu\nu}
F^{\mu\nu}$ is that we have calculated it in the Feynman-Siegel gauge. 
To see this, take level 1 fields without gauge fixing
\begin{equation}
|L1\rangle =\int d^dk \left(A_{\mu}(k)\alpha_{-1}^{\mu}c_1-i\alpha(k)c_0
\right)|k\rangle , \label{eq:AGa}
\end{equation}
and substitute it into the quadratic part of the action. The result is 
given by
\begin{eqnarray}
& &-\frac{1}{2\ap g_o^2}\langle L1|Q_B|L1\rangle = \frac{1}{g_o^2}\int d^dx
\left(-\frac{1}{2}\partial_{\mu}A_{\nu}\partial^{\mu}A^{\nu}-\frac{1}{\ap}
\alpha^2-\sqrt{\frac{2}{\ap}}\alpha\partial_{\mu}A^{\mu}\right) \nonumber
\\ &=&\frac{1}{g_o^2}\int d^dx \left(-\frac{1}{2}(\partial_{\mu}A_{\nu}
\partial^{\mu}A^{\nu}-\partial_{\mu}A_{\nu}\partial^{\nu}A^{\mu})-\frac{1}{2}
\left(\partial_{\mu}A^{\mu}+\sqrt{\frac{2}{\ap}}\alpha\right)^2\right),
\label{eq:AGb}
\end{eqnarray}
where we completed the square in the second line. 
The first term is nothing but 
$-\frac{1}{4}F_{\mu\nu}F^{\mu\nu}$ with $F_{\mu\nu}=\partial_{\mu}A_{\nu}-
\partial_{\nu}A_{\mu}$. Defining $\displaystyle B\equiv \partial_{\mu}A^{\mu}
+\sqrt{\frac{2}{\ap}}\alpha$, $B$ is the Nakanishi-Lautrup auxiliary field. 
If we integrate out the auxiliary $B$ field, we are left with only gauge 
invariant term $-\frac{1}{4}F_{\mu\nu}F^{\mu\nu}$. In the Feynman-Siegel 
gauge, which corresponds to $\alpha =0$, the second term in~(\ref{eq:AGb}) 
gives gauge-fixing term $(-1/2)(\partial_{\mu}A^{\mu})^2$, and the 
gauge-fixed action becomes of the form~(\ref{eq:AG}).

\medskip

We proceed to the cubic term. From (\ref{eq:S}) and (\ref{eq:W}), we get 
\begin{eqnarray}
w_1&\equiv& f_1(0)=-\sqrt{3}, \nonumber \\
w_2&\equiv& f_2(0)=0, \nonumber \\
w_3&\equiv& f_3(0)=\sqrt{3}, \nonumber \\
w_1^{\prime}&\equiv& f_1^{\prime}(0)=\frac{8}{3}e^{-\frac{2\pi i}{3}}
\frac{1}{(g_1(0)+1)^2}\left(\frac{1+iz}{1-iz}\right)^{-\frac{1}{3}}
\frac{1}{(1-iz)^2}\bigg|_{z=0}=\frac{8}{3}, \nonumber \\
w_2^{\prime}&\equiv& f_2^{\prime}(0)=\frac{2}{3}, \nonumber \\
w_3^{\prime}&\equiv& f_3^{\prime}(0)=\frac{8}{3}. \label{eq:AH}
\end{eqnarray}
Conformal transformations are 
\begin{eqnarray*}
f_i\circ\Phi(0)&=&\int d^dk\left\{\phi(k)f_i\circ (ce^{ikX})(0)+\frac{i}{
\sqrt{2\ap}}A_{\mu}(k)f_i\circ (c\partial X^{\mu}e^{ikX})(0)\right\} \\
&=&\int d^dk\Biggl\{(f_i^{\prime}(0))^{\ap k^2-1}\phi(k)ce^{ikX}(f_i(0)) \\
& &{}+
\frac{i}{\sqrt{2\ap}}(f_i^{\prime}(0))^{\ap k^2}A_{\mu}(k)c\partial X^{\mu}
e^{ikX}(f_i(0))\Biggr\}.
\end{eqnarray*}
Then 3-point function is
\begin{eqnarray}
& &\langle f_1\circ\Phi(0)f_2\circ\Phi(0)f_3\circ\Phi(0)\rangle \nonumber \\
&=& \int d^dk\,
d^dp\,d^dq\,w_1^{\prime\ap k^2}w_2^{\prime\ap p^2}w_3^{\prime\ap q^2} 
\nonumber \\
& & \times\Biggl\{\frac{1}{w_1^{\prime}w_2^{\prime}w_3^{\prime}}\phi(k)\phi(p)
\phi(q)\langle ce^{ikX}(w_1)ce^{ipX}(w_2)ce^{iqX}(w_3)\rangle \nonumber \\
& &-\frac{1}{2\ap w_1^{\prime}}\phi(k)A_{\nu}(p)A_{\rho}(q)\langle ce^{ikX}
(w_1)c\partial X^{\nu}e^{ipX}(w_2)c\partial X^{\rho}e^{iqX}(w_3)\rangle 
\nonumber \\
& &-\frac{1}{2\ap w_2^{\prime}}\phi(p)A_{\mu}(k)A_{\rho}(q)\langle c
\partial X^{\mu}e^{ikX}(w_1)ce^{ipX}(w_2)c\partial X^{\rho}e^{iqX}(w_3)
\rangle \nonumber \\
& &-\frac{1}{2\ap w_3^{\prime}}\phi(q)A_{\mu}(k)A_{\nu}(p)\langle c
\partial X^{\mu}e^{ikX}(w_1)c\partial X^{\nu}e^{ipX}(w_2)ce^{iqX}(w_3)
\rangle \Biggr\} \\
&=&(2\pi)^d\int d^dk\,d^dp\,d^dq\,\delta^d(k+p+q)|w_{12}|^{2\ap kp+1}
|w_{23}|^{2\ap pq+1}|w_{13}|^{2\ap kq+1}w_1^{\prime\ap k^2}w_2^{\prime\ap p^2}
w_3^{\prime \ap q^2} \nonumber \\
& & \times \Biggl\{\frac{1}{w_1^{\prime}w_2^{\prime}w_3^{\prime}}\phi(k)
\phi(p)\phi(q)+\frac{1}{w_1^{\prime}w_{23}^2}\phi(k)A_{\mu}(p)A^{\mu}(q) 
\nonumber \\ & &+\frac{1}{w_2^{\prime}w_{13}^2}\phi(p)A_{\mu}(k)A^{\mu}(q)
+\frac{1}{w_3^{\prime}w_{12}^2}\phi(q)A_{\mu}(k)A^{\mu}(p) \nonumber \\
& &+\frac{2\ap}{w_1^{\prime}}\phi(k)A_{\nu}(p)\left(-\frac{k^{\nu}}{w_{12}}
+\frac{q^{\nu}}{w_{23}}\right)A_{\rho}(q)\left(-\frac{k^{\rho}}{w_{13}}
-\frac{p^{\rho}}{w_{23}}\right) \nonumber \\
& &+\frac{2\ap}{w_2^{\prime}}\phi(p)A_{\mu}(k)\left(\frac{p^{\mu}}{w_{12}}
+\frac{q^{\mu}}{w_{13}}\right)A_{\rho}(q)\left(-\frac{k^{\rho}}{w_{13}}
-\frac{p^{\rho}}{w_{23}}\right) \nonumber \\
& &+\frac{2\ap}{w_3^{\prime}}\phi(q)A_{\mu}(k)\left(\frac{p^{\mu}}{w_{12}}
+\frac{q^{\mu}}{w_{13}}\right)A_{\nu}(p)\left(-\frac{k^{\nu}}{w_{12}}
+\frac{q^{\nu}}{w_{23}}\right)\Biggr\}, \label{eq:AI}
\end{eqnarray}
where we dropped $\phi\phi A$ and $AAA$ terms because of a twist 
symmetry~\cite{SenZw,GabZw}. A `twist operation' $\Omega$ is a combination of 
the world-sheet parity reversal and an $SL(2,\aaru)$ transformation. 
Under the action of $\Omega$, odd level fields such as $A_{\mu}$ change sign, 
while even level fields remain unchanged. 
In other words, the $\Omega$-eigenvalue of a state $|A\rangle=Ac_1|0\rangle$ 
is equal to $(-1)^{N_A}$, where $N_A$ is level of $|A\rangle$. The 3-point 
interaction terms have the following properties 
\begin{eqnarray}
(a)& & \langle A,B,C\rangle =\langle B,C,A\rangle =\langle C,A,B\rangle ,
\quad (\mathrm{cyclicity}) \nonumber \\
(b) & & \langle A,B,C\rangle =(-1)^{N_A+N_B+N_C}\langle C,B,A\rangle.
\quad (\mathrm{twist}) \label{eq:AIa}
\end{eqnarray}
The string field $\Phi$ is divided into two parts as $\Phi=\Phi^e+\Phi^o$ 
according to twist even or odd. Since we find, by using~(\ref{eq:AIa}),
\begin{eqnarray*}
\langle\Phi^e,\Phi^e,\Phi^o\rangle& \mathop{=}\limits^{(b)} &(-1)\langle
\Phi^o,\Phi^e,
\Phi^e\rangle \mathop{=}\limits^{(a)} 
-\langle\Phi^e,\Phi^e,\Phi^o\rangle =0, \\
\langle\Phi^o,\Phi^o,\Phi^o\rangle &\mathop{=}\limits^{(b)}& 
(-1)^3\langle\Phi^o,
\Phi^o,\Phi^o\rangle =0,
\end{eqnarray*}
odd level fields appear in the action neither linearly nor trilinearly. 
That is, the action involves interactions of even levels only. We conclude 
that $\phi\phi A$ and $AAA$ are absent in the action. Of course, it can be 
verified explicitly by computations.
Picking out the $\phi^3$ term, and using~(\ref{eq:AH}), it takes the form 
\[ 6\sqrt{3}\frac{3^3}{2^7}(2\pi)^d\int d^dk\ d^dp\ d^dq\,\delta^d(k+p+q)
\phi(k)\phi(p)\phi(q)F(k,p,q) \]
with
\begin{eqnarray*}
F(k,p,q)&=&\exp\biggl[\ap kp\ln 3+\ap pq\ln 3+\ap kq(\ln 3+2\ln 2) \\
& &+\ap k^2(3\ln 2-\ln 3)+\ap p^2(\ln 2-\ln 3)+\ap q^2(3\ln 2-\ln 3)\biggr].
\end{eqnarray*}
Since $F$ is multiplied by $\delta^d(k+p+q)$, we can set $q=-p-k$. Using 
this relation, $F$ can be simplified as 
\begin{eqnarray*}
F(k,p,q)&=& \exp\left(\ap (k^2+p^2+kp)\ln \frac{2^4}{3^3}\right) \\
&=& \exp\left(\ap\ln\left(\frac{4}{3\sqrt{3}}\right)(k^2+p^2+q^2)\right).
\end{eqnarray*}
Finally, the $\phi^3$ term in the action is given by
\begin{eqnarray}
S_{\phi\phi\phi}&=&-\frac{1}{3g_o^2}6\sqrt{3}\frac{3^3}{2^7}(2\pi)^d\int 
d^dk\ d^dp\ d^dq\,\delta^d(k+p+q)\exp\left(\ap 
k^2\ln\frac{4}{3\sqrt{3}}\right)\phi(k)
\nonumber \\ & &\times \exp\left(\ap p^2\ln\frac{4}{3\sqrt{3}}\right)\phi(p)
\exp\left(\ap q^2\ln\frac{4}{3\sqrt{3}}\right)\phi(q) \nonumber \\ &=& 
-\frac{1}{g_o^2}\frac{1}{3}\left(\frac{3\sqrt{3}}{4}\right)^3\int d^dx\,
\tilde{\phi}(x)^3, \label{eq:AJ}
\end{eqnarray}
where we carried out the Fourier transformation and defined
\begin{equation}
\tilde{\phi}(x)=\exp\left(-\ap\ln\frac{4}{3\sqrt{3}}\partial_{\mu}
\partial^{\mu}\right)\phi(x). \label{eq:AK}
\end{equation}
And the tilded fields for other fields are also defined in the same way 
as in (\ref{eq:AK}). $\phi AA$ terms in~(\ref{eq:AI}) can similarly be 
calculated and given by
\begin{eqnarray}
S_{\phi AA}&=&-\frac{1}{g_o^2}\frac{3\sqrt{3}}{4}\int d^dx\,\tilde{\phi}
\tilde{A}_{\mu}\tilde{A}^{\mu} \label{eq:AL} \\ & &-\frac{1}{g_o^2}
\frac{3\sqrt{3}}{8}\ap\int d^dx\,(\partial_{\mu}\partial_{\nu}\tilde{\phi}
\tilde{A}^{\mu}\tilde{A}^{\nu}+\tilde{\phi}\partial_{\mu}\tilde{A}^{\nu}
\partial_{\nu}\tilde{A}^{\mu}-2\partial_{\mu}\tilde{\phi}\partial_{\nu}
\tilde{A}^{\mu}\tilde{A}^{\nu}). \nonumber
\end{eqnarray}
We have now determined the level (1,3) truncated (in fact, it is equal to the 
level (1,2) truncated action because of the absence of $A^3$ term) 
gauge-fixed action to be\footnote{
To be honest, I failed to derive the coefficients of 
the last three terms in (\ref{eq:AL}) using purely CFT method. 
Instead, I calculated them with the help of the conservation laws 
for the current $(i/\ap)\partial X^{\mu}$, which will be explained in 
section~\ref{sec:conserve}. The results~(\ref{eq:AM}) precisely agree with 
those in~\cite{KS}.}
\begin{eqnarray}
S_{(1,2)}&=&\frac{1}{g_o^2}\int d^dx\Biggl(-\frac{1}{2}\partial_{\mu}\phi
\partial^{\mu}\phi +\frac{1}{2\ap}\phi^2-\frac{1}{3}\left(\frac{3\sqrt{3}}{4}
\right)^3\tilde{\phi}^3 \nonumber \\
& &{}-\frac{1}{2}\partial_{\mu}A_{\nu}\partial^{\mu}A^{\nu}
-\frac{3\sqrt{3}}{4}\tilde{\phi}\tilde{A}_{\mu}
\tilde{A}^{\mu} \nonumber \\ & &
-\frac{3\sqrt{3}}{8}\ap(\partial_{\mu}\partial_{\nu}\tilde{\phi}
\tilde{A}^{\mu}\tilde{A}^{\nu}+\tilde{\phi}\partial_{\mu}\tilde{A}^{\nu}
\partial_{\nu}\tilde{A}^{\mu}-2\partial_{\mu}\tilde{\phi}\partial_{\nu}
\tilde{A}^{\mu}\tilde{A}^{\nu})\Biggr). \label{eq:AM}
\end{eqnarray}
This action has an outstanding feature that the particle fields in the cubic 
terms always appear as the tilded fields. This momentum dependence has 
interesting physical effects on the spectrum, as will be discussed later.

\medskip

To be compared with the quadratic action~(\ref{eq:AG}), let's rederive the 
quadratic action making use of the operator method. This time we involve 
the fields of level 2 as well. Substituting the expansion~(\ref{eq:AE}) into 
$\langle\Phi |Q_B|\Phi\rangle$, we get 
\begin{eqnarray}
S_{\mathrm{quad}}&=&-\frac{1}{2\ap g_o^2}\langle\Phi |Q_B|\Phi\rangle 
\nonumber \\ &=& -\frac{1}{2\ap g_o^2}\int d^dk\ d^dq\Biggl(\phi(q)\phi(k)
\langle q|c_{-1}Q_Bc_1|k\rangle +A_{\mu}(q)A_{\nu}(k)\langle q|c_{-1}
\alpha_1^{\mu}Q_B\alpha_{-1}^{\nu}c_1|k\rangle \nonumber \\ & &+\frac{-i}
{\sqrt{2}}\frac{i}{\sqrt{2}}B_{\mu}(q)B_{\nu}(k)\langle q|c_{-1}\alpha_2^{\mu}
Q_B\alpha_{-2}^{\nu}c_1|k\rangle \nonumber \\ & &
+\frac{1}{2}B_{\mu\nu}(q)B_{\lambda\rho}(k)
\langle q|c_{-1}\alpha_1^{\mu}\alpha_1^{\nu}Q_B\alpha_{-1}^{\lambda}
\alpha_{-1}^{\rho}c_1|k\rangle \nonumber \\ & &+\beta_1(q)\beta_1(k)\langle
q|c_1Q_Bc_{-1}|k\rangle\Biggr) \nonumber \\ &=&-\frac{(2\pi)^d}{2\ap g_o^2}
\int d^dk\Biggl((\ap k^2-1)\phi(-k)\phi(k)+\ap k^2A_{\mu}(-k)A^{\mu}(k) 
\nonumber \\ & &+\frac{1}{2}(\ap k^2+1)2B_{\mu}(-k)B^{\mu}(k)+\frac{1}{2}
(\ap k^2+1)B_{\mu\nu}(-k)B_{\lambda\rho}(k)(\eta^{\mu\lambda}\eta^{\nu\rho}+
\eta^{\nu\lambda}\eta^{\mu\rho}) \nonumber \\ & &-(\ap k^2+1)\beta_1(-k)
\beta_1(k)\Biggr) \nonumber \\
&=&\frac{1}{g_o^2}\int d^dx\Biggl(-\frac{1}{2}\partial_{\mu}\phi\partial^{\mu}
\phi+\frac{1}{2\ap}\phi^2-\frac{1}{2}\partial_{\mu}A_{\nu}\partial^{\mu}
A^{\nu}-\frac{1}{2}\partial_{\mu}B_{\nu}\partial^{\mu}B^{\nu}-\frac{1}{2\ap}
B_{\mu}B^{\mu} \nonumber \\ & &-\frac{1}{2}\partial_{\lambda}B_{\mu\nu}
\partial^{\lambda}B^{\mu\nu}-\frac{1}{2\ap}B_{\mu\nu}B^{\mu\nu}+\frac{1}{2}
\partial_{\mu}\beta_1\partial^{\mu}\beta_1+\frac{1}{2\ap}\beta_1^2\Biggr),
\label{eq:AN}
\end{eqnarray}
where we used (\ref{eq:Qbos}) for $Q_B$, and $\langle q|k\rangle =(2\pi)^d
\delta^d(q+k)$. The first three terms precisely agree with the 
result~(\ref{eq:AG}) from the CFT method. And mass terms are correctly 
reproduced also for the level 2 fields. (The $\beta_1$ kinetic term, though, 
has the opposite sign to the other `physical' fields.)
\medskip

We will give some comments. Firstly, in calculating the Hilbert space inner 
product we needed so-called `BPZ conjugation' to obtain a bra state $\langle
\Phi |=\mathrm{bpz}(|\Phi\rangle )$. For a primary field 
$\phi(z)=\sum_{n=-\infty}^{\infty}\phi_n/z^{n+h}$ of weight $h$, 
the BPZ conjugation is defined via the inversion $\cI$ as 
\begin{eqnarray}
\mathrm{bpz}(\phi_n|0\rangle)&=&\langle 0|\mathrm{bpz}(\phi_n); \nonumber \\
\mathrm{bpz}(\phi_n)&=&\cI[\phi_n]\equiv \oint\frac{dz}{2\pi i}z^{n+h-1}
\cI\circ\phi(z) \nonumber \\ &=&\oint\frac{dz}{2\pi i}z^{n+h-1}
\left(\frac{1}{z^2}\right)^h\phi\left(-\frac{1}{z}\right)=\oint\frac{dz}
{2\pi i}z^{n-h-1}\sum_{m=-\infty}^{\infty}\phi_m(-1)^{m+h}z^{m+h} \nonumber 
\\ &=&(-1)^{-n+h}\phi_{-n}. \label{eq:AO}
\end{eqnarray}
For example, for a primary field $\partial X^{\mu}$ of weight 1, we have 
\[ \mathrm{bpz}(\alpha_{-n}^{\mu})=(-1)^{n+1}\alpha_n^{\mu}. \]

Secondly, we consider the reality condition of the string field. When the 
string field $\Phi$ is regarded as a functional of the string embeddings 
$X^{\mu}(\sigma_1,\sigma_2)$ and ghosts $b(\sigma_1,\sigma_2),\,c(\sigma_1,
\sigma_2)$, its reality condition is written as~\cite{WiSFT1}
\begin{equation}
\Phi[X^{\mu}(\sigma_1),b(\sigma_1),c(\sigma_1)]=\Phi^*[X^{\mu}(\pi-\sigma_1),
b(\pi-\sigma_1),c(\pi-\sigma_1)]. \label{eq:AOa}
\end{equation}
It corresponds to the hermiticity of the `matrix' $\Phi$. 
In terms of the state $|\Phi\rangle$, the Hermitian conjugation 
operation (denoted by $\mathrm{hc}$) alone takes the ket to the bra, so 
$\mathrm{hc}$ is combined with BPZ conjugation to define the star 
conjugation~\cite{GabZw}
\begin{equation}
*=\mathrm{bpz}^{-1}\circ\mathrm{hc}=\mathrm{hc}^{-1}\circ\mathrm{bpz}.
\label{eq:AOb}
\end{equation}
Then the reality condition reads 
\begin{equation}
|\Phi^*\rangle \equiv\mathrm{bpz}^{-1}\circ\mathrm{hc}(|\Phi\rangle)
=|\Phi\rangle . \label{eq:AOc}
\end{equation}
This condition guarantees the reality of the component fields 
$\phi,A_{\mu},B_{\mu},\ldots$ in the expansion~(\ref{eq:AE}). For instance, 
as regards the tachyon state, 
\[ \mathrm{bpz}^{-1}\circ\mathrm{hc}(\phi(k)c_1|k\rangle)=\mathrm{bpz}^{-1}
(\langle 0|e^{-ikX}\phi^*(k)c_{-1})=\phi^*(k)c_1|-k\rangle \]
should be equal to $\phi(q)c_1|q\rangle$ under momentum integration, which 
gives $\phi^*(k)=\phi(-k)$. In the position space, it becomes the reality 
condition $\phi^*(x)=\phi(x)$, as expected. The second example is the term 
$\frac{i}{\sqrt{2}}B_{\mu}\alpha_{-2}^{\mu}c_1|0\rangle$ with the momentum 
dependence ignored. Since 
\begin{equation}
\mathrm{bpz}^{-1}\circ\mathrm{hc}\left(\frac{i}{\sqrt{2}}B_{\mu}
\alpha_{-2}^{\mu}c_1|0\rangle\right)=\mathrm{bpz}^{-1}\left(\frac{-i}{
\sqrt{2}}\langle 0|B_{\mu}^*\alpha_2^{\mu}c_{-1}\right)=-\frac{i}{\sqrt{2}}
B_{\mu}^*(-\alpha_{-2}^{\mu})c_1|0\rangle , \label{eq:AOd}
\end{equation}
the factor $i$ is needed for $B_{\mu}$ to be real.

Thirdly, the $\beta_1$ kinetic term in~(\ref{eq:AN}) has the opposite sign 
to the other `physical' fields. Since the phase of $\beta_1$ in the 
expansion~(\ref{eq:AE}) has been determined (up to $\pm$) in such a way that 
$\beta_1$ should be real, we cannot selfishly redefine $\beta_1^{\prime}=i
\beta_1$. Therefore the wrong sign is not a superficial one. These 
`auxiliary' fields have the effect of cancelling the unphysical components 
of vector (tensor) fields.

We showed here the level (1,2) truncated string field theory 
action~(\ref{eq:AM}) together with the actual calculations. 
Further, the level (2,6) 
Lagrangian is presented in~\cite{KS}. To compare it with ours, we should 
set $g=2$ in the results of~\cite{KS}.

\section{Universality of the Tachyon Potential}\label{sec:univ}
Now that we have finished setting up bosonic string field theory, next we 
turn to the subject of tachyon condensation. To begin with, we show that the 
tachyon potential has the universal form~\cite{Univ} 
which is independent of the details of the 
theory describing the D-brane. We also relate 
the open string coupling $g_o$ to the D-brane tension so that the expression 
of the action 
has the form appropriate for examining the conjecture on annihilation of the 
unstable D-brane. Then, we calculate the tachyon potential and its minimum, 
and discuss the meaning of them.

\bigskip

We consider oriented bosonic open string theory on a single D$p$-brane 
extended in the $(1,\ldots,p)$-directions. Some of the directions tangential 
to the D-brane may be wrapped on non-trivial cycles. If there are noncompact 
tangential directions, we compactify them on a torus of large radii to make 
the total mass of the D-brane finite. 
Let $V_p$ denote the spatial volume of the 
D-brane. In general, such a wrapped D-brane is described by a non-trivial 
boundary conformal field theory. 
What we want to prove is that the tachyon potential defined 
below is independent of this boundary CFT.
\smallskip

Since we will be looking for a Lorentz invariant vacuum as a solution to 
the equations of 
motion derived from the string field theory action, only Lorentz scalar 
fields acquire nonvanishing vacuum expectation values. Once the tachyon 
field $\phi$ develops a nonzero vacuum 
expectation value, equations of motion require that (infinitely many) scalar 
fields of higher levels also have nonvanishing vacuum expectation values due 
to the cubic interaction terms. But since not all the scalar 
fields must be given nonzero values, we want to drop as many fields that 
need not acquire nonvanishing expectation values as possible from beginning. 
The idea is as follows: 
If some component field $\psi$ always enters the action quadratically or 
in higher order (in other words, the action contains no linear term in 
$\psi$), 
each term in the equation of motion obtained by differentiating the action 
with respect to $\psi$ involves at least 
one factor of $\psi$. This equation of motion is trivially 
satisfied by setting $\psi=0$. In finding a solution to the 
equations of motion, we are allowed to set to zero all such fields as 
$\psi$ throughout the calculation. We will often give arguments like this 
in the remainder of this paper. Now let us identify the set of such fields. 
A string field is an element of the Hilbert space $\cH^1$ of ghost number 1. 
We decompose it into two parts as $\cH^1=\cH^1_1\oplus \cH^1_2$ in the 
following manner: Let $\cH^1_1$ consist of states obtained by acting with 
the ghost oscillators $b_n,c_n$ and the matter Virasoro generators 
$L_n^{\mathrm{m}}$ 
on the $SL(2,\aaru)$ invariant vacuum $|0\rangle$. Note that $\cH^1_1$ 
contains zero momentum tachyon state $c_1|0\rangle$. $\cH^1_2$ includes 
all other states in $\cH^1$, that is, states with nonzero momentum $k$ along 
the D$p$-brane and states obtained by the action of $b_n,c_n,L_n^{\mathrm{m}}$
on the non-trivial \textit{primary} states of weight $>0$. Let us prove that 
a component of the string field $\Phi$ along 
$\cH^1_2$ never appears in the action linearly. For the nonzero momentum 
sector, it is obvious that the momentum conservation law requires the 
couplings $\langle 0|Q_B|k\rangle$ and $\langle V_3|0\rangle\otimes |0\rangle
\otimes |k\rangle$ to vanish for $k\neq 0$. Hence we focus on the zero 
momentum sector. Taking the states $|\psi\rangle\in\cH^1_2$ and $|\phi_1
\rangle,\ |\phi_2\rangle\in\cH^1_1$, we consider $S_2=\langle\phi_1|Q_B|
\psi\rangle$ and $S_3=\langle V_3|\phi_1\rangle\otimes |\phi_2\rangle\otimes
|\psi\rangle$. For these to have nonvanishing values, the ghost parts must be 
of the form $\langle c_{-1}c_0c_1\rangle$. Assuming that this condition is 
satisfied, we consider only the matter parts. Since $Q_B$ is constructed out 
of $b_n,c_n,L_n^{\mathrm{m}}$ (as is seen from~(\ref{eq:Qbos})), the matter 
part of $S_2$ can generally be written as 
\begin{eqnarray*}
S_2&=&\langle 0|L_{n_1}^{\mathrm{m}}\cdots L_{n_i}^{\mathrm{m}}|\pi\rangle \\
&=&\langle 0|\pi\rangle +\sum\langle 0|L_{m_1}^{\mathrm{m}}\cdots 
L_{m_j}^{\mathrm{m}}|\pi\rangle , 
\end{eqnarray*}
where $|\pi\rangle\in\cH^1_2$ is a primary state and $m$'s are taken to be 
all positive or all negative by using the Virasoro algebra. In either way, 
such terms vanish when $L_m^{\mathrm{m}}=(L_{-m}^{\mathrm{m}})^{\dagger}$ 
acts on $\langle 0|$ or $|\pi\rangle$ because they are primary states. And 
the only remaining term $\langle 0|\pi\rangle$, if any, also vanishes 
because $|\pi\rangle$ has nonzero conformal weight. So we have found 
$S_2=0$. In a similar notation, $S_3$ can be written as 
\[ S_3=\langle V_3|L_{-n_1}^{\mathrm{m}}\cdots L_{-n_i}^{\mathrm{m}}|0
\rangle_3\otimes L_{-m_1}^{\mathrm{m}}\cdots L_{-m_j}^{\mathrm{m}}|0
\rangle_2\otimes L_{-\ell_1}^{\mathrm{m}}\cdots L_{-\ell_k}^{\mathrm{m}}
|\pi\rangle_1. \]
Using the Virasoro conservation laws explained in section~\ref{sec:conserve}, 
we can move the Virasoro generators $L_{-n}^{\mathrm{m}},L_{-m}^{\mathrm{m}}$ 
to the 1st string Hilbert space as 
\[ S_3=\sum\langle V_3|0\rangle_3\otimes |0\rangle_2\otimes 
\overbrace{L_{k_1}^{\mathrm{m}}\cdots L_{k_a}^{\mathrm{m}}}^a|\pi\rangle_1. \]
If $k$'s are positive, $L_k^{\mathrm{m}}$ annihilates the primary state 
$|\pi\rangle_1$. If $a=0$ or $k$'s are negative, the state 
$L_{k_1}^{\mathrm{m}}\cdots L_{k_a}^{\mathrm{m}}|\pi\rangle_1$ has a strictly 
positive weight so that the 3-point coupling vanishes. From these results, 
we have also established $S_3=0$. Therefore, we can consistently truncate 
the string field $\Phi$ to lie in $\cH^1_1$ by setting the component fields 
along $\cH^1_2$ to zero \textit{all together}. We may further truncate the 
string field by appealing to the special symmetries of the 3-string vertex 
or equivalently the gluing prescription, as shown in~\cite{Trim,11238}. 
But we do not try it here.
\smallskip

For clarity, we introduce new symbols: We denote by $T$ the string field 
$\Phi$ truncated to $\cH^1_1$, and by $\tilde{S}(T)$ the cubic string field 
theory action $S(\Phi)$ truncated to $\cH^1_1$. Since the fields in $\cH^1_1$ 
have zero momenta and hence 
are independent of the coordinates on the D-brane world-volume, the 
integration over $x$ gives the ($p+1$)-dimensional volume factor $V_{p+1}$. 
So the action is written as
\begin{equation}
\tilde{S}(T)=V_{p+1}\tilde{\cL}(T)=-V_{p+1}U(T), \label{eq:AP}
\end{equation}
where we defined the tachyon potential $U(T)$ as the negative of the 
Lagrangian. By definition of $\cH^1_1$, the truncated action 
$\tilde{S}$ is entirely given by the correlation functions involving only 
the ghost fields $b , c$ and matter energy momentum tensor $T^{\mathrm{m}}$. 
In the oscillator representation, we only need the commutation relations 
among $b_n , c_n$ and the matter Virasoro algebra. Though the latter depends 
on the central charge $c$, it is now set to 26 in the critical string 
theory. Then the action, accordingly the tachyon potential as well, is 
universal in the sense that it has no room for containing 
information on the boundary CFT which describes the D-brane. More precisely, 
what is universal is part of the action except for the open string coupling 
$g_o^{-2}$. Next we establish the relation of $g_o$ to the D$p$-brane tension 
$\tau_p$ measured at the perturbative open string vacuum where 
$\langle T\rangle =0$. Though 
its derivation is outlined in~\cite{Univ} for the bosonic string case,
we will nevertheless give it in the next section with detailed string 
field theory calculations.\footnote{I would like to thank T. Takayanagi for 
detailed discussions about this point.}

\section{Mass of the D-brane}\label{sec:mass}

We consider a bosonic D$p$-brane which extends in the 
$(1,\cdots,p)$-directions. Let us see the quadratic part of the string field 
theory action which involves the following mode\footnote{Since we are 
compactifying all directions tangential to the D-brane, the momentum $k$ must 
correctly be discretized. By `integration', we mean summation over all 
possible values of $k$, giving rise to no misunderstanding.}

\begin{equation}
|\Phi\rangle =\int d^{p+1}k \ \phi^i(k)\delta^p(\vec{k})c_1\alpha^i_{-1}
e^{ik\cdot X(0)}|0\rangle ,
\end{equation}
where label $i$ denotes the directions transverse to the D-brane that are 
noncompact and flat (we assume that at least one such direction exists).
And the $\delta$-function sets to zero the space components of the momentum 
on the brane. We evaluate it with CFT prescription

\begin{eqnarray*}
S_{\mathrm{quad}}&=&-\frac{1}{2g_o^2\ap}\langle \Phi |Q_B|\Phi\rangle=
-\frac{1}{2g_o^2\ap}\langle\cI\circ\Phi(0)\oint\frac{dz}{2\pi i}j_B(z)
\Phi(0)\rangle \\ &=&\frac{1}{4g_o^2\ap{}^2}\int d^{p+1}k\ d^{p+1}q\ \phi^i(k)
\phi^j(q)\left(\frac{1}{\epsilon^2}\right)^{\ap k^2}\delta^p(\vec{k})
\delta^p(\vec{q})\Big\langle e^{ikX}c\partial X^i\left(-\frac{1}{\epsilon}
\right) \\ & & \times\oint\frac{dz}{2\pi i}(cT^{\mathrm{m}}
+bc\partial c)(z)c\partial X^je^{iqX}(\epsilon)\Big\rangle_{\epsilon\to 0} \\
&=&(2\pi)^{p+1}\frac{1}{2g_o^2}\int dk_0dq_0\phi^i(k_0)\phi^j(q_0)\delta^{ij}
q_0^2\delta(k_0+q_0)\delta^p(\vec{0})\left(\frac{1}{\epsilon}\right)^{-\ap
k_0^2-\ap k_0q_0} \\
&=&\frac{\pi V_p}{g_o^2}\int dk_0(k_0)^2\phi^i(k_0)\phi^i(-k_0) ,
\end{eqnarray*}
where we used the normalization convention
\begin{eqnarray*}
\langle\partial^2 c\partial c c(z)\rangle_{\mathrm{ghost}}&=&\langle0|2c_{-1}
c_0c_1|0\rangle=2 ,\\
\langle e^{ikX(z_1)}e^{iqX(z_2)}\rangle&=&(2\pi)^{p+1}
\delta^{p+1}(k+q)|z_1-z_2|^{2\ap k\cdot q} ,\\
V_p=\int d^p x &=& (2\pi)^p\delta^p(0) ,
\end{eqnarray*}
and minus sign in front of $k_0q_0$ in the third line 
originated from $\eta_{00}=-1$. Fourier 
transforming as $\displaystyle{\phi^i(k_0)=\int\frac{dt}{2\pi}\chi^i(t)
e^{-ik_0t}}$, the action can be written as
\begin{equation}
S_{\mathrm{quad}}=\frac{V_p}{2g_o^2}\int dt \partial_t\chi^i\partial_t
\chi^i.\label{eq:chi2}
\end{equation}
Here, $\chi^i$ is interpreted as the `location' of the D-brane in the 
transverse space, although its normalization is not correctly chosen. 
So we next fix it.
\medskip

Let's take a pair of identical D-branes whose locations are $0^i$ and $b^i$, 
respectively, in the transverse space. An open string state which stretches 
between them acquires the additional contribution $|\vec{b}|^2/(2\pi\ap)^2$ 
to the mass squared due to 
the string tension. If the D-brane at $b^i$ moves by a small amount $Y^i$, 
the mass squared of the open string state changes into
\[\frac{|\vec{b}+\vec{Y}|^2}{(2\pi\ap)^2}=\frac{|\vec{b}|^2}{(2\pi\ap)^2}+
\frac{\vec{b}\cdot\vec{Y}}{2\pi^2\ap{}^2}+\cO(Y^2). \]
So the mass squared of the state shifts by 
\begin{equation}
\frac{\vec{b}\cdot\vec{Y}}{2\pi^2\ap{}^2}, \label{eq:bY}
\end{equation}
due to the translation of the brane 
induced by the marginal mode $\chi^i$. 
\medskip

Now we describe the above situation in the string field theory language. 
As to the stretched string, any state will do, but for simplicity we take 
it to be tachyonic ground state
\[ |T\rangle=\left(\int dk_0u(k_0)U_{k_0}(z=0)\otimes \left(
	\begin{array}{cc}
	0 & 1 \\
	0 & 0
	\end{array}
\right)+\int dq_0v(q_0)V_{q_0}(z=0)\otimes\left(
	\begin{array}{cc}
	0 & 0 \\
	1 & 0
	\end{array}
\right)\right)|0\rangle,\]
where
\begin{eqnarray*}
U_{k_0}(z)&=&ce^{i\frac{b^i}{2\pi\ap}X^i}e^{ik_0X^0}, \\
V_{k_0}(z)&=&ce^{-i\frac{b^i}{2\pi\ap}X^i}e^{ik_0X^0},
\end{eqnarray*}
and the $2\times 2$ matrices tensored to the vertex operators are Chan-Paton 
factors representing two D$p$-branes. Though they involve slight abuse 
of notation because $X^i$ stands for $T$-dualized coordinate $X_L^i-X_R^i$ 
while $X^0=X_L^0+X_R^0$, no ambiguity will occur. The $L_0^{\mathrm{tot}}$ 
eigenvalue of $|T\rangle$
is $\displaystyle -\ap (k_0)^2+\frac{\vec{b}^2}{(2\pi)^2\ap}-1$, as required. 
Further, we add to $|T\rangle$ the following term as a background, 
\[|\mathrm{tr}\rangle=\chi^iP^i(z=0)\otimes\left(
	\begin{array}{cc}
	1 & 0 \\
	0 & 0
	\end{array}
\right)|0\rangle ; \quad P^i(z)=i\frac{1}{\sqrt{2\ap}}c\partial X^i(z), \]
which has the effect of moving one of the two branes. Now we are ready to do 
actual calculations. We plug $|\Phi\rangle =|\mathrm{tr}\rangle+|T\rangle$ into
the string field theory action
\begin{equation}
S=-\frac{1}{g_o^2}\left(\frac{1}{2\ap}\langle\Phi|Q_B|\Phi\rangle+\frac{1}{3}
\langle\Phi|\Phi|\Phi\rangle\right),
\end{equation}
and look at the quadratic terms in $u,v$.

The relevant terms in the quadratic part are
\begin{eqnarray*}
\langle\Phi|Q_B|\Phi\rangle&\to& \mathrm{Tr}\Bigg\langle\left(
	\begin{array}{cc}
	0 & \int dk_0u(k_0)U_{k_0}(-1/\epsilon) \\
	\int dk_0v(k_0)V_{k_0}(-1/\epsilon) & 0
	\end{array}
\right) \\ &\times& \left(\frac{1}{\epsilon^2}\right)^{-\ap (k_0)^2+\frac{b^2}
{4\pi^2\ap}-1}\partial c(0)\left(
	\begin{array}{cc}
	0 & \int dq_0u(q_0)U_{q_0}(0) \\
	\int dq_0v(q_0)V_{q_0}(0) & 0
	\end{array}
\right) \\ &\times&\left(-\ap (q_0)^2+\frac{b^2}{4\pi^2\ap}-1\right)
\Bigg\rangle_{\epsilon\to 0} \\ &=&4\pi V_p\int dk_0u(k_0)v(-k_0)
\left(-\ap k_0^2+\frac{b^2}{4\pi^2\ap}-1\right).
\end{eqnarray*}
Hence $u,v$ kinetic term in the position space becomes
\begin{equation}
S_{uv}=\frac{V_p}{g_o^2}\int dt \partial_tu \partial_t v. \label{eq:uvkin}
\end{equation}

In the cubic part we draw only $\chi u v$ term. From among $3^3=27$ 
terms, 6 terms have the combination 
$\chi u v$, but three of them vanish due to 
the Chan-Paton factors: \textit{e.g.} $\left(
	\begin{array}{cc}
	0 & 1 \\
	0 & 0
	\end{array}
\right)\left(
	\begin{array}{cc}
	1 & 0 \\
	0 & 0
	\end{array}
\right)=0$. The remaining three terms give the same contribution 
thanks to the cyclicity of 3-point vertex. Therefore we have only 
to calculate one correlator:
\begin{eqnarray*}
\langle\Phi|\Phi|\Phi\rangle &\to& 3\cdot i\sqrt{\frac{1}{2\ap}}
\chi^i\int dk_0dq_0u(k_0)v(q_0)\Bigg\langle f_1\circ (c\partial X^i)
(0)f_2\circ \left(ce^{i\frac{b^i}{2\pi\ap}X^i+ik_0X^0}\right)(0)
\\ &\times&f_3\circ \left(ce^{-i\frac{b^i}{2\pi\ap}X^i+iq_0X^0}
\right)(0)\Bigg\rangle \\ & & \!\! =\frac{3}{\pi}\sqrt{\frac{1}
{2\ap}}V_p\vec{\chi}\cdot\vec{b}\int_{\mbox{\tiny on-shell}}dt\,uv(t)
+(\mbox{off-shell}).
\end{eqnarray*}
Hence $\displaystyle S_{\chi uv}=-\frac{V_p}{g_o^2}\left(\frac{1}
{\pi\sqrt{2\ap}}\vec{\chi}\cdot\vec{b}\right)
\int_{\mbox{\tiny on-shell}}dt\,uv$. Comparing it with the kinetic term 
$S_{uv}$~(\ref{eq:uvkin}), we find that $u,v$ 
fields have gotten additional mass squared 
\begin{equation}
\Delta m^2_{uv}=\frac{1}{\pi\sqrt{2\ap}}\vec{\chi}\cdot\vec{b}.
\label{eq:muv}
\end{equation}
Since the change of the location of one of the D-branes by an 
amount $\chi^i$ caused the shift~(\ref{eq:muv}) of the mass
squared of the open string state stretched between the D-branes, 
the quantity~(\ref{eq:muv}) must be equal to the previous 
result~(\ref{eq:bY}). This consideration determines the normalization
of $\chi^i$ as 
\[\chi^i=\frac{1}{\sqrt{2}\pi\ap{}^{3/2}}Y^i.\]
Substituting it into~(\ref{eq:chi2}) leads to 
\begin{equation}
S_{\mathrm{quad}}=\frac{V_p}{4\pi^2g_o^2\ap{}^3}\int dt \,
\partial_tY^i\partial_tY^i.\label{eq:SY}
\end{equation}
This is interpreted as the kinetic energy arising from the 
collective motion of the D-brane in the nonrelativistic mechanics. 
Hence the prefactor is identified with half of the D-brane mass.
So we have obtained the D$p$-brane tension as 
\begin{equation}
\tau_p=\frac{1}{2\pi^2g_o^2\ap{}^3}. \label{eq:AQ}
\end{equation}
This formula may seem to include the wrong dependence on $\ap$, but remember 
that $g_o$ is a dimensionful parameter as we have already mentioned. 
In order for the action in~(\ref{eq:SY}) to be dimensionless, $g_o$ must have 
dimension of mass${}^{\frac{-p+5}{2}}$. 
We then define the dimensionless open string coupling $\bar{g}_o$ by
\[ g_o^2=\bar{g_o}^2\ap{}^{\frac{p-5}{2}},\]
and if we rewrite the tension in terms of $\bar{g_o}$, 
then the correct formula 
\begin{equation}
\tau_p=\frac{1}{2\pi^2\bar{g_o}^2\ap{}^{\frac{p+1}{2}}}
\end{equation}
is recovered. Incidentally, dimensionlessness of the action 
in~(\ref{eq:AG}) requires 
that the component fields $\phi,A_{\mu}$ have dimension of mass${}^2$. 
It is not surprising that the gauge field has such dimension because the 
expansion~(\ref{eq:AE}) of the string field was not appropriately normalized.

\section{Tachyon Potential in the Level Truncation Scheme}
Using the relation~(\ref{eq:AQ}), the action can be rewritten as
\begin{eqnarray*}
\tilde{S}&=&-\frac{1}{g_o^2}\left(\frac{1}{2\ap}\langle\cI\circ T(0)Q_BT(0)
\rangle +\frac{1}{3}\langle f_1\circ T(0)f_2\circ T(0)f_3\circ T(0)\rangle
\right) \\ &=& -V_{p+1}2\pi^2\ap{}^3\tau_p\left(\frac{1}{2\ap}\langle\cI\circ 
T(0)Q_BT(0)\rangle_{\mathrm{norm.}} +\frac{1}{3}\langle f_1\circ T(0)f_2\circ 
T(0)f_3\circ T(0)\rangle_{\mathrm{norm.}}\right).
\end{eqnarray*}
Generically, the correlation function includes the momentum conservation delta
 function $(2\pi)^{p+1}\delta^{p+1}(\sum k)$, but since $T\in\cH^1_1$ has no 
momentum dependence, this factor simply gives the volume $V_{p+1}$ of the 
D$p$-brane. Accordingly, the correlation functions in the second 
line are normalized such that $\langle 1\rangle_{\mathrm{matter}}=1$. In the 
rest of this chapter, we will use the symbol $\langle\cdots\rangle$ without 
`norm.' to represent the correlation function normalized in the above 
mentioned way. From the tachyon potential $U(T)=-\tilde{S}/V_{p+1}$, we 
define the following `universal function'
\begin{equation}
f(T)=\frac{U(T)}{\tau_p}=2\pi^2\ap{}^3\left(\frac{1}{2\ap}\langle\cI\circ T(0)
Q_BT(0)\rangle +\frac{1}{3}\langle f_1\circ T(0)f_2\circ T(0)f_3\circ T(0)
\rangle\right). \label{eq:AR}
\end{equation}
Total energy density coming from the tachyon potential and the D-brane 
tension is given by
\[ U(T)+\tau_p=\tau_p(1+f(T)). \]
According to the conjecture of \cite{Descent}, at the minimum $T=T_0$ of the 
tachyon potential the negative energy contribution from the tachyon 
potential exactly cancels the D-brane tension, leading to the `closed 
string vacuum' without any D-brane. In terms of $f(T)$, such a phenomenon 
occurs if 
\begin{equation}
f(T=T_0)=-1 \label{eq:AS}
\end{equation}
is true. We will investigate it by calculating $f(T)$ in the level truncation 
scheme, solving equations of motion to find $T_0$, and evaluating the minimum 
$f(T_0)$. Since the universality (independence from the boundary CFT) of the 
function $f(T)$ has already been established, we conclude that 
eq.(\ref{eq:AS}) persists for all D-branes compactified 
in arbitrary ways if we verify the relation~(\ref{eq:AS}) 
for the simplest (toroidally compactified) case. 
\medskip

\begin{figure}[htbp]
	\begin{center}
	\includegraphics{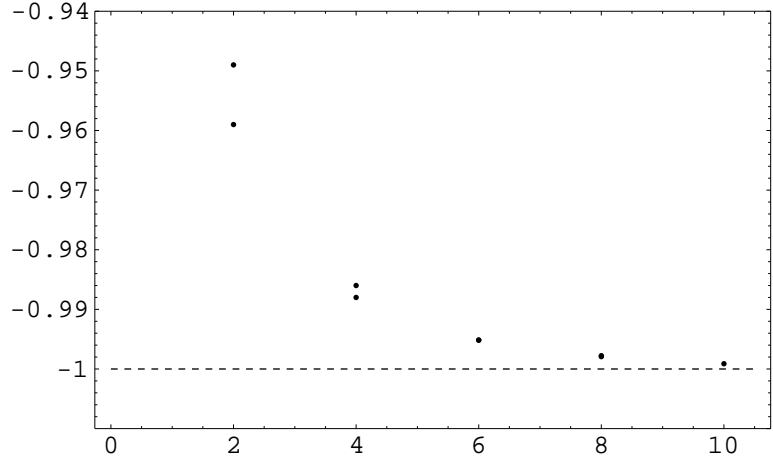}
	\end{center}
	\caption{The minimum values of the potential.}
	\label{fig:J}
\end{figure}

Now, we proceed to the actual calculations. Level (0,0) truncation, 
namely only the zero momentum tachyon state being kept, has, in fact, 
already been worked out. By picking the second and third terms out of 
the action~(\ref{eq:AM}), $f(\phi)$ is given by 
\begin{equation}
f(\phi)=2\pi^2\ap{}^3\left(-\frac{1}{2\ap}\phi^2+\frac{1}{3}\left(\frac{
3\sqrt{3}}{4}\right)^3\phi^3\right), \label{eq:AT}
\end{equation}
where we set $\tilde{\phi}=\phi$ because $\phi$ is a constant. Its minimum is 
easily found. By solving $\partial f(\phi)/\partial\phi |_{\phi_0}=0$, 
we find 
\begin{equation}
\phi_0=\left(\frac{4}{3\sqrt{3}}\right)^3\frac{1}{\ap} \quad \mathrm{and} 
\quad f(\phi_0)\simeq -0.684 . \label{eq:AU}
\end{equation}
Although we have considered only the tachyon state in the vast Hilbert space 
$\cH^1_1$, the minimum value~(\ref{eq:AU}) accounts for as much as 68\% of 
the conjectured value~(\ref{eq:AS})! We then compute corrections to it by 
including the fields of higher level. Before that, we can still restrict the 
fields by appealing to the `twist symmetry'. As mentioned earlier, the 
twist invariance requires the odd level fields to enter the action in pairs. 
Then we can set to zero \textit{all} odd level fields 
without contradicting the 
equations of motion. Hence the terms we should take into account next are the 
level 2 fields
\begin{equation}
|L2\rangle =-\beta_1c_{-1}|0\rangle +\frac{v}{\sqrt{13}}L_{-2}^{\mathrm{m}}c_1
|0\rangle . \label{eq:AV}
\end{equation}
Though the state $b_{-2}c_0c_1|0\rangle$ is also at level 2, it is excluded by
the Feynman-Siegel gauge condition~(\ref{eq:AA}). The vertex operator 
representation of the string field \textit{up to} level 2 is then given by 
\begin{equation}
T(z)=\phi c(z)-\frac{1}{2}\beta_1\partial^2c(z)+\frac{v}{\sqrt{13}}
T^{\mathrm{m}}(z)c(z). \label{eq:AW}
\end{equation}
Here we make a comment on the level truncation. 
In the level ($M,N$) truncation, 
in order for the quadratic terms (kinetic term + mass term) for the level $M$ 
fields to be included in the action, 
$N$ must be equal to or larger than $2M$. On the other hand, 
as we are using the \textit{cubic} string field theory action, $N$ 
cannot become larger than $3M$. So the possible truncation levels are 
$(M,2M) \sim (M,3M)$. In the case of~(\ref{eq:AW}) which includes fields up 
to level 2, we can obtain the potentials $f^{(4)}(T) , f^{(6)}(T)$ as 
functions of the fields $\phi , \beta_1 , v$ by substituting~(\ref{eq:AW}) 
into (\ref{eq:AR}), where the superscripts (4),(6) indicates the truncation 
level. Though we do not repeat them here, the expressions for the potential 
both at level (2,4) and at (2,6) are shown in~\cite{SenZw}. 
The fact that $\partial^2c$ is not a conformal primary 
field complicates the actual calculation. In the next section we will 
explicitly reproduce the level (2,6) potential using another method. 
By extremizing 
$f(T)$ with respect to these field variables, we find the value of the 
potential at $T=T_0$. We show the results in Table~\ref{tab:B}.
\begin{table}[htbp]
	\begin{center}
	\begin{tabular}{|c|c|}
	\hline
	level & $f(T_0)$ \\
	\hline \hline
	(0,0) & $-0.684$ \\
	\hline 
	(2,4) & $-0.949$ \\
	(2,6) & $-0.959$ \\
	\hline
	(4,8) & $-0.986$ \\
	(4,12) & $-0.988$ \\
	\hline
	(6,12) & $-0.99514$ \\
	(6,18) & $-0.99518$ \\
	\hline
	(8,16) & $-0.99777$ \\
	(8,20) & $-0.99793$ \\
	\hline
	(10,20) & $-0.99912$ \\
	\hline
	\end{tabular}
	\end{center}
	\caption{The minimum values of the potential at various truncation 
	levels.}
	\label{tab:B}
\end{table}
In this table, the results obtained by truncating at other levels are also 
quoted. 
The explicit form of level (4,8) potential is given in~\cite{SenZw}, and 
the minimum values of the potential 
at various truncation levels up to (10,20) are found 
in~\cite{MoeTay}. And these values are illustrated in Figure~\ref{fig:J}.
These results suggest that the value of the potential $f(T)$ at the extremum
$T=T_0$ converges rapidly to the conjectured one $f(T_0)=-1$ as we increase 
the truncation level. It may be somewhat surprising that the 
level (2,4) approximation, in which only three fields are included, 
gives about 95\% 
of the expected value. At $T=0$, open string degrees of freedom are living on 
the original D$p$-brane; at the new vacuum $T=T_0$, we believe that 
the D$p$-brane completely disappears together with open strings. 
Since such a brane annihilation process can be thought of as highly 
non-perturbative, we could have expected that the higher level fields have 
great influence on determination of $T_0$. But the approximated results tells 
us that this expectation is false. In fact, a few low lying modes dominate 
the solution $T_0$. And it may be said that the string field theory 
appropriately describes non-perturbative features of string dynamics.
\smallskip

In the above sentences we used the word `extremum' rather than `minimum'. 
This is because the stationary point is actually not a minimum but a saddle 
point: String field $T(z)$ in~(\ref{eq:AW}) contains the component field 
$\beta_1$. As remarked before, 
$\beta_1$ kinetic term has the wrong sign, which 
causes an unstable direction in the tachyon potential. Thus the physical 
stability is not violated, as $\beta_1$ is one of auxiliary fields.

Comparison of the results displayed in Table~\ref{tab:B} reveals the fact 
that the precision of the values of the tachyon potential at the stationary 
point is not greatly improved even if we add higher level terms to the 
action while the level of the expansion of the string field is kept 
fixed (\textit{i.e.} $(M,N)\to (M,N+n)$). 
This can clearly be seen from Figure~\ref{fig:J}. Consequently, we consider 
that the most efficient approximations are 
obtained in the level $(M,2M)$ truncation schemes. 

By the way, in \cite{SenZw} the authors are saying that they used both the 
CFT correlator method and Neumann function method to compute the three string 
vertices in the level (4,8) approximation, and that the two methods gave the 
same results. This encourages us to be sure of equivalence of these two 
approaches.
\bigskip

Though we are mainly focusing on the level truncation scheme in this chapter, 
here we briefly refer to an approach to analytical construction of the 
nonperturbative vacuum~\cite{8252}. The string field configuration $|T_0
\rangle$ representing the nonperturbative vacuum was constructed as 
\begin{equation}
|T_0\rangle=F(\alpha)\tilde{F}(b,c)|\cS\rangle , 
\end{equation}
where $|\cS\rangle$ was given a closed form expression as a squeezed (or 
coherent) state, and the form of a series $F\tilde{F}$ of creation operators 
was determined recursively. The nonperturbative vacuum found in this way 
indeed seems to coincide with the one found in the level truncation scheme. 
\smallskip

In connection with the analytical expressions for the tachyon condensate, we 
also note down the special (anomalous) symmetries that the 3-string vertex 
$\langle V_3|$ has. It was pointed out in~\cite{Hata} that $\langle V_3|$ 
is invariant under a discrete $\zetto_4$ transformation
\begin{equation}
b_{-n}\longrightarrow -nc_{-n}\ , \quad c_{-n}\longrightarrow \frac{1}{n}
b_{-n}, \label{eq:RA}
\end{equation}
which can explicitly be seen from the expressions for the Neumann 
coefficients. After that, it was shown~\cite{Trim} that this $\zetto_4$ 
symmetry is actually a discrete subgroup of the continuous $U(1)$ 
symmetry
\begin{eqnarray}
b_{-n}(\theta)&=&b_{-n}\cos\theta-nc_{-n}\sin\theta\equiv e^{\theta\cS_1}
b_{-n}e^{-\theta\cS_1}, \nonumber \\
c_{-n}(\theta)&=&c_{-n}\cos\theta+\frac{1}{n}b_{-n}\sin\theta\equiv 
e^{\theta\cS_1}c_{-n}e^{-\theta\cS_1}, \label{eq:RB}
\end{eqnarray}
where $\cS_1$ is the generator of this $U(1)$ symmetry and is defined to be 
\begin{equation}
\cS_1=\sum_{n=1}^{\infty}\left(\frac{1}{n}b_{-n}b_n-nc_{-n}c_n\right). 
\label{eq:RC}
\end{equation}
Further, if we combine $\cS_1$ with the ghost number generator
\begin{equation}
\mathcal{G}=\sum_{n=1}^{\infty}(c_{-n}b_n-b_{-n}c_n)
\end{equation}
and another generator defined as a commutator of these two generators
\begin{equation}
\cS_2\equiv \frac{1}{2}[\cS_1,\mathcal{G}]=\sum_{n=1}^{\infty}\left(
\frac{1}{n}b_{-n}b_n+nc_{-n}c_n\right), 
\end{equation}
then the set $\{\cS_1,\cS_2,\mathcal{G}\}$ turns out to satisfy the algebra 
of $SU(1,1)$,
\[ [\cS_1,\mathcal{G}]=2\cS_2\ , \quad [\cS_2,\mathcal{G}]=2\cS_1 \quad 
\mathrm{and} \quad [\cS_1,\cS_2]=-2\mathcal{G}. \]
The 3-string vertex $\langle v_3|$ is fully invariant under this $SU(1,1)$ 
transformation, where $\langle v_3|$ differs from $\langle V_3|$ in that 
the contributions from the ghost zero modes $c_0,b_0$ have been removed. 
Since we can show that $SU(1,1)$ non-singlet states need not acquire 
expectation values, the string field can further be truncated to $SU(1,1)$ 
singlets in searching for the closed string vacuum. Besides, it was pointed 
out in~\cite{11238} that the 3-string vertex $\langle V_3|$ still obeys 
some more identities and that the field expectation values obtained so far 
in the level truncation scheme actually satisfies the 
identities approximately. 
%\begin{eqnarray*}
%\langle V_3|\sum_{i=1}^3\left(L_{-n}^{\mathrm{m}(i)}-
%L_n^{\mathrm{m}(i)}\right)&=&\frac{65}{18}n(-1)^{n/2}\langle V_3|, \\
%\langle V_3|\sum_{i=1}^3\left(J_{-n}^{(i)}+J_n^{(i)}\right)&=&\Bigl[ -3
%(-1)^{n/2}+9\delta_{n,0}\Bigr]\langle V_3|,
%\end{eqnarray*}
%where we take $n$ to be even and $J$ is the ghost number current. 

Though the above symmetries 
are applicable only to the specific vertex $\langle V_3|$, by exploiting 
these symmetries we can considerably reduce the number of independent 
coefficient fields appearing in the expansion of the string field. 
And if we could find an infinite number of such constraints (symmetries) 
which relate the expectation values of various fields so that only a finite 
number of fields could vary independently, then we would obtain an exact 
solution for the nonperturbative vacuum.

\section{Conservation Laws}\label{sec:conserve}
In \cite{RasZw}, an efficient computational scheme for evaluating 
the string field theory vertex was developed. We show here the procedures 
involved in this method and also 
its application to the calculation of level (2,6) tachyon potential. In this 
method we use Virasoro-Ward identities to convert negatively moded (creation) 
Virasoro generators into linear combinations of positively moded 
(annihilation) Virasoro generators, and use the commutation relations to move 
the annihilation operators to the right. By repeating these procedures any 
correlation function eventually reaches the 3-point function $\langle c,c,c
\rangle$ of the primary field $c(z)$, which is easily computed. 
\medskip

We adopt the doubling trick, where holomorphic and antiholomorphic fields in 
the upper half plane are combined into one holomorphic field defined in the 
whole 
complex plane. One of the important ingredients here is the 
conformal transformation~(\ref{eq:Ttransf}) of the energy-momentum tensor 
with central charge $c$,
\begin{eqnarray}
\left(\frac{\partial z'}{\partial z}\right)^2T'(z')&=&T(z)-\frac{c}{12}
\{z',z\}; \label{eq:AX} \\ 
\{z',z\}&=&\frac{2\partial_z^3z'\partial_zz'-3\partial_z^2z'
\partial_z^2z'}{2\partial_zz'\partial_zz'} \nonumber .
\end{eqnarray}
And we define a holomorphic vector field $v(z)$, that is, $v(z)$ 
transforms as a 
primary field of weight $-1$ under a conformal transformation $z\to 
z^{\prime}$. $v(z)$ is further required to be holomorphic in the whole 
$z$-plane except at three points where the vertex operators are inserted. In 
particular, $v(z)$ must be nonsingular at infinity. Under the transformation 
$z^{\prime}=-1/z$, $v(z)$ transforms as $v^{\prime}(z^{\prime})=z^{-2}v(z)$. 
Regularity of $v^{\prime}(z^{\prime})$ at $z^{\prime}=0$ is equivalent to 
nonsingularity of $z^{-2}v(z)$ at $z=\infty$, which demands that $v(z)$ 
should be at most quadratic in $z$.

Keeping the properties of $T(z),v(z)$ in mind, consider the following 
correlator
\begin{equation}
\left\langle\oint_Cv(z)T(z)dz\, f_1\circ\Phi(0)f_2\circ\Phi(0)f_3\circ\Phi(0)
\right\rangle , \label{eq:AY}
\end{equation}
where $C$ is a contour which encircles the insertion points $f_1(0),f_2(0),
f_3(0)$. Since there is no singularity outside the contour, one can smoothly 
deform the contour to shrink around the point at infinity. In this process we 
need to perform 
a conformal transformation $z^{\prime}=-1/z$. But fortunately, the 
Schwarzian in~(\ref{eq:AX}) happens to vanish for such a transformation. More 
generally, the same thing occurs for any $SL(2,\shii)$ transformation 
$\displaystyle z^{\prime}=\frac{az+b}{cz+d}$. This means that if we restrict 
the transformations to $SL(2,\shii)$ ones, we can tentatively regard $T(z)$ 
as a true tensor field. Then the integrand transforms as a weight 0 primary,
\begin{equation}
v(z)T(z)dz=v^{\prime}(z^{\prime})T^{\prime}(z^{\prime})dz^{\prime} \, ; \quad 
\mathrm{if} \ \  z\equiv z^{\prime} \ \ \mathrm{mod} \ \ SL(2,\shii). 
\label{eq:AZ}
\end{equation}
By deforming the contour, (\ref{eq:AY}) vanishes identically thanks 
to~(\ref{eq:AZ}) and regularity at infinity. This can be rewritten using the 
3-point vertex $\langle V_3|$ in~(\ref{eq:Vthree}) as 
\begin{equation}
\langle V_3|\oint_Cv(z)T(z)dz|\Phi\rangle_1\otimes |\Phi\rangle_2\otimes 
|\Phi\rangle_3=0. \label{eq:BA}
\end{equation}
In fact, it must be true that
\begin{equation}
\langle V_3|\oint_Cv(z)T(z)dz=0, \label{eq:BB}
\end{equation}
because vanishing of the left hand side of~(\ref{eq:BA}) does not depend on 
any properties of $\Phi$ at all. Next we deform the contour $C$ to three small
circles $C_i$ each of which encircles the insertion point $f_i(0)$. At the 
same time, we perform the conformal transformation from global coordinate $z$ 
to local coordinate $z_i$ associated with three string world-sheets. 
Since the form of the transformation~(\ref{eq:X}) with $n=3$ is 
far from $SL(2,\shii)$ transformation, we cannot drop the Schwarzian 
term. So the equation~(\ref{eq:BB}) becomes
\begin{equation}
\langle V_3|\sum_{i=1}^3\oint_{C_i}dz_iv^{(i)}(z_i)\left(T(z_i)-\frac{c}{12}
\{f_i(z_i),z_i\}\right)=0, \label{eq:BC}
\end{equation}
where we used (\ref{eq:AX}). Three Schwarzians labeled by $i=1,2,3$ give the 
same form because $f_1(z),f_2(z),f_3(z)$ are related among each other 
through an $SL(2,\aaru)$ transformation 
$\displaystyle S(z)=\frac{z-\sqrt{3}}{1+\sqrt{3}z}$ and its inverse. They have
the following expansion around the insertion points $z_i=0$
\begin{equation}
\{f_i(z_i),z_i\}=-\frac{10}{9}+\frac{20}{9}z_i^2-\frac{10}{3}z_i^4+
\frac{40}{9}z_i^6+\cdots , \> i=1,2,3. \label{eq:BD}
\end{equation}
If we take the vector field to be $\displaystyle v(z)=\frac{z^2-3}{z}$ so 
that $v(z)$ has zeros of order one at the insertion points $\pm\sqrt{3}$ 
of the 1st and 3rd string field vertex operators and a pole of 
order one at the insertion point $0$ 
of the 2nd vertex operator, then $v^{(i)}(z_i)$'s are given by
\begin{equation}
v^{(i)}(z_i)=\left(\frac{\partial z}{\partial z_i}\right)^{-1}v(z) \qquad 
\mathrm{with} \quad z=f_i(z_i) \, ; \label{eq:BE}
\end{equation}
\begin{eqnarray*}
v^{(1)}(z_1)&=&\left(\frac{8}{3}-\frac{32\sqrt{3}}{9}z_1+\frac{248}{27}z_1^2
-\frac{1664\sqrt{3}}{243}z_1^3\right)^{-1} \\ & &\times\left( 
-\frac{16\sqrt{3}}{3}z_1+\frac{160}{9}z_1^2-\frac{1264\sqrt{3}}{81}z_1^3
\right) \\ & &\times\left(-\sqrt{3}+\frac{8}{3}z_1-\frac{16\sqrt{3}}{9}z_1^2
+\frac{248}{81}z_1^3\right)^{-1}+\cO(z_1^4) \\
&=&2z_1+\frac{20}{3\sqrt{3}}z_1^2+\frac{28}{9}z_1^3+\cO(z_1^4), \\
v^{(2)}(z_2)&=&-\frac{27}{4z_2}-4z_2+\frac{19}{36}z_2^3+\cO(z_2^5), \\
v^{(3)}(z_3)&=&2z_3-\frac{20}{3\sqrt{3}}z_3^2+\frac{28}{9}z_3^3+\cO(z_3^4).
\end{eqnarray*}
Substituting these expressions into~(\ref{eq:BC}) and using $\displaystyle 
L_n^{(i)}=\oint_{C_i}\frac{dz_i}{2\pi i}z_i^{n+1}T(z_i)$, we obtain the 
following conservation law
\begin{eqnarray}
\langle V_3|L_{-2}^{(2)}&=&\langle V_3|\left(\frac{8}{27}L_0+
\frac{80}{81\sqrt{3}}L_1+\frac{112}{243}L_2+\cdots\right)^{(1)} \nonumber \\
&+&\langle V_3|\left(-\frac{5}{54}c-\frac{16}{27}L_0+\frac{19}{243}L_2+\cdots
\right)^{(2)} \label{eq:BF} \\ &+&\langle V_3|\left(\frac{8}{27}L_0-
\frac{80}{81\sqrt{3}}L_1+\frac{112}{243}L_2+\cdots\right)^{(3)}. \nonumber
\end{eqnarray}
In calculating level (2,6) potential, the second term in~(\ref{eq:AV}) gives 
rise to, for example, the following coupling 
\begin{equation}
\langle V_3|c_1^{(1)}|0\rangle_1\otimes L_{-2}^{\mathrm{m}(2)}c_1^{(2)}|0
\rangle_2\otimes c_1^{(3)}|0\rangle_3. \label{eq:BG}
\end{equation}
Since $L_{-2}^{\mathrm{m}(2)}$ directly touches $\langle V_3|$, we can 
use~(\ref{eq:BF}) to exclude $L_{-2}^{\mathrm{m}(2)}$ though infinitely many 
zero or positive modes enter. In this case, however, there is no other 
negatively moded operator in~(\ref{eq:BG}), so only zero modes can 
contribute the nonvanishing values. Then~(\ref{eq:BG}) becomes 
\begin{eqnarray}
& &\langle V_3|\left(\frac{8}{27}(L_0^{\mathrm{m}(1)}+L_0^{\mathrm{m}(3)})
-\frac{5}{54}c^{\mathrm{m}}-\frac{16}{27}L_0^{\mathrm{m}(2)}\right)c_1^{(1)}
c_1^{(2)}c_1^{(3)}|0\rangle_3\otimes |0\rangle_2 \otimes |0\rangle_1 \nonumber
\\ &=&-\frac{5}{54}\cdot 26\cK=-\frac{65}{27}\cK , \label{eq:BH}
\end{eqnarray}
where we set 
\[\cK\equiv\langle V_3|c_1^{(1)}c_1^{(2)}c_1^{(3)}|0\rangle_3\otimes |0
\rangle_2\otimes |0\rangle_1=\left(\frac{3\sqrt{3}}{4}\right)^3 \]
as calculated in (\ref{eq:AJ}), and used the fact that central charge of the 
matter Virasoro algebra in critical 
bosonic string theory is $c^{\mathrm{m}}=26$. Due to 
the cyclic symmetry, we can do calculations in the same way when $L_{-2}$ 
appears in the 1st or 3rd Hilbert space. And by replacing $v(z)=(z^2-3)/z$ 
with $(z^2-3)z^{-k+1}$, we obtain similar conservation laws for 
$L_{-k}^{\mathrm{m}(2)}$. If we have 
the full collection of such conservation laws at hand, 
any excitation in the matter part could be removed to arrive at $\cK$. 
But we must work out also the ghost conservation laws so as to pull down the 
ghost excitations\footnote{As explained in \cite{RasZw}, in the pure ghost 
CFT the Fock space consisting of states of \textit{ghost number 1} obtained 
by acting with $b,c$ oscillators on the tachyon state $c_1|0\rangle$ agrees 
with the Verma module built on the primary $c_1|0\rangle$, that 
is, the space of states obtained as Virasoro descendants of $c_1|0\rangle$. 
Though we can dispense with the ghost conservation laws for this reason, 
we intentionally derive and use them because at the ghost number 0 sector 
there are states which cannot be written as Virasoro descendants of 
$|0\rangle$, and simply because we want to use the expressions like 
$c_{-1}|0\rangle$ rather than $L_{-2}^{\mathrm{g}}c_1|0\rangle$.}. 
\smallskip

In finding the Virasoro conservation law~(\ref{eq:BF}), the properties we 
used are 
\begin{enumerate}
	\item conformal transformation~(\ref{eq:AX}) of the energy-momentum 
	 tensor, and 
	\item the fact that the energy-momentum tensor has conformal weight 2.
\end{enumerate}
Since other information such as OPEs with various fields is irrelevant, the 
conservation laws for $b$ ghost, whose weight is 2 as well, should be 
obtained by setting the central charge $c$ to zero and $L_n^{(i)}\to 
b_n^{(i)}$ in the Virasoro conservation laws. So much for $b$ ghost.
\medskip

$c$ ghost is a primary field of weight $-1$. In this case, the previous 
`vector field $v(z)$' has to be replaced with a holomorphic field $\phi(z)$ 
of weight 2. As before, the regularity of $\phi(z)$ at infinity requires 
$\phi(z)\sim\cO(z^{-4})$ for $z\to\infty$. By inserting $\displaystyle \oint_C
\phi(z)c(z)dz$ into the 3-point vertex just as in~(\ref{eq:AY}), we find
\begin{equation}
\langle V_3|\oint_C\phi(z)c(z)dz =0. \label{eq:BI}
\end{equation}
Since both $\phi(z)$ and $c(z)$ are primary fields, we can carry out the 
conformal transformations to local coordinates without any additive term 
such as Schwarzian,
\begin{equation}
\langle V_3|\sum_{i=1}^3\oint_{C_i}\phi^{(i)}(z_i)c(z_i)dz_i=0. \label{eq:BJ}
\end{equation}
By the choice $\phi(z)=z^{-3}(z^2-3)^{-1}$, it turns out that the excitation 
$c_{-1}^{(2)}|0\rangle_2$ in the 2nd Hilbert space can be converted to 
$c_1^{(2)}|0\rangle_2$, resulting in $\cK$. To see this, expand 
$\phi^{(i)}(z_i)$ around $z_i=0$ as
\begin{eqnarray}
\phi^{(1)}(z_1)&=&\frac{4}{27}\frac{1}{z_1}+\cO(1) , \nonumber \\
\phi^{(2)}(z_2)&=&-\frac{1}{2z_2^3}+\frac{11}{54}\frac{1}{z_2}+\cO(1) , 
\label{eq:BK} \\
\phi^{(3)}(z_3)&=&\frac{4}{27}\frac{1}{z_3}+\cO(1). \nonumber
\end{eqnarray}
By expanding $c(z)=\sum_{n=-\infty}^{\infty}c_n/z^{n-1}$ and picking up 
simple poles, we get
\begin{equation}
\langle V_3|\left(\frac{8}{27}c_1^{(1)}-c_{-1}^{(2)}+\frac{11}{27}c_1^{(2)}
+\frac{8}{27}c_1^{(3)}+\cdots\right)=0. \label{eq:BL}
\end{equation}
\medskip

We are now ready to compute the level (2,6) truncated tachyon potential. 
First, we treat the quadratic term. Level 2 truncated string field in the 
Feynman-Siegel gauge is
\begin{equation}
|T\rangle =\left(\phi c_1-\beta_1c_{-1}+\frac{v}{\sqrt{13}}
L_{-2}^{\mathrm{m}}c_1\right)|0\rangle. \label{eq:BM}
\end{equation}
Corresponding bra state is found by BPZ conjugation:
\begin{equation}
\langle T|=\mathrm{bpz}(|T\rangle)=\langle 0|\left(\phi c_{-1}-\beta_1c_1
+\frac{v}{\sqrt{13}}L_2^{\mathrm{m}}c_{-1}\right). \label{eq:BN}
\end{equation}
And relevant part of the BRST operator is 
\begin{equation}
Q_B=c_0(L_0^{\mathrm{m}}-1)+c_0c_{-1}b_1+(c_0b_{-1}+2c_{-1}b_0)c_1. 
\label{eq:BO}
\end{equation}
Quadratic terms become
\begin{eqnarray}
\langle T|Q_B|T\rangle &=&(-\phi^2-\beta_1^2)\langle 0|c_{-1}c_0c_1|0\rangle
+\frac{v^2}{13}\langle 0|c_{-1}c_0c_1[L_2^{\mathrm{m}},L_{-2}^{\mathrm{m}}]|0
\rangle \nonumber \\ &=&-\phi^2-\beta_1^2+v^2, \label{eq:BP}
\end{eqnarray}
where we used 
\[ \{b_n,c_m\}=\delta_{n+m,0}, \]
\begin{equation}
[L_0^{\mathrm{m}},L_{-2}^{\mathrm{m}}]=2L_{-2}^{\mathrm{m}}, 
\label{eq:BPa} 
\end{equation}
\[[L_2^{\mathrm{m}},L_{-2}^{\mathrm{m}}]=4L_0^{\mathrm{m}}+
\frac{c^{\mathrm{m}}}{2}=4L_0^{\mathrm{m}}+13. \]

Then we turn to the cubic term. In the notation of operator formalism, 
cubic vertex is written as 
\[\cV_3\equiv \langle V_3|T\rangle_1\otimes |T\rangle_2 \otimes |T\rangle_3.\]
By substituting~(\ref{eq:BM}) and using the cyclicity, we find 
\begin{eqnarray}
\cV_3&=&\langle V_3|\Biggl(\phi^3c_1^{(1)}c_1^{(2)}c_1^{(3)}-3\beta_1\phi^2
c_1^{(1)}c_{-1}^{(2)}c_1^{(3)}+\frac{3}{\sqrt{13}}v\phi^2
L_{-2}^{\mathrm{m}(2)}c_1^{(1)}c_1^{(2)}c_1^{(3)} 
\nonumber \\ & &+3\beta_1^2\phi
c_1^{(1)}c_{-1}^{(2)}c_{-1}^{(3)}+\frac{3}{13}v^2\phi L_{-2}^{\mathrm{m}(2)}
L_{-2}^{\mathrm{m}(3)}c_1^{(1)}c_1^{(2)}c_1^{(3)}-\frac{3}{\sqrt{13}}\phi
\beta_1vL_{-2}^{\mathrm{m}(3)}c_1^{(1)}c_{-1}^{(2)}c_1^{(3)} 
\nonumber \\ & &{}-
\frac{3}{\sqrt{13}}\phi\beta_1vL_{-2}^{\mathrm{m}(1)}c_1^{(1)}c_{-1}^{(2)}
c_1^{(3)}-\beta_1^3c_{-1}^{(1)}c_{-1}^{(2)}c_{-1}^{(3)}+
\frac{v^3}{13\sqrt{13}}L_{-2}^{\mathrm{m}(1)}L_{-2}^{\mathrm{m}(2)}
L_{-2}^{\mathrm{m}(3)}c_1^{(1)}c_1^{(2)}c_1^{(3)} 
\nonumber \\ & &+\frac{3}{\sqrt{13}}
\beta_1^2vL_{-2}^{\mathrm{m}(2)}c_{-1}^{(1)}c_1^{(2)}c_{-1}^{(3)}
-\frac{3}{13}\beta_1v^2L_{-2}^{\mathrm{m}(1)}L_{-2}^{\mathrm{m}(3)}
c_1^{(1)}c_{-1}^{(2)}c_1^{(3)}\Biggr)|0\rangle_3\otimes |0\rangle_2\otimes
|0\rangle_1 \nonumber \\
&\equiv& \sum_{I=1}^{11}\cV_3^I. \label{eq:BQ}
\end{eqnarray}
To give an example, we show the detailed calculation of the fifth term.
\begin{eqnarray}
\cV_3^5&=&\frac{3}{13}v^2\phi\langle V_3|L_{-2}^{\mathrm{m}(2)}
L_{-2}^{\mathrm{m}(3)}c_1^{(1)}c_1^{(2)}c_1^{(3)}|0\rangle_3\otimes |0
\rangle_2\otimes |0\rangle_1 \nonumber \\
&=&\frac{3}{13}v^2\phi\langle V_3|\left(\frac{8}{27}L_0^{\mathrm{m}(3)}+
\frac{112}{243}L_2^{\mathrm{m}(3)}-\frac{5}{54}c^{\mathrm{m}}\right)
L_{-2}^{\mathrm{m}(3)}c_1^{(1)}c_1^{(2)}c_1^{(3)}|0\rangle_3\otimes |0
\rangle_2\otimes |0\rangle_1 \nonumber \\
&=&\frac{3}{13}v^2\phi\langle V_3|\left(\frac{8}{27}\cdot 2
L_{-2}^{\mathrm{m}(3)}+\frac{112}{243}\left(4L_0^{\mathrm{m}(3)}+
\frac{c^{\mathrm{m}}}{2}\right)-\frac{65}{27}L_{-2}^{\mathrm{m}(3)}\right)
c_1^{(1)}c_1^{(2)}c_1^{(3)}|0\rangle_3\otimes |0\rangle_2\otimes |0\rangle_1
\nonumber \\
&=&\frac{3}{13}v^2\phi\langle V_3|\left(-\frac{49}{27}\left(-\frac{5}{54}26
\right)+\frac{112}{243}\times 13\right)c_1^{(1)}c_1^{(2)}c_1^{(3)}
|0\rangle_3\otimes |0\rangle_2\otimes |0\rangle_1 \nonumber \\
&=&\frac{3}{13}\frac{7553}{729}\cK v^2\phi =\frac{581}{243}\cK v^2\phi .
\label{eq:BR}
\end{eqnarray}
In the second line, we used the conservation law~(\ref{eq:BF}) for 
$L_{-2}^{\mathrm{m}(2)}$; in the third line, the commutation 
relations~(\ref{eq:BPa}) were used; in the fourth line, we again used the 
conservation law for $L_{-2}^{\mathrm{m}(3)}$. Other ten terms can be 
evaluated in similar fashions, the result being
\begin{eqnarray*}
\cV_3&=&\Biggl(\phi^3-\frac{11}{9}\phi^2\beta_1-\frac{65}{9}
\frac{v}{\sqrt{13}}\phi^2+\frac{19}{81}\beta_1^2\phi \\ & &+\frac{581}{243}
v^2\phi+\frac{1430}{243}\frac{v}{\sqrt{13}}\beta_1\phi-\frac{1}{81}\beta_1^3
-\frac{1235}{2187}\frac{v}{\sqrt{13}}\beta_1^2 \\ & &-\frac{6391}{6561}\beta_1
v^2-\frac{62853}{19683}\frac{v^3}{\sqrt{13}}\Biggr)\cK .
\end{eqnarray*}
We can now write down 
the final expression\footnote{This result precisely agrees 
with those given in \cite{SenZw} and \cite{RasZw} if the fields in refs. are 
appropriately rescaled.} for the level (2,6) truncated tachyon potential
\begin{eqnarray}
f(T)&=&2\pi^2\ap{}^3\left(\frac{1}{2\ap}\langle T|Q_B|T\rangle+\frac{1}{3}
\cV_3\right) \label{eq:BS} \\ &=& 2\pi^2\ap{}^3\Biggl(-\frac{1}{2\ap}\phi^2
+\frac{3^3\sqrt{3}}{2^6}\phi^3-\frac{1}{2\ap}\beta_1^2+\frac{1}{2\ap}v^2
-\frac{11\cdot 3\sqrt{3}}{2^6}\phi^2\beta_1 \nonumber \\
& &{}-\frac{5\cdot 3\sqrt{39}}{2^6}\phi^2v+\frac{19\sqrt{3}}{3\cdot 2^6}
\phi\beta_1^2+\frac{581\sqrt{3}}{3^2\cdot 2^6}\phi v^2+\frac{5\cdot 11
\sqrt{39}}{3^2\cdot 2^5}\phi\beta_1v \nonumber \\ & &{}-\frac{1}{2^6\sqrt{3}}
\beta_1^3-\frac{5\cdot 19\sqrt{39}}{2^6\cdot 3^4}v\beta_1^2-\frac{6391\sqrt{3}}
{2^6\cdot 3^5}v^2\beta_1-\frac{20951\sqrt{39}}{2^6\cdot 3^5\cdot 13}v^3 
\Biggr). \nonumber
\end{eqnarray}
The simultaneous equations of motion derived from the potential $f(T)$ are 
solved by
\[T_0=(\phi\simeq 0.544/\ap , \> \beta_1\simeq -0.190/\ap , \> v\simeq 
0.202/\ap ),\]
and $f(T_0)\simeq -0.959$, as advertised earlier.

We wrote down only the conservation laws~(\ref{eq:BF}),(\ref{eq:BL}) which 
are needed to reproduce the level (2,6) potential~(\ref{eq:BS}). 
In~\cite{RasZw}, Virasoro and ghost conservation laws including operators 
of still higher modes as well as the current conservation laws are shown. 
These conservation laws, together with the commutation relations, enables us 
to avoid finding the complicated finite conformal transformations of 
non-primary fields or resorting to the explicit form~(\ref{eq:Vthree}) of 
the vertex $\langle V_3|$.

\section{Physics of the New Vacuum}\label{sec:phys}
We begin by tidying up the terminology. The usual open string vacuum with one 
or some D$p$-branes, where all fields except for those describing the 
collective motion of the D-branes have vanishing expectation values, is 
called `\textit{perturbative vacuum}'. In contrast, the new 
vacuum found in the previous sections, where various fields develop nonzero 
expectation values, is termed `\textit{nonperturbative vacuum}' for reasons 
discussed below, or `\textit{closed string vacuum}' because we believe that 
in this new vacuum the negative energy 
contribution from the tachyon potential
associated with the rolling down of the tachyon field exactly cancels the 
positive energy density (tension) of the D-brane, resulting in a true vacuum 
without any D-brane or open string. Here a problem can arise: On a single 
bosonic D-brane, the tachyon field is in the adjoint representation and hence 
is neutral under the $U(1)$ gauge group. 
Since the gauge field would remain massless even if the tachyon and other 
neutral scalar fields acquire nonzero vacuum expectation valules, we 
na\"{\i}vely expect that the gauge field would continue to exist in the 
spectrum even after the tachyon condenses. However, 
this situation can never be realized as long as the D-brane annihilation 
accompanies the tachyon condensation because the open string degrees of 
freedom, which include $U(1)$ gauge field, are fully removed together with 
the D-brane. If so, through what kind of mechanism does the $U(1)$ gauge 
field disappear? This is so-called `$U(1)$ problem' and discussed under the 
name of `Fate of $U(1)$ gauge 
field'~\cite{DK,IIB,NonBPSD,Univ,Yi,Conf,11009,12081}. 
Before meddling in this problem, let us investigate the 
properties of the `closed string vacuum' more closely.

\subsection{Effective tachyon potential}\label{sub:effpot}
\begin{figure}[htbp]
\begin{center}
	\includegraphics{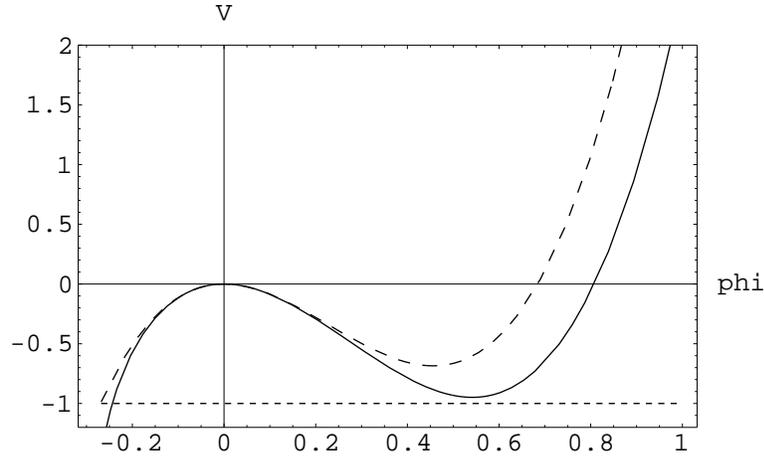}
	\caption{Effective tachyon potential at level (0,0) (dashed line) and at 
	level (2,4) (solid line).}
	\label{fig:K}
\end{center}
\end{figure}

\begin{figure}[htbp]
\begin{center}
	\includegraphics{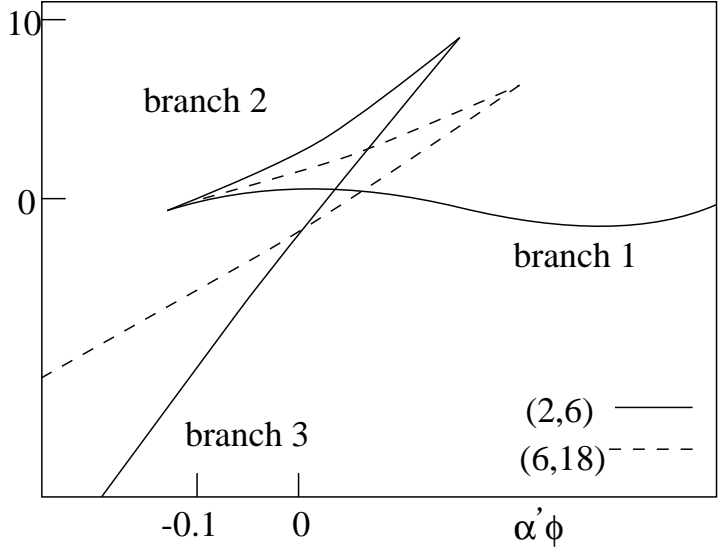}
	\caption{Structure of the effective tachyon potential.}
	\label{fig:L}
\end{center}
\end{figure}

First, we examine the structure of the effective tachyon potential. At level 
(0,0), the tachyon potential $f(\phi)$ has already been given 
in~(\ref{eq:AT}),
\begin{equation}
f(\phi)=2\pi^2\ap{}^3\left(-\frac{1}{2\ap}\phi^2+2\kappa\phi^3\right),
\label{eq:BT}
\end{equation}
where $\displaystyle \kappa\equiv\frac{1}{3!}\left(
\frac{3\sqrt{3}}{4}\right)^3\simeq 0.365$\footnote{A factor 2 is inserted 
in front of $\kappa$ in~(\ref{eq:BT}) to reconcile the definition of $\kappa$ 
here with that of \cite{KS} and \cite{MoeTay}.}. It has a minimum at 
\[ \phi_0=\left(\frac{4}{3\sqrt{3}}\right)^3\frac{1}{\ap}\simeq\frac{0.456}
{\ap}. \]
If we rescale the string field $\Phi$ back to $g_o\Phi$, in which case a 
factor of $g_o$ appears in the coefficient of every cubic interaction term 
like~(\ref{eq:O}), then the minimum occurs at $\phi_0\simeq 0.456/\ap g_o$. 
This expression suggests the nonperturbative nature of the `closed string 
vacuum'. Next, at level (2,4), the multiscalar potential can be obtained by 
dropping the last four terms in~(\ref{eq:BS}). Since the potential is 
quadratic both in $\beta_1$ and in $v$, we can eliminate these two fields 
by integrating them out exactly. The resulting effective 
tachyon potential is, in the $\ap=1$ unit, given by 
\begin{eqnarray*}
f_{\mathrm{eff}}(\phi)&=&{\frac{6\pi^2\phi^2}{256\,
     {{\left( 288 + 581\,{\sqrt{3}}\,\phi \right) }^2}\,{{\left( 432 + 786\,
{\sqrt{3}}\,\phi + 97\,{\phi^2} \right) }^2}}} \\ \nonumber & &\times 
\Bigl( -660451885056 - 4510794645504\,{\sqrt{3}}\,\phi \\ 
\nonumber & &{}- 32068942626816\,{\phi^2} - 25455338339328\,{\sqrt{3}}\,
{\phi^3} + 
27487773823968\,{\phi^4} \\ \nonumber & &
{}+54206857131636\,{\sqrt{3}}\,{\phi^5} +
24845285906980\,{\phi^6} + 764722504035\,{\sqrt{3}}\,{\phi^7} \Bigr) . 
\label{eq:BU}
\end{eqnarray*}
Its minimum occurs at $\phi_0\simeq 0.541/\ap$, which has 
increased by about 20\% 
compared to the (0,0) case. The effective potential obtained at level (0,0) 
and (2,4) is indicated in Figure~\ref{fig:K}. 
The depth of the potential is shown in Table~\ref{tab:B} and it is 
approaching the conjectured value $f_{\mathrm{eff}}(\phi_0)=-1$. The vacuum 
expectation values of $\beta_1$ and $v$ could also be obtained by 
substituting the value of $\phi_0$ to the equations of motion derived from 
the level 4 truncated multiscalar potential and solving for $\beta_1$ and 
$v$, though we do not try it. Beyond level (2,4), fields other than the 
tachyon are no longer
quadratic, so we cannot analytically proceed. Instead, we take the following 
numerical approach. In the case of level (2,6), the fields involved are $\phi,
\beta_1,v$, and the potential~(\ref{eq:BS}) 
includes cubic interactions among them. Firstly,
derive the equation of motion with respect to $\beta_1$. It becomes a 
quadratic equation for $\beta_1$, concretely given by
\begin{eqnarray}
\beta_1^2&+&\left(\frac{2^6}{\sqrt{3}\ap}-\frac{38}{3}\phi+\frac{190\sqrt{13}}
{3^4}v\right)\beta_1 \nonumber \\
&+& \left(33\phi^2-\frac{110\sqrt{13}}{9}\phi v+\frac{6391}{3^5}v^2\right)=0.
\label{eq:BV}
\end{eqnarray}
Solving (\ref{eq:BV}) for $\beta_1$ and substituting it into the equation of 
motion with respect to $v$, we get an equation containing $v$ and $\phi$. 
This equation can numerically be solved for $v$ if $\phi$ is given a 
numerical value $\phi_*$. We denote it by $v_*$. Then $\beta_{1*}$ is also 
given by solving the quadratic equation~(\ref{eq:BV}) for $\phi=\phi_* , 
v=v_*$. By plugging these values into the multiscalar potential 
$f(T)=f(\phi,\beta_1,v)$, we obtain 
one particular value $f_{\mathrm{eff}}(\phi_*)\equiv f(\phi_*,\beta_{1*},
v_*)$ of the effective tachyon potential at $\phi=\phi_*$. 
When we have repeated these 
procedures for various values of $\phi_*$, we can grasp the entire picture 
of the effective tachyon potential $f_{\mathrm{eff}}(\phi)$\footnote{The 
resulting (2,6) effective potential has changed from the (2,4) potential 
only just a little.}. The whole process given above is extended to the higher 
level case. Suppose that the $(M,N)$ truncated multiscalar potential includes 
$n$ kinds of scalar fields. Then we can write down $(n-1)$ simultaneous 
quadratic equations of motion by differentiating the multiscalar potential 
with respect to each of $(n-1)$ fields other than the 
tachyon. For a given value of $\phi=\phi_*$, $(n-1)$ simultaneous equations 
of motion can be solved to give $\psi^i=\psi^i_*$ ($2\le i\le n$, with 
$\psi^i$ 
representing the scalar fields involved in the truncated system). The value 
of the effective potential at $\phi=\phi_*$ is defined to be 
\[ f_{\mathrm{eff}}(\phi_*)=f(\phi_*,\psi^i_*). \]
But we must be careful in solving equations of motion. Since they are 
quadratic equations, each equation gives two solutions due to the branch of 
the square root. Generically, there are $2^{n-1}$ possible solutions in case 
of $n$ scalar fields, but we cannot regard all of them as appropriate ones. 
The most reliable choice is that $\phi_*=0$ gives $\psi_*^i=0$ for all $i$, 
\textit{i.e.} we should choose the branch which contains the perturbative 
vacuum in its 
profile. But it is interesting to see the branch structure in detail at level 
(2,6). As (2,6) potential includes $\beta_1 , v$ in addition to $\phi$, four 
branches can arise. However, the fourth branch behaves too strangely to be 
trustworthy, so we exclude it. The remaining three branches are drawn roughly 
in Figure~\ref{fig:L}\footnote{Since this is really a rough sketch, 
see~\cite{KS,MoeTay} for precise shapes.}. 
The branch named `branch 1' 
corresponds to the one indicated in Figure~\ref{fig:K} and actually contains 
both the perturbative vacuum and the nonperturbative vacuum. The reason why 
these branches terminate at some points is clear if we recall the origin of 
the branch structure: It originates from the square root branches of the 
quadratic equations. The number of real solutions to a quadratic equation 
$ax^2+bx+c=0$ depends on the sign of its discriminant $b^2-4ac$. In the 
region where the effective potential seems to be a two-valued function of 
$\phi$, discriminant is positive and two real solutions exist. At the point 
where two branches meet, discriminant vanishes and two solutions coincide. 
Even if we try to extend the branch beyond that point, no real solutions 
exist and we fail to do so. The existence of a critical point even on the 
branch 1 indicates that the power series expansion 
\begin{equation}
f_{\mathrm{eff}}(\phi)=2\pi^2\ap{}^3\sum_{n=2}^{\infty}c_n\kappa^{n-2}\phi^n
=2\pi^2\ap{}^3\left(-\frac{1}{2\ap}\phi^2+2\kappa\phi^3+\cdots \right) 
\label{eq:BW}
\end{equation}
of the effective potential around the perturbative vacuum $\phi=0$
has a finite radius of convergence. At level (2,6) the critical point is 
reported to appear at $\phi\simeq -0.17/\ap$~\cite{KS}\footnote{Notice 
that we set $g=2$ in~\cite{KS} for comparison with our results.}. 
In~\cite{MoeTay}, the authors pursued the movement of the critical point when 
higher level fields are taken into account, up to level (10,20). They found 
that the radius of convergence (the absolute value of $\phi$ at the critical 
point) is monotonically decreasing as the truncation level is increased, and 
is likely to converge to a finite value
\begin{equation}
r_c=-\phi_{\mathrm{crit}}\simeq\frac{0.125}{\ap} \label{eq:BX}
\end{equation}
in the limiting theory (\textit{i.e.} full string field theory with no level 
truncation approximation). Accordingly, the perturbative 
expansion~(\ref{eq:BW}) of the 
effective potential is valid only within the radius of 
convergence~(\ref{eq:BX}). We then conclude that the closed string vacuum is 
quite nonperturbative because the expectation value of the tachyon at that 
vacuum is $\phi_0\simeq 0.54/\ap$\footnote{According to \cite{MoeTay}, 
the vacuum expectation value $\phi_0\simeq 0.55/\ap$ persists at least up to 
level (10,20).} at level (2,6), which is well outside the radius of 
convergence. Hence the closed string vacuum is also called `nonperturbative 
vacuum', although it is located at the opposite side of the singularity 
($\phi_{\mathrm{crit}}<0$) as to the origin.
\smallskip

Beyond level (2,6), though many branches begin to appear, we focus 
only on three of them which correspond to the ones we encountered at level 
(2,6). In Figure~\ref{fig:L}, the branches found at level (6,18) truncation 
is illustrated as well. The branch 1 has scarecely changed its shape, which 
encourages us to expect that the branch 1 has converged to the true profile 
quite rapidly. In contrast, the branch 2 and 3 shifted substantially, though 
the qualitative behavior remains the same. In~\cite{MoeTay}  the 
branch structures at level (2,6), (4,12), (6,18) are shown, which 
suggest that the branch 2,3 are not sufficiently described by low-lying 
fields only. In any case, we are not very interested in the details of 
the branch 2,3 in searching for vacua because we know that the physically 
interpretable two vacua (perturbative vacuum with a D-brane and closed 
string vacuum) are located on the branch 1. 

Returning to the problem of the critical point on the branch 1, it is 
significant to understand the physical meaning of the critical point and 
the physics beyond that point. The singularity at $\phi_{\mathrm{crit}}$ 
may be related to the problem of the instability toward $\phi\to -\infty$ 
(it is clear from the level 0 truncated potential~(\ref{eq:BT})). If the 
range of definition of the effective tachyon potential were indeed restricted 
to some finite region, the nonperturbative vacuum could become 
\textit{the global} minimum without any instability. 

%The authors conjectured that the branch 3 eventually intersects 
%the critical point, inducing a phase transition in the effective potential. 
%terms of bosonic D-branes, if possible. 

\subsection{Small fluctuations around the nonperturbative vacuum}
\label{sub:fluc}
After we saw the static structure of the effective tachyon potential, we now 
examine the fluctuations around the nonperturbative vacuum. Then we determine 
the physical excitations there.
\medskip

If the conjecture that the D-brane disappears after tachyon condensation is 
true, there should be no physical excitations of open string degrees 
of freedom at the nonperturbative vacuum. 
In fact, such a vacuum was constructed in~\cite{Hor}, though it does not 
directly correspond to the `closed string vacuum' considered here. 
In the purely cubic 
formulation of Witten's bosonic string field theory, the way of constructing 
a class of exact solutions to equation of motion was invented in~\cite{Hor}.
By taking one particular 
solution\footnote{We do not explain the meaning of the symbols; 
see~\cite{Hor} for detail. But we note that $Q$ is an operator which has some 
common properties with the BRST charge.} $\Phi_0=Q_LI$ and expanding the 
string field $\Phi$ about this solution as $\Phi=\Phi_0+\cA$, we get the 
linearized equation of motion $Q\cA =0$. If $Q$ is appropriately chosen, 
$Q\cA =0$ implies $\cA =0$. That is to say, there are no physical 
excitations (defined to be the ones satisfying $Q\cA =0$) 
at all around the solution $\Phi_0=Q_LI$. Though this solution is 
different from the closed string vacuum, we could anyway obtain an example of 
a vacuum around which there are no physical excitations. 
\bigskip

Then let's investigate the fluctuations about the closed string vacuum, which 
is of most interest to us. First we deal with level (1,2) truncated 
action~(\ref{eq:AM}) at \textit{zero} momentum,
\begin{eqnarray}
S_{(1,2)}&=&-\frac{V_d}{g_o^2}U(\phi,A_{\mu}) ; \nonumber \\
U(\phi,A_{\mu})&=&-\frac{1}{2\ap}\phi^2+2\kappa\phi^3+\frac{3\sqrt{3}}{4}\phi
A_{\mu}A^{\mu}, \label{eq:BY}
\end{eqnarray}
where $\kappa$ is defined to be $\displaystyle \frac{1}{3!}\left(
\frac{3\sqrt{3}}{4}\right)^3$ as earlier. In this level of approximation, 
$\phi$ has vacuum expectation value $\phi_0=1/6\kappa\ap\simeq 0.456/\ap$ 
at the nonperturbative vacuum. By shifting $\phi=\phi_0+\phi^{\prime}$, the 
potential $U$ becomes 
\begin{equation}
U(\phi^{\prime},A_{\mu})=-\frac{2^{11}}{3^{10}\ap{}^3}+\frac{1}{2\ap}
\phi^{\prime 2}+2\kappa\phi^{\prime 3}+\frac{16}{27\ap}A_{\mu}A^{\mu}
+\frac{3\sqrt{3}}{4}\phi^{\prime}A_{\mu}A^{\mu}. \label{eq:BZ}
\end{equation}
The first term is a constant which determines the depth of the potential. 
The second, third and fifth terms inherit those in~(\ref{eq:BY}), but 
notice that the sign of $\phi^{\prime 2}$ term has been reversed: The tachyon
field on the original D-brane (perturbative vacuum) is no longer tachyonic 
around the nonperturbative vacuum. 
This partly shows the stability of the 
nonperturbative vacuum. And the fourth term looks like a mass term for the 
vector field $A_{\mu}$: During the process of tachyon condensation, 
mass of the (originally massless) vector field has spontaneously been 
generated! The value of mass${}^2$ is 
\[\frac{m^2}{2}=\frac{16}{27\ap}\simeq \frac{0.59}{\ap}. \]
Analyses focusing on the mass of the vector field are done using level 
truncation method in~\cite{8033}. In addition to the even level scalars 
considered so far, we should incorporate odd level vectors since they can 
mix with $A_{\mu}$ (level 1). 
For the multiple vectors, we can take two ways of analyzing mass of 
$A_{\mu}$. One is to examine the smallest eigenvalue of mass matrix of the 
form $\sum_{m,n}M_{mn}\eta_{\mu}^{(m)}\eta^{(n)\mu}$, where 
$\eta^{(m)}_{\mu}$ is the $m$-th vector field, in particular 
$\eta_{\mu}^{(1)}=A_{\mu}$. The other is to derive the effective action for 
the tachyon $\phi$ and the `massless' vector $A_{\mu}$ of the form
\begin{equation}
S_{\phi,A}=-\frac{V_p}{g_o^2}\left(U(\phi)+G(\phi)A_{\mu}A^{\mu}+\cO(A^4)
\right). \label{eq:CA}
\end{equation}
Then $G(\phi_0)$ ($\phi_0$ satisfies $U^{\prime}(\phi_0)=0$) is regarded as 
half of the mass squared of the vector field $A_{\mu}$. The results of both 
approaches are given in~\cite{8033} up to level (9,10,20) truncation, which 
means that we keep vectors up to and including level 9, scalars level 10 and 
interaction terms in the action up to and including level 20. These results 
suggest that the smallest eigenvalue of mass matrix seems to converge to a 
value in the vicinity of 0.59, while the value of $G(\phi_0)$ appears to 
approach a value near 0.62. Both results are consistently saying that the 
originally massless vector field $A_{\mu}$ has acquired nonzero 
mass after tachyon 
has condensed, whose value is approximately $m^2/2\simeq 0.6$. 
\medskip

Though the above results seem to mean that the vector mass is 
spontaneously generated during the condensation process, in the ordinary 
gauge theory the gauge invariance prohibits the mass terms of gauge fields 
from showing up. Does the spontaneous mass generation mean the breakdown of 
the gauge field theory description of the system as the tachyon condenses? 
To find a clue to the solution of this problem, 
consider the string field theory action 
expanded around the perturbative vacuum,
\begin{equation}
S=-\frac{V_p}{g_o^2}\left(-\frac{1}{2}\phi^2+2\kappa\phi^3+\frac{3\sqrt{3}}{4}
\phi A_{\mu}A^{\mu}+\ldots \right). \label{eq:CB}
\end{equation}
This action is not invariant under the standard gauge transformation
\begin{equation}
\left\{
	\begin{array}{l}
	\delta\phi =0, \\
	\delta A_{\mu}=\partial_{\mu}\Lambda ,
	\end{array}
\right. \label{eq:CC}
\end{equation}
even if the scalar $\phi$ has a \textit{vanishing}
 expectation value. Instead, it is 
invariant under the `BRST' transformation
\begin{equation}
\left\{
	\begin{array}{l}
	\delta\phi =\frac{3\sqrt{3}}{2}A^{\mu}\partial_{\mu}\Lambda +\ldots , \\
	\delta A_{\mu}=\partial_{\mu}\Lambda +\ldots ,
	\end{array}
\right. \label{eq:CD}
\end{equation}
at least for the first few terms including only $\phi$ and $A_{\mu}$. Even 
though we integrate out all massive fields to get the effective action 
$S(\phi,A_{\mu})$ for $\phi$ and $A_{\mu}$, 
it would be invariant not under~(\ref{eq:CC}), but 
under~(\ref{eq:CD}) with some modifications in higher order terms. That is, 
the `component' fields $\phi$ and $A_{\mu}$ of the string field do not have 
`gauge' symmetry of the form~(\ref{eq:CC}). This is in fact quite natural 
because the 
string field theory action is constructed in such a way that it should be 
invariant under the gauge transformation~(\ref{eq:M})
\[\delta\Phi =Q_B\Lambda +(\Phi *\Lambda-\Lambda *\Phi). \]
In the component form, it certainly gives nonlinear transformation law 
to each 
component field, never of the form~(\ref{eq:CC}). To relate the component 
fields $\phi,A_{\mu}$ to the `conventional' fields which have a 
symmetry~(\ref{eq:CC}) and live on the D-brane world-volume, 
some kind of field redefinition is needed~\cite{5085}. 
Hence the mass generation of the `component gauge field' does not necessarily 
mean that of the `conventionally defined' gauge field. 
Actually, if we carry out the following field redefinition
\begin{eqnarray}
\phi&=&\overline{\phi}+\frac{3\sqrt{3}}{4}\overline{A}_{\mu}\overline{A}^{\mu}
+\cdots , \label{eq:CI} \\
A_{\mu}&=&\overline{A}_{\mu}+\cdots , \nonumber
\end{eqnarray}
then the resulting action does not contain the cubic coupling term of the 
form $\overline{\phi}\overline{A}\overline{A}$.
As a result, the mass term for the gauge field $\overline{A}_{\mu}$ 
would not be generated even after $\overline{\phi}$ develops a 
nonzero expectation value. If this is the case, it seems to contradict the 
previous result which insists that all vectors become massive, but notice 
that the field redefinition~(\ref{eq:CI}) is defined only 
perturbatively. Considering that the closed string vacuum is highly 
nonperturbative, this redefinition may become invalid outside the radius of 
convergence and the vector field might really acquire nonzero mass. In such 
a case, the D-brane system could no longer 
be described by any low energy effective \textit{gauge} 
field theory after tachyon condenses, although we are usually assuming 
that a certain gauge theory can give a sufficiently good picture of the 
D-brane. 
%Or, it is also possible that the 
%coordinate system describing the configuration of the string field, namely 
%the set of component fields $\phi,A_{\mu},\cdots$, becomes 
%singular at the closed string vacuum, at which apparently different 
%$A_{\mu}$'s actually represent the physically equivalent 
%configurations. In this case the mass term of $A_{\mu}$ is illusory. 
%This idea was originally proposed by A. Sen in the context of noncommutative 
%soliton in~\cite{9038}(see also section~\ref{sec:NC}), 
%but it has a wider range of application.
We cannot answer which is correct to our present knowledge, 
because we need to understand 
directly the nonperturbative behavior of the fields for that purpose. Like 
this, zero momentum analyses present difficult problems, but we cannot finish 
the story without examining the kinetic term. If the kinetic term of a massive
field vanishes at the nonperturbative vacuum, its mass effectively grows to 
infinity and the field decouples from the spectrum there. 
This decoupling mechanism 
nicely fits the conjecture of annihilation of the D-brane. In fact, 
in~\cite{Univ} the standard (\textit{i.e.} two derivatives) kinetic term of 
the gauge field is shown to be absent at the nonperturbative vacuum, as long 
as the conjecture $f(T_0)=-1$ is true. Its outline is as follows. Under this 
assumption, one can show that the value of the potential at $T=T_0$ 
is independent of the classical background of the massless fields on the 
original D-brane ($T=0$), if it is a classical solution to the equations 
of motion at $T=0$. Applying it to the gauge field $A_{\mu}$, since 
a constant $F_{\mu\nu}$ is a solution of the equation of motion, the action 
($\propto f(T)$) should have no dependence on constant $F_{\mu\nu}$ at 
$T=T_0$. This means that the action does not contain the standard kinetic 
term $-\frac{1}{4}F_{\mu\nu}F^{\mu\nu}$ at the nonperturbative vacuum. 
Though this argument leaves the possibility that the action does depend on 
the terms which include derivatives of $F_{\mu\nu}$, it is conjectured that 
the effective action is altogether independent of the gauge field at the 
nonperturbative vacuum. Moreover, the authors of~\cite{0101162} obtained 
numerical supports for the expectation that the BRST invariant (physical) 
linear combination of the scalar fields $\phi,\beta_1,v$ up to level 2 loses 
its kinetic term at the nonperturbative vacuum in the level truncation 
scheme, where the `kinetic term' means the one which is quadratic in fields 
and contains at least one derivative, 
and they carried out the analysis using the derivative expansion up to 
four derivatives. In the following, however, instead of expanding the 
exponential of derivatives, 
we shall deal with the action keeping the derivative terms fully~\cite{KS}.
\bigskip

Here we will focus on the quadratic forms in the action whose zeroes 
represent the mass${}^2$ of the states. If we had a closed form expression 
$\Phi_0$ for the nonperturbative vacuum, we could examine the small 
fluctuations around it by shifting the string field as $\Phi=\Phi_0+
\widetilde{\Phi}$. But since we have found no such exact solutions up to 
now, we are compelled to analyze them in the level truncation approximation 
scheme. The quadratic part of the level (1,2) 
truncated action can be obtained from~(\ref{eq:AM}) and becomes
\begin{eqnarray}
S_{(1,2)}^{\mathrm{quad}}&=&\frac{1}{g_o^2}\int d^dx\Biggl(-\frac{1}{2}
\partial_{\mu}\phi^{\prime}\partial^{\mu}\phi^{\prime}+\frac{1}{2\ap}
\phi^{\prime 2}-\frac{1}{\ap}\tilde{\phi}^{\prime 2} \nonumber \\
& &{}-\frac{1}{2}\partial_{\mu}A_{\nu}\partial^{\mu}A^{\nu}-\frac{2^4}{3^3\ap}
\tilde{A}_{\mu}\tilde{A}^{\mu}-\frac{2^3}{3^3}\partial_{\mu}\tilde{A}_{\nu}
\partial^{\nu}\tilde{A}^{\mu}\Biggr). \label{eq:CJ}
\end{eqnarray}
From now on, we drop the prime on $\phi^{\prime}$, understanding that 
`$\phi$' represents 
a fluctuation about the nonperturbative vacuum. The zeroes of the quadratic 
form for $\phi$ in the momentum space are determined by 
\begin{equation}
-\frac{1}{2}p_{\mu}p^{\mu}+\frac{1}{2\ap}-\frac{1}{\ap}\exp\left(2\ap\ln
\frac{4}{3\sqrt{3}}p_{\mu}p^{\mu}\right)=0. \label{eq:CK}
\end{equation}
Setting $p_{\mu}p^{\mu}=-m^2$, it turns out that the left hand side 
of~(\ref{eq:CK}) is negative-definite for all real values of $m^2$. 
Therefore we found that the propagator of the field $\phi$ 
does not have any physical pole 
at the nonperturbative vacuum! It is rather surprising because 
if it were not for the exponential factor in~(\ref{eq:CK}), 
the equation would have a solution $m^2=+1/\ap$, 
which is already mentioned around the eq.(\ref{eq:BZ}). In a similar way, 
quadratic forms for $A_{\mu}$ are
\begin{eqnarray}
-\frac{1}{2}p^2-\frac{2^4}{3^3\ap}\exp\left(2\ap\ln\frac{4}{3\sqrt{3}}p^2
\right)&=&0 \quad (\mathrm{transverse} \ \ A^T_{\mu}), \label{eq:CL} \\
-\frac{1}{2}p^2-\left(\frac{2^4}{3^3\ap}+\frac{2^3p^2}{3^3}\right)\exp\left(
2\ap\ln\frac{4}{3\sqrt{3}}p^2\right)&=&0 \ \ (\mathrm{longitudinal} 
\quad A^L), \nonumber
\end{eqnarray}
where 
\[ A_{\mu}^T=A_{\mu}-\frac{\partial_{\mu}\partial^{\nu}A_{\nu}}{\partial^2}\, 
, \> A_{\mu}^L=\frac{\partial_{\mu}\partial^{\nu}A_{\nu}}{\partial^2}\, , \>
A^L=\frac{\partial^{\mu}A_{\mu}}{\sqrt{|-\partial^2|}} \]
so that $A_{\mu}=A_{\mu}^T+A_{\mu}^L \, , \> \partial_{\mu}A^{T\mu}=0 \, , \> 
\partial_{\mu}A^{L\mu}=\partial_{\mu}A^{\mu}$. For $A_{\mu}^T$, since 
there is no real pole again, the transverse vector field is not physical. 
On the other hand, the scalar field $A^L$ has a pole at $m^2=-p^2=1/\ap$. 
Though we have found a pole for $A^L$ propagator, it does not immediately 
mean the appearance of a physical state because it may be null. In any case, 
these results revealed that the states associated with $\phi$ and $A_{\mu}^T$ 
have been removed from the physical spectrum at the nonperturbative vacuum.
Thus, the number of physical states at the nonperturbative vacuum is clearly 
smaller than that at the perturbative vacuum, though it does not prove the 
conjecture  
that there are no physical excitations at all at the nonperturbative vacuum. 
\medskip

At level (2,4), the fields $B_{\mu\nu}, B_{\mu}, \beta_1$ are newly added. 
By decomposing the tensor and vectors into transverse and 
longitudinal components as in (\ref{eq:CL}), we obtain
\begin{equation}
\left.
	\begin{array}{cl}
	\mbox{2-tensor} & b^{TT}_{\mu\nu} \\
	\mbox{vector} & A_{\mu}^T; B_{\mu}^T , b_{\mu}^{TL} \\
	\mbox{scalar} & A^L; \phi, B^L, B, b^{LL}, \beta_1 ,
	\end{array}
\right. \label{eq:CM}
\end{equation}
where $b_{\mu\nu}$ and $B$ are traceless and trace part of $B_{\mu\nu}$ 
respectively. At this level, the quadratic form for $b_{\mu\nu}^{TT}$ never 
vanishes, so that the transverse tensor has no physical excitation. The pole 
for $A^L$ slightly changes the position from $m^2=1/\ap$ found at level (1,2) 
to $m^2\simeq 1.37/\ap$. Since the vectors $B^T_{\mu}$ and $b_{\mu}^{TL}$ mix 
with each other in the action, their quadratic form becomes a $2\times 2$ 
matrix. Its inverse exhibits a pole at $m^2\simeq 2.1/\ap$. In a similar way, 
the scalars $\phi, B^L, B, b^{LL}, \beta_1$ mix and their quadratic form 
matrix has three zeroes at
\[ m^2\simeq 1.3/\ap \, , \> 2.1/\ap \, , \> 15.8/\ap , \]
where the last value may be too large to be trusted. For the explicit 
expressions of the quadratic forms stated above, see~\cite{KS}. Lastly, 
$A_{\mu}^T$ quadratic form is given by 
\[ -\frac{1}{2}p^2-\frac{3^{7/2}}{2^6}\exp\left(2\ap\ln\frac{4}{3\sqrt{3}}p^2
\right)\left\{\frac{2^4}{3^2}\phi_0-\frac{2^4\cdot 11}{3^5}(\beta_1)_0
-\frac{2^4}{3^5\sqrt{13}}(7^2-2^4\ap p^2)(B)_0\right\}, \]
which has a zero at $m^2=-p^2=16.8/\ap$. The pole for $A_{\mu}^T$ has 
appeared  at level (2,4), though it was not present at level (1,2). 
But it is not clear whether or not it gives rise to a physical state. 
Concerning this vector pole, the analysis was extended to level (4,8) 
in~\cite{5085}, in which the mixing among 
$A^T_{\mu}$ and vector fields $V_{\mu}^{iT}$ at level 3 was taken into 
account. At level (3,6), it was shown that there exists at least one zero by 
using the continuity of the effective quadratic form for $A_{\mu}^T$ obtained 
by eliminating $V_{\mu}^i$'s. However, since this argument is not valid at 
level (4,8), it is not yet solved whether the pole persists after including 
higher level terms.

\medskip

So far, we have considered the spectrum of 
the low-lying fields at the nonperturbative vacuum. Although we 
could not show that there are no physical excitations at all at the 
nonperturbative vacuum, we saw that the number of physical states 
considerably decreased as compared with that at the perturbative 
vacuum. 
%The key role in removing the states was played by the 
%exponential factor $\displaystyle \exp\left(\ap\ln\frac{4}{3\sqrt{3}}p^2
%\right)$. Due to this factor, the equations whose solutions give mass${}^2$'s 
%of the states became transcendental. Around the perturbative vacuum, this 
%factor arises only in the cubic interaction terms. In this case, the 
%quadratic form is a polynomial, so the pole of the propagator appears 
%on the real axis. But by shifting $\phi=\phi_0+\phi^{\prime}$, the 
%exponential factors enter into the quadratic forms, typically of the form 
%$\phi_0\cdot \tilde{A}_{\mu}\tilde{A}^{\mu}$. The transcendental equations 
%may not necessarily have solutions, resulting in the removal of states. 
And we have found that the exponential factor $\displaystyle \exp\left(\ap
\ln\frac{4}{3\sqrt{3}}p^2\right)$, which originates ultimately from the 
structure of the cubic string vertex, does have the significant effect on the 
determination of the spectrum at the nonperturbative vacuum in the cubic 
string field theory. Let us see physical aspects of this factor. 

Around the perturbative vacuum, all fields in the cubic interaction terms are 
accompanied by $\exp\left(-\ap\ln\frac{4}{3\sqrt{3}}\partial^2\right)$. 
This implies that we can define the effective open string coupling as 
\begin{equation}
g_o^{\mathrm{eff}}(p)=\exp\left(3\ap\ln\frac{4}{3\sqrt{3}}p^2\right)g_o
\end{equation}
in momentum space. As a result, the effective coupling 
runs even at the tree-level. Since $\ln(4/3\sqrt{3})\simeq -0.26<0$, the 
effective coupling exponentially falls off as $p^2\to +\infty$. Consequently, 
the cubic string field theory is asymptotically free at the tree-level. 
In fact, any string theory should have such tree-level asymptotic 
freedom\footnote{In favor of this argument, it 
seems that the similar exponential factor also appears in the calculations 
of boundary string field theory~\cite{TTUpriv}.} for the following reason: 
The factor $\exp\left(-\ap\ln\frac{4}{3\sqrt{3}}\partial^2\right)$, 
which contains an infinite number of derivatives, smears the interactions 
at short distance and hence is indicative of the extended nature of the 
string. Such an effect should correspond to the ultraviolet cut-off of the 
string theory. Thus, the large space-like 
momentum $p^2\to +\infty$ provides good short distance behavior to the theory.
On the other hand, large time-like momentum invalidates the perturbation 
theory. In level ($M,N$) truncation, we keep only fields of level $\le M$. 
Since the level $M$ on-shell field has mass $m^2=-p^2=(M-1)/\ap$ at the 
perturbative 
vacuum, the states with mass beyond this truncation scale cannot be 
reliably treated in the perturbation theory. In the previous analyses, we 
found several poles around $m^2\sim 15/\ap$ at level (2,4), but 
these are questionable in the above sense.

\bigskip

Finally, we discuss the stability of the nonperturbative vacuum. As remarked 
before, the existence of `auxiliary' fields, whose kinetic terms have the 
wrong sign, gives rise to unstable directions in the potential, even if they 
all have positive mass squared. Actually, the stationary point corresponding 
to the nonperturbative vacuum is a saddle point. Nonetheless, we think of it 
as perturbatively stable if there exist no tachyonic modes around that point. 
Considering that the presence of the open string tachyon at the perturbative 
vacuum signals the instability of the bosonic D-branes, this criterion will 
be valid. Since every pole we have found in the above analysis is located at 
$p^2=-m^2<0$, the nonperturbative vacuum is considered to be stable at this 
level of approximation. 

Though we found the nonperturbative vacuum to be perturbatively stable, it 
may be unstable against the quantum mechanical tunnelling effect. 
This problem is very difficult to answer, so we can give no clear 
conclusions. Generically, it is hard for a system with many degrees of 
freedom to tunnel. In the string field theory, degrees of freedom involved in 
the system are an infinite number of particle fields, each of which has 
infinite degrees of freedom. Since the system has an enormous amount of 
degrees of freedom, we think that the tunnelling seldom occurs.

\subsection{Cubic string field theory around the closed string vacuum}
Thus far, we have attempted to show that there are no physical excitations of 
open string modes at the closed string vacuum in the following way: Expanding 
the string field $\Phi$ around the closed string vacuum solution $\Phi_0$ as 
$\Phi=\Phi_0+\widetilde{\Phi}$, we examine the spectrum of the fluctuation 
field $\widetilde{\Phi}$. Instead of doing so, it was proposed that starting 
by guessing the string field theory action around the closed string vacuum, 
which manifestly implement the absence of open string excitations and is 
gauge invariant, we try to construct the original D-brane as well as lump 
solutions representing lower dimensional D-branes~\cite{RSZ}. 
\medskip

We begin by the following cubic string field theory action on a D$p$-brane 
\begin{equation}
S(\Phi)=-V_{p+1}\tau_p-\frac{1}{g_o^2}\left[\frac{1}{2\ap}\int\Phi *Q_B\Phi
+\frac{1}{3}\int\Phi *\Phi *\Phi\right] , \label{eq:RSZA}
\end{equation}
where we think of the string field configuration $\Phi=0$ as representing the 
original D$p$-brane. The D-brane mass term $-V_{p+1}\tau_p$ have been added 
so that the closed string vacuum $\Phi=\Phi_0$ have vanishing energy density 
if $f(\Phi_0)=-1$~(\ref{eq:AS}) is true. As we have seen before, this action 
has gauge invariance under 
\[ \delta\Phi=Q_B\Lambda +\Phi *\Lambda -\Lambda *\Phi \]
because of the axioms
\begin{equation}
\left.
	\begin{array}{ll}
	\mbox{nilpotence} & Q_B^2=0, \\
	\mbox{odd derivation} & Q_B(A*B)=(Q_BA)*B+(-1)^AA*(Q_BB), \\
	\mbox{partial integrability} & \int Q_B(\ldots)=0, \\
	 & \int A*B=(-1)^{AB}\int B*A, 
	\end{array}
\right. \label{eq:RSZB}
\end{equation}
and the associativity of the $*$-product. Shifting the string field by the 
closed string vacuum solution $\Phi_0$ as $\Phi=\Phi_0+\widetilde{\Phi}$, 
the action is rewritten as 
\begin{eqnarray}
S(\Phi_0+\widetilde{\Phi})&=&-V_{p+1}\tau_p-\frac{1}{g_o^2}\biggl[\frac{1}{
2\ap}\int \Phi_0*Q_B\Phi_0+\frac{1}{3}\int\Phi_0*\Phi_0*\Phi_0 \nonumber \\
& &+\int\widetilde{\Phi}*\left(\frac{1}{\ap}Q_B\Phi_0+\Phi_0*\Phi_0\right)
\label{eq:RSZC} \\ & &+\frac{1}{2}\int\widetilde{\Phi}*\left(\frac{1}{\ap}
Q_B\widetilde{\Phi}+\widetilde{\Phi}*\Phi_0+\Phi_0*\widetilde{\Phi}\right)
+\frac{1}{3}\int\widetilde{\Phi}*\widetilde{\Phi}*\widetilde{\Phi}\biggr].
\nonumber 
\end{eqnarray}
Since $\Phi_0$ is a \textit{solution} to the string field equation of motion 
from~(\ref{eq:RSZA})
\[ \frac{1}{\ap}Q_B\Phi+\Phi *\Phi=0, \]
the second line vanishes. And the first line also vanishes due to the brane 
annihilation conjecture $S(\Phi_0)=0$. Hence by defining the new `BRST-like' 
operator $Q$ to be 
\begin{equation}
\frac{1}{\ap}Q\widetilde{\Phi}=\frac{1}{\ap}Q_B\widetilde{\Phi}+\Phi_0*
\widetilde{\Phi}+\widetilde{\Phi}*\Phi_0 \label{eq:RSZD}
\end{equation}
the action is written as 
\begin{equation}
S(\Phi_0+\widetilde{\Phi})\equiv S_0(\widetilde{\Phi})=-\frac{1}{g_o^2}
\left[\frac{1}{2\ap}\int\widetilde{\Phi}*Q\widetilde{\Phi}+\frac{1}{3}
\int\widetilde{\Phi}*\widetilde{\Phi}*\widetilde{\Phi}\right]. 
\label{eq:RSZE}
\end{equation}
Moreover, if we perform a field redefinition 
\[ \widetilde{\Phi}=e^K\Psi \]
with $K$ satisfying the properties 
\begin{eqnarray*}
\gh (K)&=&0 \quad (\mbox{Grassmann even}), \\
K(A*B)&=&(KA)*B+A*(KB), \quad (\langle V_3|(K^{(1)}+K^{(2)}+K^{(3)})=0), \\
\int KA*B&=&-\int A*KB, 
\end{eqnarray*}
then we have 
\begin{equation}
S_0(e^K\Psi)\equiv S_1(\Psi)=-\frac{1}{g_o^2}\left[\frac{1}{2\ap}\int\Psi *
\cQ\Psi+\frac{1}{3}\int\Psi *\Psi *\Psi\right], \label{eq:RSZF}
\end{equation}
where $\cQ=e^{-K}Qe^K$. Since the new operator $\cQ$ satisfies the 
axioms~(\ref{eq:RSZB}) with $Q_B$ replaced by $\cQ$, the action $S_1(\Psi)$ 
is invariant under the gauge transformation 
\[ \delta\Psi=\cQ\Lambda+\Psi *\Lambda-\Lambda *\Psi. \]
Noting that the configuration $\Psi=0$ corresponds to the closed string 
vacuum, the `BRST' operator $\cQ$ governs the open string dynamics there. 
To agree with the expectation that there are no open string excitations 
around the closed string vacuum, 
\[ \cQ \mbox{ must have vanishing cohomology.} \]
In addition, since the closed string vacuum created after tachyon 
condensation should contain no information about the original D-brane before 
condensation, $\cQ$ must be universal in the sense that $\cQ$ is 
independent of the details of the boundary CFT describing the original 
D-brane. This condition is satisfied if $\cQ$ is constructed purely from 
ghost operators. As examples of $\cQ$ which satisfy the above conditions, 
we consider the operators 
\begin{equation}
\cC_n=c_n+(-1)^nc_{-n} \ , \quad n=0,1,2,\cdots. \label{eq:RSZG}
\end{equation}
In fact, each of these operators has zero cohomology: Take a state 
$|\psi\rangle$ which is annihilated by $\cC_n$, 
\textit{i.e.} $\cC_n|\psi\rangle=0$. 
Since $\mathcal{B}_n=\frac{1}{2}(b_{-n}+(-1)^nb_n)$ obeys $\{\cC_n,
\mathcal{B}_n\}=1$, $|\psi\rangle$ is always expressed as 
\[ |\psi\rangle =\{\cC_n,\mathcal{B}_n\}|\psi\rangle =\cC_n(\mathcal{B}_n
|\psi\rangle), \]
which is $\cC_n$-exact. More generally, we find 
\begin{equation}
\cQ=\sum_{n=0}^{\infty}a_n\cC_n \label{eq:RSZH}
\end{equation}
with $a_n$'s constant to satisfy the required properties 
\begin{equation}
\left.
\begin{array}{cl}
(1) & \mbox{the algebraic structure (\ref{eq:RSZB}) ($Q_B$ replaced by $\cQ$)}
\\ & \ \mbox{which guarantees the gauge invariance,} \\
(2) & \mbox{vanishing cohomology,} \label{eq:RSZI} \\
(3) & \mbox{universality (not including any matter sector)}.
\end{array}
\right. 
\end{equation}
If we had a closed form expression for $\Phi_0$, we would be able to 
construct the operator $\cQ$ and see whether $\cQ$ satisfies the 
properties~(\ref{eq:RSZI}). However, since we have not yet succeeded in 
obtaining such a solution $\Phi_0$, we \textit{assume} that the kinetic 
operator $\cQ$ obtained as a result of shifting and redefining the string 
field as in~(\ref{eq:RSZF}) from the standard BRST operator $Q_B$ satisfies 
the requirements~(\ref{eq:RSZI}). 
In particular, it is important in what follows 
to note that $\cQ$ does not involve any matter operator. We will then 
justify the above postulate by showing the existence of suitable lump 
solutions representing lower dimensional D-branes. 
\medskip

\begin{figure}[htbp]
\begin{center}
	\includegraphics{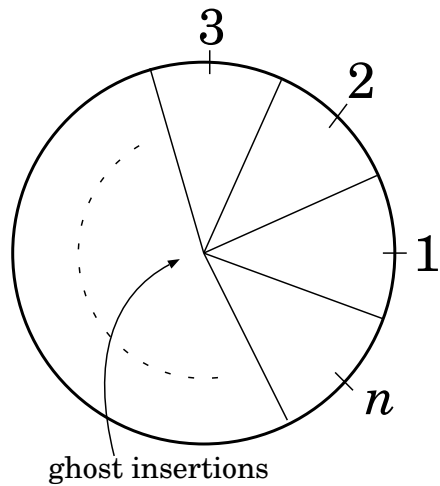}
	\caption{$n$-point amplitude with purely ghost propagator.}
	\label{fig:Pizza}
\end{center}
\end{figure}

We now fix the gauge by some condition which also does not contain any matter 
operator. Then, since both the interaction vertex and the propagator 
factorize into the matter sector and the ghost sector, so do general 
$N$-point amplitudes. The generating functional of the connected $n$-tachyon 
Green's functions takes the form 
\begin{equation}
W[J]=\sum_{n=2}^{\infty}\frac{1}{n!}\int d^{p+1}\!k_1\cdots d^{p+1}\!k_n\ 
J(k_1)\cdots J(k_n)\ \cG^{(n)}(k_1,\cdots,k_n)\delta\left(\sum_{i=1}^nk_i
\right), \label{eq:RSZJ}
\end{equation}
where $J(k)$ is the current coupled to the tachyon field $\phi(-k)$. 
$\cG^{(n)}(k_1,\cdots,k_n)\delta(\sum k_i)$ is the $n$-tachyon connected 
off-shell amplitude and is given by summing over the string Feynman diagrams. 
If we were considering the cubic string field theory with the conventional 
BRST operator $Q_B$, we would have the propagator 
\[ (c_0L_0^{\mathrm{m}})^{-1}=b_0\int_0^{\infty}dt\ e^{-tL_0^{\mathrm{m}}}, \]
which is interpreted as the integration over the propagation length $t$ of 
the intermediate strip constructing the connected world-sheet. In the present 
case of $\cQ$ consisting purely of ghost operators, however, such strips 
shrink to zero length so that we are left with the world-sheet conformally 
equivalent to a unit disk divided into $n$ wedges of angle $2\pi /n$, each 
of which corresponds to the open string world-sheet with a tachyon vertex 
operator inserted, with some ghost insertions. This situation is illustrated 
in Figure~\ref{fig:Pizza}.
As the propagator collapses, the diagram necessarily looks like tree-level. 
Aside from the ghost insertions, this gluing prescription of $n$ half-disks 
formally coincides with that of the $n$-string vertex $\int\Phi *\cdots 
*\Phi$ defined in~(\ref{eq:X}). So we have 
\begin{equation}
\cG^{(n)}(k_1,\cdots,k_n)\delta\left(\sum_{i=1}^nk_i\right)=C_n\langle
g_1^{(n)}\circ T_{k_1}(0)\cdots g_n^{(n)}\circ T_{k_n}(0)\rangle_{\mathrm{
matter}}, \label{eq:RSZK}
\end{equation}
where $T_k(z)=e^{ikX(z)}$ is the matter part of the tachyon vertex operator 
and $C_n$ is the contribution from the ghost sector which is regarded as a 
constant. Evaluating the matter sector correlator, we can completely 
determine the momentum dependence of the $n$-point amplitude as 
\[ \cG^{(n)}(k_1,\cdots,k_n)=C_n\exp\left[\left(\ln\frac{4}{n}\right)
\sum_{i=1}^n\ap k_i^2+\sum_{i.j=1 \atop i\neq j}^n\ap k_i\cdot k_j\ 
\ln \left(2 \sin\frac{\pi}{n}|i-j|\right)\right] \quad n\ge 3, \]
and $\cG^{(2)}$ is momentum independent. To obtain explicit values for 
$\cC_n$'s, we must specify the precise form of the propagator, or 
equivalently, of $\cQ$. We do not try it here. 
\medskip

We then get the tachyon effective action by Legendre-transforming $W[J]$. 
First, we define the \textit{classical field} $\varphi$ to be 
\begin{equation}
\varphi(q)\equiv\phi_c[q,J]=\frac{\delta W[J]}{\delta J(-q)}, \label{eq:RSZL}
\end{equation}
namely, $\varphi$ is the expectation value of the tachyon field $\phi$ in the 
presence of the background source $J(k)$. The effective action 
$\Gamma[\varphi]$ is defined as a functional of the classical field $\varphi$ 
via the Legendre transformation by 
\begin{equation}
\Gamma[\varphi]\equiv\int d^{p+1}\!k\ J(k)\varphi(-k)-W[J]. \label{eq:RSZM}
\end{equation}
As is well known from the particle field theory, the effective action $\Gamma[
\varphi]$ is obtained by summing over the one-particle irreducible diagrams 
and coincides with the classical action $S[\varphi]$ at the tree-level (in 
this case, `tree-level' corresponds to dropping more than three-point 
contact-type interaction vertices as in Figure~\ref{fig:Pizza}). 
By definition of $\Gamma[\varphi]$, differentiating with respect to 
$\varphi(k)$ gives 
\begin{equation}
J(-k)=\frac{\delta \Gamma[\varphi]}{\delta\varphi(k)}. \label{eq:RSZMa}
\end{equation}
What we really want to know is the expectation value of the tachyon field 
$\phi$ \textit{in the absence of} the external source: $J=0$. Then, the 
effective action is stationary, $\delta\Gamma/\delta\varphi|_{J=0}=0$. And 
the expectation value is given by 
\begin{equation}
\varphi(q)=\phi_c[q,J=0]=\frac{\delta W}{\delta J(-q)}\bigg|_{J=0}, 
\label{eq:RSZN}
\end{equation}
which is the usual 1-point function of $\phi$. 
\smallskip

\begin{figure}[htbp]
\begin{center}
	\includegraphics{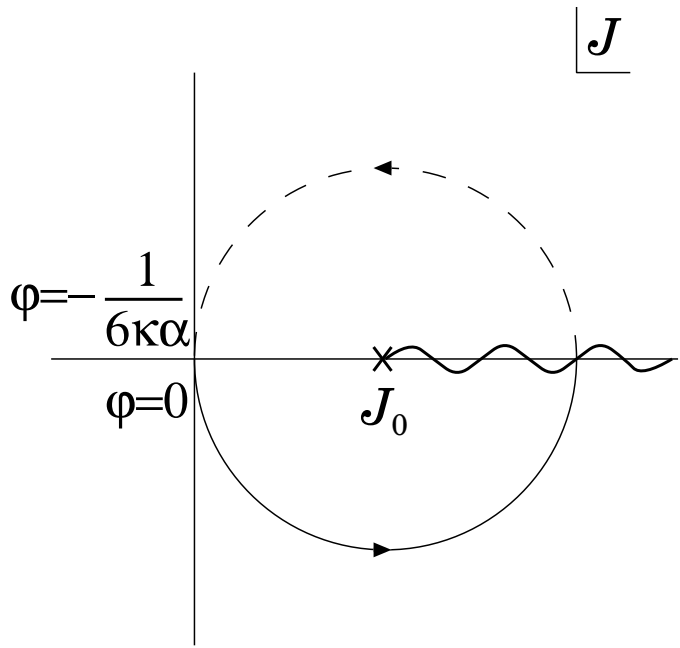}
	\caption{A branch cut on the complex $J$ plane.}
	\label{fig:RSZA}
\end{center}
\end{figure}

Now let us seek a translationally invariant (constant) solution which is to 
represent the original D$p$-brane. For this purpose, we take the external 
current to be a delta function in the momentum space, $J(k)=u\delta(k)$. 
The solution $\varphi(q)$ is obtained from~(\ref{eq:RSZN}) and 
(\ref{eq:RSZJ}) as 
\begin{eqnarray}
\varphi(q)&=&\phi_c[q,u\delta(k)\to 0] \nonumber \\
&=&\sum_{n=2}^{\infty}\frac{1}{(n-1)!}\int d^{p+1}\!k_1\cdots 
d^{p+1}\!k_{n-1}u\delta(k_1)\cdots u\delta(k_{n-1}) \nonumber \\
& &\qquad \times \cG^{(n)}(k_1,\cdots,
k_{n-1},-q)\delta^{p+1}\left(-q+\sum_{i=1}^{n-1}k_i\right)\bigg|_{u=0} 
\nonumber \\ &=&\sum_{n=1}^{\infty}\frac{u^n}{n!}C_{n+1}\bigg|_{u=0}
\delta^{p+1}(-q)\equiv F(u)\Big|_{u=0}\delta^{p+1}(q), \label{eq:RSZO}
\end{eqnarray}
where we used $\cG^{(n)}(0,\cdots,0)=C_n$. Since $F(u)$ starts, however, with 
a linear power in $u$, it na\"{\i}vely seems that $F(u=0)$ and hence the 
constant solution $\varphi|_{J=0}$ vanish. To find a clue to solve this 
problem, let us consider the level (0,0) truncated effective action with 
$\cQ=c_0$ at the tree-level. For $\Psi=\varphi c_1|0\rangle$ with $\varphi$ 
constant, the action becomes 
\[ \Gamma(\varphi)=-\frac{1}{g_o^2}\left(\frac{1}{2\ap}\varphi^2+2\kappa
\varphi^3\right). \]
Substituting it into (\ref{eq:RSZMa}), we find 
\[ -g_o^2J=\frac{1}{\ap}\varphi+6\kappa\varphi^2. \]
From this equation, we get the 1-point expectation value $\varphi$ of the 
tachyon field $\phi$ under the influence of the external field $J$ as 
\begin{equation}
\varphi=-\frac{1}{12\kappa\ap}\left(1\pm\sqrt{1-(24\kappa\ap{}^2g_o^2)J}
\right). \label{eq:RSZP}
\end{equation}
Clearly, $\varphi$ has a square root branch point in the complex $J$ plane 
at $J=1/24\kappa\ap{}^2g_o^2\equiv J_0$ as illustrated in 
Figure~\ref{fig:RSZA}.
Suppose that we initially choose the $-$ sign in front of the square root 
so that $\varphi|_{J=0}=0$. But by letting $J$ move around the branch point 
$J_0$ as $J(\theta)=J_0(1-e^{i\theta})$ from $\theta=0$ to $\theta=2\pi$, 
$\varphi$ becomes 
\[ \varphi(\theta)=-\frac{1}{12\kappa\ap}(1-e^{i\theta/2})\quad 
\longrightarrow \ \ \varphi(2\pi)=-\frac{1}{6\kappa\ap}, \]
and, of course, $J(\theta=2\pi)=0$. Further, the `solution' $\varphi=-1/
6\kappa\ap$ actually extremizes the action $\Gamma(\varphi)$ as promised. 
This result encourages us to \textit{assume} that the function $F(u)$ defined 
in~(\ref{eq:RSZO}) also has a branch point $u_0$ and that $F(u)$ takes a 
nonzero value $\varphi_0$ after $u$ goes around the branch point and comes 
back to the origin. We write this situation as 
\[ \lim_{u\to 0^{\prime}}F(u)=\varphi_0, \]
where $0^{\prime}$ denotes the origin on the suitable branch, and then 
$\varphi(q)=\varphi_0\delta(q)$. Since we are considering that we have 
reached the action~(\ref{eq:RSZF}) at the closed string vacuum 
by starting from 
the action~(\ref{eq:RSZA}) on a D$p$-brane, shifting the string field by a 
constant $\Phi_0$ representing the closed string vacuum, and performing a 
field redefinition, there \textit{must exist} a solution which represents 
the original D$p$-brane configuration. Here we identify the translationally 
invariant solution $\varphi(q)=\varphi_0\delta(q)$ as the original D-brane, 
though we cannot determine the precise value of $\varphi_0$ due to lack of 
knowledge of the ghost correlators $C_n$. 
\medskip

Next we turn to the lump solutions, but we only quote the main results here, 
leaving the details to the reference~\cite{RSZ}. We consider a tachyon 
profile of the form 
\begin{equation}
\phi_c[q,J(k)=u\psi(k)]=F(u)\chi(q)+H(u,q), \label{eq:RSZQ}
\end{equation}
where the Taylor series expansion of $H(u,q)$ in $u$ converges more rapidly 
than that of $F(u)\chi(q)$. When we take the limit $u\to 0^{\prime}$ 
described in the last paragraph, $F(u)$ goes to $\varphi_0$ as above. In this 
limit, $H(u,q)$ will become some function of $q$, 
\[ \lim_{u\to 0^{\prime}}H(u,q)=h(q), \]
possibly $h(q)=0$ if $H(u,q)$ is not singular at $u=u_0$. Then we obtain a 
classical spacetime dependent solution
\begin{equation}
\varphi(q)=\lim_{u\to 0^{\prime}}\phi_c[q,u\psi(k)]=\varphi_0\chi(q)+h(q),
\label{eq:RSZS}
\end{equation}
if we can find two functions $\psi(q)$ and $\chi(q)$, and the original 
D$p$-brane solution $\varphi_0\delta(q)$. In searching for a codimension $n$ 
lump solution on the original D$p$-brane, we try the following ansatz 
\begin{equation}
\psi(q)=K\exp\left(-\frac{\alpha q^2_{\bot}}{2}\right)\delta^{p+1-n}
(q_{{}_{/\!\!/}}), \quad \chi(q)=\gamma K\exp\left(-\frac{\beta q_{\bot}^2}{
2}\right)\delta^{p+1-n}(q_{{}_{/\!\!/}}), \label{eq:RSZT}
\end{equation}
where we have divided ($p+1$)-dimensional momentum vector $q$ into $q_{\bot}$ 
and $q_{{}_{/\!\!/}}$, the former being the momentum in the $n$-dimensional 
transverse space and the latter parallel to the $(p+1-n)$-dimensional lump. 
The better convergence property of $H(u,q)$ stated below eq.(\ref{eq:RSZQ}) 
puts severe restrictions on the functions $\psi,\chi$. After some 
manipulations, we finally get 
\begin{eqnarray}
\ln K&=&-\frac{1}{2}nA(\alpha), \label{eq:RSZU} \\
\ln\gamma&=&\frac{1}{2}n\left(-\ln\frac{2\pi}{\alpha+\beta}+B(\alpha)+2
A(\alpha)\right), \nonumber 
\end{eqnarray}
where $A$ and $B$ are functions of $\alpha$ which can be determined. Also 
$\beta$ is determined as a function of $\alpha$. Although the solutions 
seem to form a family parametrized by $\alpha$, the full solution $\varphi(q)
=\varphi_0\chi(q)+h(q)$ is independent of $\alpha$. The value of the 
effective action $\Gamma[\varphi]$ on the solution $\varphi(q)=\varphi_0
\chi(q)+h(q)$ is given by 
\begin{eqnarray}
\Gamma[\varphi_0\chi(q)+h(q)]&=&-\lim_{u\to 0^{\prime}}W[J(q)=u\psi(q)] 
\label{eq:RSZV} \\ &=&-\frac{V_{p+1-n}}{(2\pi\sqrt{\ap})^{p+1-n}}\left(
e^{nB(\alpha)/2}G_0+P_0\right)\equiv -V_{p+1-n}\cT_{p-n} \nonumber
\end{eqnarray}
with unknown constants $G_0,P_0$. \textit{Assuming} that $P_0$ vanishes, the 
ratio of the lump tensions of different dimensions is found to be 
\[\frac{\cT_{p-n-1}}{\cT_{p-n}}=2\pi\sqrt{\ap}\ e^{B(\alpha)/2}, \]
which is the expected relation if $B(\alpha)\equiv 0$. In fact, $B$ was 
numerically shown to vanish with very high accuracy, though the analytic 
proof has not yet been given. 
\medskip

To sum up, we can construct lump solutions of the gaussian form of every 
codimension if we make assumptions:
\begin{itemize}
\item the kinetic operator $\cQ$ obtained by shifting the string field 
has vanishing cohomology and is constructed purely out of ghosts,
\item the existence of the translationally invariant solution $\varphi(q)=
\varphi_0\delta(q)$ which is suitable for the original D$p$-brane. 
\end{itemize}
And more, the ratio of the lump tensions exactly agrees with that of the 
D-brane tensions if we further assume 
\begin{itemize}
\item the vanishing of the constant $P_0$ and of the function $B(\alpha)$ 
(numerically verified)
\end{itemize}
in the notation of \cite{RSZ}. To completely understand the tachyon 
condensation and the properties of the D-brane solutions within this 
formulation of string field theory, however, many things are left to be 
accomplished. Though we have seen that the ratio of the lump tensions agrees 
with the expected result, we have obtained no definite conclusions about 
the overall normalization of the brane tension. It was also pointed out 
in~\cite{RSZ} that the level truncation scheme in string field theories 
with non-standard BRST operators $\cQ$ does not seem to be successful at 
least in the standard Feynman-Siegel gauge. Hence it might be inevitable 
to determine the whole set of ghost correlators $C_n$, which in turn requires 
specifying the form of the kinetic operator $\cQ$. If it is possible, we 
could prove some of the assumptions raised in the course of the above 
analyses.

\subsection{Fate of the $U(1)$ gauge field}
In this subsection, we discuss the fate of open string modes at the closed 
string vacuum from the viewpoint of the low-energy world-volume field 
theory\footnote{We will give a related argument in the context of background 
independent open string field theory in section~\ref{sec:bsft}.}. 
\medskip

In Type II superstring theory, it was shown in~\cite{NonBPSD} and in the 
context of background independent open string field theory (see 
chapter~\ref{ch:biosfit}) that the bosonic part of the world-volume action 
for a non-BPS D$p$-brane, if the induced metric $g_{\mu\nu}$, the $U(1)$ 
gauge field strength $F_{\mu\nu}$ and the tachyon field $T$ are constant, 
takes the form\footnote{The authors of~\cite{3122,pyo,klu} proposed the 
possible forms of the action which extend the one~(\ref{eq:FateA}) to 
include the derivatives of the tachyon field.} 
\begin{equation}
S=-\tau_p\int d^{p+1}\!x\sqrt{-\det(g_{\mu\nu}+F_{\mu\nu})}U(T), 
\label{eq:FateA}
\end{equation}
where the tachyon potential $U(T)$ is assumed to vanish at the minimum 
$T=T_0$. Note that the D$p$-brane tension $\tau_p$ is proportional to the 
inverse of the closed string coupling $g_s$ so that it can be written as 
$\tau_p=C/g_s$. Since the action itself vanishes identically at the closed 
string vacuum $T=T_0$, Sen has argued that the $U(1)$ gauge field, which 
remains massless even after the neutral tachyon condenses, acts as a 
Lagrange multiplier which plays the role of removing the states charged 
under this $U(1)$. Thus the $U(1)$ gauge field on the unstable D-brane 
disappears together with the charged open string states through the 
tree-level effect in string theory. 
\medskip

According to the authors of~\cite{Conf}, however, the above argument is 
incomplete to solve the problem fully. The effective gauge coupling on the 
D$p$-brane world-volume is defined by 
\begin{equation}
\frac{1}{g_{\mathrm{eff}}}=\frac{U(T)}{g_s}=\frac{\tau_p}{C}U(T), 
\label{eq:FateB}
\end{equation}
so that the vanishing of the potential $U(T)$ at $T=T_0$ implies that the 
strong gauge coupling $g_{\mathrm{eff}}\to \infty$ invalidates the original 
gauge theory description of the D-brane world-volume theory near the closed 
string vacuum. In fact, it was argued in~\cite{Yi,Conf} that the unbroken 
$U(1)$ gauge group might be \textit{confined} via the nonperturbative dual 
Higgs mechanism. We shall describe the basic ideas and their conclusions 
below. 
\medskip

We will first consider a single D3-brane--anti-D3-brane pair in Type IIB 
theory, and later make a comment on generalizations of it. 
There exist two $U(1)$ 
gauge fields $A_{\mu}$ and $A_{\mu}^{\prime}$ living on the D3 and 
$\overline{\mathrm{D}3}$ world-volume respectively, while a complex tachyonic 
scalar field $T$ arises on strings stretched between the D3-brane and the 
$\overline{\mathrm{D}3}$-brane. Since the tachyon on the oriented 
3-$\overline{3}$ string transforms in the bifundamental representation under 
the gauge group $U(1)\times U(1)^{\prime}$, it couples to the gauge fields as 
\begin{equation}
|\partial_{\mu}T-i(A_{\mu}-A_{\mu}^{\prime})T|^2. \label{eq:FateC}
\end{equation}
From this expression, one can see that a nonvanishing vacuum expectation 
value $T=T_0=|T_0|e^{i\theta}$ of the tachyon field breaks the relative 
$U(1)$ gauge group generated by $A_{\mu}^-=A_{\mu}-A_{\mu}^{\prime}$, 
being accompanied by the usual Higgs mechanism. Since the expectation value 
$|T_0|$ is of order of the string scale at the closed string vacuum, one 
gauge field $A_{\mu}^-$ with mass of $\cO(\ap{}^{-1/2})$ decouples from the 
low-energy spectrum. On the other hand, the other linear combination 
$A_{\mu}^+=A_{\mu}+A_{\mu}^{\prime}$, which generates an overall $U(1)$, 
remains massless even after tachyon condensation and hence seems to stay 
in the low-energy spectrum. What we want to show is that the latter also 
disappears when the brane annihilation process takes place. 

The key point of realizing it is, as remarked earlier, the strong coupling 
behavior of the remaining gauge field. As the tachyon rolls down toward the 
minimum, the value of the potential $U(T)$ diminishes. When the effective 
gauge coupling $g_{\mathrm{eff}}$ has exceeded 1 according 
to~(\ref{eq:FateB}), the original description in terms of $A_{\mu}^+$ breaks 
down and we must move to the magnetic description by performing the 
electric-magnetic $S$-duality in the D3-$\overline{\mathrm{D}3}$ 
world-volume. In the special case of D3-brane, the dual gauge potential 
$\widetilde{A}^+$ is again the 1-form field with coupling $\sim 
g_{\mathrm{eff}}^{-1}$. Now let us consider a D-string stretched between 
the D3 and the $\overline{\mathrm{D}3}$. Since we are interested in the 
weak string coupling $(g_s\ll 1)$ behavior of the theory, we cannot reliably 
determine the spectrum of the D-string. Here we assume that the ground state 
of the D-string stretched between the brane-antibrane pair is represented by 
a complex tachyonic scalar $\widetilde{T}$ as it \textit{is} for $g_s\gg 1$ 
(the Type IIB $S$-duality). It is important to be aware that $\widetilde{T}$ 
is charged with respect to the dual gauge field $\widetilde{A}_{\mu}^+$ 
(or equivalently, \textit{magnetically} charged under the original gauge 
field $A_{\mu}^+$). To see this, we begin by considering the coupling of 
the original gauge fields $A_{\mu},\ A_{\mu}^{\prime}$ to the NS-NS 2-form 
field $B$ whose source is the F-string. Expanding the Born-Infeld action 
for the D3-brane, we find 
\begin{equation}
\int_{\mathrm{D}3}d^4x\sqrt{-\det(g_{\mu\nu}+F_{\mu\nu}+B_{\mu\nu})} \ \ 
\longrightarrow \ \ \int d^4x\ F_{\mu\nu}B^{\mu\nu}\propto \int_{\mathrm{D}3}
\mathop{*}\limits^4F\wedge B, \label{eq:FateD}
\end{equation}
where $F=dA$ and $\mathop{*}\limits^4$ is the Hodge dual operator in the 
(Minkowskian) 4-dimensional world-volume. Adding the contribution from the 
anti-D3-brane, the coupling becomes 
\begin{equation}
\int_{\aaru^4}\mathop{*}\limits^4F^-\wedge B=\int_{\mathrm{D}3}
\mathop{*}\limits^4F\wedge B-\int_{\overline{\mathrm{D}3}}\mathop{*}\limits^4
F^{\prime}\wedge B. \label{eq:FateE}
\end{equation}
The reason why the field strength appearing in~(\ref{eq:FateE}) is not $F^+$ 
but $F^-$ is simply a matter of convention we have determined 
in~(\ref{eq:FateC}). On the other hand, the nonvanishing field strength $F$ 
on the D3-brane induces the D-string charge through the following 
Chern-Simons coupling 
\begin{equation}
\int_{\mathrm{D}3}e^F\wedge C \ \ \longrightarrow\ \ \int_{\mathrm{D}3}F\wedge
C_{\mathit{2}}\propto \int_{\mathrm{D}3}d^4x\ \epsilon^{\mu\nu\rho\sigma}
F_{\mu\nu}C_{\rho\sigma}, \label{eq:FateF}
\end{equation}
where $C_{\mathit{2}}$ is the R-R 2-form field whose source is the D-string. 
When we add the contribution from the $\overline{\mathrm{D}3}$-brane, we 
should notice that the orientation of the anti-D3-brane is reversed to that 
of the D3-brane. Since~(\ref{eq:FateD}) and (\ref{eq:FateF}) have opposite 
parities under the orientation reversal operation due to the factor 
$\epsilon^{\mu\nu\rho\sigma}$, the total Chern-Simons coupling is found to be 
\begin{equation}
\int_{\mathrm{D}3}F\wedge C_{\mathit{2}}-\int_{\overline{\mathrm{D}3}}
F^{\prime}\wedge C_{\mathit{2}}=\int_{\aaru^4}F^+\wedge C_{\mathit{2}}=
-\int_{\aaru^4}\mathop{*}\limits^4\left(\mathop{*}\limits^4F^+\right)\wedge 
C_{\mathit{2}}, \label{eq:FateG}
\end{equation}
where `$\aaru^4$' has the same orientation as that of the D3-brane and hence 
opposite to that of the $\overline{\mathrm{D}3}$-brane. The 
expression~(\ref{eq:FateE}) tells us that the tachyon $T$ on the stretched 
open F-string is electrically charged with respect to the relative gauge 
field $A_{\mu}^-$, whereas the coupling~(\ref{eq:FateG}) indicates that the 
tachyon $\widetilde{T}$ which arises on the stretched open D-string carries 
electric charge with respect to the dualized overall gauge field 
$\widetilde{A}_{\mu}^+$ which is defined by $d\widetilde{A}^+=
\mathop{*}\limits^4F^+$. In other words, $\widetilde{T}$ carries magnetic 
charge under $A_{\mu}^+$. 
\medskip

Let us return to the condensation process. By now the expectation value of 
the `electric' tachyon $T$ has grown enough to make the effective coupling 
$g_{\mathrm{eff}}$ larger than 1, so that we may change the picture to the 
$S$-dualized magnetic discription. At this moment, the `magnetic' tachyon 
$\widetilde{T}$ from the open D-string remains zero expectation value 
$\langle\widetilde{T}\rangle =0$. But once the magnetic tachyon starts to 
condense toward some nonzero value $\widetilde{T}_0$ which together with 
$T=T_0$ minimizes the `full' tachyon potential $\mathcal{U}(T,
\widetilde{T})$, the \textit{magnetic} overall $U(1)$ is broken and the 
dual gauge field $\widetilde{A}_{\mu}^+$ which minimally couples to 
$\widetilde{T}$ acquires a mass. Though we cannot estimate the magnitude of 
$\widetilde{T}_0$, by assuming $\widetilde{T}_0$ to be of order of string 
scale the dual gauge field $\widetilde{A}_{\mu}^+$ completely disappears from 
the low-lying spectrum as well. In terms of the original gauge field 
$A_{\mu}^+$, the overall $U(1)$ gauge group was confined by the dual Higgs 
mechanism. Consequently, we are left with no world-volume gauge fields 
at all (at least from the low-energy field theoretic standpoint) after the 
tachyons have condensed to the minimum $(T_0,\widetilde{T}_0)$ where 
$\mathcal{U}(T_0,\widetilde{T}_0)=0$. 
\medskip

Although we have considered only the D3-$\overline{\mathrm{D}3}$ system thus 
far, we can extend it to the case of a D$p$-$\overline{\mathrm{D}p}$ pair 
with $p\ge 4$ by the following replacements: 
\begin{eqnarray*}
\mbox{dual gauge field (1-form) }\widetilde{A}_{\mu}^+\ \ &\longrightarrow&
\ \ \mbox{dual ($p-2$)-form potential }\widetilde{A}^+ \\
\mbox{open D-string} \ \ &\longrightarrow& \ \ \mbox{open D($p-2$)-brane} \\
\mbox{particle-like magnetic tachyon }\widetilde{T}\ \ &\longrightarrow& \ \ 
(p-3)\mbox{-dimensional tachyonic object `$\widetilde{T}$'}.
\end{eqnarray*}
Though the generalization of the Higgs mechanism to antisymmetric tensor 
fields of higher rank is possible, we do not precisely know how to deal with 
the magnetically charged \textit{extended} tachyonic (\textit{i.e.} negative 
`tension squared') objects which will appear as boundaries of suspended 
D($p-2$)-branes. But for the special case of $p=4$, both the F-string and the 
D2-brane stretched between the D4-brane and the anti-D4-brane are described 
as the same objects from $M$-theoretical point of view. That is to say, the 
F-string is an M2-brane wrapped on the circle in the eleventh direction while 
the D2-brane is a transverse M2-brane, both of which are suspended between 
an M5-brane and an anti-M5-brane. A single phenomenon in the eleven 
dimensional picture, namely the Higgsing of the 2-form gauge fields living on 
the M5-brane and $\overline{\mathrm{M}5}$-brane induced by the condensation 
of the tachyonic `string' (boundary of an M2-brane), looks like the 
perturbative Higgs mechanism (F1) and the nonperturbative dual Higgs 
mechanism (D2) from the Type IIA viewpoint. Even if the above argument 
justifies the $p=4$ case, it cannot be generalized to other cases 
unfortunately. Anyway, we hope that the confinement mechanism works well 
for arbitrary $p$, involving $p\le 2$ (see~\cite{Conf} for details). 

Further, the confinement mechanism can surely be generalized to the system 
of $N$ D$p$-$\overline{\mathrm{D}p}$ pairs, as suggested in~\cite{Conf}. 
And the generalization to the non-BPS D$p$-brane case can be achieved by 
modding out the D$p$-$\overline{\mathrm{D}p}$ system by $(-1)^{F_L}$, 
where $F_L$ denotes the left-moving part of the spacetime fermion number. 
\medskip

As a piece of evidence that the confinement takes place in reality, 
let us see briefly that the confined electric flux tube can be identified 
with the macroscopic fundamental string. To begin with, we consider a 
D0-$\overline{\mathrm{D}0}$ pair. Just like the relative $U(1)$ gauge field 
$A^-$ in the D$p$-$\overline{\mathrm{D}p}$ case, the transverse scalars 
representing the relative motion of the D-particle and the anti-D-particle 
become massive after the tachyon condensation, so we concentrate on the other 
massless scalars $\phi^i$ $(1\le i\le 9)$ which represent the center of mass 
motion. Note that there are no `dual' objects on the 0-brane world-line. The 
Lagrangian for the scalars is given by 
\begin{equation}
L=-\tau_0U(T)\sqrt{1-\sum_{i=1}^9(\partial_0\phi^i)^2}. \label{eq:FateH}
\end{equation}
The canonical momentum conjugate to $\phi^i$ is 
\begin{equation}
p_i=\frac{\partial L}{\partial\dot{\phi}^i}=\tau_0U(T)\frac{\dot{\phi}^i}{
\sqrt{1-\sum(\dot{\phi}^j)^2}}, \label{eq:FateI}
\end{equation}
and the Hamiltonian is found to be 
\begin{equation}
H=p_i\dot{\phi}^i-L=\sqrt{\sum(p_i)^2+\tau_0^2U(T)^2}. \label{eq:FateJ}
\end{equation}
It looks like the energy of a relativistic particle of mass $\tau_0U(T)$. 
From now on, we focus on a single direction, say $x^1$, and compactify it 
on a circle of radius $R$. Then the momentum~(\ref{eq:FateI}) is discretized 
as \[ p_1=\frac{n}{R}. \]
Because of the assumption $U(T_0)=0$, the Hamiltonian after tachyon 
condensation takes the form 
\begin{equation}
H_0=\frac{|n|}{R}. \label{eq:FateK}
\end{equation}
Since the D0-$\overline{\mathrm{D}0}$ pair disappears after the condensation, 
it is natural to consider that the momentum $p_1$ is carried by a massless 
mode on the closed string. It is surprising that the Hamiltonian remains 
nontrivial even if the Lagrangian itself vanishes at $T=T_0$. Here we 
consider taking the $T$-duality in the $x^1$-direction. Before tachyon 
condensation, the system becomes the D-string--anti-D-string pair, and 
since the transverse scalar $\phi^1$ is altered into a gauge field component 
$A_1$ by the $T$-duality, the canonical momentum $p_1\sim\partial_0\phi^1$ is 
now regarded as an electric flux $F_{01}$ along the D-string world-volume. 
And the value of the flux is quantized as 
\[ F_{01}=\frac{nR}{\ap}, \]
where $\ap/R$ is the radius of the compactified $x^1$-direction in the 
$T$-dualized geometry. 
After tachyon condensation, the D-string and the anti-D-string disappear in 
pairs, while the momentum $n/R$ carried by the closed string is transferred 
to the fundamental closed string wrapped $n$ times on the circle in the 
$x^1$-direction. To sum up, the electric flux on the 
D1-$\overline{\mathrm{D}1}$ world-volume has, through the tachyon 
condensation, changed itself into the macroscopic fundamental closed string 
with the corresponding winding number. Moreover, its energy can be written 
in the form 
\begin{equation}
H_0=|F_{01}|=\frac{|n|}{2\pi\ap}\cdot 2\pi R.
\end{equation}
From this expression, we can see that the tension of the confined flux string 
precisely agrees with that of the $n$ fundamental strings. 

We have seen that the fundamental closed string can be constructed as the 
confined electric flux tube in a rather simplified 
setting. Concerning the problem of producing the fundamental closed strings 
at the closed string vacuum in open string theory, more elaborate arguments 
are given in~\cite{Conf,9061,10240,12081,0101213}. 

\chapter{Superstring Field Theory}

After the bosonic open string field theory was constructed, many attempts to 
apply these techniques to open superstring theory have been made. In this 
chapter, we will discuss cubic superstring field theory by Witten (and its 
variant) and Wess-Zumino-Witten--like superstring field theory by 
Berkovits, and see their applications to the problem of tachyon condensation. 

\section{Superconformal Ghosts and Picture}
In the superstring case, there exist new world-sheet fields $\psi^{\mu}, 
\beta,
\gamma$ due to world-sheet supersymmetry\footnote{We restrict our arguments 
to Type II superstring theory, so the two-dimensional field theory on 
the closed string world-sheet has an $\cN =(1,1)$ superconformal symmetry.}, 
in addition to $X^{\mu}, b, c$ which already exist on the bosonic string 
world-sheet. Though we explicitly write only the holomorphic 
(left-moving) side of 
various world-sheet fields almost everywhere, there is also the 
corresponding antiholomorphic side which will be denoted with tilde like 
$\tilde{\psi}^{\mu}, \tilde{b}$ (except for $\bar{\partial}X^{\mu}(\bar{z})$).
The energy-momentum tensor $T(z)$ is modified in the superstring case as
\begin{eqnarray*}
T^{\mathrm{m}}(z)&=&-\frac{1}{\ap}\partial X^{\mu}\partial X_{\mu}-\frac{1}{2}
\psi^{\mu}\partial\psi_{\mu}, \\
T^{\mathrm{g}}(z)&=&(\partial b)c-2\partial(bc)+(\partial\beta)\gamma-\frac{3}
{2}\partial(\beta\gamma).
\end{eqnarray*}
The supercurrent $G(z)$, the superpartner of the energy-momentum tensor, is 
defined by 
\begin{eqnarray*}
G^{\mathrm{m}}(z)&=&i\sqrt{\frac{2}{\ap}}\psi^{\mu}\partial X_{\mu}(z), \\
G^{\mathrm{g}}(z)&=&-\frac{1}{2}(\partial\beta)c+\frac{3}{2}\partial(\beta c)
-2b\gamma.
\end{eqnarray*}
See section \ref{sec:Conv} or \cite{Pol} for their mode expansions and  
commutation relations among them. According to~\cite{FMS}, we `bosonize' the 
superconformal ghosts $\beta, \gamma$ as 
\begin{eqnarray}
\beta(z)&=&e^{-\phi(z)}\partial\xi(z), \nonumber \\
\gamma(z)&=& \eta(z)e^{\phi(z)}. \label{eq:DA}
\end{eqnarray}
The newly defined fields $\xi, \eta$ are fermionic and $e^{n\phi}$ is also 
defined to be fermionic if $n$ is odd. So the products appearing 
in~(\ref{eq:DA}) are bosonic, just as $\beta, \gamma$ are. And their 
orderings are determined such that the $\beta\gamma$ OPE
\[ \beta(z)\gamma(w)\sim -\frac{1}{z-w} \]
is preserved by the bosonization. In fact, using the following OPE
\[ \xi(z)\eta(w)\sim \frac{1}{z-w} \, , \quad 
\phi(z)\phi(w)\sim -\log (z-w) ,\]
one can verify 
\begin{eqnarray*}
\beta(z)\gamma(w)=e^{-\phi(z)}\partial\xi(z)\eta(w)e^{\phi(w)}&\sim& 
\partial_z\frac{1}{z-w}\exp\left( +\log (z-w)\right):e^{-\phi(z)}e^{\phi(w)}:
\\ &\sim& -\frac{1}{z-w}.
\end{eqnarray*}
As is clear from the definition (\ref{eq:DA}), the $\beta\gamma$ system can 
be bosonized without using zero mode of $\xi$, which we denote 
by $\displaystyle \xi_0
=\oint\frac{dz}{2\pi i}z^{-1}\xi(z)$. We define a ``small" Hilbert space 
to be the one which does not contain the $\xi$ zero mode. In contrast, a 
Hilbert space containing also $\xi_0$ is called a ``large" Hilbert space. 
In~\cite{FMS}, it was shown that it is possible to do all calculations in a 
``small" Hilbert space in the first-quantized superstring theory. In the 
string field theory context too, the $*$ operation 
(gluing) can consistently be defined within the ``small" Hilbert 
space~\cite{WiSFT1}. Here, we collect 
the properties of the fields considered above in Table~\ref{tab:D}.
\begin{table}[htbp]
\begin{center}
	\begin{tabular}{|l||c|c||c|c||c|c||c|c|c||c|c|}
	\hline
	holomorphic field & $\partial X^{\mu}$ & $\psi^{\mu}$ & $b$ & $c$ 
	& $\beta$ & $\gamma$ & $e^{\ell\phi}$ & $\xi$ & $\eta$ & $T$ & $G$ \\
	\hline \hline
	conformal weight $h$ &1&1/2&2&$-1$&3/2&$-1/2$&$-\frac{1}{2}\ell^2-\ell$
	&0&1&2&3/2 \\
	\hline
	ghost number $\gh$ &0&0& $-1$ & $+1$ & $-1$ & $+1$ &0& $-1$ & $+1$ &0&0 \\
	\hline
	picture number $\pic$ &0&0&0&0&0&0& $\ell$ & $+1$ & $-1$ &0&0 \\
	\hline
	world-sheet statistics &B&F&F&F&B&B& ${F (\ell:\mathrm{odd}) \atop B 
	(\ell:\mathrm{even})}$  &F&F&B&F \\
	\hline
	\end{tabular}
	\caption{Some properties of the fields on an $\cN=1$ superstring 
	 world-sheet.}
	\label{tab:D}
\end{center}
\end{table}

Next we explain the concept of `picture'~\cite{FMS}. The $\phi$ current is 
defined to be 
\begin{equation}
j^{\phi}=-\partial\phi(z) \label{eq:DC}
\end{equation}
so that $e^{\ell\phi}$ has $\phi$-charge $\ell$. A vertex operator of 
$\phi$-charge $\ell$ is said to be in the $\ell$-\textit{picture}. The 
pictures of $\xi,\eta$ are conventionally defined to take $\beta,\gamma$ 
ghosts in the 0-picture. Every state has some (actually, an infinitely 
many countable 
number of) equivalent representations in terms of 
vertex operators with different 
picture numbers. For example, the massless open string vector state in the 
Neveu-Schwarz sector 
\begin{equation}
A_{\mu}(k)\psi^{\mu}_{-1/2}|0;k\rangle_{NS} \label{eq:DD}
\end{equation}
is equivalently represented by the following vertex operators 
\begin{eqnarray}
\cV^{(-1)}(z)&=&A_{\mu}(k)\psi^{\mu}(z)e^{ikX(z)}c(z)e^{-\phi(z)}, 
\label{eq:DE} \\ \cV^{(0)}(z)&=&\frac{A_{\mu}(k)}{\sqrt{2\ap}}(i\partial_t
X^{\mu}+2\ap k_{\nu}\psi^{\nu}\psi^{\mu})e^{ikX}c(z), \label{eq:DF}
\end{eqnarray}
where the superscript ($\ell$) on $\cV$ denotes the picture number. The 
fact that the vertex operators of the superstring have a variety of forms is 
accounted for by the following argument: In bosonic string theory, consider 
an $n$-point amplitude on a disk, 
\begin{equation}
\cA_n=\frac{1}{g_o^2}\int\frac{\cD X\cD g}{V_{\mathrm{diff\times Weyl}}}
\exp(-\cS^{\mathrm{m}})\prod_{i=1}^n\int_{\partial\Sigma}d\sigma_i
\sqrt{g(\sigma_i)}\cV_i(\sigma_i). \label{eq:DG}
\end{equation}
We can use the degrees of freedom of diffeomorphism$\times$Weyl 
transformations on the world-sheet to fix the world-sheet metric $g(\sigma)$ 
to some specific fiducial metric $\hat{g}(\sigma)$, eliminating the $\cD g$ 
integral. But there still remain some symmetries that are not fixed by the 
choice of metric, which we call the \textit{conformal Killing group}. The 
conformal Killing group of the disk is known to $PSL(2,\aaru)=SL(2,\aaru)/
\zetto_2$. To remove these degrees of freedom completely, we fix three vertex 
operators to the specific positions $\hat{\sigma}_i$ 
on the disk boundary $\partial\Sigma$. 
If we carry out the above procedures using the Faddeev-Popov method, we reach 
\begin{eqnarray}
\cA_n&=& \frac{1}{g_o^2}\int\cD X\cD b\cD c\exp(-\cS^{\mathrm{m}}
-\cS^{\mathrm{g}}) \label{eq:DH} \\ & &\times \left(\prod_{i=1}^3\sqrt{\hat{g}
(\hat{\sigma}_i)}c(\hat{\sigma}_i)\cV_i(\hat{\sigma}_i)\right)\left(
\prod_{i=4}^n\int_{\partial\Sigma}d\sigma_i\sqrt{\hat{g}(\sigma_i)}\cV_i
(\sigma_i)\right).
\end{eqnarray}
On the way, fermionic $b,c$ ghosts were introduced in terms of which the 
Faddeev-Popov determinant was rewritten. Note that each fixed vertex 
operator ($i=1,2,3$) is accompanied by the $c$-ghost, whereas the unfixed 
vertex operators ($i=4,\ldots ,n$) are integrated without $c$-ghosts. In the 
case of superstring, its world-sheet coordinates include the anticommuting 
variables $\theta_i,\bar{\theta}_i$ in the superspace representation. Two of 
them are fixed in a similar way, 
resulting in the superconformal ghosts $\beta,\gamma$. 
In the bosonic case, $c$ was inserted to fill in three $c$ zero modes 
$c_1,c_0,c_{-1}$, making the fermionic integral nonvanishing ($\int dc\, c=1,
\int dc\, 1=0$). On the other hand, in the superstring case $\gamma$ ghost is 
bosonic, hence the two insertions must be of the form $\delta(\gamma)$. 
Fortunately, $\delta(\gamma)$ is easily bosonized, 
\begin{equation}
\delta(\gamma)=e^{-\phi}. \label{eq:DHa}
\end{equation}
Since the original (before introducing the Faddeev-Popov ghosts) vertex 
operators are constructed purely from matters ($X^{\mu},\psi^{\mu}$), the two 
$\theta$-fixed vertex operators are in the $-1$-picture due to~(\ref{eq:DHa}),
and the remaining $(n-2)$ vertex operators are in the 0-picture: This is the 
origin of the vertex operators of different picture numbers. As a result, the 
insertions have total $\phi$-charge $-2$, which is actually needed to obtain 
a nonzero amplitude. Let us see this explicitly. The $\phi$-current is 
defined in~(\ref{eq:DC}) to be $j^{\phi}=-\partial\phi$. Consider the 
following amplitude 
\begin{equation}
\left\langle\oint_C\frac{dz}{2\pi i}j^{\phi}(z)(\mbox{some insertions})
\right\rangle_{S^2}=(\mbox{total $\phi$-charge})\biggl\langle(\mbox{some 
insertions})\biggr\rangle_{S^2}, \label{eq:DI}
\end{equation}
where the contour $C$ encircles \textit{all} of the inserted vertex operators 
counterclockwise. Suppose that we are treating the holomorphic side of the 
closed string or adopting the doubling trick in the open string theory to be 
able to consider the sphere amplitude. Then we can move the contour $C$ to 
shrink around the point at infinity. But we must perform a conformal 
transformation $z\to u=1/z$ at that time. From the OPE 
\begin{eqnarray*}
T^{\phi}(z)j^{\phi}(0)&=&\left(-\frac{1}{2}\partial\phi\partial\phi-\partial^2
\phi\right)_z(-\partial\phi)_0 \\ &\sim& \frac{2}{z^3}+\frac{1}{z^2}j^{\phi}
(0)+\frac{1}{z}\partial j^{\phi}(0),
\end{eqnarray*}
we find that the current $j^{\phi}$ is not a primary field. Accordingly, it 
transforms under the conformal transformation $z\to u$ as 
\begin{equation}
(\partial_uz)j^{\phi}(z)=j^{\phi}(u)-\frac{\partial_u^2z}{\partial_uz}.
\label{eq:DJ}
\end{equation}
Substituting $u=1/z$, the inhomogeneous term becomes $\displaystyle 
-\frac{\partial_u^2z}{\partial_uz}=\frac{2}{u}$. Integrating both side 
of~(\ref{eq:DJ}) with respect to $u$ along the contour $D$ which encircles 
$u=0$ \textit{clockwise}, 
\begin{equation}
\oint_D\frac{dz}{du}\frac{du}{2\pi i}j^{\phi}(z(u))=\oint_D\frac{du}{2\pi i}
j^{\phi}(u)+2\oint_D\frac{du}{2\pi i}\frac{1}{u}=0-2. \label{eq:DK}
\end{equation}
The last equation holds because there are no operators inside the contour $D$ 
and $j^{\phi}$ is holomorphic (the 1st term), and $D$ is clockwise (the 2nd 
term). In the most left hand side we can convert the variable of integration 
from $u$ to $z$. But then the contour must be seen from the point of view of 
$z$, which encircles $z=\infty$ \textit{counterclockwise}. It is nothing but 
the contour $C$ appearing in~(\ref{eq:DI}). Composing (\ref{eq:DI}) and 
(\ref{eq:DK}), we conclude that
\[(\mbox{total $\phi$-charge of the inserted vertex operators})=-2 \]
must be true if we are to have nonvanishing amplitudes on the sphere.
\medskip

Since we have the following state-operator correspondences
\begin{equation}
\left.
	\begin{array}{lc}
	\mbox{tachyon} & |\Omega\rangle_{NS}\cong c e^{-\phi}, \\
	\mbox{massless NS} & \psi^{\mu}_{-1/2}|\Omega\rangle_{NS}
	\cong \psi^{\mu} ce^{-\phi}, \\
	\mbox{massless R} & |\vec{s}\rangle_R\cong ce^{-\phi/2}\Theta_{\vec{s}},
	\end{array}
\right. \label{eq:DL}
\end{equation}
we define the `natural' picture as 
\[
\begin{array}{lc}
\mbox{Neveu-Schwarz sector} & -1 \\
\mbox{Ramond sector} & -1/2.
\end{array}
\]
In (\ref{eq:DL}), $\Theta_{\vec{s}}$ is the \textit{spin field} $\displaystyle
\exp\left[i\sum_{a=0}^4s_aH^a\right]$, where $s_a=\pm\frac{1}{2}$ and $H^a$ 
are the bosonized form of $\psi^{\mu}$'s
\begin{equation}
\frac{1}{\sqrt{2}}(\pm\psi^0+\psi^1)=e^{\pm iH^0}\, , \quad 
\frac{1}{\sqrt{2}}(
\psi^{2a}\pm i\psi^{2a+1})=e^{\pm iH^a} \, (a=1,2,3,4). \label{eq:DM}
\end{equation}
We will focus on the Neveu-Schwarz sector for a while. If we think of 
$-1$-picture vertex operators as natural and fundamental representation, the 
vertex operators in the 0-picture can be obtained by acting on the 
$-1$-picture vertex operators with the following 
`picture-changing operator'\footnote{We use the unconventional symbol $\cX$ 
for the picture-changing operator to distinguish it from the embeddings 
$X^{\mu}$ of the world-sheet. Usually picture-changing operator is also 
denoted by $X$.} 
\begin{eqnarray}
\cX(z)&\equiv&\{Q_B,\xi(z)\}=\oint\frac{d\zeta}{2\pi i}j_B(\zeta)\xi(z) 
\nonumber \\ &=& c\partial\xi +e^{\phi}G^{\mathrm{m}}+e^{2\phi}b\partial\eta
+\partial (e^{2\phi}b\eta), \label{eq:DN}
\end{eqnarray}
where $j_B(z)$ is the BRST current
\begin{eqnarray}
j_B(z)&=&cT^{\mathrm{m}}+\gamma G^{\mathrm{m}}+\frac{1}{2}(cT^{\mathrm{g}}
+\gamma G^{\mathrm{g}}) \nonumber \\
&=& c(T^{\mathrm{m}}+T^{\eta\xi}+T^{\phi})+\eta e^{\phi}G^{\mathrm{m}}+
bc\partial c-\eta\partial\eta e^{2\phi}b , \label{eq:DO} \\
T^{\eta\xi}&=&(\partial\xi)\eta \> , \quad T^{\phi}=-\frac{1}{2}\partial\phi
\partial\phi-\partial^2\phi. \nonumber
\end{eqnarray}
Though the picture-changing operator may look like a BRST-exact object at 
first sight, it is not a trivial element because 
the field $\xi(z)$ does not exist 
in the ``small" Hilbert space. But $\cX$ itself \textit{does} exist in the 
``small" Hilbert space. This fact can easily be seen from the 
expression~(\ref{eq:DO}) of the BRST current which shows that $\cX$ 
depends on $\xi$ only through its derivative. The properties of $\cX$ are 
\[\mbox{conformal weight 0, \quad ghost number 0, \quad picture number } 
+1. \]
General $n$-point tree-level amplitudes are obtained as follows. Take all 
vertex operators in the natural $-1$-picture, and insert $(n-2)$ 
picture-changing operators as well on the sphere\footnote{As mentioned before,
we use the doubling trick in the open string context.} so that the whole 
operator has $-2$-picture. We can show that the resulting amplitude is 
independent of the positions at which the picture-changing operators are 
inserted if all of the external states are on-shell, \textit{i.e.} 
represented by BRST invariant vertex operators. For that purpose, consider
\begin{equation}
\cA\equiv\left\langle\cX(z_2)\prod_{i=1}^n\cV_i\right\rangle-\left\langle
\cX(z_1)\prod_{i=1}^n\cV_i\right\rangle=\left\langle\left\{Q_B,
\int_{z_1}^{z_2}dz\, \partial\xi(z)\right\}\prod_i\cV_i\right\rangle. 
\label{eq:DP}
\end{equation}
Though $\cX=\{Q_B,\xi\}$ is not a BRST commutator, the difference $\cX(z_1)
-\cX(z_2)$ \textit{can be} written as a commutator even in the ``small" 
Hilbert space. In this case, we can express $Q_B$ as an integral of $j_B$ 
and freely deform the integration contour. By the usual contour argument, 
\begin{eqnarray*}
\cA&=&\left\langle\int\limits_{z_1}^{z_2}dz\oint\limits_C\frac{d\zeta}{2\pi i}
j_B(\zeta)\partial\xi(z)\prod_i\cV_i\right\rangle \\
&=&+\left\langle\int\limits_{z_1}^{z_2}dz\,\partial\xi(z)\oint\limits_{-C}
\frac{d\zeta}{2\pi i}j_B(\zeta)\prod_i\cV_i\right\rangle \\
&=&\left\langle\int\limits_{z_1}^{z_2}dz\,\partial\xi(z)\sum_{i=1}^n\cV_1
\cdots \{Q_B,\cV_i\}\cdots\cV_n\right\rangle=0,
\end{eqnarray*}
where we used the BRST invariance of the vertex operators 
in the last equality, and the original 
contour $C$ encircles the position $z$ of $\partial\xi$. In the third line, 
$C$ was decomposed into small circles encircling each vertex operator. 
$\cA=0$ shows that the amplitude does not depend on the position where the 
picture-changing operator is inserted. The same is true when more than one 
picture-changing operators are inserted because $\cX=\{Q_B,\xi\}$ is a 
manifestly BRST invariant object 
just like the on-shell vertex operators. Note that 
the conclusion holds only if all of the external states are on-shell. Once we 
abandon this assumption to go to the off-shell calculations in string field 
theory context, we must put picture-changing operators on the common 
interaction point, namely the center of the disk representing the interaction.
When we take the limit where a picture-changing operator coincides with a 
vertex operator, the picture-changing operation takes place and the 
vertex operator changes its picture from $-1$ to 0. 
Since there is no obstacles to prevent more picture-changing 
operators from acting on the 0-picture vertex operators, we can 
construct still higher picture vertex operators at our own will. 
If the picture-raising 
operation is extended this way, we want to have the `picture-lowering' 
operation as well. 
It can be achieved by acting the following 
`inverse picture-changing operator' 
\begin{equation}
Y=c\ \partial\xi \ e^{-2\phi} \label{eq:DQ}
\end{equation}
on the vertex operators.
It is the inverse operator of $\cX$ in the sense that
\begin{equation}
\lim_{z\to w}\cX(z)Y(w)=\lim_{z\to w}Y(z)\cX(w)=1. \label{eq:DR}
\end{equation}
By making use of $\cX$ and $Y$, we will arrive at vertex operators of 
arbitrary integer picture number. So far, we have concentrated on the 
Neveu-Schwarz sector, but we can similarly define the picture-changing 
operations for the Ramond sector vertex operators. They differ from the 
Neveu-Schwarz vertex operators in that the 
picture numbers of the Ramond sector vertex operators are half-integer valued.

\section{Witten's Cubic Superstring Field Theory 
\\ and its Problems}\label{sec:cubicsuper}
The cubic open superstring field theory action proposed by Witten 
in~\cite{WiSFT1,WiSFT2} is a straightforward extension of the cubic bosonic 
open string field theory action introduced in the last chapter. However, it is
rather complicated due to the existence of the Ramond sector states and the 
concept of picture.
\medskip

We first consider the Neveu-Schwarz sector and take the string field $A$ to 
have the ghost number $+1$ and the picture number $-1$ (natural picture). If 
we assume the same action as that of the bosonic cubic open string field 
theory
\begin{equation}
S_?=\int\left(A*Q_BA+\frac{2}{3}A*A*A\right) \label{eq:DS}
\end{equation}
with exactly the same definitions of $\int$ and $*$, the second term turns 
out to vanish because it has the wrong value $-3$ of $\phi$-charge. It can 
easily be remedied by inserting the picture-changing operator $\cX$ only in 
the second term. For further extension, however, we purposely modify both the 
$*$ and $\int$ operations. Let's define the following new operations, 
\begin{eqnarray}
A\star B&=&\cX (A*B) , \nonumber \\
\oint A&=&\int Y A, \label{eq:DT}
\end{eqnarray}
where $\cX$ and $Y$ are inserted at the string midpoint $\sigma=\pi/2$. If we 
use these symbols, the action for the Neveu-Schwarz sector can be written as 
\begin{equation}
S_{NS}=\oint \left(A\star Q_BA+\frac{2g_o}{3}A\star A\star A\right). 
\label{eq:DU}
\end{equation}
Moreover, in order for the gauge parameter to form a closed subalgebra under 
the `star-product', we \textit{must} use the $\star$ operation defined above.
A gauge transformation of a string field $A$ is 
\[ \delta A=Q_B \Lambda +\ldots , \]
where $\Lambda$ is a gauge parameter and $\ldots$ represents the nonlinear 
terms. Since the string field $A$ has $(\gh =+1 , \pic =-1)$ and $Q_B$ has 
$(\gh =+1 , \pic =0)$, the gauge parameter $\Lambda$ has $(\gh =0 , 
\pic =-1)$. If we want the product of two gauge parameters $\Lambda_1 , 
\Lambda_2$ to have the same ghost and picture number as that of 
each of the original gauge 
parameters, we must assign $(\gh =0 , \pic =+1)$ to the `star'-product. 
That is just the property of $\star =\cX\cdot*$. In accordance with this, 
the `integration' operation should also be modified to $\oint=\int Y$. 
\medskip

In any case, a gauge invariant cubic superstring field theory action for the 
Neveu-Schwarz sector 
was constructed, \textit{at least formally}. Next we take the 
Ramond sector into account. As the product of two Ramond sector gauge 
parameters is thought to be in the Neveu-Schwarz sector, we must consider the 
combined Ramond-Neveu-Schwarz string field. We denote by $M=(A,\psi)$ a 
combined state, where $A$ is a Neveu-Schwarz state and $\psi$ is a Ramond 
state. From the state-operator correspondence~(\ref{eq:DL}), we assign 
\[ \left(\gh =+1 , \pic =-\frac{1}{2}\right) \]
to the Ramond sector state. We define the product of two string fields 
$M_1,M_2$ of the combined system by
\begin{equation}
M_1\hat{\star}M_2=(A_1,\psi_1)\hat{\star}(A_2,\psi_2)=\Bigl(A_1\star A_2+
\psi_1*\psi_2 \> , \> A_1\star\psi_2+\psi_1\star A_2 \Bigr), \label{eq:DV}
\end{equation}
where $*$ is the usual $*$-product, $\star$ is the modified product defined 
in~(\ref{eq:DT}) and we denoted the new product by $\hat{\star}$\footnote{
Under the $\hat{\star}$-product, the set of gauge parameters for the combined 
system certainly forms a closed subalgebra.}. We can easily see that the 
product $M_1\hat{\star}M_2$ has the ghost number $+2$ and picture number $(-1,
-1/2)$ for (Neveu-Schwarz, Ramond) state. And a new integration operation for 
the combined system is defined simply by
\begin{equation}
\int\!\!\!\int (A,\psi)=\oint A, \label{eq:DW}
\end{equation}
that is, we take out only the Neveu-Schwarz state\footnote{The integral of a 
Ramond sector string field must be zero because of Lorentz invariance.} and 
integrate it using $\oint$ defined in~(\ref{eq:DT}). Note that all of the 
axioms we saw in section~\ref{sec:cubic} are obeyed by the products $\star, 
\hat{\star}$ and integrations $\oint,\int\!\!\int$. By putting these objects 
and operations together, we can write down a gauge invariant action for the 
combined Ramond-Neveu-Schwarz system, 
\begin{equation}
S_{RNS}=\int\!\!\!\int\left(M\hat{\star}Q_BM+\frac{2g_o}{3}M\hat{\star}M
\hat{\star}M\right). \label{eq:DX}
\end{equation}
This is invariant under the following gauge transformation
\[ \delta M=Q_B \Lambda+g_o(M\hat{\star}\Lambda -\Lambda\hat{\star}M), \]
though the proof is quite formal and actually has some problems, as shown 
later. We can rewrite the action~(\ref{eq:DX}) 
in terms of $*$-product and $\int$, the result being 
\begin{equation}
S_{RNS}=\int\left(A*Q_BA+Y\psi*Q_B\psi+\frac{2g_o}{3}\cX A*A*A+2g_oA*\psi 
*\psi\right). \label{eq:DY}
\end{equation}
If we set to zero the Ramond sector string field $\psi$, it correctly 
reproduces the action~(\ref{eq:DU}) of only the Neveu-Schwarz sector. The 
quadratic part of the action~(\ref{eq:DY}) can be altered into the standard 
structure if we impose the Feynman-Siegel gauge condition $b_0A=b_0\psi =0$. 
For the Neveu-Schwarz part, we can set $A=b_0A^{\prime}$ in the Feynman-Siegel 
gauge, so
\[ S_{NS}^{\mathrm{quad}}=\int A*Q_BA=\langle A^{\prime}|b_0Q_Bb_0|A^{\prime}
\rangle=\langle A^{\prime}|\{b_0,Q_B\}b_0|A^{\prime}\rangle=\langle A^{\prime}
|L_0^{\mathrm{tot}}b_0|A^{\prime}\rangle . \]
If we extract the ghost zero modes from $A^{\prime}$ as $|A^{\prime}\rangle=
|\tilde{A}\rangle\otimes c_0|\downarrow\rangle$, then 
\begin{equation}
S_{NS}^{\mathrm{quad}}=\langle\downarrow|c_0b_0c_0|\downarrow\rangle\langle
\tilde{A}|\tilde{L}_0^{\mathrm{tot}}|\tilde{A}\rangle, \label{eq:DZ}
\end{equation}
where tilded objects do not include ghost zero modes $c_0,b_0$ at all. 
Since the factor $\langle\downarrow|c_0b_0c_0|\downarrow\rangle$ 
simply gives 1, this 
is the standard form of the gauge fixed action. For the Ramond part, we must 
appropriately handle the zero modes of $\beta,\gamma$ as well. At the 
linearized level, it can be shown that we can carry out the gauge 
transformation $\delta\psi=Q_B\chi$ such that the transformed $\psi$ 
satisfies the gauge conditions 
\[ b_0\psi=\beta_0\psi=0 . \]
It was shown in~\cite{WiSFT2} that under these gauge conditions the quadratic 
action for the Ramond sector becomes 
\begin{equation}
S_R^{\mathrm{quad}}=\langle\tilde{\psi}|\tilde{G}_0^{\mathrm{tot}}|
\tilde{\psi}\rangle, \label{eq:EA}
\end{equation}
where $\tilde{\psi}, \tilde{G}_0^{\mathrm{tot}}$ do not include $b,c,\beta,
\gamma$ zero modes. Since its derivation is rather complicated, we leave it to
ref.~\cite{WiSFT2}, but we point out an essential part. While the 
`Klein-Gordon operator' $L_0^{\mathrm{tot}}$ appeared through 
$L_0^{\mathrm{tot}}=\{Q_B,b_0\}$ in the Neveu-Schwarz sector, the 
`Dirac operator' $G_0^{\mathrm{tot}}$ arises as $G_0^{\mathrm{tot}}=[Q_B,
\beta_0]$ in the Ramond sector, which is suitable for the spacetime fermions.
\medskip

So far, we have seen the Witten's construction of superstring field theory. 
The action (\ref{eq:DX}) or (\ref{eq:DY}) has the formal gauge invariance and 
is a natural extension of the bosonic string field theory cubic action. 
It also reproduces the correct propagators for the Neveu-Schwarz and Ramond 
sector fields. Important roles in constructing the theory 
are played by the 
picture-changing operators $\cX$ and $Y$ among other things. These are 
necessary for the construction of the nonvanishing Chern-Simons-like cubic 
action in superstring theory. However, it was pointed out in~\cite{Wendt} 
that just these picture-changing operators bring about various problems. 
Let's see them in some detail. 

\begin{figure}[htbp]
	\begin{center}
	\includegraphics{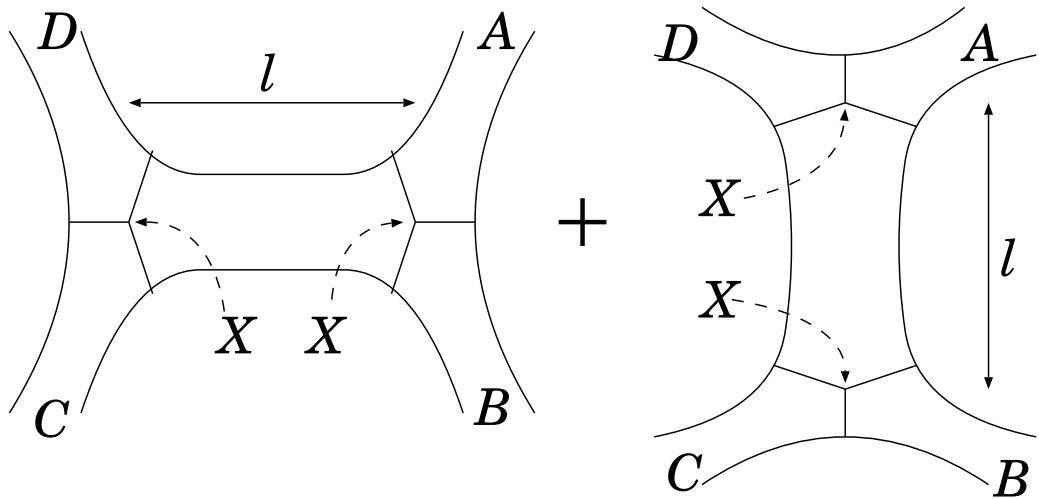}
	\end{center}
	\caption{The two diagrams representing the 4-boson scattering.}
	\label{fig:AA}
\end{figure}

\begin{figure}[htbp]
	\begin{center}
	\includegraphics{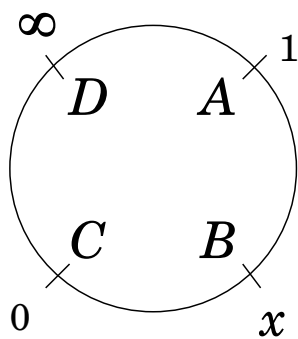}
	\end{center}
	\caption{4-point disk amplitude.}
	\label{fig:AB}
\end{figure}

The 4-point amplitudes for massless states are studied in~\cite{Wendt}. Since 
there are no quartic interaction terms in the Witten's action, a 4-point 
amplitude consists of two field theory diagrams labelled as $s$- and 
$t$-channel. We consider here the four-boson (\textit{i.e.} Neveu-Schwarz 
sector) amplitude. As $A*A*\psi$ interaction is absent, the intermediate 
states are also in the Neveu-Schwarz sector. The diagrams we consider are 
shown in Figure~\ref{fig:AA}. 
Note that the picture-changing operator $\cX$ is inserted at each interaction 
point. Through a suitable conformal transformation these two diagrams together
give a conventional 4-point disk amplitude indicated in Figure~\ref{fig:AB}. 
The positions of three vertex operators corresponding to the external states 
C,A,D are fixed to $0,1,\infty$ respectively 
by using degrees of freedom of the conformal Killing group 
$PSL(2,\aaru)$. The integration over the `propagation length' $\ell$ in 
Figure~\ref{fig:AA} is taken the place of by 
the $x$-integration $\int_0^1dx$ in 
Figure~\ref{fig:AB}. At $\ell =0$, however, two inserted picture-changing 
operators collide. Let us see the operator product of two $\cX$'s. If we move 
to the ``large" Hilbert space, this can be written as 
\begin{equation}
\cX(z)\cX(0)=\oint_C\frac{d\zeta}{2\pi i}j_B(\zeta)\xi(z)\cX(0), 
\label{eq:cXcX}
\end{equation} 
where $C$ encircles only $z$. But since $\cX$ is BRST-invariant, the contour 
can be deformed into the one which encircles 
both $z$ and 0. Then we can take $\xi\cX$ OPE in 
advance, which gives
\[ \xi(z)\cX(0)\sim-\frac{1}{z^2}be^{2\phi}(0)+\cO(z^{-1}). \]
To find the OPE~(\ref{eq:cXcX}), 
we need to take operator product of $j_B(\zeta)$ 
with $be^{2\phi}(0)$, and to pick 
up a simple pole $\zeta^{-1}$ by the contour integration. Though the 
resulting operator may take a complicated form, it certainly is a nonsingular 
operator. Therefore we find the most singular term in $\cX(z)\cX(0)$ behaves 
as $z^{-2}\times (\mbox{regular operator})$. Since $\cX(z)\cX(w)$ diverges as 
$(z-w)^{-2}$ in the limit $z\to w$, this collision 
poses a problem. The 
explicit expression of the divergent amplitude will be as 
follows. $\ell=0$ 
corresponds to $x=\frac{1}{2}$, so that the four-point amplitude 
regularized by an infinitesimal quantity $\epsilon$ 
becomes~\cite{Wendt}
\begin{equation}
\cA_s+\cA_t=\int\limits_0^1dx\left\langle\cV_{-1}^D(\infty)\cV_{-1}^A(1)
\cV_0^B(x)\cV_0^C(0)
\right\rangle-\left(\frac{\pi}{2\epsilon}+\cO(\epsilon)
\right)F_{DABC}\left(\frac{1}{2}\right) \label{eq:EC}
\end{equation}
with
\[ F_{DABC}(y)=\left\langle\cV_{-1}^D(\infty)\cV_{-1}^A(1)\cV_{-1}^B(y)
\cV_{-1}^C(0)\right\rangle_{\mathrm{matter}}, \]
where the subscripts represent the picture number. The first term 
in~(\ref{eq:EC}) coincides with the result of the first-quantized string 
theory. The second term, arising from the contact interaction at $\ell=0$, 
diverges in the limit $\epsilon\to 0$. Furthermore, the infinitesimal gauge 
invariance is violated for the same reason: Under the gauge transformation
\[ \delta A=Q_B\lambda +g_o\cX (A*\lambda -\lambda *A) \]
of the Neveu-Schwarz string field $A$, the variation of the second term 
in~(\ref{eq:DU}) includes the collision of two $\cX$'s. Because of this 
divergence, the gauge invariance at order $g_o^2$ is problematic. We can 
avoid this problem at order $g_o^2$ by introducing a four-boson counterterm 
into the action. After regularizing as 
\[ \delta A=Q_B\lambda +g_oe^{-\epsilon L_0}\cX (A*\lambda -\lambda *A)
+(\mbox{appropriate } \cO(g_o^2) \mbox{ term}), \]
the gauge variation of the following counterterm\footnote{Notice that the 
inclusion of $\xi,\tilde{\xi}$(right-mover) suggests that we should work in 
the ``large" Hilbert space, though we will not pursue it here.}
\begin{equation}
S_4=g_o^2\int(A*A)*\left\{\xi\left(\frac{\pi}{2}\right)-\tilde{\xi}\left(
\frac{\pi}{2}\right)\right\}e^{-\epsilon L_0}\cX\left(\frac{\pi}{2}\right)
(A*A) \label{eq:ED}
\end{equation}
cancels the variation of the cubic term at order $g_o^2$, \textit{if} $A$ is 
BRST invariant. Fortunately, this counterterm also exactly cancels both 
the divergent and finite part of the contact term in~(\ref{eq:EC}). Though 
this method may seem to solve the problems 
at order $g_o^2$, more counterterms 
are needed to restore gauge invariance to higher orders in $g_o$. And since 
the contact interaction and its divergence already exist at the 
tree-level, the counterterm must be infinite even in the classical action, 
making the action ill-defined unless it is regularized in some way. 
\medskip

Aside from the problems mentioned above, this formalism seems not to be 
suitable for the study of tachyon condensation. We now describe the attempts 
to find the `closed string vacuum' in the cubic superstring field theory .

First of all, we must seek the `tachyonic' state. Since we are discussing 
superstring theory, usually the tachyon state is projected out by the GSO 
projection operator $\frac{1+e^{\pi iF}}{2}$, where $F$ measures the 
world-sheet fermion number\footnote{The word `world-sheet fermion number' is 
imprecise because we assign $e^{\pi iF}=+1$ to anticommuting ghosts $b,c$ and 
$e^{\pi iF}=-1$ to commuting ghosts $\beta,\gamma$. It is more appropriate to 
call $F$ the world-sheet spinor number since $b,c$ have integer spins 
(weights) while $\beta,\gamma$ have half-integer spins.} of a state. This is 
true for Type II 
closed superstrings and open superstrings on \textit{BPS} D-branes. 
But if we consider non-BPS D-branes or D-brane anti D-brane 
system~\cite{Sen},
GSO($-$) sectors, in which the open string tachyon lives, appear in the 
theory. In fact, such tachyonic modes bring about instability of 
configurations with no spacetime supersymmetry. Here we consider a single 
non-BPS D-brane, on which both GSO($+$) and GSO($-$) sectors are present. 
In this sense, we may say that the open superstring system on a non-BPS 
D-brane resembles the bosonic open string system on a bosonic D-brane. 
Since we are looking for Lorentz invariant vacua, we set to zero all Ramond 
states (spacetime spinors). Note that $e^{\pi iF}$ must be multiplicatively 
conserved at every string vertex. This means that the GSO($-$) Neveu-Schwarz 
string field $A_-$ must appear in pairs in the cubic interaction term. Let's 
consider level 0 truncation, namely, keep only the tachyon state. Since the 
tachyon state lives in the GSO($-$) sector, the string field becomes\footnote{
Additional structures will be given later, which do not matter here.}
\begin{equation}
A_+=0 \> , \quad A_-(z)=t\cdot ce^{-\phi}(z), \label{eq:EE}
\end{equation}
where $t$ represents the tachyon field. Though we used the symbol $\phi$ for 
the tachyon field in chapter~\ref{ch:sft}, $\phi$ is already reserved for 
bosonization of the superconformal ghosts, so we cannot help using another 
symbol $t$. Because of 
the conservation of $e^{\pi iF}$, the cubic action 
for the Neveu-Schwarz sector can include only terms of the form
\begin{equation}
(\mathrm{a}) A_+*Q_BA_+\quad(\mathrm{b}) A_-*Q_BA_-\quad(\mathrm{c})
A_+*A_+*A_+\quad(\mathrm{d}) A_+*A_-*A_- . \label{eq:EEa}
\end{equation}
In particular, $A_-^3$ is absent. In the above list, all but $(\mathrm{b})$ 
vanish due to $A_+=0$. Then it is clear that the level 0 truncated tachyon 
potential is purely quadratic and has no minimum. It is disappointing because 
in bosonic string field theory we saw that 
even at level 0 approximation the minimum value 
of the tachyon potential 
produced about 68\% of the expected answer. One may have a faint hope that the
higher level fields could change the shape of the effective potential 
drastically into the `double well' 
form which has two minima at $t=\pm t_0$. 
Such a calculation was carried out 
in~\cite{4112}, where the multiscalar 
potential was calculated up to level (2,4) and the effective tachyon potential
up to level (1,2). We simply quote the results without details about actual 
calculations. At level $(\frac{1}{2},1)$ and (1,2), the fields other than the 
tachyon $t$ can be integrated out exactly to give the effective tachyon 
potentials
\begin{eqnarray}
U_{(\frac{1}{2},1)}(t)&=&-\frac{t^2}{2}-\frac{81}{32}t^4, \label{eq:EF} \\
U_{(1,2)}(t)&=&-\frac{t^2}{2}-\frac{81}{32}\frac{t^4}{1-16t^2}. \label{eq:EG}
\end{eqnarray}
Since the sign of the $t^4$ term is minus, the 
effective potential became 
even steeper when we took higher level fields into account 
and have no minimum
as ever. To make matters worse, even singularities have 
arisen in the level 
(1,2) potential. Seeing the changes of the effective potential, we expect that
the inclusion of higher modes would never give rise to any minimum of the 
effective potential, which breaks the hope given a little while ago. Here we 
state an empirical law that the contributions from the higher level modes do 
not change the qualitative behavior of the effective potential obtained 
at the lowest 
level, and at most make the bottom of the potential deeper. The results 
that support this statement were obtained in the bosonic D-brane case and just 
above, and will be obtained later in this chapter.
This observation is already pointed out in~\cite{BSZ}.

Anyway, the Witten's original proposal for superstring field theory seems to 
be rejected by the calculation of the tachyon potential as well as the 
problems of the contact term divergences.

\section{Modified Cubic Superstring Field Theory}\label{sec:yy}
To overcome the contact term divergences, a slightly different formalism was 
proposed~\cite{PTY,Med,Zub} within the framework of cubic superstring field 
theory. Main ideas are that we put the string fields in the 0-picture and use 
the `double-step' picture-changing operator. Again we will concentrate on the 
Neveu-Schwarz sector. Then the action takes the form
\begin{equation}
S_{NS}=-\frac{1}{g_o^2}\left(\frac{1}{2\ap}\int Y_{-2}\cA *Q_B\cA +\frac{1}{3}
\int Y_{-2}\cA *\cA *\cA\right), \label{eq:EH}
\end{equation}
where $\int$ and $*$ are the most basic ones without any insertion of 
the picture-changing operators. 
$Y_{-2}$ is a \textit{double-step inverse 
picture-changing operator} which satisfies
\begin{equation}
\lim_{z\to w}Y_{-2}(z)\cX(w)=\lim_{z\to w}\cX(z)Y_{-2}(w)=Y(w), \label{eq:EI}
\end{equation}
and inserted at the string midpoint. And $\cA$ is the Neveu-Schwarz string 
field in the 0-picture. In this formalism we regard these 0-picture states as 
most fundamental, so that the $-1$-picture string field $A$ in the last 
section can be obtained by acting $Y$ on $\cA$,
\begin{equation}
A^{(-1)}=Y\cdot\cA^{(0)}, \label{eq:EJ}
\end{equation}
where we tentatively put the superscripts representing the picture number 
for clarity. \textit{If} $Y$ and $\cX$ were always invertible, namely they 
had no non-trivial kernel, then $\cA^{(0)}$ would be written as $\cA^{(0)}
=\cX \cdot A^{(-1)}$ and the action would become 
\[ S_{NS}=-\frac{1}{g_o^2}\left(\frac{1}{2\ap}\int A*Q_BA+\frac{1}{3}\int\cX 
A*A*A\right) \]
because of the relation~(\ref{eq:EI}). It has precisely gone back to the 
Witten's action~(\ref{eq:DU}) of the Neveu-Schwarz string field. But the 
assumption of invertibility is not quite correct, as we will see explicitly 
later. Therefore the 0-picture formalism can give different results from those
of Witten's. In fact, it is a field in $\ker Y$ that lets the tachyon 
potential have a non-trivial mininum. The original authors insisted that with 
this formalism we are free from the contact term 
problems~\cite{PTY,Med}. We 
leave the details about it to references, but can illustrate it as follows. 
Even though the 4-boson amplitude still contains two picture-changing 
operators at two 3-string vertices, the Neveu-Schwarz propagator involves one 
factor of $Y_{-2}^{-1}$ (though quite inexact) which cancels one of two 
$Y_{-2}$'s, resulting in no collision. Further, gauge 
variations have no colliding $Y_{-2}$'s. Next we will construct $Y_{-2}$ 
explicitly.

$Y_{-2}$ is required to satisfy that 
\renewcommand{\labelenumi}{(\theenumi)}
\renewcommand{\theenumi}{\roman{enumi}}
\begin{enumerate}
	\item BRST invariance $[Q_B,Y_{-2}]=0$,
	\item Lorentz invariance, in particular $Y_{-2}$ does not depend on 
	momentum,
	\item $Y_{-2}$ has conformal weight 0,
\end{enumerate}
in addition to the crucial requirement~(\ref{eq:EI}). Two possible candidates 
are
\begin{itemize}
\item[$\cdot$] chiral one \cite{Med,Zub}
\begin{equation}
Y_{-2}(z)=-4e^{-2\phi(z)}-\frac{16}{5\ap}e^{-3\phi}c\partial\xi\psi_{\mu}
\partial X^{\mu}(z), \label{eq:EJa}
\end{equation}
\item[$\cdot$] nonchiral one \cite{PTY}
\begin{equation}
Y_{-2}(z,\bar{z})=Y(z)\tilde{Y}(\bar{z}). \label{eq:EK}
\end{equation}
\end{itemize}
The latter includes both holomorphic and antiholomorphic fields in the upper 
half plane, and $Y(z)$ is the usual inverse picture-changing 
operator~(\ref{eq:DQ}). When we use the doubling trick, the nonchiral one 
becomes
\begin{equation}
Y_{-2}(z,\bar{z})=Y(z)Y(\bar{z}), \label{eq:EL}
\end{equation}
that is, the same operators $Y$ are inserted at the two points conjugate to 
each other with respect to the disk boundary (the real axis, in this case). 
In this sense, $Y_{-2}$ is not truly local, but \textit{bilocal}. We think of 
different choices of $Y_{-2}$ as defining \textit{different theories}. The 
above two versions of $Y_{-2}$, though both theories give the same $N$-point 
tree-level amplitudes \textit{on-shell}, can lead to entirely different 
theories \textit{off-shell}. In fact, it was shown in~\cite{Uro} that these 
two theories have different off-shell supersymmetry transformations in the 
\textit{free} (quadratic) level, though they are equivalent on-shell. 
Moreover, it was found that two quadratic actions written in terms of the 
component fields take completely different forms, so are equations of motion. 
Thus it is still unknown whether these two theories arising from two possible 
choices of $Y_{-2}$ are equivalent or not. Anyway, calculation of the tachyon 
potential in the level truncation scheme 
was carried out using the \textit{bilocal} 
operator~(\ref{eq:EL}) in~\cite{11117}. 
Now we will refer to it.
\smallskip

The action (\ref{eq:EH}) can be represented in terms of the CFT correlators 
as in chapter~\ref{ch:sft},
\begin{eqnarray}
S_{NS}&=&-\frac{1}{g_o^2}\Biggl(\frac{1}{2\ap}\left\langle Y_{-2}(i,
\bar{\imath})\cI\circ\cA(0)Q_B\cA(0)\right\rangle \nonumber \\
& &{}+\frac{1}{3}\left\langle Y_{-2}(i,\bar{\imath})f_1\circ\cA(0)f_2\circ
\cA(0)f_3\circ\cA(0)\right\rangle\Biggr) \nonumber \\
&\equiv& -\frac{1}{g_o^2}\left(\frac{1}{2\ap}\llk Y_{-2}|\cA,Q_B\cA\rangle +
\frac{1}{3}\llk Y_{-2}|\cA,\cA,\cA\rangle\right), \label{eq:EM}
\end{eqnarray}
where the insertion point $i=h^{-1}(0)$ of $Y_{-2}$\footnote{We implicitly 
used the fact that $Y_{-2}$ is a primary field of conformal weight 0, so that 
$f\circ Y_{-2}(z)=Y_{-2}(f(z))$.} is the common interaction point among two 
or three strings, labelled by $Q$ in Figure~\ref{fig:F}. The definitions of 
$\cI$ and $f_i$ are exactly the same as those in (\ref{eq:Yb}), (\ref{eq:W}), 
(\ref{eq:S}). The correlators are evaluated in the ``small" Hilbert space. 
On a non-BPS D-brane, we must generalize the action~(\ref{eq:EM}) to include 
GSO($-$) sector as explained before. By the conservation of $e^{\pi iF}$ the 
possible terms are the same as in (\ref{eq:EEa}). If we demand that $\cA_+Q_B
\cA_+$ and $\cA_+\cA_+\cA_+$ terms precisely take the form~(\ref{eq:EM}) and 
that the total action be invariant under the gauge transformation
\begin{eqnarray}
\delta\cA_+&=&Q_B\Lambda_++\frac{1}{\ap}[\cA_+,\Lambda_+]-\frac{1}{\ap}\{\cA_-
,\Lambda_-\}, \nonumber \\
\delta\cA_-&=&Q_B\Lambda_-+\frac{1}{\ap}[\cA_-,\Lambda_+]+\frac{1}{\ap}\{
\cA_+,\Lambda_-\}, \label{eq:EN}
\end{eqnarray}
then the action is determined to be
\begin{eqnarray}
S_{NS}^{\mbox{\scriptsize{non-BPS}}}&=&-\frac{1}{g_o^2}\Biggl(\frac{1}{2\ap}
\llk Y_{-2}|
\cA_+,Q_B\cA_+\rangle+\frac{1}{3}\llk Y_{-2}|\cA_+,\cA_+,\cA_+\rangle 
\nonumber \\ & &{}-\frac{1}{2\ap}\llk Y_{-2}|\cA_-,Q_B\cA_-\rangle+\llk 
Y_{-2}|\cA_+,\cA_-,\cA_-\rangle \Biggr), \label{eq:EO}
\end{eqnarray}
where we used the cyclicity properties
\begin{eqnarray}
\llk Y_{-2}|\cA_+,Q_B\cA_+\rangle &=& \llk Y_{-2}|Q_B\cA_+,\cA_+\rangle ,
\nonumber \\
\llk Y_{-2}|\cA_-,Q_B\cA_-\rangle &=& -\llk Y_{-2}|Q_B\cA_-,\cA_-\rangle ,
\nonumber \\
\llk Y_{-2}|\cA_+,\cA_-,\cA_-\rangle &=& -\llk Y_{-2}|\cA_-,\cA_+,\cA_-
\rangle = \llk Y_{-2}|\cA_-,\cA_-,\cA_+\rangle , \nonumber \\
\llk Y_{-2}|\cA_1,\ldots ,\cA_n\rangle &=& e^{2\pi ih_n}\llk Y_{-2}|\cA_n,
\cA_1,\ldots ,\cA_{n-1}\rangle , \label{eq:EP}
\end{eqnarray}
where $h_n$ is conformal weight of the vertex operator $\cA_n$. For more 
details, see~\cite{11117}. Now we expand the string field in a basis of the 
Hilbert space. The Hilbert space we should consider here is the 
`universal subspace' $\cH_1^{1,0}$, where the superscripts 1,0 indicate that 
it contains states of ghost number 1 and picture number 0, and the subscript 
1 means that the states are constructed by acting $L_n^{\mathrm{m}},
G_n^{\mathrm{m}}, b_n, c_n, \beta_n, \gamma_n$ on the oscillator vacuum 
$|\Omega\rangle=c_1|0\rangle , \> |0\rangle$ being the $SL(2,\aaru)$ invariant
vacuum. That is, non-trivial primary states and states with nonzero momenta 
are truncated out. The proof that this gives 
a consistent truncation of the theory is the same 
as in the bosonic case in section~\ref{sec:univ}. We will find in this 
0-picture formalism there are much more auxiliary fields as compared with 
the Witten's $-1$-picture formalism. And again we impose on the string field 
the Feynman-Siegel gauge condition
\[ b_0\cA_{\pm}=0, \]
so that we can exclude the states including $c_0$, \textit{if} they have 
nonzero $L_0^{\mathrm{tot}}$ eigenvalues. Recall that for a state of 
$L_0^{\mathrm{tot}}=0$ we cannot perform a suitable gauge transformation 
leading to the Feynman-Siegel gauge. We can easily find the state of lowest 
$L_0^{\mathrm{tot}}$ eigenvalue satisfying the above conditions. It is 
$c_1|0\rangle\cong c(0)$, which takes the same form as the zero momentum 
tachyon in the bosonic string theory. Hence $L_0^{\mathrm{tot}}(c_1|0\rangle)
=-1(c_1|0\rangle)$. Surprisingly, this state is obviously annihilated by the 
action of $Y=c\partial \xi e^{-2\phi}$. That is to say, such a state does not 
exist in a theory based on the $-1$-picture states. Let's seek the 
next-to-lowest lying state. It is $\gamma_{1/2}|0\rangle\cong\gamma(0)=
\eta e^{\phi}$, whose $L_0^{\mathrm{tot}}$ eigenvalue is $-1/2$. And since 
\begin{equation}
\lim_{z\to 0}Y(z)\gamma(0)=\lim_{z\to 0}c\partial\xi e^{-2\phi}(z)\eta 
e^{\phi}(0)=-ce^{-\phi}(0), \label{eq:EQ}
\end{equation}
this state just corresponds to the zero momentum tachyon in the $-1$-picture 
theory. Therefore, we had found the `second' tachyonic state $c(z)$ in the 
0-picture theory which is more tachyonic than the usual tachyon $\eta e^{\phi}
\sim ce^{-\phi}$. Here we define the level of the component field to be $h+1$,
where $h$ is the conformal weight of the associated vertex operator, so that 
the second tachyon should be at level 0. As the level 0 tachyon state $c(z)$ 
lies in the GSO($+$) sector, one may think that this formalism does not 
reproduce the spectrum found on a BPS D-brane in the first-quantized 
superstring theory. But when one calculates the quadratic part 
$\llk Y_{-2}|\cA_+,Q_B\cA_+\rangle$ of the action for level 0 truncated 
GSO($+$) sector string field $\cA_+(z)=\int d^{p+1}k\, u(k)c(z)e^{ikX(z)}$, 
he finds the kinetic term of $u$ field to be absent. As we have already seen 
in the bosonic string field theory, the standard kinetic terms of the form 
$(\partial_{\mu}\psi)^2$ have arisen through the 
contribution from $Q_B\to c_0
L_0^{\mathrm{m}}=c_0(\ap p^2+m^2)$. In this case, however, $Y_{-2}$ supplies 
two factors of $c$'s, and two more $c$'s come from two $\cA_+$'s. So the 
contribution from $j_B^1=cT^{\mathrm{m}}$ term vanishes because $bc$ ghost 
number does not satisfy the condition for the correlator to have a nonzero 
value. Instead, the term including $j_B^6=\partial\eta\, \eta e^{2\phi}b$ 
remains nonzero, which gives only the mass term. We then conclude that the $u$
field does not represent any physical degrees of freedom. Hence the existence 
of $u$-tachyon in GSO($+$) sector does not contradict the first-quantized 
theory. In~\cite{Uro}, the quadratic part of the action for low-lying 
component fields was calculated also for the different choice~(\ref{eq:EJa}) 
of $Y_{-2}$. According to that, there is nonvanishing kinetic term for $u$, 
which is produced roughly through $uu\langle e^{-2\phi}c(\ap k^2-1)c\partial 
c\rangle$. But since this term has the wrong sign as opposed to the `physical'
fields, it does not contribute to the physical spectrum too. 

\smallskip

Incidentally, this auxiliary tachyon can be used to construct a non-trivial 
vacuum even on a \textit{BPS} D-brane~\cite{Aref}, though it is not related 
to the tachyon condensation under consideration. Since this superstring field 
theory contains a tachyonic field $u$ in the GSO($+$) sector, there can be 
solutions to the equations of motion where some (scalar) fields develop 
nonvanishing vacuum expectation values on a BPS D-brane. Such a solution was 
actually found in~\cite{Aref}, in which the gauge vector field in the 
Neveu-Schwarz sector becomes massive through the Higgs mechanism as in 
section~\ref{sec:phys} while the gaugino in the Ramond sector remains 
massless, leading to the spontaneous breakdown of supersymmetry in this new 
vacuum. 

\medskip

\begin{figure}[htbp]
	\begin{center}
	\includegraphics{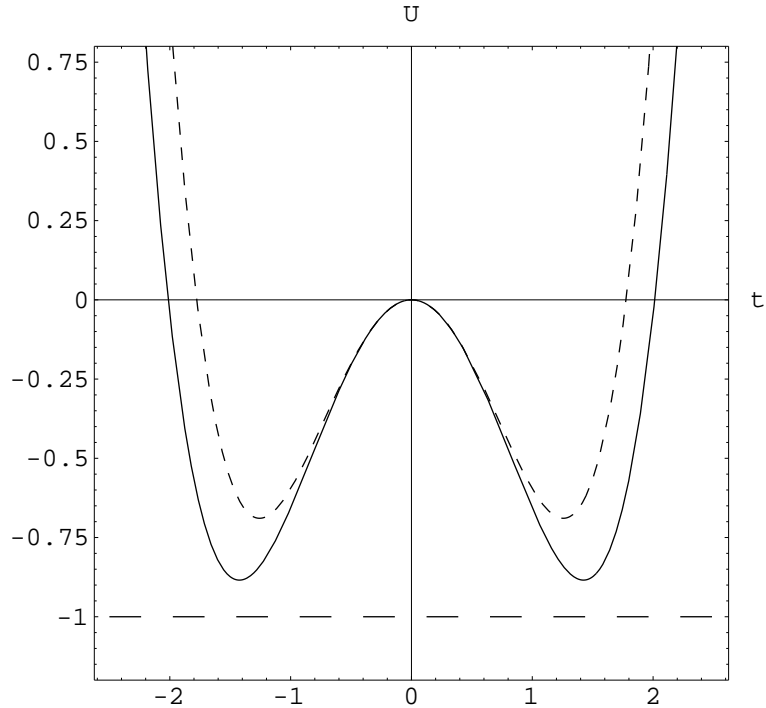}
	\end{center}
	\caption{Effective tachyon potentials at level 
	$(\frac{1}{2},1)$ (dashed 
	line) and at level (2,4) (solid line).}
	\label{fig:AC}
\end{figure}

Now we examine the level $(\frac{1}{2},1)$ truncated tachyon potential. The 
string field is 
\[\cA_+(z)=u c(z) \> , \quad \cA_-(z)=te^{\phi}\eta(z), \]
and we use the full cubic action~(\ref{eq:EO}). As $Y_{-2}$ insertions 
together with handling of $\xi,\eta,\phi$ rather complicate the actual 
calculations, we do not explicitly show the detailed processes but simply 
write down the results. With some proper normalization, one gets
\[ S_{NS}^{\mbox{\scriptsize{non-BPS}}}=\frac{V_{p+1}}{g_o^2}\left(
\frac{1}{\ap}u^2 -\frac{9}{16}ut^2+\frac{1}{4\ap}t^2\right). \]
A $u^3$ term is absent for the same reason as the absence of 
$u$-kinetic term. And using the same method as that in section~\ref{sec:mass} 
the tension of the non-BPS D$p$-brane is 
determined\footnote{I will make comments 
concerning this point at the end of this section.} in~\cite{11117} to be
\begin{equation}
\tilde{\tau}_p=\frac{2\sqrt{2}}{\bar{g}_o^2\pi^2\ap{}^{\frac{p+1}{2}}},
\label{eq:ES}
\end{equation}
where $\bar{g}_o$ is the dimensionless open string coupling and related to 
$g_o$ as $g_o^2=\ap{}^{\frac{p-5}{2}}\bar{g}_o^2$. Putting these results 
together, we finally obtain the level $(\frac{1}{2},1)$ truncated multiscalar 
tachyon potential
\begin{equation}
U(t,u)=-\frac{S_{NS}^{\mbox{\scriptsize{non-BPS}}}}{V_{p+1}}=\tilde{\tau}_p
\frac{\pi^2\ap{}^3}{2\sqrt{2}}\left(-\frac{1}{\ap}u^2+\frac{9}{16}ut^2-
\frac{1}{4\ap}t^2\right). \label{eq:EW}
\end{equation}
Since there is no $u^3$ term, we can exactly integrate out the level 0 
auxiliary tachyon $u$ to obtain 
the effective $t$-tachyon potential 
\begin{equation}
U^{(1)}_{\mathrm{eff}}(t)=\tilde{\tau}_p\frac{\pi^2\ap{}^3}{2\sqrt{2}}\left(
\frac{3^4\ap}{2^{10}}t^4-\frac{1}{4\ap}t^2\right). \label{eq:EX}
\end{equation}
Note that the coefficient in front of $t^4$ is now \textit{positive}, which 
means the effective potential has two non-trivial extrema 
other than $t=0$. 
As we are considering only two fields $u$ and $t$, we find the auxiliary 
tachyonic field $u$, which does not exist in the original $-1$-picture 
theory, has the effect of letting the tachyon potential be of the double-well 
form. The minima of the potential occur at $t=\pm t_0$; $t_0=8\sqrt{2}/9\ap$, 
and the depth of the potential is 
\begin{equation}
U^{(1)}_{\mathrm{eff}}(t_0)=-\frac{\pi^2\ap{}^3}{2\sqrt{2}}\frac{2^4}{
3^4\ap{}^3}\tilde{\tau}_p\simeq -0.689\tilde{\tau}_p. \label{eq:EY}
\end{equation}
Now that we found the level $(\frac{1}{2},1)$ truncated tachyon potential 
reproduces about 69\% of the conjectured value, we want to extend it to 
higher level. According to the empirical law, 
we expect that the shape of the effective 
potential should keep the form of a double-well, and the value of the 
potential at the bottom should approach $-\tilde{\tau}_p$. It is actually 
achieved in~\cite{11117}. Since the level 1 and level $3/2$ fields are odd 
under a $\zetto_2$ twist symmetry, the next truncation level we should 
consider is (2,4). At level 2, the Neveu-Schwarz string field has six states 
in the 0-picture, there being twice as many fields as those 
at the same $L_0$ eigenvalue in the $-1$-picture 
formalism. After integrating out $u$ and the 
level 2 fields, the resulting (2,4) effective tachyon potential is reported 
to be
\begin{equation}
U^{(4)}_{\mathrm{eff}}=\tilde{\tau}_p\frac{\pi^2\ap{}^3}{2\sqrt{2}}\left(
\frac{86929\ap}{2^{10}\cdot 3^4\cdot 17}t^4-\frac{1}{4\ap}t^2\right), 
\label{eq:EZ}
\end{equation}
whose minima occur at $t_0=72\sqrt{34}/\sqrt{86929}\ap$, and 
\[ U^{(4)}_{\mathrm{eff}}(t_0)=-\frac{\pi^2\ap{}^3}{2\sqrt{2}}\frac{2^4\cdot 
3^4\cdot 17}{86929\ap{}^3}\tilde{\tau}_p\simeq -0.884\tilde{\tau}_p. \]
These results agree with our expectations. The shapes of the 
potentials~(\ref{eq:EX}), (\ref{eq:EZ}) are shown in Figure~\ref{fig:AC}. 

So far we have seen how the modified cubic superstring field theory conquers 
the problems by which the Witten's original cubic superstring field theory 
was confronted, for example, the contact term divergences and the absence of 
the minimum of the tachyon potential. However, there remain some difficulties 
in this formalism too. For instance, states that are in reality unphysical 
also solve the equation of motion 
$Y_{-2}(Q_B\cA+\cA *\cA)=0$. 
if they are in the kernel of the 
double-step inverse picture-changing operator $Y_{-2}$. 
And the physical meaning of the different choices of $Y_{-2}$ is 
not clear. At least, the two theories truncated to the low level fields have 
completely different properties off-shell, for example, equations of motion 
and supersymmetry transformations. It remains open whether these two theories 
are equivalent after including higher level contributions fully. For these 
and other reasons, it seems that not everyone believes this 0-picture 
formalism really solves all problems about the construction of superstring 
field theory. 

\bigskip

Before closing this section, I'd like to make an 
objection against one of the results. 
In~\cite{11117}, the authors insist that they get the tension $\tau_p$ of 
the \textit{BPS} D$p$-brane by the method explained in 
section~\ref{sec:mass}, 
and then obtain the tension $\tilde{\tau}_p$ of the \textit{non-BPS} 
D$p$-brane via the well-known formula $\tilde{\tau}_p=\sqrt{2}\tau_p$. But 
I think it is the tension $\tilde{\tau}_p$ of the non-BPS D-brane that we 
obtain by seeing the coefficient of the kinetic term $(\partial_tY)^2$ for 
the correctly normalized translation mode in the string field theory action, 
because in this method we ought to get the tension of the `D-brane' 
represented by the external Chan-Paton factor $\left(
	\begin{array}{cc}
	1 & 0 \\
	0 & 0
	\end{array}
\right)$ irrespective of the property of \textit{the} brane. The information 
whether the brane is BPS or non-BPS should be reflected in the way the string 
fields are GSO-projected. In fact, this superstring field theory cannot 
describe \textit{both} BPS and non-BPS D$p$-branes. For definiteness, let's 
consider the situation where we are interested in the tachyon condensation on 
a non-BPS D3-brane. Since it lives in Type IIA superstring theory, no BPS 
D3-brane exists. And further, we cannot simply $T$-dualize to alter IIA theory
into IIB theory because the relation between the open string coupling $g_o$ 
and the closed string coupling $g_c$ may depend on the details of the 
boundary CFT describing the D-brane, for the relation $\tilde{\tau}_p=\sqrt{2}
\tau_p$ holds when written in terms of closed string coupling constant $g_c$. 
In such circumstances, it is impossible to obtain the tension $\tau_3$ of the 
BPS D3-brane from the superstring field theory on a non-BPS D3-brane. If we 
assume our opinion mentioned above is true, $\tilde{\tau}_p$ does not include 
the factor of $\sqrt{2}$ in~(\ref{eq:ES}). Were it not for the factor, the 
negative energy contribution at the bottom of the tachyon potential would 
exceed 100\% of the conjectured value at level (2,4). Since the minimum value 
of the potential monotonically decreases as we include higher level 
modes\footnote{Since the level truncation processes set to zero the fields 
$\psi^i$ of higher levels, we are looking for a minimum in the restricted 
subspace represented by the set $\{\psi^i=0\}$ in the level truncation scheme.
Therefore the string field configuration which minimizes the 
potential at some level does \textit{not} minimize the 
potential found when higher modes are included 
in general.}, the 
expected exact cancellation between the non-BPS D-brane 
tension and the tachyon potential can never be realized. For this reason, 
I think the results displayed above needs to be reexamined.

\section{Berkovits' Open Superstring Field Theory}\label{sec:berk}
As another approach to superstring field theory, here we introduce the 
formalism which has been developed by N. Berkovits. To understand the 
profound structure of the theory we should read through a series of 
papers~\cite{twisted,Uniqueness,Top,sP,New}. We do not have enough space to 
review the whole structure of the theory, so we will content ourselves with 
describing only part of it necessary for the study of tachyon condensation. 

\bigskip

To begin with, we embed $\cN =1$\footnote{As remarked earlier, we write only 
the holomorphic side, though we also have the equivalent antiholomorphic 
side.} critical ($c^{\mathrm{m}}=15$) Ramond-Neveu-Schwarz superstring into a 
critical ($c^{\mathrm{m}}=6$) $\cN=2$ superstring~\cite{Uniqueness}. In 
general, an $\cN=2$ superstring contains a set of generators ($T,G^+,G^-,J$) 
which satisfies the following $\cN=2$ superconformal algebra
\begin{eqnarray}
T(z)T(0)&\sim&\frac{c}{2z^4}+\frac{2}{z^2}T(0)+\frac{1}{z}\partial T(0),
\nonumber \\ T(z)G^{\pm}(0)&\sim&\frac{3}{2z^2}G^{\pm}(0)+\frac{1}{z}
\partial G^{\pm}(0), \nonumber \\ T(z)J(0)&\sim&\frac{1}{z^2}J(0)+\frac{1}{z}
\partial J(0), \nonumber \\ G^+(z)G^+(0)&\sim&G^-(z)G^-(0)\sim 0, \nonumber \\
G^+(z)G^-(0)&\sim&\frac{2c}{3z^3}+\frac{2}{z^2}J(0)+\frac{1}{z}\left(2T(0)+
\partial J(0)\right), \nonumber \\ J(z)G^{\pm}(0)&\sim&\pm\frac{1}{z}G^{\pm}
(0), \nonumber \\ J(z)J(0)&\sim&\frac{c}{3z^2}. \label{eq:SCA}
\end{eqnarray}
We further extend it to have a `small' $\cN=4$ superconformal symmetry by 
introducing additional generators. We list world-sheet superconformal algebras
in Table~\ref{tab:E}.
\begin{table}[htbp]
	\begin{center}
	\begin{tabular}{|l|c|c|c|c|c|}
	\hline
	world-sheet SUSY & supercurrent & spin 1 current & spin 
	$\frac{1}{2}$ & $c^{\mathrm{g}}$ & symmetry \\ \hline \hline
	$\cN=0$ (bosonic) & --- & --- & --- & $-26$ & --- \\ \hline
	$\cN=1$ & $G$ & --- & --- & $-15$ & --- \\ \hline
	$\cN=2$ & $G^{\pm}$ & $J$ & --- & $-6$ & $U(1)$ \\ \hline 
	$\cN=4$ small & $G^{\pm},\tilde{G}^{\pm}$ & $J,J^{++},J^{--}$ & --- & 12 &
	$SU(2)$ \\ \hline
	$\cN=4$ big & $G^{\pm},\tilde{G}^{\pm}$ & seven $J$'s & four & 0 & 
	$SU(2)\times SU(2) \times U(1)$ \\ \hline
	\end{tabular}
	\caption{Classification of the world-sheet superconformal algebras. 
	$c^{\mathrm{g}}$ is the ghost central charge, and `symmetry' is the one 
	generated by the spin 1 currents. Spin is equal to conformal weight 
	because of holomorphicity.}
	\label{tab:E}
	\end{center}
\end{table}
From this table, we find that two additional supercurrents $\tilde{G}^{\pm}$ 
and two additional spin 1 currents $J^{++},J^{--}$ are required. By writing 
$J=\partial H$, we can define two currents as 
\begin{equation}
J^{++}=e^H \> , \quad J^{--}=e^{-H}, \label{eq:FA}
\end{equation}
where $H(z)$ obeys the following OPE
\[ H(z)H(0)\sim \frac{c}{3}\log z . \]
Three spin 1 currents $(J^{++}=e^H, J=\partial H, J^{--}=e^{-H})$ form an 
$SU(2)$ algebra. Using $J^{++}$ and $J^{--}$, 
$\tilde{G}^{\pm}$ are defined by
\begin{eqnarray}
\tilde{G}^-(z)&\equiv& \oint\frac{d\zeta}{2\pi i}J^{--}(\zeta)G^+(z)
\equiv [J_0^{--}, G^+(z)], \nonumber \\
\tilde{G}^+(z)&\equiv& \oint\frac{d\zeta}{2\pi i}J^{++}(\zeta)G^-(z)
\equiv [J_0^{++}, G^-(z)]. \label{eq:FB}
\end{eqnarray}
By definition, $(G^+,\tilde{G}^-)$ and $(\tilde{G}^+,G^-)$ transform as 
doublets under the $SU(2)$. The $\cN=2$ algebra together with the $SU(2)$ 
algebra defined above generates the $\cN=4$ superconformal algebra, which 
contains four fermionic supercurrents $G^{\pm},\tilde{G}^{\pm}$~\cite{Top}. 
Since the operator products of $U(1)$ generator $J$ with the other $SU(2)$ 
generators $J^{++},J^{--}$ become
\[J(z)J^{\pm\pm}(0)\sim\frac{\pm c/3}{z}J^{\pm\pm}(0), \]
if we take a critical $(c\equiv c^{\mathrm{m}}=6)$ $\cN=2$ superstring then 
the superscripts $\pm$ on the generators correctly represent the charges under
the $U(1)$ generated by $J=\partial H$. 
\medskip

Now that we see the critical $\cN=2$ superconformal algebra can be extended to
the $\cN=4$ superconformal algebra, we wish to explicitly construct the 
generators of $\cN=4$ algebra in terms of the fields appearing in the $\cN=1$ 
Ramond-Neveu-Schwarz superstring. This can be done by embedding the critical 
$\cN=1$ Ramond-Neveu-Schwarz superstring into an $\cN=2$ superconformal 
algebra~\cite{Uniqueness}. In the combined matter-ghost Ramond-Neveu-Schwarz 
superstring, we allow the ghosts to have arbitrary conformal weights 
parametrized by one parameter $\lambda$ as 
\begin{equation}
h_b=\lambda \> , \quad h_c=1-\lambda \> , \quad h_{\beta}=\lambda-\frac{1}{2}
\> , \quad h_{\gamma}=\frac{3}{2}-\lambda. \label{eq:FC}
\end{equation}
These are so chosen as to preserve $h_b+h_c=h_{\beta}+h_{\gamma}=1$ 
and to be 
able to combine $b$ and $\beta$ into a two-dimensional superfield $B=\beta+
\theta b$, where $\theta$ is the fermionic coordinate of the superspace. 
In this case, the ghost part of the energy-momentum tensor is written as 
\begin{equation}
T^{\mathrm{g}}(\lambda)=(\partial b)c-\lambda\partial (bc)+(\partial\beta)
\gamma-\frac{2\lambda -1}{2}\partial (\beta\gamma). \label{eq:FD}
\end{equation}
The central charge of (\ref{eq:FD}) can be determined by seeing the most 
singular term of $T^{\mathrm{g}}T^{\mathrm{g}}$ OPE, the result being 
\begin{equation}
c^{\mathrm{g}}=9-12\lambda . \label{eq:FE}
\end{equation}
Of course, the `standard' ghosts are recovered by choosing $\lambda =2$. 
Looking at the form of $T^{\mathrm{g}}(\lambda)$, one finds 
\begin{equation}
T^{\mathrm{g}}(2)=T^{\mathrm{g}}(3/2)+\frac{1}{2}\partial (-bc-\xi\eta) 
+\frac{1}{2}\partial (\xi\eta-\beta\gamma). \label{eq:FF}
\end{equation}
If we define 
\begin{equation}
J=-bc-\xi\eta \> , \quad T=T^{\mathrm{m}}+T^{\mathrm{g}}(3/2)\> , \quad 
J_2=\xi\eta-\beta\gamma \> , \label{eq:FG}
\end{equation}
then one can easily verify
\begin{eqnarray}
J(z)J(0)&\sim&\frac{6}{3z^2}, \nonumber \\
T(z)T(0)&\sim&\frac{6}{2z^4}+\frac{2}{z^2}T(0)+\frac{1}{z}\partial T(0),
\label{eq:FH}
\end{eqnarray}
because the matter central charge $c^{\mathrm{m}}$ of the critical $\cN=1$ 
Ramond-Neveu-Schwarz superstring is 15 and $c^{\mathrm{g}}$~(\ref{eq:FE}) 
is equal to $-9$ when $\lambda =3/2$. By constructing two spin $3/2$ 
supercurrents as~\cite{twisted}
\begin{equation}
G^-=b \> , \quad G^+=j_B+\partial(c\xi\eta)+\partial^2c, \label{eq:FI}
\end{equation}
we find $(T,G^+,G^-,J)$ to satisfy the $\cN=2$ superconformal algebra with 
$c=6$. Since weights of the ghosts are shifted by $1/2$, $G^{\pm}$ 
in~(\ref{eq:FI}) actually have spin (weight) $3/2$. From Table~\ref{tab:E}, 
the central charge $c^{\mathrm{m}}$ of the matter system of the $\cN=2$ 
\textit{critical} superstring must be 
equal to 6, so we have gotten the 
\textit{matter} system for the $\cN=2$ critical superstring from the combined 
\textit{matter-ghost} system of the critical $\cN=1$ Ramond-Neveu-Schwarz 
superstring~\cite{Uniqueness}. A straightforward method for calculating 
scattering amplitudes for the critical $\cN=2$ superstring is introducing 
$\cN=2$ ghosts $(b,c,\beta^{\pm},\gamma^{\pm},\eta,\xi)$, constructing $\cN=2$
BRST charge, computing correlation functions of BRST-invariant vertex 
operators, and integrating them over the moduli of $\cN=2$ super-Riemann 
surfaces. But it was shown in~\cite{Top} that there is a simpler method for 
that purpose. It involves extending the $\cN=2$ superconformal algebra 
obtained in~(\ref{eq:FG}), (\ref{eq:FI}) to an $\cN=4$ algebra by the 
prescription mentioned before 
and then twisting the resulting $\cN=4$ algebra 
by the $U(1)$ current $J$. That is to say, we define a new energy-momentum 
tensor by 
\begin{equation}
T^{\mathrm{twist}}(z)\equiv T(z)+\frac{1}{2}\partial J(z), \label{eq:FJ}
\end{equation}
where $T$ and $J$ are defined in~(\ref{eq:FG}). In this `new theory',
an operator $\cO$ of `original' weight $h$ and $U(1)$ charge $q$ will have the
new weight $h^{\prime}\neq h$. This can be seen by the following OPE, 
\begin{eqnarray*}
T^{\mathrm{twist}}(z)\cO(0)&=&T(z)\cO(0)+\frac{1}{2}\partial J(z)\cO(0) \\
&\sim&\frac{h}{z^2}\cO(0)+\frac{1}{z}\partial \cO(0)+\frac{1}{2}\partial 
\left(\frac{q}{z}\cO(0)\right) \\ &=& \frac{h-\frac{1}{2}q}{z^2}\cO(0)
+\frac{1}{z}\partial\cO(0), 
\end{eqnarray*}
where we assumed $\cO$ to be a primary field 
for simplicity. From this, we can read off the new weight
\begin{equation}
h^{\prime}=h-\frac{1}{2}q. \label{eq:FK}
\end{equation}
So the various charged fields under the $U(1)$ change their weights according 
to~(\ref{eq:FK}) in the twisted theory. We now summarize in Table~\ref{tab:F} 
the weights (spins) of the fields appearing in the $\cN=4$ superconformal 
algebra.
\begin{table}[htbp]
\begin{center}
	\begin{tabular}{|c||c|c|c|c|c|c|c|c|}
	\hline
	 & $T$ & $G^+$ & $G^-$ & $\tilde{G}^+$ & $\tilde{G}^-$ & $J^{++}$ & $J$ 
	& $J^{--}$ \\ \hline
	before twist & 2 & 3/2 & 3/2 & 3/2 & 3/2 & 1 & 1 & 1 \\ \hline
	after twist & 2 & 1 & 2 & 1 & 2 & 0 & 1 & 2 \\ \hline
	\end{tabular}
	\caption{Spins of the fields before and after twisting.}
	\label{tab:F}
\end{center}
\end{table}
By definition~(\ref{eq:FJ}) of the twisted energy-momentum tensor, the OPE 
among two $T^{\mathrm{twist}}$'s obeys the usual Virasoro algebra with 
\textit{zero} central charge. So one can hope that in the twisted theory 
we need not introduce $\cN=2$ ghosts and complications arising from them. 
Though the story is not such simple, it was shown in~\cite{Top} that $\cN=2$ 
physical vertex operators and scattering amplitudes can be computed 
\textit{without} introducing $\cN=2$ ghosts after twisting. 
We will not pursue this point any more. Here, we write 
down the full set of $\cN=4$ superconformal generators (after twisting) in 
terms of Ramond-Neveu-Schwarz variables,
\begin{eqnarray}
T^{\mathrm{twist}}&=&T^{\mathrm{m}}+T^{\mathrm{g}}(\lambda =2)+\frac{1}{2}
\partial (\beta\gamma-\xi\eta ), \nonumber \\
G^+&=&j_B+\partial (c\xi\eta )+\partial^2c, \nonumber \\
G^-&=&b, \nonumber \\
J^{++}&=&-c\eta , \nonumber \\
J&=& -bc-\xi\eta , \nonumber \\
J^{--}&=&-b\xi , \label{eq:FL} \\
\tilde{G}^+(z)&=&\oint\frac{d\zeta}{2\pi i}J^{++}(\zeta)G^-(z)=\eta(z),
\nonumber \\
\tilde{G}^-(z)&=&\oint\frac{d\zeta}{2\pi i}J^{--}(\zeta)G^+(z)
=\mathop{\mathrm{Res}}\limits_{\zeta\to z}j_B(z)J^{--}(\zeta) \nonumber \\
&=&[Q_B,b\xi(z)]=(T^{\mathrm{m}}+T^{\mathrm{g}}(\lambda =2))\xi(z)
-b\cX(z), \nonumber
\end{eqnarray}
where $j_B$ is the standard $\cN=1$ BRST current of Ramond-Neveu-Schwarz 
superstring. Since the $U(1)$ current $J=-bc-\xi\eta$ is essentially the 
ghost number current, conformal weights of the ghosts $b,c,\beta,\gamma$ have 
been brought back to the `standard' weights represented 
by $\lambda =2$ after twisting. And 
notice that the $\xi$ zero mode explicitly appears in several generators 
forming $\cN=4$ superconformal algebra. This means we cannot help doing 
calculations in the ``large" Hilbert space.
\medskip

Next let's consider the $\cN=2$ vertex operators and correlation functions 
among them. As we are in the ``large" Hilbert space, an $\cN=2$ vertex 
operator $\Phi$ is constructed by using the corresponding $\cN=1$ 
Ramond-Neveu-Schwarz vertex operator $A$ as 
\begin{equation}
\Phi(z)=\> :\xi A(z): , \label{eq:FM}
\end{equation}
where $A$ has ghost number $+1$ and is in the natural $-1$-picture. From now 
on we will abbreviate the normal ordering symbol 
$:\ldots :$ as usual. Since both 
the ghost number and the picture number of $\Phi$ are zero, $\Phi$ is 
bosonic (Grassmann even). In particular, $\Phi$ is neutral under the $U(1)$ 
generated by the current $J=-bc-\xi\eta$ which belongs to the $\cN=2$ algebra.
Simple $N$-point functions $\langle \Phi(z_1)\cdots \Phi(z_N)\rangle$ on the 
sphere\footnote{Doubling trick.} among them, however, turn out to vanish, 
because the $U(1)$ current $J$ carries nonzero anomaly. By exactly the same 
argument as below eq.~(\ref{eq:DI}), one can find any correlator 
vanishes unless total $U(1)$ charge of the inserted vertex operators is equal 
to $+2$. Or this fact can easily be verified as follows. It is well known that
the field $c$ of conformal weight $-1$ has three zero modes $c_1,c_0,c_{-1}$, 
so three insertions of $c$ is necessary for the path integral to have a 
nonvanishing value. Similarly, the field $\xi$ of conformal weight 0 has one 
zero mode $\xi_0$, which requires one insertion of 
$\xi$ in the integrand of 
path integral. To sum up, we must include factors of $ccc\xi$, the sum of 
whose $U(1)$ charges is $+3-1=+2$, which agrees with the previously mentioned 
result. Thus, the simplest nonvanishing amplitude can be written as 
\begin{equation}
\langle\Phi(z_1)G_0^+\Phi(z_2)\tilde{G}_0^+\Phi(z_3)\rangle, \label{eq:FN}
\end{equation}
where $G_0^+\Phi(z)$ is defined by $\displaystyle \oint_z\frac{d\zeta}{2\pi i}
G^+(\zeta)\Phi(z)$, and similarly for $\tilde{G}_0^+\Phi(z)$. Note that both 
$G^+,\tilde{G}^+$ are primary fields of conformal weight 1 after twisting. 
Three vertex operator insertions at $z_1,z_2,z_3$ are needed to fix the 
conformal Killing group $PSL(2,\aaru)$. Recall that 
both $G_0^+$ and 
$\tilde{G}_0^+$ have one unit of $U(1)$ charge. 
Since we first want 
to consider the on-shell amplitudes, we should determine the physical 
condition for the $\cN=2$ vertex operator $\Phi$. It 
turns out that the suitable physical condition is written as
\begin{equation}
G_0^+\tilde{G}_0^+\Phi(z)=0, \label{eq:FO}
\end{equation}
because through the following computations 
\begin{eqnarray*}
G_0^+\tilde{G}_0^+\Phi(z)&=&\oint\frac{dx}{2\pi i}j_B(x)\oint\frac{dy}{2\pi i}
\eta(y)\xi(z)A(z) \\ &=& \{Q_B,A(z)\}
\end{eqnarray*}
(\ref{eq:FO}) is equivalent to the usual on-shell condition $Q_B|A\rangle=0$. 
Note that since $A$ lives in the ``small" Hilbert space, $\eta$-$A$ 
contraction never gives a simple pole $1/(y-z)$. Moreover, it is important to 
be aware that 
\begin{equation}
\{G_0^+,\tilde{G}_0^+\}=0 \> , \quad (G_0^+)^2=(\tilde{G}_0^+)^2=0. 
\label{eq:FP}
\end{equation}
One can see that the 3-point amplitude~(\ref{eq:FN}) is invariant under the 
`gauge' transformation 
\begin{equation}
\delta\Phi =G_0^+\Lambda_1+\tilde{G}_0^+\Lambda_2 \label{eq:FQ}
\end{equation}
if all vertex operators are on-shell. It is easily verified by 
using~(\ref{eq:FO}), (\ref{eq:FP}) and contour deformations. And the gauge 
transformation~(\ref{eq:FQ}) also preserves the 
`equation of motion'~(\ref{eq:FO}) thanks 
to~(\ref{eq:FP}). So we should physically identify two vertex operators which 
are related through the gauge transformation~(\ref{eq:FQ}), namely $\Phi_1
\cong\Phi_2$ if there exist $\Lambda_1,\Lambda_2$ such that $\Phi_1=\Phi_2+
G_0^+\Lambda_1+\tilde{G}_0^+\Lambda_2$. In fact, the `pure gauge' state of the
form $\Phi=G_0^+\Lambda_1+\tilde{G}_0^+\Lambda_2$ 
corresponds to the following $\cN=1$ vertex operator 
\[ A=\eta_0\Phi=\tilde{G}_0^+(G_0^+\Lambda_1+\tilde{G}_0^+\Lambda_2)=G_0^+
(-\tilde{G}_0^+\Lambda_1)=Q_B(-\eta_0\Lambda_1), \]
which is nothing but a BRST-exact\footnote{Though 
$\Lambda_1$ itself does not exist in the ``small" Hilbert space, $\eta_0
\Lambda_1$ does.} state in the $\cN=1$ sense. 
In the end, we can construct the physical Hilbert space in 
terms of $\cN=2$ vertex operators $\{\Phi\}$ in a similar way as in $\cN=1$ 
case: A physical ``large" Hilbert space consists of states that are 
annihilated by $G_0^+\tilde{G}_0^+=Q_B\eta_0$ modulo the gauge 
transformation~(\ref{eq:FQ}).

\medskip

As a next step, we want to generalize the above on-shell formalism to an 
off-shell free string field theory. The constraints on the form of the action 
are that it should reproduce the equation of motion~(\ref{eq:FO}) and it 
should incorporate the gauge invariance~(\ref{eq:FQ}). An obvious guess is 
\begin{equation}
S_{\mathrm{quad}}=\int\Phi *G_0^+\tilde{G}_0^+\Phi , \label{eq:FR}
\end{equation}
where $\int$ and $*$ are the same gluing operations as in the cubic bosonic 
string field theory. $\Phi$ is now promoted to an off-shell 
string field, though the substance is almost the same. But there is a subtle 
point in the reality condition of the string field. Under the Hermitian 
conjugation, $J$ flips its sign as $(bc)^{\dagger}=cb=-(bc)$ and so on. 
Hence the reality condition 
\[\mathrm{hc}(\Phi[X(\sigma)])=\Phi[X(\pi-\sigma)] \]
must be accompanied by the $SU(2)$ rotation 
\[ J\to -J \> , \quad J^{++}\to J^{--} \> , \quad J^{--} \to J^{++} \> . \]

\medskip

At last, we have come to the position to extend the free action~(\ref{eq:FR}) 
to include interactions. For the action to produce 
the correct 3-point on-shell 
amplitude, it is shown that the cubic action must be of the form
\begin{equation}
S_{\mathrm{cubic}}=\int\biggl((G_0^+\Phi)*(\tilde{G}_0^+\Phi)+\Phi*\{(G_0^+
\Phi),(\tilde{G}_0^+\Phi)\}\biggr) \label{eq:FS}
\end{equation}
up to numerical coefficients. But it is impossible to extend the gauge 
transformation~(\ref{eq:FQ}) to a non-linear form under which 
$S_{\mathrm{cubic}}$ is invariant. This fact forces us to add quartic or even 
higher order terms to the cubic action $S_{\mathrm{cubic}}$. As a clue to 
seek the correct form of the action, let's look at the Wess-Zumino-Witten 
model
\begin{equation}
S_{WZW}=\mathrm{Tr}\int d^2z\left[(g^{-1}\partial g)(g^{-1}\bar{\partial}g)
-\int_0^1dt(\hat{g}^{-1}\partial_t\hat{g})[(\hat{g}^{-1}\partial\hat{g}),
(\hat{g}^{-1}\bar{\partial}\hat{g})]\right], \label{eq:FT}
\end{equation}
where $g$ takes value in some group $G$ and $\hat{g}$ 
is a $G$-valued function extended 
to depend also on $t\in [0,1]$, 
whose boundary values are constrained such that 
$\hat{g}(t=0)=I \> , \quad \hat{g}(t=1)=g$. By setting $g=e^{\varphi}$ and 
expanding in $\varphi$, we get from $S_{WZW}$ the correct cubic action of the 
form~(\ref{eq:FS}) with the anticommutator replaced by a commutator. Expecting
more successes, we mimic it and assume the following string field theory 
action
\begin{eqnarray}
S&=&\frac{1}{2}\int\Biggl[\left(e^{-\Phi}G_0^+e^{\Phi}\right)\left(e^{-\Phi}
\tilde{G}_0^+e^{\Phi}\right) \nonumber \\
& &{}-\int_0^1dt\left(e^{-t\Phi}\partial_te^{t\Phi}\right)\left\{\left(e^{-t
\Phi}G_0^+e^{t\Phi}\right),\left(e^{-t\Phi}\tilde{G}_0^+e^{t\Phi}\right)
\right\}\Biggr]. \label{eq:FU}
\end{eqnarray}
Here, the products and integral among the string fields are defined by the 
Witten's gluing prescription of the strings. This action has many advantages 
for our purpose. Firstly, 
\begin{itemize}
	\item the equation of motion derived from the action~(\ref{eq:FU}) is 
\end{itemize}
\begin{equation}
\tilde{G}_0^+\left(e^{-\Phi}G_0^+e^{\Phi}\right)=0 , \label{eq:FV}
\end{equation}
whose linearized version correctly reproduces $\tilde{G}_0^+G_0^+\Phi=0$. 
In terms of the Ramond-Neveu-Schwarz variables, (\ref{eq:FV}) is written as
\begin{equation}
\eta_0\left(e^{-\Phi}Q_Be^{\Phi}\right)=0 . \label{eq:FW}
\end{equation}
Secondly,
\begin{itemize}
	\item the action~(\ref{eq:FU}) has the nonlinear gauge invariance under 
\end{itemize}
\begin{equation}
\delta e^{\Phi}=(G_0^+\Lambda_1)e^{\Phi}+e^{\Phi}(\tilde{G}_0^+\Lambda_2)
=(Q_B\Lambda_1)e^{\Phi}+e^{\Phi}(\eta_0\Lambda_2), \label{eq:FX}
\end{equation}
where $\Lambda_1,\Lambda_2$ are gauge parameters of ghost number $-1$. The 
proof of this gauge invariance can be found in~\cite{BSZ}. Of course, the 
linearized gauge transformation of~(\ref{eq:FX}) takes the form $\delta\Phi =
G_0^+\Lambda_1+\tilde{G}_0^+\Lambda_2$, as desired. Thirdly, 
\begin{itemize}
	\item this action accurately reproduces the on-shell amplitudes found in 
	the first-quantized superstring theory.
\end{itemize}
As remarked above, the action~(\ref{eq:FU}) gives the correct answer for the 
3-point on-shell scattering amplitude. Moreover, the on-shell 4-point 
\textit{tree-level} 
amplitude was computed in~\cite{Eche}. There are three types of 
diagrams which contribute to the 4-point tree 
amplitude. We call them $s$-channel 
$(\cA_s)$, $t$-channel $(\cA_t)$ and quartic $(\cA_q)$ amplitudes 
respectively. The sum $\cA_s+\cA_t$ of the two diagrams which include two 
cubic vertices is shown to contain a surface term (contact term) of the same 
kind as the one found in~(\ref{eq:EC}), but it is \textit{finite} in this 
case because there are no dangerous operators such as colliding 
picture-changing operators. And still, it was found that the quartic 
contribution $\cA_q$ exactly \textit{cancels} the finite contact term. When 
we take the four external states to be on-shell, \textit{i.e.} $G_0^+
\tilde{G}_0^+\Phi=0$, the tree-level 4-point amplitude $\cA_s+\cA_t+\cA_q$ 
computed from the second-quantized superstring field theory 
action~(\ref{eq:FU}) precisely agrees with the first-quantized result, 
without any finite or divergent contact term. Generally speaking, 
\begin{itemize}
	\item the superstring field theory action~(\ref{eq:FU}) does not suffer 
	from the contact term divergence problems.
\end{itemize}

\medskip

I'd like to close this section with some comments which 
are not directly related to our route to the problem 
of tachyon condensation. 
There are two other examples of a critical $\cN=2$ superstring, though we do 
not depict them in detail. One of them is the self-dual string describing 
self-dual Yang-Mills theory in $D=(2,2)$~\cite{OV}. And the other is the 
`hybrid' description of the modified Green-Schwarz superstring in 
four-dimensional superspace combined with the $c=9, \cN=2$ superconformal 
field theory which describes a six-dimensional compactification 
manifold~\cite{GS}. In the latter example, by expressing the action in terms 
of Green-Schwarz-like variables,
\begin{itemize}
	\item it can be made manifestly $SO(3,1)$ super-Poincar\'e invariant.
\end{itemize}

\section{Open Superstring Field Theory \\ on Various D-Branes}
In the last section, the basic formalism of Berkovits' open superstring field 
theory was reviewed. We write here the superstring field theory action 
again, with a slightly different notation:
\begin{eqnarray}
S_D&=&\frac{1}{2g_o^2}\bllk\left(e^{-\Phi}Q_Be^{\Phi}\right)\left(e^{-\Phi}
\eta_0e^{\Phi}\right) \nonumber \\
& &{}-\int_0^1dt\left(e^{-t\Phi}\partial_te^{t\Phi}\right)\left\{\left(e^{-t
\Phi}Q_Be^{t\Phi}\right),\left(e^{-t\Phi}\eta_0e^{t\Phi}\right)
\right\}\brrk . \label{eq:Berk}
\end{eqnarray}
The meaning of $\llk\ldots\rrk$ will be explained below. Though we did not 
specify what properties the Grassmann even string field $\Phi$ has, it does 
correspond to a state in the GSO($+$) Neveu-Schwarz sector. It is not yet 
known how to incorporate the Ramond sector states. But in order to look for a 
Lorentz invariant vacuum produced by the tachyon condensation, we can ignore 
the Ramond sector states because they represent the spacetime spinors. 
In spite of this fact, we strongly hope that we will succeed in including the 
Ramond sector into this formalism because in that case we can examine whether 
the spacetime supersymmetry is restored when we construct BPS D-branes via 
tachyon condensation as kinks on a non-BPS D-brane or as vortices on a D-brane 
anti D-brane pair. Aside from the problems on the Ramond sector,
we must also consider the GSO($-$) Neveu-Schwarz sector 
for the tachyon condensation, since the only tachyonic 
state lives there. On a non-BPS D-brane, it is convenient to introduce 
\textit{internal} Chan-Paton factors. With the help of them, the algebraic 
structures obeyed by the GSO($+$) string fields are 
preserved even after 
introducing the GSO($-$) string fields. Moreover, the multiplicative 
conservation of $e^{\pi iF}$ is guaranteed by simply taking the trace. 
On a D-brane anti D-brane pair, in addition to the internal ones 
\textit{external} Chan-Paton factors representing two branes must be 
tensored too. These affairs will be discussed in this section.

\subsection{On a BPS D-brane}
We begin by explaining the meaning of $\llk\ldots\rrk$ in~(\ref{eq:Berk}) in 
the case of a BPS D-brane, namely, $\Phi$ is in the GSO($+$) sector. By 
expanding the exponentials in a formal power series, the action is decomposed 
into the sum of the various $n$-point (possibly infinite) 
vertices $\llk A_1\ldots A_n\rrk$ with 
some vertex operators $A_i$. But we must take great care not to change the 
order of operators. Recall that in section~\ref{sec:eva} we have defined an 
arbitrary $n$-point vertex $\int\Phi *\cdots *\Phi$ as the $n$-point CFT 
correlator $\langle f_1\circ\Phi(0)\ldots f_n\circ\Phi(0)\rangle$. Now we 
adopt the same definition for $\llk\ldots\rrk$, that is, 
\begin{equation}
\llk A_1\ldots A_n\rrk=\langle f_1^{(n)}\circ A_1(0)\ldots f_n^{(n)}\circ 
A_n(0)\rangle \label{eq:FY}
\end{equation}
where 
\begin{eqnarray}
f_k^{(n)}(z)&=&h^{-1}\circ g_k^{(n)}(z), \nonumber \\
h^{-1}(z)&=&-i\frac{z-1}{z+1}, \nonumber \\
g_k^{(n)}(z)&=&e^{\frac{2\pi i}{n}(k-1)}\left(\frac{1+iz}{1-iz}\right)^{
\frac{2}{n}}, \label{eq:FYa} \\
& & 1\le k\le n. \nonumber
\end{eqnarray}
Of course, we can use $g_k^{(n)}$ (unit disk representation) instead of 
$f_k^{(n)}$ (upper half plane representation) in~(\ref{eq:FY}). One thing we 
must keep in mind is that the correlator is evaluated in the ``large" Hilbert 
space. So the correlator is normalized such that 
\begin{equation}
\left\langle\xi(z)\frac{1}{2}c\partial c\partial^2c(w)e^{-2\phi(y)}
\right\rangle=\langle 0|\xi_0c_1c_0c_{-1}e^{-2\phi(0)}|0\rangle =1.
\label{eq:FZ}
\end{equation}
In the most left hand side the correlator is independent of $z,w,y$ because 
they supply only zero modes. When we consider the GSO($-$) sector too, vertex 
operators can have half-integer-valued conformal weights. In this case, some 
ambiguity arises in the definition of the conformal transformation 
\[ f\circ\varphi(0)=(f^{\prime}(0))^h\varphi(f(0)) \]
for a primary field. Here we unambiguously define the phase of it. 
For $f_k^{(n)}(z)$, since one finds 
\[f_k^{(n)\prime}(0)=\frac{2}{n}\sec^2\frac{\pi (k-1)}{n}, \]
we can simply define 
\begin{equation}
f_k^{(n)}\circ\varphi(0)=\bigg|\left(\frac{2}{n}\right)^h\sec^{2h}
\frac{\pi (k-1)}{n}\bigg| \> \varphi(f_k^{(n)}(0)). \label{eq:GA}
\end{equation}
For $g_k^{(n)}(z)$, $g_k^{(n)\prime}(0)$ contains a factor of $i$, which we
define to be $i=e^{\pi i/2}$. Then 
\[ g_k^{(n)\prime}(0)=\frac{4i}{n}e^{\frac{2\pi i}{n}(k-1)}=\frac{4}{n}
e^{2\pi i(\frac{k-1}{n}+\frac{1}{4})}. \]
Finally we set
\begin{equation}
g_k^{(n)}\circ\varphi(0)=\bigg|\left(\frac{4}{n}\right)^h\bigg| \> e^{2\pi ih
(\frac{k-1}{n}+\frac{1}{4})}\varphi(g_k^{(n)}(0)). \label{eq:GB}
\end{equation}
\medskip

Now we enumerate the algebraic properties the correlator $\llk\ldots\rrk$ 
satisfies. Let $A_i$'s denote arbitrary vertex operators, whereas 
$\Phi,\Phi_i$ represent the string fields in GSO($+$) Neveu-Schwarz sector, 
\textit{i.e.} Grassmann even vertex operators of ghost number 0.
\begin{itemize}
\item Cyclicity properties
\end{itemize}
\begin{eqnarray}
\llk A_1\ldots A_{n-1}\Phi\rrk &=&\llk\Phi A_1\ldots A_{n-1}\rrk , 
\label{eq:GCa} \\
\llk A_1\ldots A_{n-1}(Q_B\Phi)\rrk &=& -\llk(Q_B\Phi)A_1\ldots A_{n-1}\rrk ,
\label{eq:GCb} \\
\llk A_1\ldots A_{n-1}(\eta_0\Phi)\rrk &=& -\llk(\eta_0\Phi)A_1\ldots A_{n-1}
\rrk. \label{eq:GCc}
\end{eqnarray}
Since the similar relations hold for the string fields in the GSO($-$) sector,
we postpone proving them until we introduce the GSO($-$) sector. 
\begin{itemize}
\item Anticommutativity
\end{itemize}
\begin{equation}
\{Q_B,\eta_0\}=0 \> , \quad Q_B^2=\eta_0^2=0. \label{eq:GG}
\end{equation}
\begin{itemize}
\item Leibniz rules
\end{itemize}
\begin{eqnarray}
Q_B(\Phi_1\Phi_2)&=&(Q_B\Phi_1)\Phi_2+\Phi_1(Q_B\Phi_2), \nonumber \\
\eta_0(\Phi_1\Phi_2)&=&(\eta_0\Phi_1)\Phi_2+\Phi_1(\eta_0\Phi_2). 
\label{eq:GH}
\end{eqnarray}
\begin{itemize}
\item Partial integrability
\end{itemize}
\begin{equation}
\llk Q_B(\ldots)\rrk=\llk\eta_0(\ldots)\rrk =0. \label{eq:GI}
\end{equation}
We have already seen (\ref{eq:GG}) in the form of (\ref{eq:FP}). Because
the GSO($+$) string field $\Phi_1$ is Grassmann even, there are no minus 
signs in~(\ref{eq:GH}). (\ref{eq:GI}) follow from the fact that both $Q_B$ and
$\eta_0$ are the integrals of the primary fields $j_B$ and $\eta$, 
respectively, of conformal weight 1, and that the 
integration contours can be deformed to shrink around infinity.
\medskip

These are the basic structures of the theory defined by the 
action~(\ref{eq:Berk}). In generalizing it to the theory on a non-BPS D-brane,
we must also include the GSO($-$) sector without spoiling these basic 
structures. We will discuss below how to do that in an appropriate way.

\subsection{On a non-BPS D-brane}
GSO($-$) states are represented by the Grassmann odd vertex operators and have
half-integer weights if the states have zero momenta. This fact can be seen 
from the following examples, 
\[\left.
	\begin{array}{ccccc}
	\mbox{state} & \mbox{sector} & \mbox{vertex operator} & 
	\mbox{Grassmannality} & \mbox{weight} \\
	\hline
	\mbox{tachyon} & \mbox{NS}- & \xi ce^{-\phi} & \mbox{odd} & -1/2 \\
	\mbox{gauge field} & \mbox{NS}+ & \xi\psi^{\mu}ce^{-\phi} & \mbox{even} & 
	0
	\end{array}
\right. \]
where the vertex operators in the ``large" Hilbert space are obtained by 
combining~(\ref{eq:DL}) with (\ref{eq:FM}). Clearly the Leibniz 
rules~(\ref{eq:GH}) do not hold true for Grassmann odd operators. To remedy 
this point, we introduce \textit{internal} Chan-Paton matrices and trace over 
them. Concretely, we multiply vertex operators in the GSO($+$) sector by the 
$2\times 2$ identity matrix $I$ and ones in the GSO($-$) sector by the Pauli 
matrix $\sigma_1$, so that the complete string field becomes
\begin{equation}
\widehat{\Phi}=\Phi_+\otimes I+\Phi_-\otimes\sigma_1, \label{eq:GJ}
\end{equation}
where the subscripts $\pm$ denote not the Grassmannality but the $e^{\pi iF}$ 
eigenvalue. In this case, however, these two happen to coincide. In order to 
recover~(\ref{eq:GH}), $Q_B$ and $\eta_0$ should be tensored by a matrix which
anticommutes with $\sigma_1$. For that purpose, it turns out that $\sigma_3$ 
plays a desired role. So we define 
\begin{equation}
\widehat{Q}_B=Q_B\otimes \sigma_3 \> , \quad 
\widehat{\eta}_0=\eta_0\otimes\sigma_3.
\label{eq:GK}
\end{equation}
In the rest of this chapter, we always regard the 
hat on an operator as meaning that the operator contains a $2\times 2$ 
internal Chan-Paton matrix. When the vertex operators have internal 
Chan-Paton factors in them, we modify the definition of 
the the correlator $\llk\ldots\rrk$ as 
\begin{equation}
\llk \widehat{A}_1\ldots \widehat{A}_n\rrk =\mathrm{Tr}
\langle f_1^{(n)}\circ\widehat{A}_1
(0)\ldots f_n^{(n)}\circ\widehat{A}_n(0)\rangle , \label{eq:GL}
\end{equation}
where the trace is taken over the internal Chan-Paton matrices. With these 
definitions, basic properties satisfied by the GSO($+$) string field hold 
even for the combined system including both GSO($+$) and GSO($-$) sectors. 
That is, 
\begin{itemize}
\item cyclicity properties
\begin{eqnarray}
\llk\widehat{A}_1\ldots\widehat{A}_{n-1}\widehat{\Phi}\rrk &=& \llk 
\widehat{\Phi}\widehat{A}_1\ldots\widehat{A}_{n-1}\rrk , \label{eq:GMa} \\
\llk\widehat{A}_1\ldots\widehat{A}_{n-1}(\widehat{Q}_B\widehat{\Phi})\rrk 
&=&-\llk(\widehat{Q}_B\widehat{\Phi})\widehat{A}_1\ldots\widehat{A}_{n-1}
\rrk , \label{eq:GMb} \\
\llk\widehat{A}_1\ldots\widehat{A}_{n-1}(\widehat{\eta}_0\widehat{\Phi})\rrk 
&=&-\llk(\widehat{\eta}_0\widehat{\Phi})\widehat{A}_1\ldots\widehat{A}_{n-1}
\rrk , \label{eq:GMc}
\end{eqnarray}
\item anticommutativity
\begin{equation}
\{\widehat{Q}_B,\widehat{\eta}_0\}=0 \> , \quad (\widehat{Q}_B)^2=
(\widehat{\eta}_0)^2=0, \label{eq:GMd}
\end{equation}
\item Leibniz rules
\begin{eqnarray}
\widehat{Q}_B(\widehat{\Phi}_1\widehat{\Phi}_2)&=&(\widehat{Q}_B
\widehat{\Phi}_1)\widehat{\Phi}_2+\widehat{\Phi}_1(\widehat{Q}_B
\widehat{\Phi}_2), \nonumber \\
\widehat{\eta}_0(\widehat{\Phi}_1\widehat{\Phi}_2)&=&(\widehat{\eta}_0
\widehat{\Phi}_1)\widehat{\Phi}_2+\widehat{\Phi}_1(\widehat{\eta}_0
\widehat{\Phi}_2), \label{eq:GMe}
\end{eqnarray}
\item partial integrability
\begin{equation}
\llk\widehat{Q}_B(\ldots)\rrk = \llk\widehat{\eta}_0(\ldots)\rrk =0. 
\label{eq:GMf}
\end{equation}
\end{itemize}
Now we begin with proving (\ref{eq:GCa}) and (\ref{eq:GMa}). Since the 
Chan-Paton matrices obey the cyclicity symmetry without any sign factor 
under the trace, we will put the Chan-Paton matrices aside and focus on the 
unhatted vertex operators. Consider an $SL(2,\aaru)$\footnote{
This is because $T$ maps the interior of the unit disk to itself in a 
one-to-one manner and maps any point on the disk boundary to some point on the
boundary of the transformed disk. This rotation of the disk is the simplest 
transformation in the whole set of the conformal Killing group $PSL(2,\aaru)$ 
of the disk.} transformation $T(z)=e^{2\pi i/n}z$ 
in the unit disk (as opposed to the 
upper half plane) representation of the disk. 
By $SL(2,\aaru)$ invariance
of the CFT correlator on the disk, the left hand side of~(\ref{eq:GCa}) can 
be written as 
\begin{equation}
\langle (g_1\circ A_1)\cdots(g_{n-1}\circ A_{n-1})(g_n\circ\Phi)\rangle =
\langle(T\circ g_1\circ A_1)\cdots(T\circ g_{n-1}\circ A_{n-1})(T\circ g_n
\circ \Phi)\rangle , \label{eq:GD}
\end{equation}
where we abbreviate the superscript $(n)$ on $g_k^{(n)}$ and the argument (0) 
of $(g_k\circ A_k)(0)$. From the definition~(\ref{eq:FYa}) of $g_k$, we find 
$T\circ g_k=g_{k+1}$. Though $T\circ g_n(z)=T^n\circ g_1(z)$ and $T^n(z)=
e^{2\pi i}z$, it acts non-trivially on the fields of non-integer weights. In 
fact, for a primary field $\varphi(z)$ of conformal weight $h$, 
\begin{equation}
T^n\circ \varphi(z)=\left(\frac{d}{dz}(e^{2\pi i}z)\right)^h\varphi(T^n(z))
=e^{2\pi ih}\varphi(z) \label{eq:GE}
\end{equation}
shows that $T^n$ gives a non-trivial factor $e^{2\pi ih}$. Keeping this in 
mind, we move the last factor $(T\circ g_n\circ\Phi)$ to the extreme 
left. If it is Grassmann even, \textit{i.e.} $\Phi$ is in the GSO($+$) sector,
we can do so trivially. But if $\Phi$ is Grassmann odd, the sign factor can 
arise. Since the product of $n$ operators in the correlator~(\ref{eq:GD}) 
must be Grassmann even, the product of other $(n-1)$ operators must also be 
Grassmann odd when $\Phi$ is odd. Therefore in moving $(T\circ g_n\circ\Phi)$ 
to the left an extra minus sign actually 
arises. But remember the factor $e^{2\pi ih}$
arising from $T^n$, which is $+1$ for GSO($+$) states and $-1$ for GSO($-$) 
states. So these two sign factors precisely cancel each other, which enables 
us to write~(\ref{eq:GD}) as 
\begin{equation}
\langle (g_1\circ\Phi))(g_2\circ A_1)\cdots(g_n\circ A_{n-1})\rangle =\llk
\Phi A_1\ldots A_{n-1}\rrk , \label{eq:GF}
\end{equation}
which completes the proof of (\ref{eq:GCa}), (\ref{eq:GMa}). When we replace 
$\Phi$ by $Q_B\Phi$ or $\eta_0\Phi$, the contribution $T^n\to e^{2\pi ih}$ 
remains true since $Q_B$ or $\eta_0$ does not affect the conformal weight 
(they are integrals of currents of weight 1). $Q_B$ or $\eta_0$ does, however,
change the statistics, so that the factor $(-1)$ always appears rather than 
cancels. Thus we have found (\ref{eq:GCb}), (\ref{eq:GCc}), (\ref{eq:GMb}),
(\ref{eq:GMc}) to be the case. In the above lines, we implicitly assumed that 
the operators have zero momenta so that $\ap p^2$ in~(\ref{eq:L0mo}) does not 
contribute to the conformal weight. Even in the case of general weight 
$h\in\aaru$, the cyclicity relations (\ref{eq:GCa})$\sim$(\ref{eq:GCc}), 
(\ref{eq:GMa})$\sim$(\ref{eq:GMc}) still hold if we put a cyclicity axiom that
we pick up a phase factor $e^{-2\pi ih}$ when 
moving $(T\circ g_1\circ\Phi)$ to the 
extreme left~\cite{BSZ}. The proof of (\ref{eq:GMd}) and (\ref{eq:GMf}) is 
exactly the same as that of (\ref{eq:GG}), (\ref{eq:GI}). To prove 
(\ref{eq:GMe}), it is sufficient to be aware that $Q_B$ anticommutes with 
Grassmann odd $\Phi_{1,-}$ \textit{and} $\sigma_3$ also anticommutes with 
$\sigma_1$, cancelling $-1$ among each other.
\medskip

The action~(\ref{eq:Berk}) almost needs not to be modified, as long as we use 
(\ref{eq:GL}) as the definition of $\llk\ldots\rrk$. But since the trace over 
the internal Chan-Paton matrices supplies an extra factor of 2, we must 
divide the action by an overall factor 2 to compensate for it. Thus the open 
superstring field theory action on a non-BPS D-brane is given by 
\begin{eqnarray}
\hspace{-2cm}
S_{\tilde{D}}&=&\frac{1}{4g_o^2}\bllk\left(e^{-\widehat{\Phi}}\widehat{Q}_B
e^{\widehat{\Phi}}\right)\left(e^{-\widehat{\Phi}}\widehat{\eta}_0
e^{\widehat{\Phi}}\right) \nonumber \\
& &{}-\int_0^1dt\left(e^{-t\widehat{\Phi}}\partial_te^{t\widehat{\Phi}}\right)
\left\{\left(e^{-t\widehat{\Phi}}\widehat{Q}_Be^{t\widehat{\Phi}}\right),
\left(e^{-t\widehat{\Phi}}\widehat{\eta}_0e^{t\widehat{\Phi}}\right)\right\}
\brrk . \label{eq:Berko}
\end{eqnarray}
We want the action to have gauge invariance under
\begin{equation}
\delta e^{\widehat{\Phi}}=\left(\widehat{Q}_B\widehat{\Lambda}_1\right)
e^{\widehat{\Phi}}+e^{\widehat{\Phi}}\left(\widehat{\eta}_0
\widehat{\Lambda}_2\right) \label{eq:GN}
\end{equation}
as in (\ref{eq:FX}). What Chan-Paton structure should we assign to 
$\widehat{\Lambda}_i$? Since $\Lambda_i$ were Grassmann odd (ghost number 
$-1$) in the GSO($+$) sector, $\widehat{\Lambda}_i$ should anticommute with 
$\widehat{Q}_B$ and $\widehat{\eta}_0$. Moreover, $(\widehat{Q}_B\widehat{
\Lambda}_1)$ and $(\widehat{\eta}_0\widehat{\Lambda}_2)$ must have the same 
Chan-Paton structure as that (\ref{eq:GJ}) of the string field. One can see 
that 
\begin{equation}
\widehat{\Lambda}_i=\Lambda_{i,+}\otimes \sigma_3+\Lambda_{i,-}\otimes 
i\sigma_2 \label{eq:GO}
\end{equation}
has the desired properties. As for the second requirement,
\begin{eqnarray*}
\widehat{Q}_B\widehat{\Lambda}_1&=&Q_B\Lambda_{1,+}\otimes\sigma_3\sigma_3
+Q_B\Lambda_{1,-}\otimes i\sigma_3\sigma_2 \\
&=&(Q_B\Lambda_{1,+})\otimes I+(Q_B\Lambda_{1,-})\otimes\sigma_1 ,
\end{eqnarray*}
and similarly for $\widehat{\eta}_0\widehat{\Lambda}_2$. Also we find the 
first requirement to be satisfied by~(\ref{eq:GO}) because
\[ j_B\Lambda_{i,+}=-\Lambda_{i,+}j_B \quad \mathrm{and} \quad 
\sigma_2\sigma_3=-\sigma_3\sigma_2. \]
Note that the GSO($+$) gauge parameter $\Lambda_{i,+}$ is Grassmann odd, while
the GSO($-$) $\Lambda_{i,-}$ is Grassmann even. In this case, Grassmannality 
fails to coincide with $e^{\pi iF}$ eigenvalue.
\medskip

For explicit calculations, it is useful to expand the action~(\ref{eq:Berko}) 
in a formal power series in $\widehat{\Phi}$. It can be arranged in the form
\begin{equation}
S_{\tilde{D}}=\frac{1}{2g_o^2}\sum_{M,N=0}^{\infty}\frac{(-1)^N}{(M+N+2)!}
\left({M+N \atop N}\right)\bllk\left(\widehat{Q}_B\widehat{\Phi}\right)
\widehat{\Phi}^M\left(\widehat{\eta}_0\widehat{\Phi}\right)\widehat{\Phi}^N
\brrk \label{eq:Berkov}
\end{equation}
thanks to the cyclicity.

\subsection{On a D-brane--anti-D-brane pair}
\begin{figure}[htbp]
	\begin{center}
	\includegraphics{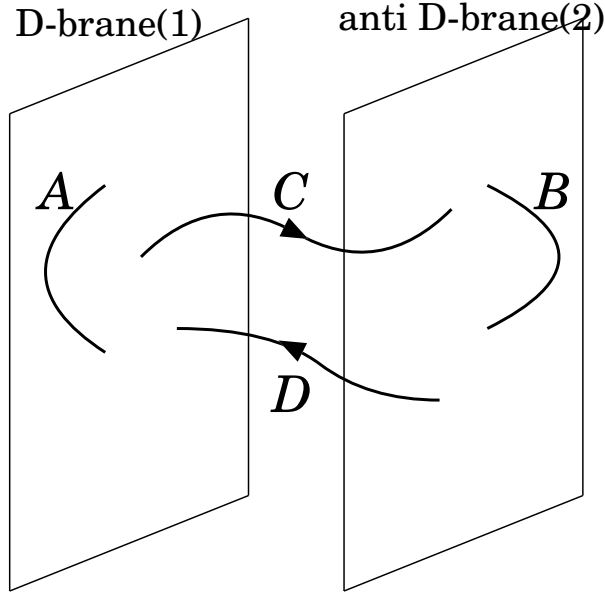}
	\end{center}
	\caption{Strings on a brane-antibrane pair.}
	\label{fig:AD}
\end{figure}

In order to deal with the D-brane anti D-brane system, we further introduce 
the \textit{external} Chan-Paton factors which resembles the conventional 
ones. Though each of the two branes is the BPS object, the GSO($-$) sector 
appears from the strings stretched between the brane and 
the antibrane. Hence we 
still continue to use the internal Chan-Paton factors introduced in the 
non-BPS D-brane case to preserve the algebraic structures. Consider the 
strings on the brane-antibrane pair as in Figure~\ref{fig:AD}. 
Though we depict two branes in Figure~\ref{fig:AD} as if they are separated 
from each other, one should think of these two branes as being coincident. 
Four kinds of strings labeled by A,B,C,D 
are represented by the following external 
Chan-Paton matrices 
\[A:\left(
	\begin{array}{cc}
	1 & 0 \\
	0 & 0
	\end{array}
\right) \> , \> B:\left(
	\begin{array}{cc}
	0 & 0 \\
	0 & 1
	\end{array}
\right) \> , \> C:\left(
	\begin{array}{cc}
	0 & 1 \\
	0 & 0
	\end{array}
\right) \> , \> D:\left(
	\begin{array}{cc}
	0 & 0 \\
	1 & 0
	\end{array}
\right). \]
States arising from strings represented by the diagonal Chan-Paton factors $A$
and $B$, or alternatively by $I$ and $\sigma_3$, live in the GSO($+$) sector. 
In contrast, states from off-diagonal strings $C$ and $D$, or $\sigma_1$ and 
$\sigma_2$, live in the GSO($-$) sector. Therefore we can write the complete 
string field as 
\begin{equation}
\widehat{\Phi}=(\Phi_+^1\otimes I+\Phi_+^2\otimes\sigma_3)\otimes I+
(\Phi_-^3\otimes\sigma_1+\Phi_-^4\otimes\sigma_2)\otimes\sigma_1. 
\label{eq:GP}
\end{equation}
Notice that here we adopt the notation
\[(\mbox{vertex operator})\otimes(\mbox{external Chan-Paton matrix})\otimes
(\mbox{internal Chan-Paton matrix}), \]
which is different from the one $(\mbox{v.o.})\otimes(\mbox{int.})\otimes
(\mbox{ext.})$ used in~\cite{BSZ}. And similarly define 
\begin{eqnarray}
\widehat{Q}_B&=&Q_B\otimes I\otimes\sigma_3 \> , \quad \widehat{\eta}_0=\eta_0
\otimes I\otimes \sigma_3 , \label{eq:GQ} \\
\widehat{\Lambda}_i&=&(\Lambda^1_{i,+}\otimes I+\Lambda^2_{i,+}\otimes 
\sigma_3)\otimes\sigma_3 \nonumber \\
& &{}+(\Lambda^3_{i,-}\otimes\sigma_1+\Lambda^4_{i,-}
\otimes\sigma_2)\otimes i\sigma_2, \label{eq:GR} \\
\llk\widehat{A}_1\ldots\widehat{A}_n\rrk &=&\mathrm{Tr}_{\mathrm{ext}}\otimes
\mathrm{Tr}_{\mathrm{int}}\langle f_1^{(n)}\circ\widehat{A}_1(0)\ldots
f_n^{(n)}\circ\widehat{A}_n(0)\rangle . \label{eq:GS}
\end{eqnarray}
These definitions guarantee that the basic properties 
(\ref{eq:GMa})$\sim$(\ref{eq:GMf}) hold true also in this case. Cyclicity is 
clear because external Chan-Paton factors satisfy this property under 
$\mathrm{Tr}_{\mathrm{ext}}$ and the other parts are the same as in the 
non-BPS D-brane case. (\ref{eq:GMd}), (\ref{eq:GMf}) are as true as ever. 
Leibniz rules hold because the external Chan-Paton factors attached to 
$Q_B,\eta_0$ are the identity element $I$, which causes no sign factor. The 
action takes the same form as in~(\ref{eq:Berko}). To determine the 
normalization factor, we require the action to reproduce the 
action~(\ref{eq:Berk}) on a BPS D-brane when the string field takes the form
$\displaystyle \widehat{\Phi}=\Phi_+\otimes\left(
	\begin{array}{cc}
	1 & 0 \\
	0 & 0
	\end{array}
\right)\otimes I$, \textit{i.e.} string is 
constrained on one D-brane.
In this case, the double trace gives a factor
\[ \mathrm{Tr}_{\mathrm{ext}}\left(
	\begin{array}{cc}
	1 & 0 \\
	0 & 0
	\end{array}
\right)^n\times \mathrm{Tr}_{\mathrm{int}}I^m=2 \]
for any $n,m\ge 1$. So we find that the following action on a brane-antibrane 
pair
\begin{eqnarray}
\hspace{-2cm}
S_{D\bar{D}}&=&\frac{1}{4g_o^2}\bllk\left(e^{-\widehat{\Phi}}\widehat{Q}_B
e^{\widehat{\Phi}}\right)\left(e^{-\widehat{\Phi}}\widehat{\eta}_0
e^{\widehat{\Phi}}\right) \nonumber \\
& &{}-\int_0^1dt\left(e^{-t\widehat{\Phi}}\partial_te^{t\widehat{\Phi}}\right)
\left\{\left(e^{-t\widehat{\Phi}}\widehat{Q}_Be^{t\widehat{\Phi}}\right),
\left(e^{-t\widehat{\Phi}}\widehat{\eta}_0e^{t\widehat{\Phi}}\right)\right\}
\brrk , \label{eq:Berkovi}
\end{eqnarray}
where $\llk\ldots\rrk$ includes the double trace (\ref{eq:GS}), contains the 
action~(\ref{eq:Berk}) on a BPS D-brane as a special case with the correct 
normalization.

\section{Some Preliminaries to the Tachyon Potential}
Here we state the conjecture about the tachyon condensation, though we already
used it in section~\ref{sec:yy}. On a non-BPS D-brane, there is a tachyonic 
mode in the GSO($-$) sector. Since the tachyon field is real-valued, the 
situation is much similar to the bosonic D-brane case. That is, the tachyon 
potential has (at least) a minimum, where the tension of the non-BPS D-brane 
is exactly canceled by the negative energy density from the tachyon potential.
And the minimum represents the `closed string vacuum' without any D-brane. 
For a brane-antibrane pair, the tachyonic mode arises on strings stretched 
between the D-brane and the anti D-brane. Since such strings are represented 
by two Chan-Paton sectors $C,D$ in Figure~\ref{fig:AD} in the oriented 
Type II theory, the tachyon field necessarily takes value in the complex 
number. Then the infinitely degenerate continuous minima of the tachyon 
potential form a vacuum manifold which has the topology of a circle. 
When the tachyon condenses homogeneously in the whole 
spacetime, the D-brane and the anti D-brane completely annihilate in pairs, 
resulting in the usual `closed string vacuum'. In this case, the negative 
energy contribution from the minimum of the tachyon potential exactly cancels 
the sum of the tensions of the D-brane and the anti D-brane. If the 
condensation happens in a topologically non-trivial way, namely $\pi_1(S^1)
\neq 0\in\zetto$, there remain some lower dimensional (anti) D-branes and the 
topological charge becomes the Ramond-Ramond charge of the resulting D-branes.
We will, however, not consider the latter case in this chapter. 
\bigskip

To investigate the above conjectures, we must express the open string coupling
$g_o$ in terms of the tension of the brane under consideration. This can be 
done by the method mentioned in section~\ref{sec:mass}. 
We outline it here, giving some comments in connection with the 
remarks at the end of section~\ref{sec:yy}.

For the brane-antibrane system, we do not have to consider the strings 
stretched between the D-brane and the anti D-brane because both of these 
BPS D-branes have the same tension. So it is sufficient to measure the 
tension of one of them. As in section~\ref{sec:mass}, we consider the 
following string field which represents the massless translation mode of 
a single BPS D-brane, 
\begin{equation}
\Phi(z)=\int d^{p+1}\!k\ \phi^i(k)\delta^p(\vec{k})\xi ce^{-\phi}\psi^i
e^{ikX}(z), \label{eq:GT}
\end{equation}
where $i$ denotes the noncompact flat directions transverse to the 
D$p$-brane. Substituting it into the action~(\ref{eq:Berk}) for a BPS 
D-brane, the quadratic part of it becomes 
\begin{equation}
S_{\mathrm{quad}}=\frac{V_p}{2g_o^2}\int dt(\partial_t\chi^i)^2, \label{eq:GU}
\end{equation}
where $\displaystyle \chi^i(t)=\int dk_0\phi^i(k_0)e^{ik_0t}$, 
which is interpreted as 
the location of the D-brane up to a normalization factor, is the Fourier 
transform of $\phi^i(k_0)$. To determine the correct normalization of 
$\chi^i$ in terms of the `physical length' $Y^i$, we compare the mass${}^2$ 
of the string state stretched between two identical BPS D$p$-branes 
separated by $\vec{b}$ found in the string field 
theory with that of the first-quantized string theory. 
The vertex operators representing the ground state on the stretched string 
are, in the oriented theory, given by 
\begin{eqnarray}
U_{k_0}(z)&=&\xi ce^{-\phi}\epsilon_i\psi^ie^{i\frac{b^i}{2\pi}X^i}e^{ik_0X^0}
(z)\otimes\left(
	\begin{array}{cc}
	0 & 1 \\
	0 & 0
	\end{array}
\right) \label{eq:GV} \\
V_{k_0}(z)&=&\xi ce^{-\phi}\epsilon_i\psi^ie^{i\frac{b^i}{2\pi}X^i}e^{ik_0X^0}
(z)\otimes\left(
	\begin{array}{cc}
	0 & 0 \\
	1 & 0
	\end{array}
\right) \nonumber
\end{eqnarray}
Plugging the string field 
\begin{equation}
\Phi(z)=\int dk_0\Bigl(u(k_0)U_{k_0}(z)+v(k_0)V_{k_0}(z)\Bigr)+P(z), 
\label{eq:puvp}
\end{equation}
where the background 
\[ P(z)=\chi^i\xi ce^{-\phi}\psi^i(z)\otimes \left(
	\begin{array}{cc}
	1 & 0 \\ 0 & 0
	\end{array}
\right) \]
represents the translation of one of the two branes by an amount of $\chi^i$, 
into the action~(\ref{eq:Berk}), the mass term for the $u,v$ fields is 
generated through the cubic coupling term and the value of the mass is found 
to be~\cite{BSZ} 
\[ m^2=-\frac{1}{\sqrt{2}\pi}\vec{b}\cdot\vec{\chi}. \]
Comparing it to\footnote{Note that we are setting $\ap =1$.} 
\[ \frac{|\vec{b}+\vec{Y}|^2}{(2\pi)^2}-\frac{\vec{b}^2}{(2\pi)^2}\simeq 
\frac{1}{2\pi^2}\vec{b}\cdot\vec{Y}, \]
we obtain the correct normalization of the translation mode $\vec{\chi}$ as 
\[ \vec{\chi}=-\frac{1}{\sqrt{2}\pi}\vec{Y}. \]
Then we can read off the tension of the BPS D$p$-brane from the quadratic 
action~(\ref{eq:GU}) with $\vec{\chi}$ rewritten in terms of $\vec{Y}$ as 
\begin{equation}
\tau_p=\frac{1}{2\pi^2g_o^2}. \label{eq:GW}
\end{equation}

For the non-BPS D-brane system, the situation is very similar. In fact, 
we find from the following argument that no more calculations are needed. 
We take the string field representing the translation of a single 
non-BPS D$p$-brane to be 
\begin{equation}
\widehat{\Psi}(z)=\int d^{p+1}\!k\ \phi^i(k)\delta^p(\vec{k})\xi ce^{-\phi}
\psi^ie^{ikX}(z)\otimes I, 
\end{equation}
which differs from~(\ref{eq:GT}) only in that $\widehat{\Psi}$ contains the 
internal Chan-Paton matrix $I$. When we substitute $\widehat{\Psi}$ into 
the action~(\ref{eq:Berko}) for a non-BPS D-brane, we obtain the same 
quadratic action as~(\ref{eq:GU}) because the trace over the internal 
Chan-Paton matrices precisely cancels the additional overall normalization 
factor $\frac{1}{2}$ in~(\ref{eq:Berko}). Since there exist GSO($+$) sector 
states on non-BPS D-branes, if we take the GSO($+$) lowest states as 
representing the stretched string states, the string field we should 
consider becomes
\[ \widehat{\Psi}=\Phi\otimes I, \]
where $\Phi$ is given in~(\ref{eq:puvp}). Again the internal Chan-Paton 
matrices, together with the trace over them, lead to no difference from 
the results obtained in the BPS D-brane case. 
Therefore, we have found that the 
non-BPS D$p$-brane tension should also be given by 
\begin{equation}
\tau_p=\frac{1}{2\pi^2g_o^2}. \label{eq:GY}
\end{equation}
The fact that the D$p$-brane tension is given by the same formula both for 
a BPS~(\ref{eq:GW}) and for a non-BPS~(\ref{eq:GY}) D-brane does not mean 
that they have physically the same tension because the definition of $g_o$ 
should be different. In particular, we must not multiply the 
formula~(\ref{eq:GY}) for a non-BPS D$p$-brane tension by the famous 
factor $\sqrt{2}$. 
\medskip

Now that we have obtained the relation between the open string coupling and 
the D-brane tension, next we argue that the tachyon potential is given by the 
same calculations for both the non-BPS D-brane and the brane-antibrane system.
Though the tachyon potential has the form of `mexican hat' or `wine bottle' in
the case of a brane-antibrane pair so that it has infinitely degenerate 
minima, we can identify all those points when looking for a spacetime 
independent vacuum state. In such a case, we can take the complex tachyon 
field to lie on the (positive) real axis, and in particular it is sufficient 
to consider the external Chan-Paton sector $\sigma_1$ in~(\ref{eq:GP}). This 
prescription can be extended to all the massive states in the GSO($-$) sector,
so that we can completely drop the external Chan-Paton sector $\sigma_2$. 
And since the D-brane and the anti D-brane must be coincident to vanish in 
pairs, we will also ignore the external Chan-Paton sector $\sigma_3$. Then the
string field becomes
\begin{equation}
\widehat{\Phi}=\Phi_+\otimes I\otimes I+\Phi_-\otimes\sigma_1\otimes\sigma_1.
\label{eq:GZ}
\end{equation}
Since under the trace over the internal Chan-Paton factors GSO($-$) states 
always appear in even numbers, the \textit{external} Chan-Paton factors with 
the trace over them 
always give an overall factor Tr${}_{\mathrm{ext}}I=2$. Thus if 
we take the trace over the external Chan-Paton factors in advance, the 
action~(\ref{eq:Berkovi}) for the brane-antibrane pair is, 
in appearance, reduced to twice 
the action~(\ref{eq:Berko}) for a non-BPS D-brane. This multiplication factor 
2 means that the depth of the tachyon potential has doubled, which enables the
negative energy density from the tachyon potential to exactly cancel the sum 
of the tensions of the brane-antibrane pair. Recall here 
that the tension of the BPS (anti) D-brane and 
that of the non-BPS D-brane are commonly represented by 
$1/2\pi^2g_o^2$ though the definitions of $g_o$ in these two theories are 
different. In sum, if we can show that the negative energy contribution from 
the tachyon potential at the bottom of the well exactly cancels the tension of
the D-brane for a non-BPS D-brane case, it automatically follows that also 
for a brane-antibrane pair the sum of the tensions of the brane and the 
antibrane is exactly canceled by the tachyon potential. This is why hereafter
we will concentrate on the open superstring field theory on a non-BPS D-brane 
represented by the action~(\ref{eq:Berko}). 
\bigskip

Let us prepare the level expansion of the (tachyonic) string field 
$\widehat{\Phi}$ on a non-BPS D-brane. The truncated 
Hilbert space, denoted by 
$\cH_1$, we should consider for the analysis of tachyon potential is the 
`universal' subspace of the ``large" Hilbert space $\cH_L$. $\cH_1$ consists 
of states which can be obtained by acting on the oscillator vacuum 
$|\Omega\rangle=\xi ce^{-\phi}(0)|0\rangle$ with matter super-Virasoro 
generators $G_r^{\mathrm{m}}, L_n^{\mathrm{m}}$ and ghost oscillators $b_n,
c_n,\beta_r,\gamma_r$, and have the same ghost and picture 
numbers as that of $|\Omega\rangle$. In terms of the corresponding $\cN=2$ 
vertex operators in the ``large" Hilbert space, all of them must have 
$\gh(\Phi)=\pic(\Phi)=0$. Just as in the bosonic string case, $\cH_1$ does not
contain states with nonzero momentum or non-trivial primary states. It can be 
shown in the same way as in section~\ref{sec:univ} that restricting the string
field to this subspace $\cH_1$ gives 
a consistent truncation of the theory in searching the 
tachyon potential for solutions to equations of motion. Since the 
$L_0^{\mathrm{tot}}$ eigenvalue (weight) of the zero momentum tachyon state 
$|\Omega\rangle$ is $-1/2$, we define the level of a component field of the 
string field to be $(h+\frac{1}{2})$, where $h$ is conformal weight of the 
vertex operator associated with the component field. Then the tachyon state
$|\Omega\rangle$ is at level 0. 
\smallskip

Here we consider the gauge fixing. Linearized gauge transformation takes the 
following form
\begin{equation}
\delta\widehat{\Phi}=\widehat{Q}_B\widehat{\Lambda}_1+\widehat{\eta}_0
\widehat{\Lambda}_2. \label{eq:HA}
\end{equation}
Using the first term, we can reach the gauge $\displaystyle b_0\widehat{\Phi}=
\oint\frac{dz}{2\pi i}zb(z)\widehat{\Phi}(0)=0$ for states of nonzero 
$L_0^{\mathrm{tot}}$ eigenvalue. One can prove it in the same way as in 
section~\ref{sec:sample}. By the second term in ~(\ref{eq:HA}), we can further
impose on $\widehat{\Phi}$ the following gauge condition
\begin{equation}
\xi_0\widehat{\Phi}(0)\equiv \oint\frac{dz}{2\pi i}\frac{1}{z}\xi(z)
\widehat{\Phi}(0)= \> :\xi\widehat{\Phi}(0): \> =0. \label{eq:HB}
\end{equation}
We can easily prove it. If $\widehat{\xi}_0\widehat{\Phi}=\widehat{\Omega}
\neq 0$, we perform the linearized gauge transformation 
$\widehat{\Phi}^{\prime}=\widehat{\Phi}-\widehat{\eta}_0\widehat{\Omega}$. 
Here, for $\widehat{\Omega}$ to be qualified as a gauge parameter, 
$\widehat{\xi}_0$ must be defined to be $\widehat{\xi}_0=\xi_0\otimes
\sigma_3$. Then it follows 
\[\widehat{\xi}_0\widehat{\Phi}^{\prime}=\widehat{\xi}_0\widehat{\Phi}-
\widehat{\xi}_0\widehat{\eta}_0(\widehat{\xi}_0\widehat{\Phi})=
\widehat{\xi}_0\widehat{\Phi}-\{\widehat{\xi}_0,\widehat{\eta}_0\}
(\widehat{\xi}_0\widehat{\Phi})=0. \] 
Note that we can reach the gauge~(\ref{eq:HB}) without any limitation. 
We simply assume that the `modified Feynman-Siegel gauge' $b_0\widehat{\Phi}
=\xi_0\widehat{\Phi}=0$ is good even nonperturbatively. 
\medskip

Now we list the low-lying states in Table~\ref{tab:G}. 
\begin{table}[htbp]
\begin{center}
	\begin{tabular}{|c|c|c|c|c|c|}
	\hline
	$L_0^{\mathrm{tot}}$ & level & $e^{\pi iF}$(GSO) & twist & state & 
	vertex operator \\ \hline
	$-1/2$ & 0 & $-$ & $+$ & $|\Omega\rangle=\xi_0c_1e^{-\phi(0)}|0\rangle$ & 
	$\widehat{T}=\xi ce^{-\phi}\otimes\sigma_1$ \\ \hline
	0 & 1/2 & + & $-$ & $c_0\beta_{-1/2}|\Omega\rangle$ & $c\partial c\>\xi
	\partial\xi e^{-2\phi}\otimes I$ \\ \hline
	1/2 & 1 & $-$ & $-$ & $\beta_{-1/2}\gamma_{-1/2}|\Omega\rangle$ & $\xi c
	\partial\phi e^{-\phi}\otimes\sigma_1$ \\ \hline
	 & & & & $c_{-1}\beta_{-1/2}|\Omega\rangle$ & $\widehat{A}=c\partial^2c
	\xi\partial\xi e^{-2\phi}\otimes I$ \\ 
	1 & 3/2 & + & + & $b_{-1}\gamma_{-1/2}|\Omega\rangle$ & $\widehat{E}=\xi
	\eta\otimes I$ \\
	 & & & & $G_{-3/2}^{\mathrm{m}}|\Omega\rangle$ & $\widehat{F}=\xi
	G^{\mathrm{m}}ce^{-\phi}\otimes I$ \\ \hline
	 & & & & $b_{-1}c_{-1}|\Omega\rangle$ & $\xi\partial^2ce^{-\phi}\otimes
	\sigma_1$ \\ & & & & $\beta_{-1/2}\gamma_{-3/2}|\Omega\rangle$ & $\xi
	\partial\xi\eta ce^{-\phi}\otimes \sigma_1$ \\ 3/2 & 2 & $-$ & + & 
	$\beta_{-3/2}\gamma_{-1/2}|\Omega\rangle$ & $\xi c(\partial\phi)^2
	e^{-\phi}\otimes\sigma_1$ \\ & & & & $(\beta_{-1/2})^2(\gamma_{-1/2})^2
	|\Omega\rangle$ & $\xi c\partial^2\phi e^{-\phi}\otimes\sigma_1$ \\ 
	 & & & & $L_{-2}^{\mathrm{m}}|\Omega\rangle$ & $\xi T^{\mathrm{m}}c
	e^{-\phi}\otimes \sigma_1$ \\ \hline
	\end{tabular}
	\caption{Zero momentum low-lying Lorentz scalar states in the ``large" 
	Hilbert space.}
	\label{tab:G}
\end{center}
\end{table}
The tachyonic ground state is $|\Omega\rangle=\xi_0c_1e^{-\phi(0)}|0\rangle$, 
and the positive level states are obtained by acting on $|\Omega\rangle$ with 
negatively moded oscillators, making sure that the states should have the 
correct ghost number. At first sight, the level 1/2 state $c_0\beta_{-1/2}|
\Omega\rangle$ seems to contradict the Feynman-Siegel gauge condition, but 
note that this state has $L_0^{\mathrm{tot}}=0$ so that it cannot be gauged 
away. These states are mapped to the vertex operators in a definite way, and 
the vertex operator representations are also shown in Table~\ref{tab:G}. 
As an example, let us see the state $c_{-1}\beta_{-1/2}|\Omega\rangle$. 
In terms of the vertex operators, it corresponds to
\begin{eqnarray*}
\hspace{-0.5cm}
(c_{-1}\beta_{-1/2})(\xi_0c_1e^{-\phi(0)}|0\rangle)&\cong&\oint\frac{dz_1}{
2\pi i}\frac{1}{z_1^3}c(z_1)\oint\frac{dz_2}{2\pi i}\beta(z_2)\cdot \xi c
e^{-\phi}(0) \\ &=&\oint\frac{dz_1}{2\pi i}\frac{1}{z_1^3}\left(\frac{1}{2!}
z_1^2\partial^2c(0)\right)\xi c(0)\oint\frac{dz_2}{2\pi i}(-\partial\xi(z_2))
e^{-\phi(z_2)}e^{-\phi(0)} \\ &=&-\frac{1}{2}\partial^2c\xi c(0)\oint
\frac{dz_2}{2\pi i}\partial\xi(z_2)e^{-\ln z_2}:e^{-\phi(z_2)}e^{-\phi(0)}: \\
&=&-\frac{1}{2}c\partial^2c \> \xi\partial\xi e^{-2\phi}(0).
\end{eqnarray*}
In section~\ref{sec:berk} we constructed $\cN=2$ vertex operators in ``large" 
Hilbert space from $\cN=1$ vertex operators in the natural $-1$-picture as 
$\Phi= \> :\xi A:$. In fact, this construction guarantees the second gauge 
condition~(\ref{eq:HB}). Since $\xi$ is a primary field of conformal weight 0,
this $\xi$-multiplication operation, or equivalently acting on a state with 
$\xi_0$, does not affect the weight or level.
\smallskip

We can still consistently truncate the string field by appealing to a 
$\zetto_2$ twist symmetry under which the action is invariant while the 
component fields have the following twist charge
\begin{eqnarray}
(-1)^{h+1} &\mathrm{if}& h\in\zetto, \nonumber \\
(-1)^{h+\frac{1}{2}} &\mathrm{if}& h\in\zetto+\frac{1}{2}, \label{eq:HC}
\end{eqnarray}
where $h$ is the conformal weight of the vertex operator with which the 
component field is associated. In other words, $h=(\mathrm{level})-
\frac{1}{2}$. Though we can prove the twist invariance in a similar (but 
more complicated) way to the proof of cyclicity~(\ref{eq:GMa}), we leave 
it to the reference~\cite{BSZ}. Since the twist-odd fields must always enter 
the action in pairs, these fields can be truncated out without contradicting 
equations of motion. Explicitly speaking, we keep only the fields of level 
$0,\frac{3}{2},2,\frac{7}{2},4,\frac{11}{2},\ldots$ which are twist-even.

\section{Computation of the Tachyon Potential}
Now we turn to the action in the level $(M,N)$ truncation scheme. Though the 
action~(\ref{eq:Berko}), or equivalently~(\ref{eq:Berkov}), is non-polynomial,
\textit{i.e.} it contains arbitrarily higher order terms in $\widehat{\Phi}$, 
only a finite number of terms can contribute nonvanishing values to the action
at a given finite level. Since any component field other than the tachyon has 
strictly positive level, it is sufficient to show that each term in the 
action contains only a finite number of tachyon fields. For a CFT correlator
in the ``large" Hilbert space to have a non-vanishing value, the insertion 
must have 
\[bc \mbox{-ghost number}: +3 \> , \quad \xi\eta \mbox{-ghost number}: -1 
\> , \quad 
\phi\mbox{-charge}: -2\] 
as in~(\ref{eq:FZ}). Because the tachyon vertex operator $T=\xi ce^{-\phi}$ 
has $-1$ unit of $\phi$-charge, 
if an infinite number of $T$'s are inserted, then 
infinitely many vertex operators of a positive $\phi$-charge must also be 
inserted inside the correlator for $\phi$-charge to add up to $-2$, but 
such a term does not exist at any finite level.
\medskip

Now we shall see the form of the action at each truncation level. At 
level (0,0), $\widehat{\Phi}=t\widehat{T}$ always supplies $-1$ unit of 
$\phi$-charge. While $\widehat{\eta}_0$ has no $\phi$-charge, $\widehat{Q}_B$
is divided according to $\phi$-charge into three parts
\begin{equation}
Q_B=Q_0+Q_1+Q_2, \label{eq:HD}
\end{equation}
where subscripts denote their $\phi$-charge, as is clear 
from~(\ref{eq:DO}). But when acting on $\widehat{\Phi}=t\widehat{T}$, 
it becomes
\begin{equation}
\widehat{Q}_B\widehat{\Phi}(0)=\oint\frac{dz}{2\pi i}j_B(z)\cdot t\xi c
e^{-\phi}(0)\otimes\sigma_3\sigma_1=-it\left(\frac{1}{2}\xi c\partial c
e^{-\phi}+\eta e^{\phi}\right)(0)\otimes\sigma_2, \label{eq:HE}
\end{equation}
which includes only terms of $\phi$-charge $-1$ or $+1$. Incidentally,
\begin{equation}
\widehat{\eta}_0\widehat{\Phi}(0)=\oint\frac{dz}{2\pi i}\eta(z)t\xi ce^{-\phi}
(0)\otimes\sigma_3\sigma_1=itce^{-\phi}(0)\otimes\sigma_2. \label{eq:HF}
\end{equation}
To sum up, the term of the form $\left(\widehat{Q}_B\widehat{\Phi}\right)
\left(\widehat{\eta}_0\widehat{\Phi}\right)\widehat{\Phi}^N$ has 
$\phi$-charge $-N-1\pm 1$. For this to become equal 
to $-2$, $N$ must be 0 or 2,
which means the level (0,0) truncated action takes the form $S_{(0,0)}=at^2+
bt^4$, as we will see explicitly. Then we move to level $(\frac{3}{2},3)$ or 
(2,4) truncation.\footnote{Because of the twist symmetry, the first 
non-trivial correction to the (0,0) potential comes from level $\frac{3}{2}$ 
fields.} Since the positive level fields enter the action always linearly or
quadratically at this level of approximation, we see how many 
tachyon vertex operators must be inserted inside the correlator for 
$\phi$-charge to be $-2$. We find from Table~\ref{tab:G} that $\widehat{A}$ 
has $\phi$-charge $-2$, $\widehat{E}$ has 0, and all the others have $-1$. 
Hence the term of the form $\llk Q_2 \eta_0E^2T^N\rrk$ can contain the 
largest number of $\Phi$ because $T$ has a negative $\phi$-charge. Since the 
$\phi$-charge of the above term is $2-N$, $N$ must be 4. Then we 
conclude that it suffices to consider the terms in the 
action~(\ref{eq:Berkov}) up to sixth order in the string field 
$\widehat{\Phi}$ at level $(\frac{3}{2},3)$ or (2,4). Using the cyclicity 
and the twist symmetry~\cite{BSZ}
\begin{eqnarray}
& &\llk\widehat{\Phi}_1\ldots\left(\widehat{Q}_B\widehat{\Phi}_k\right)\ldots
\left(\widehat{\eta}_0\widehat{\Phi}_l\right)\ldots\widehat{\Phi}_n\rrk 
\nonumber \\ &=&
(-1)^{n+1}\left(\prod_{i=1}^n\Omega_i\right)\llk\widehat{\Phi}_n\ldots\left(
\widehat{\eta}_0\widehat{\Phi}_l\right)\ldots\left(\widehat{Q}_B
\widehat{\Phi}_k\right)\ldots\widehat{\Phi}_1\rrk
\end{eqnarray}
for the twist even fields ($\Omega_i=+1$), the action up to $\widehat{\Phi}^6$
is written as
\begin{eqnarray}
\hspace{-0.5cm}
S_{\tilde{D}}^{(6)}&=&\frac{1}{2g_o^2}\bllk\frac{1}{2}(\widehat{Q}_B
\widehat{\Phi})(\widehat{\eta}_0\widehat{\Phi})+\frac{1}{3}(\widehat{Q}_B
\widehat{\Phi})\widehat{\Phi}(\widehat{\eta}_0\widehat{\Phi})+\frac{1}{12}
(\widehat{Q}_B\widehat{\Phi})\left(\widehat{\Phi}^2(\widehat{\eta}_0
\widehat{\Phi})-\widehat{\Phi}(\widehat{\eta}_0\widehat{\Phi})\widehat{\Phi}
\right) \nonumber \\ & &{}+\frac{1}{60}(\widehat{Q}_B\widehat{\Phi})\left(
\widehat{\Phi}^3(\widehat{\eta}_0\widehat{\Phi})-3\widehat{\Phi}^2(
\widehat{\eta}_0\widehat{\Phi})\widehat{\Phi}\right) \nonumber \\ & &{}+
\frac{1}{360}(\widehat{Q}_B\widehat{\Phi})\left(\widehat{\Phi}^4(
\widehat{\eta}_0\widehat{\Phi})-4\widehat{\Phi}^3(\widehat{\eta}_0
\widehat{\Phi})\widehat{\Phi}+3\widehat{\Phi}^2(\widehat{\eta}_0\widehat{\Phi}
)\widehat{\Phi}^2\right)\brrk \label{eq:HH} \\ 
&\equiv& -V_{p+1}\tau_pf_6(\widehat{\Phi})=-\frac{V_{p+1}}{2\pi^2g_o^2}
f_6(\widehat{\Phi}), \nonumber 
\end{eqnarray}
where we used (\ref{eq:GW}).
\medskip

At last, we shall show the detailed calculations of the tachyon potential.
First, let us compute the pure tachyon contribution (\textit{i.e.} level 
(0,0)) to the tachyon potential~\cite{NS}. In the quadratic term, however, 
we also include the momentum dependence for future use. For the 
following string field
\begin{equation}
\widehat{\Phi}(z)=\int d^{p+1}k\> t(k)\xi ce^{-\phi}e^{ikX}(z)\otimes\sigma_1,
\label{eq:HI}
\end{equation}
it follows that 
\begin{eqnarray}
\widehat{Q}_B\widehat{\Phi}(z)&=&i\int d^{p+1}k\> t(k)\left[\left(\ap k^2-
\frac{1}{2}\right)\xi c \partial ce^{-\phi}-\eta e^{\phi}\right]e^{ikX}(z)
\otimes\sigma_2, \nonumber \\
\widehat{\eta}_0\widehat{\Phi}(z)&=&i\int d^{p+1}k\> t(k)ce^{-\phi}e^{ikX}
(z)\otimes\sigma_2. \label{eq:HK}
\end{eqnarray}
The quadratic part of the action~(\ref{eq:HH}) becomes
\begin{eqnarray}
\hspace{-0.5cm}
S_{\tilde{D}}^{\mathrm{quad}}&=&\frac{1}{4g_o^2\ap}\llk(\widehat{Q}_B
\widehat{\Phi})(\widehat{\eta}_0\widehat{\Phi})\rrk \nonumber \\
&=&-\frac{1}{4g_o^2\ap}\int d^{p+1}kd^{p+1}q\> t(k)t(q)\left(\ap k^2-
\frac{1}{2}\right) \nonumber \\ & &\times\mathrm{Tr}_{\mathrm{int}}\langle
\cI\circ(\xi c\partial ce^{-\phi}e^{ikX})(0)(ce^{-\phi}e^{iqX})(0)\otimes I
\rangle \nonumber \\ & & \nonumber \\
&=&-\frac{(2\pi)^{p+1}}{2g_o^2\ap}\int d^{p+1}k\> t(k)
t(-k)\left(\ap k^2-\frac{1}{2}\right)\left\langle\xi\left(-\frac{1}{\epsilon}
\right)\frac{1}{2}c\partial c\partial^2c\left(-\frac{1}{\epsilon}\right)
e^{-2\phi(\epsilon)}\right\rangle \nonumber \\
&=&\frac{1}{g_o^2}\int d^{p+1}x\left(-\frac{1}{2}\partial_{\mu}t\partial^{\mu}
t+\frac{1}{4\ap}t^2\right), \label{eq:HL}
\end{eqnarray}
where we tentatively restored $\ap$ so that we can clearly see the tachyon 
mass $-1/2\ap$, but at any other place we set $\ap=1$. And we find that the 
standard normalization of the kinetic term can be obtained with the 
normalization convention~(\ref{eq:FZ}). 

The quartic term of the action for zero momentum tachyon is 
\begin{eqnarray}
S_{\tilde{D}}^{\mathrm{quartic}}&=&\frac{1}{24g_o^2}\left(\llk(\widehat{Q}_B
\widehat{\Phi})\widehat{\Phi}\widehat{\Phi}(\widehat{\eta}_0\widehat{\Phi})
\rrk-\llk(\widehat{Q}_B\widehat{\Phi})\widehat{\Phi}(\widehat{\eta}_0
\widehat{\Phi})\widehat{\Phi}\rrk\right) \nonumber \\
&=&-\frac{1}{12g_o^2}\Bigl\{\langle (f_1^{(4)}\circ(Q_2\Phi))(f_2^{(4)}\circ
\Phi)(f_3^{(4)}\circ\Phi)(f_4^{(4)}\circ(\eta_0\Phi))\rangle \nonumber \\
& &{}+\langle (f_1^{(4)}\circ(Q_2\Phi))(f_2^{(4)}\circ\Phi)(f_3^{(4)}\circ
(\eta_0\Phi))(f_4^{(4)}\circ\Phi)\rangle\Bigr\} \nonumber \\
&=&-\frac{V_{p+1}}{2g_o^2}t^4. \label{eq:HM}
\end{eqnarray}
Calculations of the correlators above are straightforward, though they are 
tedious. Combining~(\ref{eq:HL}) and (\ref{eq:HM}), the tachyon potential at 
level (0,0) approximation is found to be
\begin{equation}
f^{(0,0)}(t)=-S_{\tilde{D}}^{(0,0)}\bigg/\left(\frac{V_{p+1}}{2\pi^2g_o^2}
\right)=\pi^2\left(-\frac{t^2}{2}+t^4\right), \label{eq:HN}
\end{equation}
which has two minima at $\displaystyle t=\pm t_0=\pm\frac{1}{2}$ and the 
minimum value is 
\begin{equation}
f^{(0,0)}(\pm t_0)=-\frac{\pi^2}{16}\simeq -0.617. \label{eq:HO}
\end{equation}
For the conjecture on the 
non-BPS D-brane annihilation to be true, the `universal 
function' $f(\Phi)$ must satisfy $f(\Phi_0)=-1$ as in the bosonic case, 
where $\Phi_0$ represents the exact configuration of the string field at the 
minimum. Although we have taken only the tachyon field into account, the 
minimum value~(\ref{eq:HO}) reproduces as much as 62\% of the conjectured 
value. In particular, as opposed to the result from Witten's cubic superstring
field theory, the tachyon potential found here takes the double-well form 
and has two minima even at level (0,0) approximation.
\medskip

\begin{figure}[htbp]
	\begin{center}
	\includegraphics{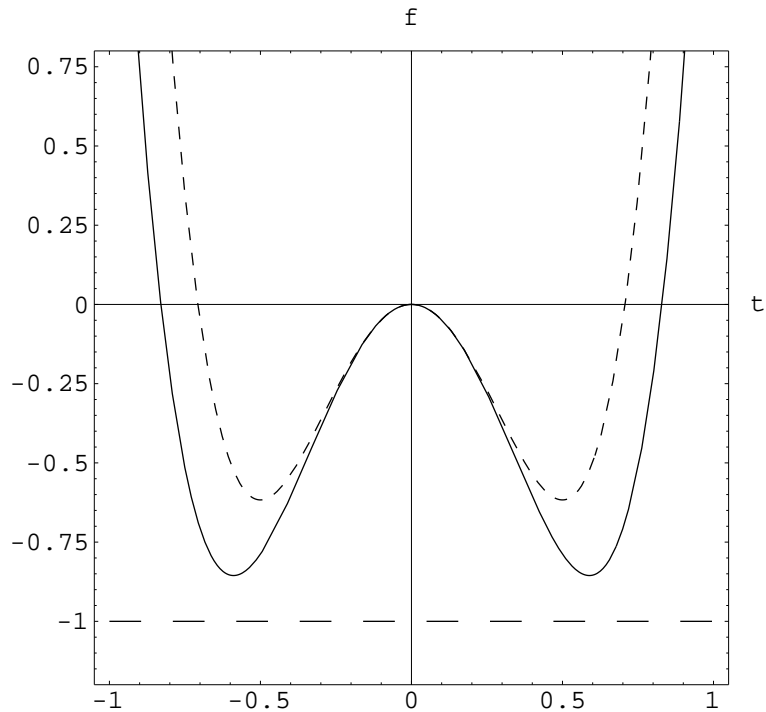}
	\end{center}
	\caption{The effective tachyon potential at level (0,0) (dashed line) 
	and $(\frac{3}{2},3)$ (solid line).}
	\label{fig:AE}
\end{figure}

Though we do not show details here, the computations of the tachyon potential 
have been extended to higher levels. We will quote the results at level 
$(\frac{3}{2},3)$~\cite{BSZ} and at level (2,4)~\cite{3220,4015}. Since at
this level of approximation, as remarked earlier, the strictly positive 
level fields are included only quadratically, these fields can exactly be 
integrated out to give the effective tachyon potential. At level 
$(\frac{3}{2},3)$, the effective potential is given by~\cite{BSZ}
\begin{equation}
f_{\mathrm{eff}}^{(\frac{3}{2},3)}(t)=-4.93t^2\frac{1+4.63t^2+3.21t^4-9.48t^6
-11.67t^8}{(1+1.16t^2)(1+2.48t^2)^2}, \label{eq:HP}
\end{equation}
whose minimum appears at $t=\pm t_0\simeq \pm 0.589$ and 
\[ f_{\mathrm{eff}}^{(\frac{3}{2},3)}(\pm t_0)\simeq -0.854. \]
At level (2,4), though the two results slightly differ from each other, the 
minimum value is reported to be
\[ f_{\mathrm{eff}}^{(2,4)}(\pm t_0)\simeq\left\{
	\begin{array}{ccc}
	-0.891 & \mathrm{in} & \cite{3220} \\
	-0.905 & \mathrm{in} & \cite{4015} 
	\end{array}
\right\} \sim -0.9. \]
We do not write down the expression of the effective potential given 
in~\cite{4015} because it is quite lengthy. The form of the effective 
potential is illustrated in Figure~\ref{fig:AE}.
\bigskip

The results from successive level truncation approximations show that the 
minimum value of the tachyon potential is approaching the conjectured value 
$-1$ as we include fields of higher levels, though it converges less rapidly 
than in the case of bosonic string theory. Not only the minimum value but 
also the shape of the effective tachyon potential possesses the desired 
properties. For example, it has no singularities, as is seen from the 
expression~(\ref{eq:HP}). And at least up to level (2,4), the potential is 
bounded below, which is expected in the superstring theory. From the fact 
that the tachyon potential on a non-BPS D-brane has a $\zetto_2$ symmetry 
$f(t)=f(-t)$, we can obtain the form of the tachyon potential on a 
brane-antibrane pair by rotating the double-well in Figure~\ref{fig:AE} 
around the $f$-axis. The resulting potential has infinitely degenerate 
minima represented by $|t|=t_0$, which have the topology of a circle 
as a whole. As we have discussed before,  
it is already guaranteed that at a minimum of this potential 
the brane and the antibrane are annihilated in 
pairs, resulting in the `closed string vacuum' without any D-brane if we 
assume that $f(\Phi_0)=-1$ is true on a non-BPS D-brane and that the 
condensation takes place in a spatially homogeneous way. To realize more 
complicated situations in this formalism of superstring field theory, 
deeper understandings about the convergence property of the level truncation 
scheme and the symmetries of interaction vertices, including accidental ones, 
seem to be necessary. And it would be interesting to examine the fluctuations 
of the open string degrees of freedom around the `closed string vacuum' 
and to show that there are no physical excitations there.
\medskip

Open superstring field theory was also applied to the D0/D4 system with a 
Neveu-Schwarz $B$-field~\cite{7235}, in which case the unstable system decays 
to a BPS state rather than the vacuum. Turning on a $B$-field in the spatial 
directions of the D4-brane, a tachyonic mode arises in the spectrum of the 
0-4 string~\cite{SW}. This is an indication that the system of a single 
D0-brane and a D4-brane in the presence of the $B$-field is unstable toward 
the formation of the D0-D4 bound state. Therefore the depth of the tachyon 
potential should represent the mass defect in forming the bound state. This 
tachyon potential was calculated~\cite{7235} in the level (0,0) approximation 
and the minimum value of the potential was compared with the expected result, 
though the agreement was not very precise. Note that in the large $B$-field 
limit with Pf$(2\pi\ap B)>0$ the original D0/D4 system effectively looks like 
the system of a D0-$\overline{\mathrm{D}0}$ pair so that the tachyon 
potential in this case reduces to the one~(\ref{eq:HN}) we have derived. 
\smallskip

It is important to see whether the Wess-Zumino-Witten--like superstring field 
theory formulated by Berkovits can successfully be applied to other systems 
as well which is not directly described by the first-quantized superstring 
theory (such as the above example).

\chapter{D-brane~as~a~Tachyonic~Lump \\ in String Field Theory} 
\label{ch:lump}
In bosonic string field theory, which contains D$p$-branes of all dimensions 
up to $p=25$, it was conjectured in~\cite{Descent} that a D$(p-1)$-brane can 
be obtained as a tachyonic lump solution on a D$p$-brane. In this chapter we 
will examine it by using level truncation method of open string field theory. 
We expect that we can learn much more about the structure of the open string 
field theory by investigating such dynamical phenomena than by calculating 
the static tachyon potential because the characteristic factor of 
$e^{\partial^2}$ can have non-trivial effects.

\section{Unstable Lumps in the Tachyonic Scalar Field Theory}
\label{sec:HK}
Since we have no method to deal with the full string field up to now, we 
resort to the level expansion analyses as in the previous chapters. The 
bosonic string field theory action on a D$(p+1)$-brane truncated to level 
(0,0) has already been given in~(\ref{eq:AM}) by
\begin{equation}
S_{(0,0)}=2\pi^2\ap{}^3\tau_{p+1}\int d^{p+2}\!x\left(-\frac{1}{2}\partial_M
\phi\partial^M\phi+\frac{1}{2\ap}\phi^2-2\kappa\tilde{\phi}^3\right),
\label{eq:IA}
\end{equation}
where $M=(0,1,\ldots,p+1)$, $\displaystyle \kappa=\frac{1}{3!}\left(\frac{
3\sqrt{3}}{4}\right)^3$ and 
\[ \tilde{\phi}(x)=\exp\left(-\ap\ln\frac{4}{3\sqrt{3}}\partial^2
\right)\phi(x). \]
Although we know the higher derivative terms in $\tilde{\phi}$ play 
important roles in determining the spectrum at the nonperturbative vacuum, 
as the first approximation we simply set $\tilde{\phi}=\phi$, thinking of the 
higher derivative terms as small corrections to it. This assumption is true if
$\phi(x)$ does not intensively fluctuate. In the next section we will discuss 
the analysis in which we keep the derivative terms fully. 
\medskip

\begin{figure}[htbp]
	\begin{center}
	\includegraphics{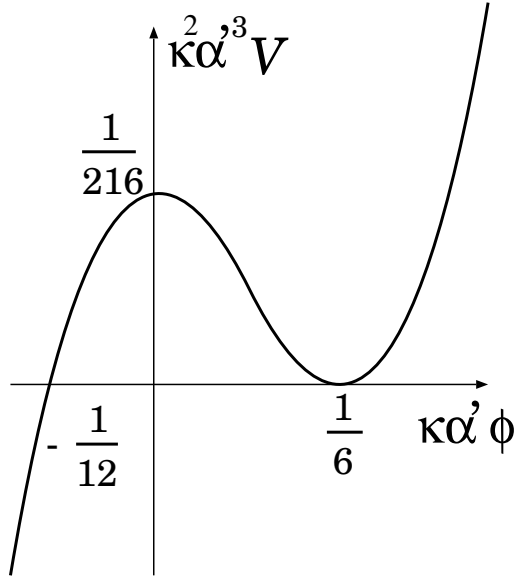}
	\end{center}
	\caption{Cubic potential.}
	\label{fig:AF}
\end{figure}

We begin with a codimension 1 lump on a D$(p+1)$-brane. We will use the 
following notation, $x^0=t, x^{p+1}=x, \vec{y}=(x^1,\ldots,x^p)$ and 
$\mu=(0,1,\ldots,p)$. Using these symbols, we rewrite the action~(\ref{eq:IA})
with tildes removed as 
\begin{equation}
S=2\pi^2\ap{}^3\tau_{p+1}\int d^{p+1}\!y\ dx\left(-\frac{1}{2}\partial_{\mu}
\phi\partial^{\mu}\phi-\frac{1}{2}\left(\frac{\partial\phi}{\partial x}
\right)^2-V(\phi)\right), \label{eq:IB}
\end{equation}
where $V(\phi)$ is the tachyon potential,
\begin{equation}
V(\phi)=-\frac{1}{2\ap}\phi^2+2\kappa\phi^3+\frac{1}{216\kappa^2\ap{}^3}.
\label{eq:IC}
\end{equation}
The last constant term was added by hand in order for the value of the 
potential to vanish at the bottom $\phi=\phi_0=1/6\kappa\ap$. In such a case, 
a configuration of $\phi(x^M)$ which sufficiently rapidly asymptotes to 
$\phi_0$ can support a lower dimensional brane of the finite energy density. 
The shape of $V(\phi)$ is illustrated in Figure~\ref{fig:AF}. 
Here we look for a solution $\overline{\phi}(x)$ which depends only on 
$x=x^{p+1}$. In this case, the equation of motion obtained by varying the 
action~(\ref{eq:IB}) with respect to $\phi$ is 
\begin{equation}
\frac{d^2\overline{\phi}(x)}{dx^2}=V^{\prime}(\overline{\phi}) 
\quad \longrightarrow \quad
\frac{d}{dx}\left[\frac{1}{2}\left(\frac{d\overline{\phi}}{dx}\right)^2\right]
=\frac{d}{dx}V(\overline{\phi}). \label{eq:ID}
\end{equation}
If we read $\overline{\phi}$ and $x$ as the position of a particle and the 
time 
variable respectively, the above equation can be regarded as the equation of 
motion for the particle in a potential $-V$. The particle, which leaves 
$\overline{\phi}=\phi_0=1/6\kappa\ap$ toward left at a 
`time' $x=-\infty$ with 
no initial velocity, reaches $\phi=-1/12\kappa\ap$ at $x=0$ for convenience 
and comes back to $\overline{\phi}=\phi_0$ at $x=\infty$ because there is 
no friction force and $V(\phi_0)=V(-1/12\kappa\ap)=0$. In exactly the same 
way, a lump solution which satisfies the boundary conditions 
\begin{equation}
\lim_{x\to\pm\infty}\overline{\phi}(x)=\phi_0 \> , \quad \overline{\phi}(x=0)
=-\frac{1}{12\kappa\ap} \label{eq:IE}
\end{equation}
can be constructed by integrating~(\ref{eq:ID}) as 
\[x=\int_{\overline{\phi}(0)}^{\overline{\phi}(x)}\frac{d\phi^{\prime}}
{\sqrt{2V(\phi^{\prime})}}=\int_{-1/2}^{6\kappa\ap\overline{\phi}}
\frac{\sqrt{3\ap}d\varphi}{\sqrt{2\varphi^3-3\varphi^2+1}}, \]
which can be solved for $\overline{\phi}$,
\begin{equation}
\overline{\phi}(x)=\frac{1}{6\kappa\ap}\left(1-\frac{3}{2}\sech^2
\left(\frac{x}{2\sqrt{\ap}}\right)\right). \label{eq:IF}
\end{equation}
Expanding $\phi(x^M)$ around the lump solution $\overline{\phi}(x)$ as 
$\phi(x^M)=\overline{\phi}(x)+\varphi(x,y^{\mu})$, the action becomes
\begin{eqnarray}
S&=&2\pi^2\ap{}^3\tau_{p+1}\int d^{p+1}\!y\ dx\biggl[\biggl\{-\frac{1}{2}
\left(\frac{d\overline{\phi}}{dx}\right)^2-V(\overline{\phi})\biggr\}-
\frac{1}{2}\partial_{\mu}\varphi\partial^{\mu}\varphi \nonumber \\
& &{}-\frac{1}{2}\varphi\left(-\frac{\partial^2\varphi}{\partial x^2}
+V^{\prime\prime}(\overline{\phi})\varphi\right)-2\kappa\varphi^3\biggr].
\label{eq:IG}
\end{eqnarray}
The first two terms in $\{\ldots\}$ represent the energy density of the lump. 
Using~(\ref{eq:ID}) and (\ref{eq:IF}), we can carry out the $x$-integration 
to find
\begin{eqnarray}
S^0&=&-2\pi^2\ap{}^3\tau_{p+1}\int d^{p+1}\!y\ dx\left(\frac{d\overline{\phi}}
{dx}\right)^2 \label{eq:IH} \\
&=&-\left(\frac{2^{13}\pi}{5\cdot 3^8}\right)2\pi\sqrt{\ap}\tau_{p+1}
\int d^{p+1}\!y \equiv -\cT_p\int d^{p+1}\!y. \nonumber
\end{eqnarray}
$\cT_p$ defined above is the tension of the lump. In string theory, the 
following relation holds between the D$p$-brane tension $\tau_p$ and the 
D$(p+1)$-brane tension $\tau_{p+1}$,
\[ \tau_p=2\pi\sqrt{\ap}\tau_{p+1}. \]
According to (\ref{eq:IH}), the tension $\cT_p$ of the lump satisfies 
\begin{equation}
\cT_p\simeq 0.784\cdot 2\pi\sqrt{\ap}\tau_{p+1} . \label{eq:II}
\end{equation}
Since 0.784 is rather close to 1, this result seems to support the conjecture 
that the tachyonic lump on a D$(p+1)$-brane represents a D$p$-brane. But 
remember that in the level (0,0) truncated action~(\ref{eq:IB}) 
D$(p+1)$-brane 
tension is not $\tau_{p+1}$, but $\displaystyle \tau_{p+1}^{(0,0)}=
\frac{\pi^2}{108\kappa^2}\tau_{p+1}=\frac{2^{12}\pi^2}{3^{10}}\tau_{p+1}
\simeq 0.684\tau_{p+1}$. Indeed, we saw in chapter~\ref{ch:sft} that the 
depth of the tachyon potential at level (0,0) is about 68\% of the D-brane 
tension. If we compare $\cT_p$ with this value, the relation becomes
\begin{equation}
\cT_p=\frac{18}{5\pi}2\pi\sqrt{\ap}\tau_{p+1}^{(0,0)}\simeq 1.146\cdot
2\pi\sqrt{\ap}\tau_{p+1}^{(0,0)}. \label{eq:IJ}
\end{equation}
In any case, the lump solution (\ref{eq:IF}) reproduces a relatively 
close value to the expected D$p$-brane tension, though we are ignoring higher 
derivative terms or higher level fields. To obtain more precise description, 
let us consider such corrections.
\smallskip

At first, we see the derivative corrections. Substituting 
\[\tilde{\phi}(x^M)=\exp\left(-\ap\ln\frac{4}{3\sqrt{3}}\partial_M\partial^M
\right)\phi(x^M)\simeq\phi(x^M)-\ap\ln\frac{4}{3\sqrt{3}}\partial_M\partial^M
\phi(x^N) \]
into the action~(\ref{eq:IA}) and keeping only two derivatives, we find 
\begin{equation}
S_{(0,0)}=2\pi^2\ap{}^3\tau_{p+1}\int d^{p+2}\!x\left(-\frac{1}{2}f(\phi)
\partial_M\phi\partial^M\phi+\frac{1}{2\ap}\phi^2-2\kappa\phi^3
-\frac{1}{216\kappa^2\ap{}^3}\right), \label{eq:IK}
\end{equation}
where 
\begin{equation}
f(\phi)=1+24\ap\kappa\ln\left(\frac{4}{3\sqrt{3}}\right)\phi. \label{eq:IL}
\end{equation}
Note that $f(\phi)$ vanishes at $\phi\simeq 0.436/\ap <\phi_0\simeq 
0.456/\ap$. This fact is consistent with the conjecture that the standard 
kinetic terms of all fields vanish at the nonperturbative vacuum $\phi=
\phi_0$. By varying~(\ref{eq:IK}) with respect to $\phi$, we get the equation
\[ \frac{1}{2}f(\phi)\left(\frac{d\overline{\phi}}{dx}\right)^2-V(
\overline{\phi})=0. \]
Its solution can formally be written as 
\begin{equation}
x=\int_{\overline{\phi}(0)}^{\overline{\phi}(x)}\frac{\sqrt{f(\phi^{\prime})}
d\phi^{\prime}}{\sqrt{2V(\phi^{\prime})}}, \label{eq:IM}
\end{equation}
but since $f(\phi^{\prime})$ ceases to be positive at $\phi^{\prime}\simeq 
0.436/\ap$ as remarked above, the solution cannot be extended beyond some 
critical value $x_c$ which corresponds to $\overline{\phi}(x_c)=0.436/\ap$. 
In~\cite{2117}, the lump solution $\overline{\phi}(x)$ was obtained by 
solving~(\ref{eq:IM}) numerically, and then the tension of the lump was 
calculated by substituting the solution 
$\overline{\phi}(x)$ into the action~(\ref{eq:IK}) 
and carrying out the $x$-integral from $-x_c$ to $x_c$. The result is 
\begin{equation}
\cT_p\simeq 0.702\cdot 2\pi\sqrt{\ap}\tau_{p+1}. \label{eq:IN}
\end{equation}
Though the leading derivative correction has decreased the tension from the 
value~(\ref{eq:II}) obtained in the zeroth approximation, the difference 
between these values is small. This suggests that ignoring the higher 
derivative terms in $\tilde{\phi}$ was a good approximation in constructing a 
lump of codimension 1.

Secondly, we consider the corrections from level 2 fields $\beta_1,B_{\mu}$ 
and $B_{\mu\nu}$. Here again we set $\tilde{\beta}_1=\beta_1,\tilde{B}_{\mu}
=B_{\mu}, \tilde{B}_{\mu\nu}=B_{\mu\nu}$, and further drop the cubic 
derivative interaction terms. The relevant terms can be found in the appendix 
B of~\cite{KS}. In the notation of it, we need $\cL^{(0)}+\cL^{(2)}+
\cL_0^{(4)}$ with $A_{\mu}=0$. The basic strategy is as follows: Write down 
the action, derive the equations of motion by varying the action with respect 
to level 2 fields, solve them in the presence of the tachyon lump background 
found in level (0,0) approximation, substitute the solutions into the 
action and evaluate it to find the corrections to the lump tension. 
Since we have to fall back on the numerical analysis, we quote the results 
shown in~\cite{2117} with no detailed explanation. Adding level-2 
corrections to the derivative-corrected result~(\ref{eq:IN}), we eventually 
get the following value for the lump tension
\[\cT_p\simeq 0.82\cdot 2\pi\sqrt{\ap}\tau_{p+1}. \]
As compared to the zeroth approximation (\ref{eq:II}), we have found a small 
correction in the right direction. From these results, it may seem that a 
tachyonic lump solution of codimension 1 on a D$(p+1)$-brane gives 
a rather good approximation of a D$p$-brane. Now, how about lumps of 
codimension more than one? We discuss them in the next paragraph.
\medskip

\begin{figure}[htbp]
	\begin{center}
	\includegraphics{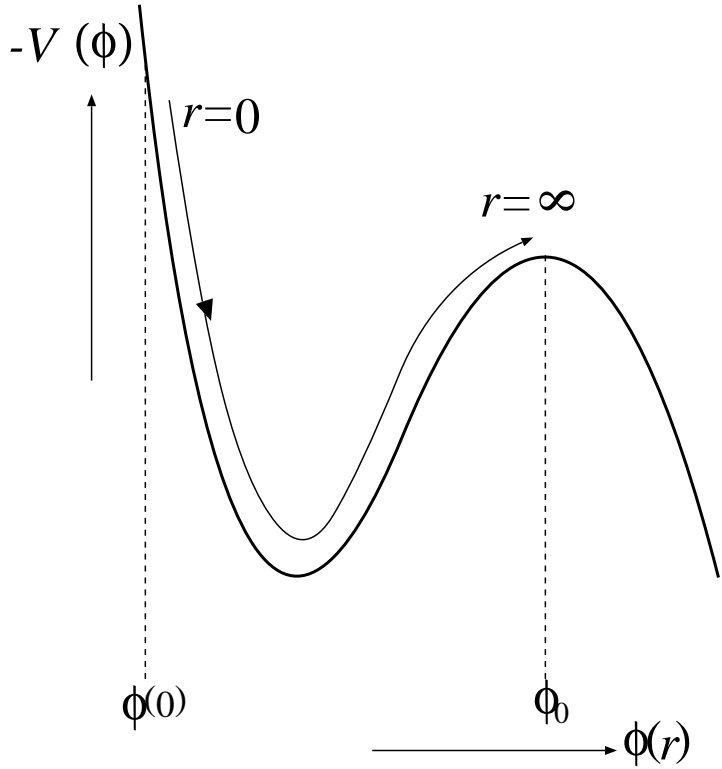}
	\end{center}
	\caption{Mechanical analogy of configuration $\phi(r)$.}
	\label{fig:AG}
\end{figure}

In general, the equation of motion derived from the action~(\ref{eq:IA}) 
takes the following form
\[ \partial_M\partial^M\phi-V^{\prime}(\phi)=0, \]
where $V(\phi)$ is defined in~(\ref{eq:IC}). Here we look for static 
configurations with $(q+1)$-dimensional translation symmetry, which we want 
to identify with D$q$-brane. In addition, we impose on them the spherical 
symmetry in the $(p+1-q)$-dimensional transverse space. Then the above 
equation of motion becomes 
\begin{equation}
\frac{\partial^2\phi}{\partial r^2}+\frac{p-q}{r}\frac{\partial\phi}{\partial
r}-V^{\prime}(\phi)=0, \label{eq:IO}
\end{equation}
where $r(\ge 0)$ is the radial coordinate in the transverse space. 
If we require 
a lump solution to have finite energy and to be smooth at $r=0$, 
the solution must obey the following boundary conditions
\begin{equation}
\lim_{r\to\infty}\phi(r)=\phi_0 \> , \quad \frac{\partial \phi}{\partial r}
\bigg|_{r=0}=0. \label{eq:IP}
\end{equation}
By the mechanical analogy of motion of a particle, 
the `motion' $\phi(r)$ can be depicted as follows: A particle which departs 
from $\phi(0)$ at time $r=0$ with zero initial velocity (the second boundary 
condition) rolls down the potential $-V(\phi)$ and reaches $\phi=\phi_0$ 
at $r=\infty$ (the first boundary condition) with zero final velocity. This 
situation is illustrated in Figure~\ref{fig:AG}.
But notice the existence of the second term in~(\ref{eq:IO}), which is 
nonzero when the value $p+1-q$ of codimension is more than 1. Since this term 
acts as a friction force when the particle moves, the particle cannot arrive 
at $\phi=\phi_0$ unless it is initially 
given more potential energy which compensates 
for the energy loss. In~\cite{2117}, for each value of $p+1-q$ an appropriate 
initial `position' $\phi(0)$ was found in such a way 
that the energy loss due to the 
friction is exactly equal to the difference $|V(\phi(0))-V(\phi_0)|$ of the 
potential energy. Then, $\phi(r)$ correctly asymptotes to $\phi_0$ 
at $r=\infty$. The results are shown in Table~\ref{tab:H}.
\begin{table}[htbp]
	\begin{tabular}{|c|c|c|c|c|c|c|c|}
	\hline
	codimension $p+1-q$ & 1 & 2 & 3 & 4 & 5 & 6 & $\ge$ 7 \\
	\hline
	$\ap \phi(0)$ & $-0.23$ & $-0.64$ & $-1.5$ & $-3.5$ & $-11.5$ & $-10^7$ & 
	no solution \\ \hline
	$\cT_q/(2\pi\sqrt{\ap})^{p+1-q}\tau_{p+1}$ & 0.78 & 0.81 & 0.72 & 0.59 & 
	0.30 & 0.003 & no solution \\ \hline
	\end{tabular}
	\caption{Initial value $\phi(0)$ and tension $\cT_q$ of the 
	$(q+1)$-dimensional lump for each value of codimension.}
	\label{tab:H}
\end{table}
From these results, we can read the feature that the tension of the 
$(q+1)$-dimensional lump approximately reproduces the D$q$-brane tension for 
sufficiently small values of codimension. But for $p+1-q=5$ and 6 the tensions
are too small, and once the value of codimension reaches or goes beyond 7 we 
cannot even find any lump solution. This occurs because the particle cannot 
climb to the top of the hill due to the strong friction force even if we 
initially put the particle to an infinitely high place. In the language of 
string field theory, the approximation where we ignored the higher derivative 
terms in $\tilde{\phi}$ was not good. Since the magnitude of the 
`initial' value $\phi(0)$ 
increases together with the increase of the value of codimension $p+1-q$, 
$\phi(0)$ necessarily intensively fluctuates around $r=0$, which invalidates 
the approximation $\tilde{\phi}\simeq\phi$. In fact, we will see in the next 
section that the lump of any codimension can be constructed if we completely 
take the higher derivatives into account. 
\bigskip

So far we have focused our attention on the tension of the lump. Here, we turn
to the fluctuation spectrum on the lump. As we saw that the approximation 
$\tilde{\phi}\simeq\phi$ is good only for small values of codimension, 
we consider the codimension 1 lump only. We begin by the 
action~(\ref{eq:IG}) with the lump tension term dropped, 
\begin{equation}
S=2\pi^2\ap{}^3\tau_{p+1}\int d^{p+1}\!y dx \left[-\frac{1}{2}\partial_{\mu}
\varphi\partial^{\mu}\varphi-\frac{1}{2}\varphi\left(-\frac{\partial^2\varphi}
{\partial x^2}+V^{\prime\prime}(\overline{\phi})\varphi\right)-2\kappa
\varphi^3\right]. \label{eq:IQ}
\end{equation}
If we use (\ref{eq:IC}) and (\ref{eq:IF}), we can write $V^{\prime\prime}
(\overline{\phi})$ explicitly as
\begin{equation}
V^{\prime\prime}(\overline{\phi})=-\frac{1}{\ap}+12\kappa\overline{\phi}
=\frac{1}{\ap}\left(1-3\sech^2\left(\frac{x}{2\sqrt{\ap}}\right)\right).
\label{eq:IR}
\end{equation}
Expanding the fluctuation field $\varphi(x,y)$ in an orthonormal complete 
basis $\{\xi_n(x)\}$ as 
\begin{equation}
\varphi(x,y)=\sum_n\phi_n(y)\xi_n(x), \label{eq:IS}
\end{equation}
and using (\ref{eq:IR}), the second term in the action~(\ref{eq:IQ}) has the 
following structure,
\begin{equation}
-\frac{1}{2}\sum_{nm}\phi_n(y)\phi_m(y)\xi_n(x)\left[-\frac{d^2}{dx^2}\xi_m(x)
+\frac{1}{\ap}\left(1-3\sech^2\left(\frac{x}{2\sqrt{\ap}}\right)\right)
\xi_m(x)\right]. \label{eq:IT}
\end{equation}
Hence, if $\xi_m(x)$ solves the following eigenvalue equation
\begin{equation}
-\frac{d^2}{dx^2}\xi_m(x)+\frac{1}{\ap}\left(1-3\sech^2\frac{x}{2\sqrt{\ap}}
\right)\xi_m(x)=M_m^2\xi_m(x), \label{eq:IU}
\end{equation}
the fields $\phi_n(y)$ living on the lump have definite mass squared $M_n^2$. 
The eigenvalue equation~(\ref{eq:IU}) is interpreted as the Schr\"{o}dinger 
equation for wavefunctions $\xi_m(x)$ of a particle in the potential 
$V^{\prime\prime}(\overline{\phi})$ generated by the tachyon lump background 
$\overline{\phi}(x)$. This equation is exactly solvable. As a matter of fact, 
this equation belongs to a family of exactly solvable reflectionless 
potentials labeled by a parameter $\ell$. To see this, define a new variable 
$u=x/2\sqrt{\ap}$ and rewrite~(\ref{eq:IU}) as 
\begin{equation}
-\frac{d^2}{du^2}\xi_m(u)+(9-12\sech^2u)\xi_m(u)=(4\ap M_m^2+5)\xi_m(u). 
\label{eq:IV}
\end{equation}
This equation is the $\ell =3$ case of the general equations 
\begin{equation}
-\frac{d^2}{du^2}\xi_m(u)+(\ell^2-\ell(\ell+1)\sech^2u)\xi_m(u)=E_m(\ell)
\xi_m(u). \label{eq:IW}
\end{equation}
It is known that the eigenvalue equation~(\ref{eq:IW}) is exactly solvable 
and that its eigenfunctions can be written in terms of Legendre polynomials 
($\ell\in\zetto$) or hypergeometric functions ($\ell\not\in\zetto$). 
For details, see~\cite{8227} and the references therein. Resticting our 
attention to the 
$\ell=3$ case~(\ref{eq:IV}), its solutions as well as the way to 
solve it are found in~\cite{8227}. As the latter is a technical issue, we 
simply write down the solutions here. There are three bound states\footnote{
Here we use the notation which is different from that of~\cite{8227}: $\xi_-
(\mathrm{here})=\xi_0(\mathrm{there})\> , \quad \xi_0(\mathrm{here})=
\eta_0(\mathrm{there})$ and $\xi_+(\mathrm{here})=\xi_1(\mathrm{there})$.} 
\begin{eqnarray}
\xi_-(x)&=&\sqrt{\frac{15}{32\sqrt{\ap}}}\sech^3\left(\frac{x}{2\sqrt{\ap}}
\right) \quad \mathrm{with} \quad M^2=-\frac{5}{4\ap}, \label{eq:IX} \\
\xi_0(x)&=&\sqrt{\frac{15}{8\sqrt{\ap}}}\tanh\left(\frac{x}{2\sqrt{\ap}}
\right)\sech^2\left(\frac{x}{2\sqrt{\ap}}\right)  
\quad \mathrm{with} \quad M^2=0, \label{eq:IY} \\
\xi_+(x)&=&\sqrt{\frac{3}{32\sqrt{\ap}}}\left(5\sech^3\left(\frac{x}{
2\sqrt{\ap}}\right)-4\sech\left(\frac{x}{2\sqrt{\ap}}\right)\right) 
\label{eq:IZ} \\ & & \quad \mathrm{with} \quad M^2=\frac{3}{4\ap}, \nonumber 
\end{eqnarray}
where the subscripts are intended to represent 
the signs of their mass squared. Thus we have a 
tachyon, a massless scalar and a massive scalar state on the lump. The 
masslessness of the field associated with $\xi_0(x)$ can be seen from the 
form of the wavefunction $\xi_0(x)\sim\partial_x\overline{\phi}(x)$, which 
is the derivative of the lump profile. These eigenfunctions are normalized 
such that 
\begin{equation}
\int_{-\infty}^{\infty}dx\ \xi_n(x)\xi_m(x)=\delta_{nm}. \label{eq:IZa}
\end{equation}
In addition, there is the continuum state labeled by momentum $k$. We do not 
write down the explicit form of it here 
because it does not participate in the essential part of our 
discussion. We note, however, that its mass squared is given 
by $\displaystyle M(k)^2=\frac{1}{\ap}+\frac{k^2}{4}$. To sum up, the 
fluctuation field $\varphi(x,y)$ is expanded as 
\begin{equation}
\varphi(x,y)=\phi_-(y)\xi_-(x)+\phi_0(y)\xi_0(x)+\phi_+(y)\xi_+(x)+
(\mathrm{continuum}). \label{eq:JA}
\end{equation}
Substituting it into the action~(\ref{eq:IQ}), we find the terms quadratic 
in the lump field $\phi_n(y)$ to be 
\begin{eqnarray}
S_{\mathrm{quad}}&=&2\pi^2\ap{}^3\tau_{p+1}\int d^{p+1}\!y \biggl[-\frac{1}{2}
(\partial_{\mu}\phi_-)^2+\frac{5}{8\ap}\phi_-^2-\frac{1}{2}(
\partial_{\mu}\phi_0)^2 \nonumber \\
& & \qquad {}-\frac{1}{2}(\partial_{\mu}\phi_+)^2-\frac{3}{8\ap}\phi_+^2
+(\mathrm{continuum})\biggr], \label{eq:JB}
\end{eqnarray}
due to the orthonormality condition~(\ref{eq:IZa}). That is to say, a 
tachyonic scalar field $\phi(x^M)$ on a D($p+1$)-brane induces a 
tachyon $\phi_-$, a massless scalar $\phi_0$, a massive scalar $\phi_+$ and 
continuum fields on the lump world-volume. 
\bigskip

Leaving investigations about tachyon condensation on the lump in this field 
theory model to later chapter, here we consider whether the 
($p+1$)-dimensional lump has appropriate features to be identified with the 
D$p$-brane. We have seen above that on the lump world-volume a tachyonic 
scalar $\phi_-$ and a massless scalar $\phi_0$ arise from the original 
tachyonic scalar field $\phi(x^M)$. Then, how does the 
massless gauge field $A_{\mu}$ on a D($p+1$)-brane look like on the lump 
world-volume? For this purpose, we consider the level (1,2) truncated 
action~(\ref{eq:AM}) with $\phi$ replaced by $\overline{\phi}$ and the 
tildes removed. The relevant parts are
\begin{eqnarray}
S_{(1,2)}&=&2\pi^2\ap{}^3\tau_{p+1}\int d^{p+2}\!x\biggl(-\frac{1}{2}
\partial_MA_N\partial^MA^N-\frac{3\sqrt{3}}{4}\overline{\phi}A_MA^M 
\label{eq:JC}
\\ & &{}-\frac{3\sqrt{3}}{8}\ap \Bigl(\partial_M\partial_N\overline{\phi}
A^MA^N+\overline{\phi}\partial_MA^N\partial_NA^M-2\partial_M\overline{\phi}
\partial_NA^MA^N\Bigr)\biggr). \nonumber 
\end{eqnarray}
Though $A_{p+1}$ is thought to represent the translation mode of the lump in 
the $x^{p+1}$-direction in the conventional D-brane picture, 
we have already interpreted the massless scalar 
$\phi_0$ as representing such a mode. Hence we take the 
`axial gauge' $A_{p+1}=0$ here. 
And we also impose on $A_M$ the Lorentz-like gauge condition 
$\partial_MA^M=0$. 
From these conditions and the fact that $\overline{\phi}$ 
depends only on $x=x^{p+1}$, the last three terms in~(\ref{eq:JC}) entirely 
vanish. Here we decompose $A_{\mu}(x,y)$ into the product of $x$-dependent 
part and $y$-dependent part as in~(\ref{eq:IS}), but since we only want to 
see whether there is a massless gauge field on the lump world-volume, we 
simply consider a single mode 
\begin{equation}
A_{\mu}(x,y)=\eta(x)\cA_{\mu}(y) \label{eq:JD}
\end{equation}
without a summation over the complete set. 
$\eta(x)$ is normalized such that $\displaystyle
\int_{-\infty}^{\infty}dx\ \eta(x)^2=1$. Plugging~(\ref{eq:JD}) into 
(\ref{eq:JC}), we find 
\begin{eqnarray}
S_{(1,2)}&=&2\pi^2\ap{}^3\tau_{p+1}\int d^{p+1}\!y\biggl[-\frac{1}{2}
\partial_{\mu}\cA_{\nu}\partial^{\mu}\cA^{\nu} \nonumber \\
& &{}-\frac{1}{2}\cA_{\mu}\cA^{\mu}\int dx\ \eta(x)\left(-\frac{d^2}{dx^2}
\eta(x)+\frac{3\sqrt{3}}{2}\overline{\phi}(x)\eta(x)\right)\biggr].
\label{eq:JE}
\end{eqnarray}
Therefore we need to consider the following eigenvalue equation
\begin{equation}
-\frac{d^2}{dx^2}\eta(x)+\frac{16}{27\ap}\left(2-3\sech^2\frac{x}{2\sqrt{\ap}}
\right)\eta(x)=M^2\eta(x), \label{eq:JF}
\end{equation}
and to see whether the above equation has a zero eigenvalue. Arranging this 
equation in the following way,
\begin{equation}
-\frac{d^2}{du^2}\eta+\left(4.897-\frac{64}{9}\sech^2u\right)\eta=(4\ap M^2
+0.156)\eta, \label{eq:JG}
\end{equation}
we find it belongs to the class~(\ref{eq:IW}) as a special case $\ell\simeq
2.213$. Though we do not write the exact solution in terms of a hypergeometric
function for this value of $\ell$, we can read the mass of the vector field 
without knowing the detailed expression of the eigenfunction. 
Since the left hand side of~(\ref{eq:JG}) vanishes for the ground 
state wavefunction $\eta_0(u)$ as in~(\ref{eq:IV}), 
we obtain $M_0^2\simeq -0.039/\ap$. This seems 
to be consistent with the value $m^2\simeq -0.06/\ap$ obtained through a 
numerical calculation in~\cite{2117}. So we have found an approximately 
massless vector field on the lump. By the way, we have seen that the 
(approximately) massless gauge field $\cA_{\mu}$ on the lump world-volume 
arises in such a way that the `bulk' gauge field $A_{\mu}$ on the 
D($p+1$)-brane is trapped by the Schr\"{o}dinger potential 
generated by the tachyonic lump background $\overline{\phi}(x)$ to form a 
bound state. The authors of~\cite{2117} pointed out that the above mechanism 
is reminiscent of the Randall-Sundrum mechanism for gravity.
\medskip

We conclude this section with a summary. By neglecting higher derivative 
terms in $\tilde{\phi}$, we obtained the exact codimension one lump solution 
$\overline{\phi}(x)$ in the level (0,0) approximation. The tension $\cT_p$ 
of the lump reproduced the value near the D$p$-brane tension $\tau_p$.
Derivative corrections and higher level corrections were not so large, 
which suggested that the zeroth approximation taken above was pretty good. 
But it was hard to construct a lump with the large value of codimension in 
this level of approximation. The analysis of small fluctuations on the 
codimension 1 lump showed that the spectrum of the light fields was 
nearly the same as the one 
expected for a bosonic D$p$-brane: a tachyon $(m^2=-5/4\ap)$, a massless 
scalar representing a translational mode of the lump, an approximately 
massless gauge field. We hope that the lump solution correctly describes a 
bosonic D$p$-brane in the full string field theory.

\section{Lumps in Pure Tachyonic String Field Theory}\label{sec:ptsft}
In this section we explain the analysis made in~\cite{3031} in which we keep 
the higher derivative terms exactly while the string field is truncated to  
level 0. This is called `pure tachyonic string field theory'~\cite{8101}. 
Since the analysis relies on a purely numerical approach, 
we can describe only the outline of the strategy and the results. Then we 
discuss the physical meanings of the results.
\medskip

In \cite{3031}, the authors explain why the approach of keeping higher 
derivative terms fully and truncating the string field to lower level fields 
is regarded as a good approximation. Its outline is as follows: Consider a 
scattering amplitude for four tachyons with nonzero momenta. One finds that 
a factor of $(3\sqrt{3}/4)^{-2}$ plays a significant role in reconciling 
the result calculated in the level truncation approximation with the exact 
result. An important fact is that the factor $(3\sqrt{3}/4)^{-2}$ does 
originate from the exponential $\exp(-\ap\ln(4/3\sqrt{3})\partial_{\mu}
\partial^{\mu})$ of derivative. Even if we discard all the fields other than
the lowest one, the difference between the approximated result and the exact 
one is about only 
20\% as long as we keep the derivatives exactly. So they concluded that it was
advantageous to keep the higher derivatives exactly and 
to truncate the string field to lower levels.
\medskip

Now we explain the actual calculations. Performing a field redefinition
\begin{equation}
\varphi=\tilde{\phi}=\exp\left(-\ap\ln\frac{4}{3\sqrt{3}}\partial_{\mu}
\partial^{\mu}\right)\phi, \label{eq:JH}
\end{equation}
the action can be rewritten as 
\begin{eqnarray}
S_{(0,0)}&=&2\pi^2\ap{}^3\tau_{25}\int d^{26}\!x \biggl(\frac{1}{2\ap}\varphi
(\ap\partial^2+1)\exp\left(2\ap\ln\frac{4}{3\sqrt{3}}\partial^2\right)\varphi
\nonumber \\ & &
-2\kappa\varphi^3-\frac{1}{216\kappa^2\ap{}^3}\biggr). \label{eq:JI}
\end{eqnarray}
Though we will consider $(p+1)$-dimensional lumps on a D25-brane, 
we can easily generalize it to a D-brane of arbitrary dimensionality. 
The equation of motion in momentum space is given by
\begin{equation}
\left(\frac{1}{\ap}-\vec{p}\,{}^2\right)\exp\left(-2\ln\frac{4}{3\sqrt{3}}\ap
\vec{p}\,{}^2\right)\varphi(\vec{p})=6\kappa\int d^{25-p}k\ \varphi(\vec{k})
\varphi(\vec{p}-\vec{k}), \label{eq:JJ}
\end{equation}
where $\vec{p},\vec{k}$ are momentum vectors in the $(25-p)$-dimensional 
transverse space. We impose on $\varphi$ a spherical symmetry 
in the transverse space. From here on, we employ numerical methods. 
Firstly, we make an 
initial guess $\varphi_0(\vec{p})$. Then we can calculate the right hand side 
of equation~(\ref{eq:JJ}), 
\begin{equation}
G_0(\vec{p})\equiv\int d^{25-p}\!k\ \varphi_0(\vec{k})\varphi_0(\vec{p}-
\vec{k})=\int\frac{d^{25-p}\!x}{(2\pi)^{25-p}}\varphi_0^2(\vec{x})e^{-i
\vec{p}\cdot\vec{x}}. \label{eq:JK}
\end{equation}
We wrote the final expression because it is technically difficult to compute 
the convolution directly in the momentum space. By Fourier-transforming 
$\varphi_0(\vec{k})$ to $x$-space, squaring it and inverse 
Fourier-transforming back to $p$-space, $G_0(\vec{p})$ is more easily 
computed. Secondly, $G_0(\vec{p})$ can then be used to compute the tachyon 
field $\varphi_1(\vec{p})$ by
\begin{equation}
\varphi_1(\vec{p})=\frac{6\kappa G_0(\vec{p})}{(\frac{1}{\ap}-\vec{p}\,{}^2)}
\exp\left(2\ln\frac{4}{3\sqrt{3}}\ap\vec{p}\,{}^2\right). \label{eq:JL}
\end{equation}
Thirdly, we convolve $\varphi_1(\vec{p})$ to obtain $\displaystyle 
G_1(\vec{p})=\int d^{25-p}\!k\ \varphi_1(\vec{k})\varphi_1(\vec{p}-\vec{k})$. 
All we have to do is to repeat these processes recursively. If we denote the 
$i$-th convolved function by $G_i(\vec{p})$, we should evaluate the error 
$\displaystyle \mathcal{E}_i=\int d^{25-p}\!p |G_i(\vec{p})-
G_{i-1}(\vec{p})|$, and we stop the calculation when $\mathcal{E}_i$ becomes 
sufficiently small. In this way, we get an approximate configuration 
$\varphi(\vec{p})=\varphi_i(\vec{p})$ of the tachyon field representing a 
$(p+1)$-dimensional lump. The tension of the lump is calculated by 
substituting $\varphi=\varphi_i$ into the action~(\ref{eq:JI}) and comparing 
it with $\displaystyle -\cT_p\int d^{p+1}\!x$. We quote the results 
from~\cite{3031} and show them in Table~\ref{tab:I}.
\begin{table}[htbp]
\begin{center}
	\begin{tabular}{|c|c|c|c|c|c|c|c|}
	\hline
	codimension & 1 & 2 & 3 & 4 & 5 & 6 & 7 \\ \hline
	$\ap\phi(0)$ & $-0.63$ & $-0.86$ & $-1.19$ & $-1.63$ & $-2.26$ & $-3.15$ &
	$-4.46$ \\ \hline
	$\cT_p/\tau_p$ & 0.706 & 0.725 & 0.857 & 0.917 & 1.064 & 1.253 & 1.640 \\
	\hline
	\end{tabular}
	\caption{Initial value $\phi(0)$ and the ratio of the tension $\cT_p$ of 
	the lump to the D$p$-brane tension $\tau_p=(2\pi\sqrt{\ap})^{25-p}
	\tau_{25}$.}
	\label{tab:I}
\end{center}
\end{table}
As opposed to the previous results shown in Table~\ref{tab:H}, there seem to 
exist lumps of codimension equal to or more than 7. But the ratio of the 
tension of the lump to the D$p$-brane tension takes a larger value for 
larger $p$. In addition, the ratio for the codimension one lump, which we 
think is most reliable, is only 71\%. These facts suggest that the 
contributions from higher level fields must be included before we conclude 
that the tachyonic lumps can be identified with D-branes. 

\section{Modified Level Truncation Scheme}
\subsection{String field and truncation}
In previous sections we saw two extreme treatments of higher derivatives in 
constructing tachyonic lumps. One was to ignore the derivatives and to set 
$\tilde{\phi}=\phi$, and then we added derivative terms as corrections. The 
other was to keep higher derivatives exactly while we approximated the string 
field by truncating higher level terms. In this section we will introduce a 
compromise between these two. If we use this new method, we will find that the
discrepancy between the tension of the lump and that of the lower-dimensional 
D-brane is only about 1\%! Before that, we look at the space-dependent string 
field configurations. 
\medskip

For simplicity, we begin by considering lumps of codimension 
one.\footnote{After we explain the procedures in this case, we will 
generalize them to lumps of codimension more than one.} We denote by $x$ the 
coordinate of the direction in which we will construct lump-like 
configurations, and by $X$ the corresponding scalar field on the string 
world-sheet. The remaining 25-dimensional manifold $\cM$ is described by a 
conformal field theory of central charge 25. When we consider a D$p$-brane 
as the original (\textit{i.e.} before tachyon condensation) configuration, 
its world-volume is labeled by coordinates $x$ and $x^0,x^1,\ldots ,x^{p-1}$ 
along $\cM$. To keep the total mass of the D$p$-brane finite, we compactify 
all spatial directions tangent to the D-brane. While $(p-1)$ directions 
represented by $(x^1,\ldots ,x^{p-1})$ are wrapped on an arbitrary 
$(p-1)$-cycle of $\cM$, $x$ is compactified on a circle of radius $R$, namely 
$x\sim x+2\pi R$. As in section~\ref{sec:mass}, we assume that there is at 
least one non-compact flat direction in $\cM$ so that the tension of the 
D$p$-brane can be written in terms of the open string coupling constant. 
In this case, we get again
\begin{equation}
\tau_p=\frac{1}{2\pi^2g_o^2\ap{}^3} \label{eq:JM}
\end{equation}
because the situation is exactly the same. The dynamics of open strings on 
this D$p$-brane is described by the following boundary conformal field theory 
(matter part)
\[ \mathrm{CFT}(X)\oplus\mathrm{CFT}(\cM). \]
Letting $L_n^X,L_n^{\cM},L_n^{\mathrm{g}}$ denote the Virasoro generators of 
$\mathrm{CFT}(X),\mathrm{CFT}(\cM)$, and the ghost system respectively, 
the total Virasoro generators of the system is 
\begin{equation}
L_n^{\mathrm{tot}}=L_n^X+L_n^{\cM}+L_n^{\mathrm{g}}. \label{eq:JN}
\end{equation}

For the CFT$(\cM)$ part, we can consistently truncate the Hilbert space of 
ghost number 1 to the `universal subspace' $\cH_1^{\cM ,1}$ by the same 
argument as in the spacetime independent tachyon condensation. Here 
$\cH_1^{\cM ,1}$ includes neither state with nonzero momentum 
$(p^0,p^1,\ldots ,p^{p-1})$ nor non-trivial primary of CFT$(\cM)$. For the 
CFT$(X)$, however, we encounter some complications. Since the lump we are 
seeking is not invariant under the translation in the $x$-direction, 
we must include nonzero momentum 
modes in the string field expansion. In a situation where the string fields 
contain states $|k\rangle =e^{ikX(0)}|0\rangle$ with nonzero momentum along 
$x$, the `1-point' function of the CFT$(X)$ primary $\varphi$ does not 
necessarily vanish,
\[ \langle V_3|(\varphi(0)|k_1\rangle_1)\otimes |k_2\rangle_2\otimes |k_3
\rangle_3\sim \langle\varphi(z_1)e^{ik_1X(z_1)}e^{ik_2X(z_2)}e^{ik_3X(z_3)}
\rangle\neq 0. \]
This is why we have to include primary states of CFT$(X)$ too. Fortunately, 
we find that for nonzero $k$ apparently non-trivial primary states can 
actually be written as Virasoro descendants of the trivial primary. To show 
this, we consider a basis of states with $x$-momentum $k=n/R$, where $n$ is 
some integer. It is obtained by acting on $e^{inX(0)/R}|0\rangle$ with the 
oscillators $\alpha_{-m}^X$. We denote the whole space spanned by such a 
basis by $\cW_n$. And we build the Verma module $\cV_n$ 
on the primary $e^{inX(0)/R}|0
\rangle$, that is, the set obtained by acting on the primary with the 
Virasoro generators $L_{-m}^X$ of CFT$(X)$. If we 
can show $\cW_n=\cV_n$, any state in $\cW_n$ can be written as a Virasoro 
descendant of the unique primary, so there are no non-trivial primary 
states. $\cW_n$ agrees with $\cV_n$ if there are no null states in 
the spectrum. When the following equation
\begin{equation}
\frac{n}{R}=\frac{p-q}{2} \label{eq:JO}
\end{equation}
holds for some integers $p,q$, null states can appear~\cite{MSZ}. Therefore, 
for nonzero $n$, we can avoid introducing non-trivial primary states by 
choosing the radius $R$ of the circle such that the equation~(\ref{eq:JO}) 
can never be satisfied for integers $n,p,q$. But for $n=0$, we cannot help 
including new non-trivial primary states such as $\alpha_{-1}^X|0\rangle$. 
We divide these \textit{zero momentum} primary states into two sets $\cP_+,
\cP_-$ according to the behavior under the reflection $X\to -X$. That is, 
\begin{equation}
\cP_+=\{ |\varphi_e^i\rangle=\varphi_e^i(0)|0\rangle_X\} \> , \quad 
\cP_-=\{ |\varphi_o^i\rangle=\varphi_o^i(0)|0\rangle_X\}, \label{eq:JP}
\end{equation}
where $\varphi_e^i(z)\to \varphi_e^i(z)\> , \ \varphi_o^i(z)\to -\varphi_o^i
(z)$ under $X\to -X$, and the subscript $X$ of $|0\rangle_X$ is added to 
emphasize that this `vacuum' state does not contain any contribution from 
CFT($\cM$) or ghost sector. For example, the state $\alpha_{-1}^X|0\rangle_X$ 
exists in $\cP_-$ while $\alpha_{-2}^X\alpha_{-2}^X|0\rangle_X$ belongs to 
$\cP_+$. And the trivial primary $|0\rangle_X$ itself is contained in 
$\cP_+$. Further, we take linear combinations of $e^{inX(0)/R}|0\rangle$ so 
that they are combined into 
eigenstates of the reflection. It can easily be done by 
\begin{eqnarray}
\cos\left(\frac{n}{R}X(0)\right)|0\rangle&=&\frac{1}{2}\left(e^{inX(0)/R}+
e^{-inX(0)/R}\right)|0\rangle\equiv\frac{1}{2}(|n/R\rangle+|-n/R\rangle),
\nonumber \\
\sin\left(\frac{n}{R}X(0)\right)|0\rangle&=&\frac{1}{2i}\left(e^{inX(0)/R}-
e^{-inX(0)/R}\right)|0\rangle \label{eq:JQ} \\
&=&\frac{1}{2i}(|n/R\rangle-|-n/R\rangle).
\nonumber
\end{eqnarray}
To sum up, the Hilbert space we should consider is constructed by acting on 
the primary states~(\ref{eq:JP}) or (\ref{eq:JQ}) with the matter Virasoro 
generators $L_{-m}^X,L_{-m}^{\cM}$ and the ghost oscillators $b_{-m},c_{-m}$. 
It includes neither the states with nonzero $\cM$-momentum nor non-trivial 
primaries of CFT($\cM$). At present, as far as CFT$(X)$ is concerned, all 
possible states are included in this Hilbert space. For $n=0$ (zero momentum) 
we keep all the primary states~(\ref{eq:JP}). For $n\neq 0$, it was shown 
that the Verma module $\cV_n$ spans the whole space $\cW_n$ if we 
choose a suitable value of $R$. 

Here, we make an exact consistent truncation of this Hilbert space. In order 
to find a one-lump solution along the circle labeled by $x$, we impose a 
symmetry under $x\to -x$ on the solution. Then, since the component fields 
associated with the states which are odd under the reflection must enter the 
action in pairs to respect the symmetry, such fields can consistently be set 
to zero. As the Virasoro generators are invariant under the reflection, we 
can eventually remove the odd primary states $|\varphi_o^i\rangle$ and $\sin
(nX(0)/R)|0\rangle$. Moreover, we can restrict the component fields to 
the ones which are even under the twist symmetry. Twist eigenvalue is given 
by $(-1)^{N+1}$, where $N$ is the eigenvalue of the oscillator number operator 
$\widehat{N}$ (its definition will explicitly be given below). For example, 
$c_1L_{-1}^X\cos(nX(0)/R)|0\rangle$ can be removed. 

After all, we need to consider the following Hilbert space, denoted by 
$\widehat{\cH}$, in discussing a codimension one lump. On the `primary' states
\begin{eqnarray}
& &\cP_+^{\prime}=\{c_1\varphi_e^i(0)|0\rangle\} \quad (\mbox{zero momentum}) 
\quad \mathrm{and} \nonumber \\
& &\left\{c_1\cos\left(\frac{n}{R}X(0)\right)|0\rangle\right\}_{n=1}^{\infty},
\label{eq:JR}
\end{eqnarray}
act with the oscillators
\begin{eqnarray}
&\mathrm{CFT}(X)& \quad L_{-1}^X,L_{-2}^X,L_{-3}^X,\ldots ,\nonumber \\
&\mathrm{CFT}(\cM)& \quad L_{-2}^{\cM},L_{-3}^{\cM}, \ldots , \label{eq:JS} \\
&\mathrm{ghost}& \quad (c_0),c_{-1},c_{-2},\ldots ;\ b_{-1},b_{-2},\ldots, 
\nonumber 
\end{eqnarray}
where $|0\rangle$ is the $SL(2,\aaru)$ invariant vacuum of the matter-ghost 
CFT. The reason why we did not include $L_{-1}^{\cM}$ is that it always 
annihilates the primary states with zero $\cM$-momentum. If we employ the 
Feynman-Siegel gauge $b_0|\Phi\rangle=0$, we still remove the states which 
include $c_0$ if $L_0^{\mathrm{tot}}\neq 0$.
\medskip

Let us take a glance at the zero momentum primaries. Due to the twist 
symmetry, we only need to consider the level 0,2,4,$\ldots$ states. At level 
2, possible states are $\alpha_{-2}^X|0\rangle,\ \alpha_{-1}^X\alpha_{-1}^X
|0\rangle$. As the former belongs to $\cP_-$, we can exclude it. The latter 
can be written as $L_{-2}^X|0\rangle$, which is a Virasoro descendant of the 
trivial primary $|0\rangle$. Therefore there are no non-trivial even 
primaries at level 2. At level 4, there are five possible states 
$\alpha_{-4}^X|0\rangle,\ \alpha_{-3}^X\alpha_{-1}^X|0\rangle,\ \alpha_{-2}^X
\alpha_{-2}^X|0\rangle,\ \alpha_{-2}^X\alpha_{-1}^X\alpha_{-1}^X|0\rangle,\ 
(\alpha_{-1}^X)^4|0\rangle$. Since the 1st and 4th are odd primaries, there 
remain three even primary states. On the other hand, the available Virasoro 
descendants are $L_{-4}^X|0\rangle,\ L_{-2}^XL_{-2}^X|0\rangle$ because 
$L_{-1}^X$ annihilates $|0\rangle$ through $L_{-1}^X|0\rangle\sim 
\alpha_{-1}^Xp|0\rangle=0$. As we have only two Virasoro descendants, one 
state must be added as a non-trivial primary to form a complete set at level 
4. Although we have seen that the first non-trivial even primary appears at 
level 4 in an \textit{ad hoc} way, more systematic approach can be found 
in~\cite{MSZ}.
\bigskip

Thus far, we have not used any approximation scheme. Here we introduce the 
modified version of level truncation. Before we incorporate 
the nonzero momentum modes, just as in the previous chapters, 
the level of a state was defined to be the 
sum of the level numbers of the creation operators acting on the oscillator 
vacuum $|\Omega\rangle=c_1|0\rangle$. Namely, if we define the number operator
\begin{equation}
\widehat{N}=\sum_{n=1}^{\infty}\alpha_{-n}^{\mu}\alpha_{\mu n}+
\sum_{n=-\infty}^{\infty}n\ \ca c_{-n}b_n\ca -1 \label{eq:JT}
\end{equation}
and $\widehat{N}|\Phi_i\rangle=N_i|\Phi_i\rangle$, the level of the state 
$|\Phi_i\rangle$ was $N_i-(-1)$. But note that $L_0^{\mathrm{tot}}$ can be 
written as $L_0^{\mathrm{tot}}=\ap p^2+\widehat{N}$. From this expression, 
once we include the nonzero momentum modes in the string field expansion it 
is natural to generalize the definition of the level of the state $|\Phi_i
\rangle$ as 
\begin{equation}
(L_0^{\mathrm{tot}} \mbox{ eigenvalue of } |\Phi_i\rangle)-(-1)=\ap p^2
+N_i-(-1). \label{eq:JU}
\end{equation}
Of course, $-1$ is the $L_0^{\mathrm{tot}}$ eigenvalue of the zero momentum 
tachyon $c_1|0\rangle$. The level of a component field, denoted by $m$, 
is defined to be equal to the level of the state 
with which the component field is 
associated, and the level of a term in the action, $n$, is defined to be the 
sum of the levels of the fields included in the term, as before. Then we can 
define the level $(M,N)$ approximation for the action to be the one in which 
we keep only those fields of level $m\le M$ and those terms of level $n\le N$ 
in the action. This approximation scheme based on the new 
definition~(\ref{eq:JU}) of the level is called \textit{modified level 
truncation}. 

Now let us see low-lying states in the modified sense. The `tachyon' state 
gives rise to an infinite tachyon tower 
\begin{equation}
|T_n\rangle=c_1\cos\left(\frac{n}{R}X(0)\right)|0\rangle \ , \quad 
\mathrm{level}=\frac{\ap n^2}{R^2}. \label{eq:JV}
\end{equation}
The two states $c_{-1}|0\rangle,\ L_{-2}^{\mathrm{m}}c_1|0\rangle$ which used 
to be at level 2 are now 
\begin{eqnarray}
|U_n\rangle&=&c_{-1}\cos\left(\frac{n}{R}X(0)\right)|0\rangle, \nonumber \\
|V_n\rangle&=&L_{-2}^Xc_1\cos\left(\frac{n}{R}X(0)\right)|0\rangle, 
\label{eq:JW} \\
|W_n\rangle&=&L_{-2}^{\cM}c_1\cos\left(\frac{n}{R}X(0)\right)|0\rangle, 
\nonumber
\end{eqnarray}
all of which are at level$=2+\ap n^2/R^2$. Recalling that $L_{-1}^X$ does not 
annihilate the state of $n\neq 0$, we find additional states
\begin{equation}
|Z_n\rangle=L_{-1}^XL_{-1}^Xc_1\cos\left(\frac{n}{R}X(0)\right)|0\rangle, 
\quad \mathrm{level}=2+\frac{\ap n^2}{R^2}. \label{eq:JX}
\end{equation}
Making use of these states, the string field is expanded as 
\begin{eqnarray}
|\Phi\rangle&=&\sum_{n=0}^{\infty}t_n|T_n\rangle+\sum_{n=0}^{\infty}u_n
|U_n\rangle+\sum_{n=0}^{\infty}v_n|V_n\rangle \label{eq:JY} \\
& &{}+\sum_{n=0}^{\infty}w_n|W_n\rangle+\sum_{n=1}^{\infty}z_n|Z_n\rangle
+\cdots. \nonumber 
\end{eqnarray}
When we fix the expansion level $(M,N)$, the largest value of $n$ (discrete 
momentum) we should keep depends on the radius $R$ of the circle. 

\subsection{Action and the lump tension}
Next, we turn to the action on a D$p$-brane. Since the string field does not 
have $\cM$-momentum, the action always contains an overall volume factor
$V_p=\int_{\cM}dt\ d^{p-1}\!x$. Using the relation~(\ref{eq:JM}), the action 
can be written as 
\begin{eqnarray}
S(\Phi)&=&-V_p2\pi^2\ap{}^3\tau_p\left(\frac{1}{2\ap}\langle\cI\circ\Phi(0)
Q_B\Phi(0)\rangle+\frac{1}{3}\langle f_1\circ\Phi(0)f_2\circ\Phi(0)f_3
\circ\Phi(0)\rangle\right) \nonumber \\
&\equiv&-V_p\tau_p\cdot 2\pi Rf(\Phi), \label{eq:JZ}
\end{eqnarray}
where $\langle\ldots\rangle$ is normalized such that 
\begin{equation}
\langle e^{inX/R}\rangle_{\mathrm{matter}}=2\pi R\delta_{n,0} \> , \quad 
\left\langle\frac{1}{2}\partial^2c\partial c\ c\right\rangle_{\mathrm{ghost}}
=1. \label{eq:KA}
\end{equation}
Since the $X$-momentum is discretized, it is normalized using Kronecker 
delta. As in section~\ref{sec:HK}, we add a constant term to the action so 
that the energy density vanishes at the bottom of the potential. This can be 
executed by adding (minus) the mass of the D$p$-brane, $-2\pi RV_p\tau_p$. 
Then the action becomes
\begin{equation}
S^{\prime}(\Phi)=-2\pi RV_p\tau_p(f(\Phi)+1). \label{eq:KB}
\end{equation}
We denote by $\Phi_0$ the string field configuration representing the 
spacetime independent closed string vacuum we dealt with in 
chapter~\ref{ch:sft}. Since the function $f(\Phi)$ is normalized 
in~(\ref{eq:JZ}) such that $f(\Phi_0)=-1$ if the brane annihilation conjecture
is true, $S^{\prime}(\Phi_0)$ actually vanishes. But considering the fact 
that we have to rely on the level truncation approximation to draw the 
results from string field theory to date, we should replace the expected 
exact D$p$-brane mass $+2\pi RV_p\tau_p$ by $-2\pi RV_p\tau_pf_{(M,N)}
(\Phi_0)$, where the subscript $(M,N)$ represents the level of approximation 
used to compute the action. Then the mass-shifted 
action in $(M,N)$ truncation is given by 
\begin{equation}
S^{\prime}_{(M,N)}(\Phi)=2\pi RV_p\tau_p(f_{(M,N)}(\Phi_0)-f_{(M,N)}(\Phi)).
\label{eq:KC}
\end{equation}
If we find a lump solution $\Phi=\Phi_{\ell}$ which extremizes the action, by 
substituting $\Phi_{\ell}$ into the above action one can write the tension 
$\cT_{p-1}$ of the codimension 1 lump as 
\begin{equation}
S^{\prime}_{(M,N)}(\Phi_{\ell})=2\pi RV_p\tau_p(f_{(M,N)}(\Phi_0)-f_{(M,N)}
(\Phi_{\ell}))\equiv -V_p\cT_{p-1}. \label{eq:KD}
\end{equation}
The conjecture about the tachyonic lump is that the tension $\cT_{p-1}$ of 
the lump actually coincides with the tension $\tau_{p-1}=2\pi\sqrt{\ap}\tau_p$
of the D$(p-1)$-brane. So we need to work out the ratio 
\begin{equation}
r^{(2)}\equiv \frac{\cT_{p-1}}{2\pi\sqrt{\ap}\tau_p}=\frac{R}{\sqrt{\ap}}
(f_{(M,N)}(\Phi_{\ell})-f_{(M,N)}(\Phi_0)) \label{eq:KE}
\end{equation}
for various values of $R$ and see whether the ratio $r^{(2)}$ takes a value 
near 1 irrespective of the values of $R$. For comparison, we can consider 
another ratio
\begin{equation}
r^{(1)}=\frac{R}{\sqrt{\ap}}(f_{(M,N)}(\Phi_{\ell})+1), \label{eq:KF}
\end{equation}
which is obtained by replacing $f_{(M,N)}(\Phi_0)$ 
in~(\ref{eq:KE}) with the expected value $f(\Phi_0)=-1$ while $f_{(M,N)}
(\Phi_{\ell})$ remains the approximate value.

\subsection{Lump solutions}
Here we explain the procedures for finding 
a lump solution as well as its tension 
for a fixed value of radius $R$ and fixed truncation level. We choose 
\[ R=\sqrt{3\ap} \quad \mathrm{and} \quad \mbox{level (3,6)}. \]
At this level, the expansion~(\ref{eq:JY}) of the string field becomes
\begin{eqnarray}
|\Phi_{(3)}\rangle &=&\sum_{n=0}^3t_n|T_n\rangle+\sum_{n=0}^1u_n|U_n\rangle
+\sum_{n=0}^1v_n|V_n\rangle \nonumber \\ & &\quad {}+\sum_{n=0}^1w_n|
W_n\rangle+z_1|Z_1\rangle. \label{eq:KG}
\end{eqnarray}
Substituting the above expansion into the action~(\ref{eq:JZ}) and 
calculating the CFT correlators, we can write down the action in terms of 
the component fields appearing in~(\ref{eq:KG}). Though we can directly 
compute the conformal transformations of the vertex operators and the 
correlators among them, by using the conservation laws explained 
in section~\ref{sec:conserve} we can reduce them to the simplest form 
\begin{equation}
\langle V_3|c_1^{(1)}c_1^{(2)}c_1^{(3)}|n_3/\sqrt{3}\rangle_3\otimes
|n_2/\sqrt{3}\rangle_2\otimes |n_1/\sqrt{3}\rangle_1=\left(\frac{3\sqrt{3}}
{4}\right)^{3-(n_1^2+n_2^2+n_3^2)\ap /R^2}. \label{eq:KH}
\end{equation}
On the way to the final expression for a lump, 
no new techniques are required. Since the 
full expression is quite lengthy and is not illuminating, 
we will not write it down. The explicit expression of the 
potential $\cV(\Phi)=f(\Phi)/2\pi^2\ap{}^3$ at level (3,6) can be found 
in~\cite{MSZ}. At any rate, assume that we now have the level (3,6) 
truncated action at hand. By solving numerically the equations of motion 
obtained by varying the action with respect to the 11 component fields 
appearing in~(\ref{eq:KG}), we want to find a lump solution $\Phi_{\ell}$. 
But, in general, the action or the potential $f_{(3,6)}(\Phi)$ has many 
extrema, so the minimizing algorithm may converge to an unwanted solution. 
Besides, it may converge to the global minimum, the closed string vacuum. 
Nevertheless, 
we can avoid these undesirable solutions by starting the numerical 
algorithm with a suitable initial values because we already have the rough 
estimation for the shape of the lump we are looking for. In this way, we 
obtain a set of the expectation values of the component fields representing 
the lump. Putting these values into~(\ref{eq:KD}), we finally find the 
tension of the lump. If we repeat the above procedures for different values 
of $R$ and different truncation levels, we can use them to see whether the 
conjecture is true. 
\medskip

From here on, we will quote the results from~\cite{MSZ} each time they are 
needed in order to discuss the properties of the lump. First of all, 
let us see how fast the tension of the lump converges to the conjectured 
value (D$(p-1)$-brane tension) as the truncation level increases. See 
Table~\ref{tab:K}.
\begin{table}[htbp]
\begin{center}
	\begin{tabular}{|c|c|c|c|c|c|}
	\hline
	level & $(\frac{1}{3},\frac{2}{3})$ & $(\frac{4}{3},\frac{8}{3})$ & 
	(2,4) & $(\frac{7}{3},\frac{14}{3})$ & (3,6) \\ \hline
	$r^{(1)}$ & 1.32 & 1.25 & 1.11 & 1.07 & 1.06 \\ \hline
	$r^{(2)}$ & 0.774 & 0.707 & 1.024 & 0.984 & 0.994 \\ \hline
	\end{tabular}
	\caption{The ratio of the lump tension to the D$(p-1)$-brane tension 
	for $R=\sqrt{3}$.}
	\label{tab:K}
\end{center}
\end{table}
We find that the value of $r^{(1)}$, or equivalently $f_{(M,N)}(\Phi_{\ell})$,
monotonically decreases as more fields are included. We saw the similar 
tendency in the case of the minimum value $f_{(M,N)}(\Phi_0)$ of the 
tachyon potential, and it is in fact natural because we 
are increasing the number 
of the adjustable parameters when seeking a minimum. Anyway, the value of 
$r^{(1)}$ seems to converge to some value in the vicinity of 1, as expected. 
Whereas the value of $r^{(1)}$ differs from 1 by 6\% at level (3,6), the 
value of $r^{(2)}$ is converging to the expected value even more rapidly: 
about 0.6\%! The value of $r^{(2)}$ oscillates because not only $f_{(M,N)}
(\Phi_{\ell})$ but also $f_{(M,N)}(\Phi_0)$ varies with the truncation level 
and $r^{(2)}$~(\ref{eq:KE}) is determined by the difference between them, but 
$r^{(2)}$ certainly provides a more accurate answer than $r^{(1)}$. From the 
above results, we have gotten the numerical evidence that the modified level 
truncation scheme has a good convergence property.

\begin{figure}[htbp]
	\begin{center}
	\includegraphics{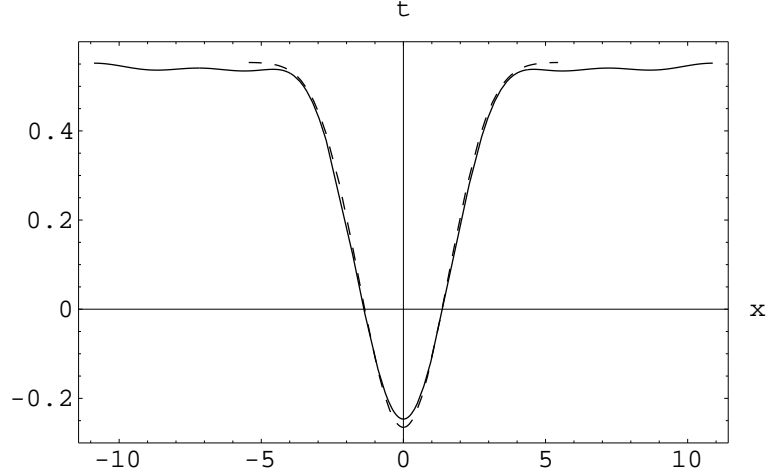}
	\end{center}
	\caption{Tachyon profile $t(x)$ for $R=\sqrt{3}$ (dashed line) and 
	$R=\sqrt{12}$ (solid line).}
	\label{fig:BA}
\end{figure}

Second, we consider several values of the radius and construct a lump 
solution for each value. In~\cite{MSZ}, the following values are chosen,
\[ R=\sqrt{\frac{35}{2}}, \quad \sqrt{12}, \quad \sqrt{\frac{15}{2}},\quad 
\sqrt{3},\quad \sqrt{\frac{11}{10}}. \]
For these values, the relation~(\ref{eq:JO}) never holds so that non-trivial 
primary states need not to be added in the nonzero momentum sectors. The 
tension of the lump for each value of $R$ is given in Table~\ref{tab:L}.
\begin{table}[htbp]
\begin{center}
	\begin{tabular}{|c|c|c|c|c|c|}
	\hline
	radius $R$ & $\sqrt{\frac{35}{2}}$ & $\sqrt{12}$ & $\sqrt{\frac{15}{2}}$ 
	& $\sqrt{3}$ & $\sqrt{\frac{11}{10}}$ \\ \hline
	level & $(\frac{72}{35},\frac{144}{35})$ & $(\frac{25}{12},\frac{25}{6})$ 
	& $(\frac{32}{15},\frac{64}{15})$ & (3,6) & 
	$(\frac{40}{11},\frac{80}{11})$ \\ \hline
	$r^{(1)}$ & 1.239 & 1.191 & 1.146 & 1.064 & 1.022 \\ \hline
	$r^{(2)}$ & 1.024 & 1.013 & 1.005 & 0.994 & 0.979 \\ \hline
	$\sigma/\sqrt{\ap}$ & 1.545 & 1.541 & 1.560 & 1.523 & 1.418 \\ \hline
	\end{tabular}
	\caption{The lump tension and thickness at various radii.}
	\label{tab:L}
\end{center}
\end{table}
The value of $r^{(1)}$ seems to be converging to 1 as the truncation level 
is increased (in this setting the truncation level is higher for a smaller 
value of $R$). Though the value of $r^{(1)}$ is too large for large radii at 
this level, $r^{(2)}$ provides a pretty good value over the whole range of 
radius. Typically, it differs from 1 only by $1\sim 3$\%. The fact that the 
tension of the lump is independent of the radius of the compactification 
circle supports the identification between the lump solution and 
the D($p-1$)-brane 
because the compactification in the directions perpendicular to the D-brane 
does not affect the tension of the D-brane and the lump should have the same 
property if it is to be identified with the D-brane. Since we have 
got expectation values 
of the `tachyon' fields $t_n$ for each radius, we can plot the tachyonic lump 
profile
\begin{equation}
t(x)=\sum_{n=0}t_n\cos\frac{nx}{R} \label{KI}
\end{equation}
as a function of $x$. The expectation values of $t_n$ are shown 
in Table~\ref{tab:M}, and the tachyon profiles are plotted in 
Figure~\ref{fig:BA} only for $R=\sqrt{3}$ and $R=\sqrt{12}=2\sqrt{3}$ because 
all five profiles are as similar as we cannot distinguish one another.
\begin{table}[htbp]
\begin{center}
	\begin{tabular}{|c|c|c|c|c|c|}
	\hline
	$R$ & $\sqrt{35/2}$ & $\sqrt{12}$ & $\sqrt{15/2}$ & $\sqrt{3}$ & 
	$\sqrt{11/10}$ \\ \hline
	$t_0$ & 0.424556 & 0.401189 & 0.363333 & 0.269224 & 0.0804185 \\ \hline
	$t_1$ & $-0.218344$ & $-0.255373$ & $-0.308419$ & $-0.394969$ & 
	$-0.317070$ \\ \hline
	$t_2$ & $-0.176679$ & $-0.190921$ & $-0.194630$ & $-0.125011$ & 
	$-0.00983574$ \\ \hline
	$t_3$ & $-0.132269$ & $-0.122721$ & $-0.0849552$ & $-0.0142169$ 
	& --- \\ \hline
	$t_4$ & $-0.0830114$ & $-0.0575418$ & $-0.0248729$ & --- & --- \\ \hline 
	$t_5$ & $-0.0409281$ & $-0.0210929$ & --- & --- & --- \\ \hline
	$t_6$ & $-0.0178687$ & --- & --- & --- & --- \\ \hline
	\end{tabular}
	\caption{The values of the tachyon fields representing a lump solution.}
	\label{tab:M}
\end{center}
\end{table}
We can then measure the size (thickness) of the lump by fitting the lump 
profile with a Gaussian curve of the form
\[ G(x)=a+b\ e^{-\frac{x^2}{2\sigma^2}} \]
for each value of radius. The results for the value of $\sigma$ are shown 
in Table~\ref{tab:L}. Surprisingly, the value of $\sigma$ seems to be 
independent of the chosen radius $R$. That is to say, if the lump can be 
identified with the D($p-1$)-brane, string field theory predicts that the 
D-brane has a thickness of order of the string scale $\sqrt{\ap}$ 
irrespective of the radius of compactification, at least at this level of 
approximation. It is not clear up to now whether this situation persists 
even after we include higher level modes or the D-brane is still described 
as an object of no thickness in full string field theory. Or, it may be 
possible that a non-trivial field redefinition relates a delta function 
profile (no thickness) to a Gaussian-like profile of a finite size. 

\subsection{Pure tachyonic string field theory revisited}
In section~\ref{sec:ptsft}, we described the numerical analysis using 
`pure tachyonic string field theory' with the action
\begin{equation}
S_0=2\pi^2\ap{}^3\tau_p\int d^{p+1}\!x \left(-\frac{1}{2}\partial_{\mu}\phi
\partial^{\mu}\phi+\frac{1}{2\ap}\phi^2-2\kappa\tilde{\phi}^3\right). 
\label{eq:KJ}
\end{equation}
If we compactify the direction $x^p$ in which we eventually form a lump-like 
configuration on a circle of radius $R$, this system can be reconsidered from 
the point of view of the modified level truncation. When we are interested in 
a codimension one lump so that the field $\phi$ depend only on $x^p$, the 
compactified version of the action~(\ref{eq:KJ}) is obtained by substituting 
the following string field 
\[ |T\rangle=\sum_{n=0}^{\infty}t_nc_1\cos\left(\frac{n}{R}X^p(0)\right)|0
\rangle \]
into the cubic string field theory action~(\ref{eq:JZ}). In this case, the 
calculations involved are not difficult and can be done exactly. The result 
is obtained in~\cite{MSZ} as 
\begin{eqnarray}
S_0&=&-2\pi RV_p\ 2\pi^2\ap{}^3\tau_p\Biggl(-\frac{1}{2\ap}t_0^2-\frac{1}{4}
\sum_{n=1}^{\infty}\left(\frac{1}{\ap}-\frac{n^2}{R^2}\right)t_n^2 \nonumber 
\\ & &{}+\frac{1}{3}K^3t_0^3+\frac{1}{2}\sum_{n=1}^{\infty}t_0t_n^2
K^{3-\frac{2n^2\ap}{R^2}}+\frac{1}{4}\sum_{n=1}^{\infty}t_n^2t_{2n}
K^{3-\frac{6n^2\ap}{R^2}} \nonumber \\ & &{}+\frac{1}{2}\sum_{n=1}^{\infty}
\sum_{m>n}^{\infty}t_nt_mt_{m+n}K^{3-\frac{2\ap}{R^2}(n^2+m^2+nm)}\Biggr),
\label{eq:KK}
\end{eqnarray}
where $\displaystyle K=\frac{3\sqrt{3}}{4}$. Truncating the string field and 
the action to some finite level, and solving the equations of motion obtained 
by differentiating the action with respect to the component fields $t_n$, we 
would obtain a lump solution for some finite radius 
$R$ at some finite truncation 
level. If we could take the limit level $\to \infty$ followed by the radius 
$R\to\infty$, the result for the codimension one lump displayed in 
section~\ref{sec:ptsft} would be reproduced. 
\medskip

To be compared with it, let us consider the following $\phi^3$ scalar field 
theory
\begin{equation}
S_0^{\prime}=2\pi^2\ap{}^3\tau_p\int d^{p+1}\!x \left(-\frac{1}{2}
\partial_{\mu}\phi\partial^{\mu}\phi+\frac{1}{2\ap}\phi^2-2\kappa\phi^3
\right). \label{eq:KL}
\end{equation}
It differs from (\ref{eq:KJ}) in that $\tilde{\phi}^3$ is replaced with 
$\phi^3$. We assume again that $\phi$ depends only on $x^p$ and that 
$x^p$-direction is compactified on a circle of radius $R$. By expanding 
$\phi(x^p)$ in cosine-Fourier series as 
\[ \phi(x^p)=\sum_{n=0}^{\infty}t_n\cos\left(\frac{nx^p}{R}\right), \]
the action (\ref{eq:KL}) is rewritten in terms of $\{t_n\}$, as 
in~(\ref{eq:KK}). But in this case the coefficient $\displaystyle K^{3-
\frac{\ap}{R^2}(l^2+m^2+n^2)}$ of the cubic term $t_lt_mt_n$ is replaced 
by $K^3$. This difference has a great influence on the convergence property 
of the theories~(\ref{eq:KJ}), (\ref{eq:KL}). Since $K>1$, contributions 
from $t_n$ for large $n$ are exponentially suppressed due to the factor 
$K^{-\ap n^2/R^2}$ in the pure tachyonic string field theory~(\ref{eq:KJ}) or 
(\ref{eq:KK}). On the other hand, there are no such suppression factors in 
the $\phi^3$ theory~(\ref{eq:KL}). From this fact, it is clear that the 
calculations in the pure tachyonic string field theory converge much more 
rapidly than in $\phi^3$ theory and that the pure tachyonic string field 
theory is more suitable for the (pure tachyonic) level truncation 
scheme~\cite{8101}.

\section{Higher Codimension Lumps}
It was discovered in~\cite{8053,8101} that lump solutions of codimension 
$d\ge 2$ can be realized in the modified level truncation scheme of bosonic 
open string field theory. Since the procedures here are almost the same as 
in the case of codimension one lump, we only point out the differences 
between them. First of all, a circle of radius $R$ labeled by $x=x^p$ is 
replaced by a $d$-dimensional torus $T^d$ whose coordinates are $(x^{p-d+1},
x^{p-d+2},\cdots, x^p)$. We take the compactification length for each of $d$ 
directions to be the same value $2\pi R$, namely $x^i\sim x^i+2\pi R$ for 
$p-d+1\le i \le p$. Then the spacetime is divided into $T^d\times \cM$, 
where $\cM$ is a $(26-d)$-dimensional Minkowskian manifold. The original 
D$p$-brane is fully wrapped on $T^d$, and a lump solution localized 
on $T^d$ is 
conjectured to represent a D($p-d$)-brane. Secondly, some of the oscillators 
and states have to be modified on $T^d$. The Virasoro generator $L_n^X$ of 
CFT$(X)\equiv$ CFT$(T^d)$ should be understood as $L_n^X=\sum_{i=p-d+1}^p
L_n^{X^i}$. Basis states~(\ref{eq:JV})$\sim$(\ref{eq:JX}) must be altered into
\[ |T_{\vec{n}}\rangle=c_1\cos\left(\frac{\vec{n}\cdot \vec{X}(0)}{R}\right)
|0\rangle \quad \mathrm{with} \quad \mathrm{level}=\frac{\ap\vec{n}^2}{R^2} \]
and so on, where $\vec{n}=(n^{p-d+1},\cdots,n^p)$. When $d\ge 2$, there 
exist $(d-1)$ new zero momentum primaries~\cite{8053}
\begin{equation}
|S^i\rangle=(\alpha_{-1}^p\alpha_{-1}^p-\alpha_{-1}^i\alpha_{-1}^i-\alpha_0^p
\alpha_{-2}^p+\alpha_0^i\alpha_{-2}^i)|0\rangle , \quad p-d+1\le i\le p-1
\label{eq:KM}
\end{equation}
at level 2. Thirdly, the action is written as 
\[ S(\Phi)\equiv -V_{p-d+1}\tau_p(2\pi R)^df(\Phi) \]
with the normalization
\[ \langle e^{i\vec{n}\cdot \vec{X}/R}\rangle_{\mathrm{matter}}=(2\pi R)^d
\delta_{\vec{n},\vec{0}}. \]
Then the tension of the codimension $d$ lump is given by
\[ \cT_{p-d}=-\frac{S_{(M,N)}^{\prime}(\Phi_{\ell})}{V_{p-d+1}}=(2\pi R)^d
\tau_p(f_{(M,N)}(\Phi_{\ell})-f_{(M,N)}(\Phi_0)) \]
\textit{cf}. (\ref{eq:KD}). The ratio of the lump tension to the 
D($p-d$)-brane tension, whose conjectured value is 1, becomes
\begin{equation}
r^{(2)}=\frac{\cT_{p-d}}{(2\pi\sqrt{\ap})^d\tau_p}=\left(\frac{R}{\sqrt{\ap}}
\right)^d(f_{(M,N)}(\Phi_{\ell})-f_{(M,N)}(\Phi_0)). \label{eq:KN}
\end{equation}
\medskip

Now we quote the results for $2\le d\le 6$ from~\cite{8053}. There, the 
radius $R$ is set to $\sqrt{3\ap}$ and the solutions are restricted to the 
ones which have discretized rotational symmetry, namely the permutations 
among $X^i$ and the reflections $X^i\to -X^i$. The values of the 
ratio~(\ref{eq:KN}) at various truncation levels are listed in 
Table~\ref{tab:N}.
\begin{table}[htbp]
\begin{center}
	\begin{tabular}{|c|c|c|c|c|c|c|}
	\hline
	$d$ & $(\frac{1}{3},\frac{2}{3})$ & $(\frac{2}{3},\frac{4}{3})$ & (1,2) &
	$(\frac{4}{3},\frac{8}{3})$ & $(\frac{5}{3},\frac{10}{3})$ & (2,4) \\ 
	\hline
	1 & 0.774 & & & 0.707 & & 1.024 \\ \hline
	2 & 1.34 & 0.899 & & 0.838 & 0.777 & 1.1303 \\ \hline
	3 & 2.32 & 1.42 & 1.07 & 1.03 & 0.931 & 1.3277 \\ \hline
	4 & 4.02 & 2.37 & 1.57 & 1.27 & 1.17 & 1.6384 \\ \hline
	5 & 6.96 & 4.01 & 2.47 & 1.78 & 1.50 & 2.0901 \\ \hline
	6 & 12.06 & 6.82 & 4.04 & 2.69 & 2.06 & 2.6641 \\ \hline
	\end{tabular}
	\caption{The ratio $r^{(2)}$ for the lump of codimension $d$.}
	\label{tab:N}
\end{center}
\end{table}
For $d=2$, no new field appears at level 1 because $\vec{n}^2/3$ cannot 
become 1 for integer $\vec{n}=(n_1,n_2)$. So the entry remains a blank. 
One may find that the value of $r^{(2)}$ suddenly increases at level (2,4). 
This is because the non-tachyon fields $u_0,v_0,w_0$ appear at this level so 
that they bring about qualitative changes in the action. To see how such 
non-tachyon fields affect the value of $r^{(2)}$ in more detail, we need to 
extend the results to still higher levels. For sufficiently small values of 
$d$, the modified level truncation scheme seems to have a good convergence 
property. For large values of $d$, however, the truncation to low-lying 
fields does not give accurate answers. In particular, the tension of the 
lump is overestimated for larger value of $d$. This trend is reminiscent of 
the results we saw in section~\ref{sec:ptsft}.

In~\cite{8101}, codimension two lumps are studied in a similar way. There it 
was shown that the size of the codimension two lump is independent of the 
radius $R$, the same thing being true of the codimension one lump. In 
addition, codimension 2 lump solutions were found also in the pure tachyonic 
string field theory and $\phi^3$ theory. They approximately had the same 
shape as the lump found in the string field theory at level (2,4). The better 
convergence property of the pure tachyonic string field theory discussed 
earlier was actually verified in this case. 

\section{Tachyonic Lumps and Kinks in Superstring Theory}\label{sec:kink}
In this section, we will briefly consider spacetime dependent configurations 
of the tachyon field on a non-BPS D-brane of Type II superstring theory. We 
should notice that in the supersymmetric case the tachyonic lump of 
codimension more than one cannot be produced in the standard field theory 
setting because of the Derrick's theorem: solitons in scalar 
field theory are energetically unstable against shrinking to zero size if 
their codimension is more than one. One of the assumptions in proving this 
theorem is that the scalar potential should be bounded from below. We 
cannot apply this theorem to the lump solutions in bosonic string field 
theory because the tachyon potential is not bounded below. This is why we 
have actually found the lump solutions of codimension more than one in 
previous sections in spite of the no-go theorem. But in superstring theory 
we have the tachyon potential of a double-well form, which is bounded below. 
Among the configurations of codimension 1, we can construct a kink solution 
because the tachyon potential has doubly 
degenerate minima. In fact, it is conjectured that the tachyonic kink on a 
non-BPS D$p$-brane represents a BPS D($p-1$)-brane~\cite{cycle,Hora}. 
And since a lump solution can be considered as a kink--anti-kink 
pair, we will focus our attention on the kink configuration and 
give a rough estimation of the tension of the kink. 
\bigskip

Let us consider the effective field theory for the tachyon. As the first 
approximation, we only take into account the potential term $V(t)$ and the 
standard kinetic term $\partial_{\mu}t\partial^{\mu}t$. The kinetic term has 
been computed in~(\ref{eq:HL}). In this case, the action is written as
\begin{equation}
S=2\pi^2\tilde{\tau}_p\int d^{p+1}\!x\left(-\frac{1}{2}\partial_{\mu}t
\partial^{\mu}t-V(t)\right), \label{eq:KO}
\end{equation}
\[ V(t)=\frac{1}{2\pi^2}(f(t)-f(t_0)), \]
where $f(t)$ is given by (\ref{eq:HN}) at level (0,0) or by (\ref{eq:HP}) at 
level $(\frac{3}{2},3)$. And $\tilde{\tau}_p$ denotes the tension of a 
non-BPS D$p$-brane. The tachyon field $t$ depends only on $x=x^p$. 
The equation of motion derived from the action~(\ref{eq:KO}) is 
\begin{equation}
\frac{d^2t(x)}{dx^2}=V^{\prime}(t(x)). \label{eq:KP}
\end{equation}
We impose on the solution $t=\overline{t}(x)$ the following boundary 
conditions
\[ \lim_{x\to\pm\infty}\overline{t}(x)=\pm t_0 \ , \quad \overline{t}(0)=0. \]
The equation of motion (\ref{eq:KP}) can be integrated to give 
\begin{equation}
x=\int_0^{\overline{t}(x)}\frac{dt'}{\sqrt{2V(t')}}=\pi\int_0^{\overline{t}
(x)}\frac{dt'}{\sqrt{f(t')-f(t_0)}}. \label{eq:KQ}
\end{equation}
Substituting the solution $\overline{t}(x)$ into the action~(\ref{eq:KO}), 
its value can be evaluated as 
\begin{eqnarray}
S(\overline{t})&=&2\pi^2\tilde{\tau}_pV_p\int_{-\infty}^{\infty}dx(-2
V(\overline{t}))=-2\tilde{\tau}_pV_p\int_{-\infty}^{\infty}dx(
f(\overline{t}(x))-f(t_0)) \nonumber \\ &=&-2\pi\tilde{\tau}_pV_p
\int_{-t_0}^{t_0}dt'\sqrt{f(t')-f(t_0)}\equiv -V_p\cT_{p-1},
\label{eq:KR}
\end{eqnarray}
where we have denoted the tension of the kink by $\cT_{p-1}$. Since 
$\tilde{\tau}_p$ is the tension of a non-BPS D$p$-brane, the tension 
$\tau_{p-1}$ of a BPS D$(p-1)$-brane is given by\footnote{Here we set 
$\ap=1$.}
\[ \tau_{p-1}=2\pi \frac{\tilde\tau_p}{\sqrt{2}}. \]
Then we should consider the ratio 
\begin{equation}
r=\frac{\cT_{p-1}}{\tau_{p-1}}=\sqrt{2}\int_{-t_0}^{t_0}dt'
\sqrt{f(t')-f(t_0)}, \label{eq:KS}
\end{equation}
whose conjectured value is 1. This ratio can be calculated at each truncation 
level. At level (0,0), 
\[ r_{(0,0)}=\sqrt{2}\int_{-1/2}^{1/2}dt\frac{\pi}{4}\sqrt{(1-4t^2)^2}=
\frac{\sqrt{2}}{6}\pi\simeq 0.74, \]
while at level $(\frac{3}{2},3)$ numerical method gives
\[ r_{(\frac{3}{2},3)}\simeq 1.03. \]
Although these values are surprisingly close to 1, there are many derivative 
corrections arising from higher order terms we have ignored. 
And we do not know 
whether such corrections are sufficiently small. Hence we should regard 
these agreements as accidental. 
Applying the modified level truncation scheme to this system after 
compactifying the $x^p$-direction, we could investigate the kink solution 
on a non-BPS D$p$-brane more accurately. 

\section{Validity of the Feynman-Siegel gauge}\label{sec:Siegel}
In this section we shall examine whether the Feynman-Siegel gauge is a good 
gauge choice at the nonperturbative level or not. We adopt the following 
criterion: If a nonpertubative solution, such as the closed string vacuum, 
found \textit{in} the Feynman-Siegel gauge solves the full set of equations 
of motion derived from the gauge invariant (\textit{i.e.} 
before gauge-fixing) action, then we can say that the gauge choice was good. 
For that purpose, we expand the string field as 
\begin{equation}
|\Phi\rangle=\sum_a\phi_a|\Phi_a\rangle+\sum_n\phi_n|\Phi_n\rangle, 
\label{eq:RD}
\end{equation}
where $|\Phi_a\rangle$'s satisfy the Feynman-Siegel gauge condition $b_0|
\Phi_a\rangle=0$ while $|\Phi_n\rangle$'s not. That is, every $|\Phi_n
\rangle$ contains a $c_0$ mode. Let $S(\phi_a,\phi_n)$ denote the full gauge 
invariant action obtained by substituting the expansion~(\ref{eq:RD}) into 
the string field theory action. Since the closed string vacuum solution and 
the lump solution we have found in the Feynman-Siegel gauge $(\phi_n\equiv 
0)$ using a level truncation approximation were obtained by solving the 
equations of motion
\begin{equation}
\frac{\delta S(\phi_b,0)}{\delta\phi_a}=0, \label{eq:RF}
\end{equation}
the remaining set of equations of motion we should consider is 
\begin{equation}
\frac{\delta S(\phi_a,\phi_m)}{\delta\phi_n}\Bigg|_{\phi_m=0}\equiv 
L_n+Q_n=0, 
\label{eq:RE}
\end{equation}
where we denoted by $L_n$ and $Q_n$ the contributions from 
linear and quadratic 
terms respectively to the left hand side of the equation~(\ref{eq:RE}). 
Since each of $L_n$ and $Q_n$ is not small, cancellation between these two 
terms must occur for the equation~(\ref{eq:RE}) to be true. We can measure 
the degree of the cancellation by forming the ratio $r_n=(L_n+Q_n)/L_n$. 
This ratio was numerically calculated in~\cite{Hata} for the closed string 
vacuum solution (in~\cite{Hata}, the equation~(\ref{eq:RE}) was interpreted 
as the BRST invariance of the solution), and 
in~\cite{0101014} for the codimension 1 lump solution. 
For the former, the value of the ratio was found to be about 1\% using the 
expectation values of the fields obtained in the level (10,20) approximation. 
For the latter, the values are typically 10-20\% at level (2,6). Though this 
is not so small, the cancellation is indeed realized so that 
we can expect the result 
will be improved if we include higher level fields. These results suggest 
that the Feynman-Siegel gauge is a valid gauge choice in searching for the 
closed string vacuum and the codimension 1 lump solution. 

\chapter{Background Independent Open String Field Theory}\label{ch:biosfit}
Around the year '92--'93, a different formulation of 
open string field theory, called background independent open string field 
theory, was proposed \cite{WiBI1,
WiBI2,Sh1,Sh2,LiWi}. It was recently shown \cite{GerSh,KMM1,KMM2} that this 
theory is well suited for the study of tachyon condensation 
in that only the tachyon field (not the whole string field) 
acquires the nonvanishing 
vacuum expectation value, so we can proceed without any approximation 
scheme such as the level truncation method. This fact means that 
we can obtain some exact results about the tachyon physics:
the tachyon potential and the description of 
the D-branes as tachyonic solitons. In this chapter we first explain 
the formulation of the theory briefly and then see the applications to 
the tachyon condensation in both cases of bosonic string and superstring.

\section{Background Independent Open String Field Theory}
In this theory, the configuration space of the open string field 
is regarded as the `space of all two-dimensional world-sheet 
field theories' on the disk~\cite{WiBI1}. The world-sheet action is given by 
\begin{equation}
\cS=\cS_0+\int_0^{2\pi}\frac{d\theta}{2\pi}\cV(\theta),\label{eq:wsSv}
\end{equation}
where $\cS_0$ stands for the open-closed string conformal 
background~(\ref{eq:wsS}). $\cV(\theta)$ contains interaction terms 
on the disk boundary\footnote{In this sense, this theory is also called 
`boundary string field theory'.} (parametrized by $\theta$) and is not 
conformal. We will consider the bosonic string for a while.

$\cV$ is a local operator constructed from $X,b,c$, and we can expand it as
\begin{equation}
\cV=T(X)+A_{\mu}(X)\partial_{\theta}X^{\mu}+B_{\mu\nu}(X)\partial_{\theta}
X^{\mu}\partial_{\theta}X^{\nu}+\cdots.
\end{equation}
If $\cV$ involves terms with equal to or more than two $\theta$-derivatives, 
which correspond to the massive states in the first-quantized open string 
Hilbert space, then the theory may be ill-defined as a two-dimensional field 
theory because such terms represent non-renormalizable interactions 
on the world-sheet. 
Consequently, we must introduce an ultraviolet cut-off parameter in 
general~\cite{LiWi}, but in the special case of tachyon condensation we have 
only to deal with the relevant perturbation $T(X)$, so we can avoid the 
problem of ultraviolet divergences here.

Interaction $\cV$, which consists of local operators with ghost number 0, 
is considered to be defined from a certain more `fundamental' 
operator $\cO$ as~\cite{WiBI1} \[ \cV=b_{-1}\cO ,\]
where $\cO$ has ghost number +1. This can be seen from the fact that the 
vertex operator associated with the massless gauge boson is of the form 
$c\partial X^{\mu} e^{ik\cdot X}$, with ghost number 1. If $\cV$ purely 
consists of matters $(X)$ without any ghosts, the above equation is 
equivalent to \[ \cO=c\cV.\]
From here on we will consider this case.
\smallskip

Letting $\cV_i$ denote the basis elements for operators of ghost number 0, 
the boundary interaction $\cV$ is written as
\begin{equation}
\cV=\sum_it^i\cV_i,
\end{equation}
where coefficients $t^i$ are couplings on the world-sheet theory, which are 
regarded as fields from the spacetime point of view. Gathering the pieces 
above, the spacetime action $S$ is defined through the Batalin-Vilkovisky 
formalism by
\begin{equation}
dS=\frac{K}{2}\int_0^{2\pi}\frac{d\theta}{2\pi}\int_0^{2\pi}\frac{d
\theta^{\prime}}{2\pi}\langle d\cO(\theta)\{Q_B,\cO(\theta^{\prime})\}
\rangle_{\cV} \label{eq:dS}
\end{equation}
namely,
\begin{equation}
\frac{\partial S}{\partial t^i}=\frac{K}{2}\int\frac{d\theta}{2\pi}\int
\frac{d\theta^{\prime}}{2\pi}\langle\cO_i(\theta)\{Q_B,\cO(\theta^{\prime})
\}\rangle_{\cV}, \label{eq:flying}
\end{equation}
where $Q_B$ is the BRST charge, and $K$ is a normalization factor that is kept
 undetermined here. $\langle\cdots\rangle_{\cV}$ is \textit{un}normalized 
correlation function in the two-dimensional world-sheet field theory evaluated 
with the full (\textit{perturbed}) world-sheet action~(\ref{eq:wsSv}).

\section{World-sheet Renormalization Group}

Although the theory is conformal in the interior of the disk, the boundary 
interaction can have arbitrary relevant (tachyonic) or irrelevant (massive) 
operators. Due to this non-conformal character, 
the coupling constants $t^i$ 
flow under the change of the scale. When we write 
down the spacetime action of the background independent open string field 
theory, it is useful to make use of the $\beta$-functions for various 
couplings. Here, we describe only those things that have to do with the study 
of tachyon condensation. For detailed discussion about world-sheet 
renormalization group, see~\cite{HKM,10247}.
\medskip

First, for a \textit{primary} field $\cV_i$ of conformal weight $h_i$ we have 
\begin{equation}
\{Q_B,c\cV_i\}=(1-h_i)c\partial c\cV_i. \label{eq:Andr}
\end{equation}
And we define the `metric' of the space of couplings $t^i$ as 
\begin{equation}
G_{ij}(t)=2K\int \frac{d\theta d\theta^{\prime}}{(2\pi)^2}\sin^2\frac{\theta
-\theta^{\prime}}{2}\langle \cV_i(\theta)\cV_j(\theta^{\prime})\rangle_{\cV},
\end{equation}
where $\sin^2$ factor comes from the ghost 3-point correlator. Then the 
spacetime action~(\ref{eq:flying}) is reexpressed as
\begin{equation}
\frac{\partial S}{\partial t^i}=-\sum_j(1-h_j)t^jG_{ij}(t). \label{eq:A}
\end{equation}
The metric $G_{ij}$ has the property that it is positive definite and 
invertible in a unitary theory. Let's consider the renormalization group (RG) 
flow in the space of coupling constants, 
\[x\frac{dt^i}{dx}=-\beta^i(t),\]
where $x$ is a distance scale. If we expand the $\beta$-function in the 
couplings, $\beta$ is, in general, of the form
\begin{equation}
\beta^i(t)=(h_i-1)t^i+\cO(t^2).
\end{equation}
But one can perform reparametrization in the space of couplings in such a way 
that, at least locally, $\beta$-functions are exactly \textit{linear} in $t$. 
This has an 
analogy in the general relativity context: In any curved spacetime, one can 
perform an appropriate coordinate transformation to reach the locally flat 
space. With such a choice of couplings, the action~(\ref{eq:A}) can be 
written as
\begin{equation}
\frac{\partial S}{\partial t^i}=\beta^jG_{ij}(t) \label{eq:kurikomi}
\end{equation}
(summation over $j$ is implicit). 
The coordinate system determined in this way is well-suited for the 
study of tachyon condensation. We see below how the linearity of $\beta$ 
simplifies the analyses.

In almost all of the remainder of this chapter, we focus on the boundary 
perturbation of the special form
\begin{equation}
T(X)=a+\sum_{i=1}^{26}u_iX_i^2, \label{eq:Tau}
\end{equation}
where $a$ and $u_i$ are couplings, and for simplicity we take the target space 
to be Euclidean flat 26-dimensional spacetime, but there is no essential 
difference from Minkowski spacetime. Since $X_i(\theta)$ itself is not a conformal primary 
field, we cannot use the formula~(\ref{eq:Andr}) for a primary field $\cV_i$ as it stands. 
Using 
\[ T^{\mathrm{m}}=-\frac{1}{4\ap}\sum_j\partial X_j\partial X_j \qquad \mathrm{and} \qquad 
X_i(z)X_j(w)\sim -2\ap\eta_{ij}\log (z-w), \]
one finds 
\begin{eqnarray*}
\{Q_B,cT(X)\}&=&a\{Q_B,c\}+\sum_{i=1}^{26}u_i\{Q_B,cX_i^2\} \\ &=&\left(a+2\ap
\sum_{i=1}^{26}u_i\right)c\partial c+c\partial c\sum_{i=1}^{26}u_iX_i^2.
\end{eqnarray*}
From this expression, we can read off the $\beta$-functions\footnote{I'd like to thank 
O. Andreev for pointing this out.} as
\begin{equation}
\beta^a=-a-2\ap\sum_{i=1}^{26}u_i \, ,\quad \beta^{u_i}=-u_i\, ,\qquad \beta^j=0 \ 
\mbox{ for any other coupling}
\end{equation}
in the above choice of couplings. From these $\beta$-functions, we find that 
nonzero $a$ and $u_i$ flow to infinity in the infrared limit. That is, 
$a=\infty$ and $u_i=\infty$ is the infrared fixed point. 
And of more importance is 
that the couplings never mix with each other. So if we take $T=a+\sum u_i
X_i^2$ at the starting point, other terms such as $X_i^4$ do not appear 
through the RG flow. Therefore, it can make sense that we calculate the 
partition function only with the perturbation~(\ref{eq:Tau}). 

But things have not been completed. It is not clear whether we could 
consistently set to zero the excited open string modes ({\it i.e.} 
non-tachyonic modes). In fact, cubic string field theory forced us to take 
infinitely many scalar fields into account. Fortunately, it is shown that 
all couplings for higher modes can be set to zero without contradicting 
equations of motion.

Here the proof. Expanding the spacetime action in the couplings as 
\begin{equation}
S(a,u,\lambda^i)=S^0(a,u)+S^1_i(a,u)\lambda^i+S^2_{ij}(a,u)\lambda^i
\lambda^j+\cdots \, ,\label{eq:Sexp}
\end{equation}
where we denoted collectively by $\lambda^i$ 
all couplings other than $a,u$, we 
want to show $S^1_i(a,u)=0$. First we suppose
\begin{equation}
0\neq S^1_i(a,u)=\frac{\partial S}{\partial \lambda^i}\Bigg|_{\lambda=0}
=\beta^jG_{ij}\Big|_{\lambda=0}. \label{eq:Si}
\end{equation}
Non-degeneracy of the metric $G_{ij}$ implies that $\beta^j(a,u,\lambda^i
=0)\neq 0$. This means that some couplings $\lambda^j$ are newly generated 
by the RG flow when we turn on $a$ and $u$. However, since the theory remains 
free (namely quadratic) even after including the 
perturbation~(\ref{eq:Tau}), such couplings should not be created 
by the RG flow. Then we 
conclude that the supposition~(\ref{eq:Si}) was false, or $S^1_i(a,u)=0$. 
Looking at the action~(\ref{eq:Sexp}) with $S^1_i=0$, couplings $\lambda^i$ 
enter $S$ quadratically or with higher powers, so the equations of motion 
$\partial S/\partial \lambda^i=0$ are trivially satisfied by all $\lambda^i=0$.
{\it q.e.d.}

An important conclusion can be drawn: In the `closed string vacuum', only the 
tachyon field develops nonzero vacuum expectation value, while all other 
fields have vanishing ones.

\section{Spacetime Action}
Since we saw that it was sufficient to examine the particular 
perturbation~(\ref{eq:Tau}) for the study of tachyon condensation, hereafter 
we concentrate on it. It will be shown in section~\ref{sec:app} that 
the spacetime action $S(a,u_i)$ is given by 
\begin{eqnarray}
S&=&K\left(\sum_i2\ap u_i-\sum_iu_i\frac{\partial}{\partial u_i}+1+a\right)
Z^{\mathrm{m}},\label{eq:SKZ}\\
Z^{\mathrm{m}}&=&e^{-a}\prod_{i=1}^{26}Z_1(2\ap u_i),\\
Z_1(z)&=&\sqrt{z}e^{\gamma z}\Gamma(z), \label{eq:Z1}
\end{eqnarray}
where $Z^{\mathrm{m}}$ is the matter part of the partition function on the 
disk. We want to rewrite the above spacetime action $S$ expressed in terms 
of $a,u$ in such a way that the action $S$ is written as a functional of 
the tachyon field $T(X)$. 
We insist that the action, when we keep terms with up to two derivatives, 
should be given by 
\begin{equation}
\tilde{S}=K\int\frac{d^{26}x}{(2\pi \ap)^{13}}\Big(\ap e^{-T}\partial_{\mu}
T\partial^{\mu}T+(T+1)e^{-T}\Big). \label{eq:Stach}
\end{equation}
The reason is as follows: Plugging~(\ref{eq:Tau}) into~(\ref{eq:Stach}) and 
carrying out the Gaussian integrals, we find 
\begin{equation}
\tilde{S}=K\left[(a+1+13)e^{-a}\frac{1}{\prod_{i=1}^{26}\sqrt{2\ap u_i}}
+2\ap e^{-a}\frac{\sum_{j=1}^{26}u_j}{\prod_{i=1}^{26}\sqrt{2\ap u_i}}
\right].
\end{equation}
This agrees with the expression of~(\ref{eq:SKZ}) for small $u_i$. An 
intuitive argument that justifies this comparison is that $T(X)$ with large 
$u_i$ corresponds to an intensively fluctuating tachyon profile so that this 
affects higher derivative terms, therefore it is sufficient to see $u\to 0$ 
behavior now. 
\medskip

So far, the normalization constant $K$ is not fixed. Although $K$ is shown 
to be related to the D25-brane tension $\tau_{25}$, some more preparations 
are needed. Now we simply give the result 
\begin{equation}
\frac{K}{(2\pi\ap)^{13}}=\tau_{25}, \label{eq:Ktension}
\end{equation}
and postpone the proof until later.

\section{Tachyon Condensation}\label{sec:bsft}
\subsection*{\underline{brane annihilation}}
\begin{figure}[htbp]
	\begin{center}
	\includegraphics{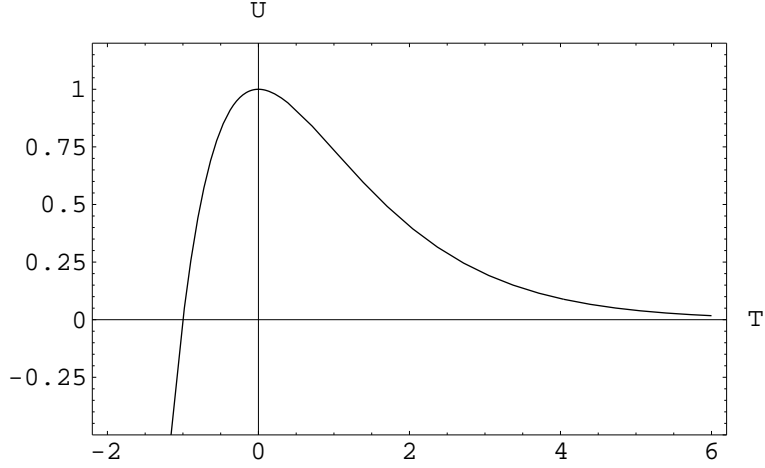}
	\end{center}
	\caption{The exact tachyon potential.}
	\label{fig:POT}
\end{figure}

We first survey the spacetime-independent tachyon field, $T=a$. The spacetime 
action~(\ref{eq:Stach}) for $T=a$ becomes 
\begin{eqnarray}
\tilde{S}&=&\tau_{25}V_{26}U(a),\label{eq:SVU} \\
U(T)&=&(T+1)e^{-T},\label{eq:pot}
\end{eqnarray}
where we used~(\ref{eq:Ktension}) and $V_{26}$ is 26-dimensional volume $
\int d^{26}x$. The form of the tachyon potential $U(T)$ is illustrated in 
Figure~\ref{fig:POT}. As the potential $U(T)$ is exact, we can give the 
following arguments confidently.
\medskip

$U(T)$ has two extrema at $T=0$ and $T=+\infty$. $T=0$ corresponds to the 
original D25-brane, and its energy density \textit{is} 
exactly equal to the known 
D25-brane tension $\tau_{25}$. On the other hand, $T=+\infty$ is thought of 
as the `closed string vacuum', with vanishing energy. So we have proved using 
background independent open string field theory that 
the negative energy contribution from the tachyon potential precisely cancels 
the D25-brane tension. One may feel it dubious that the stable vacuum is at 
infinity. But the actual distance in the field space is finite because the 
metric there is $e^{-T}$ (it can be read from the kinetic term in the action),
the situation being not so strange.
\smallskip

Incidentally, in~\cite{KMM1} it is argued that the possibility of decaying 
to $T\to -\infty$ reflects the closed string tachyon instability. As the 
argument there seems not to be based on a firm footing, we will not give any 
further comment about it.

\subsection*{\underline{Lower dimensional branes 
(two-derivative approximation)}}
The equations of motion following from the action~(\ref{eq:Stach}) is 
\[ 2\ap \partial_{\mu}\partial^{\mu}T-\ap \partial_{\mu}T\partial^{\mu}T
+T=0. \]
Substituting $T=a+\sum_{i=1}^{26}u_iX_i^2$, we get 
\[ \left(a+4\ap \sum_{i=1}^{26}u_i\right)+\sum_{i=1}^{26}X_i^2u_i
(1-4\ap u_i)=0. \]
It is solved by $u_i=0$ or $u_i=1/4\ap$ and $a=-4\ap \sum u_i=-n$, 
where $n$ is the number of nonzero $u_i$'s. In the spacetime language, 
such solutions represent codimension $n$ solitons, to be identified 
with D($25-n$)-branes. As a check, let us see the energy density of the 
solution. If we substitute the solution
\begin{equation}
T(X)=-n+\frac{1}{4\ap}\sum_{i=1}^nX_i^2 \label{eq:uap}
\end{equation}
into the action~(\ref{eq:Stach}), its value is given by 
\begin{equation}
\tilde{S}=V_{26-n}\cdot e^n(4\pi\ap )^{n/2}\tau_{25}\equiv V_{26-n}
\mathcal{T}_{25-n}.
\end{equation}
So the ratio 
\begin{equation}
\frac{\mathcal{T}_{25-n}}{\tau_{25}}=\left(\frac{e}{\sqrt{\pi}}2\pi 
\sqrt{\ap}\right)^n \label{eq:dis}
\end{equation}
does not agree with the expected answer for the D-brane tensions, 
but the discrepancy is 
attributed to the approximation of ignoring higher derivative terms. 
Using the exact action, we would obtain the correct results, as will be shown.

Note that the soliton~(\ref{eq:uap}) is spherically symmetric in the 
directions transverse to the brane and asymptotes to the 
`closed string vacuum' $T=+\infty$.

\subsection*{\underline{Lower dimensional branes (exact treatment)}}
Extending the previous method to higher derivative terms, namely, expand the 
action~(\ref{eq:SKZ}) to higher orders in $u_i$ and determine the successive 
terms in the action, is impractical. If we were able to do the above 
processes infinitely, we would get the exact result in principle, but 
it is of course impossible. Instead, we proceed on another route.

From the previous arguments, we know that we can consistently take the 
boundary perturbations $\cV=T=a+\sum uX^2$ with no other couplings, and 
that no new couplings are generated under the RG flow. Therefore all we have 
to do is to minimize the exact action~(\ref{eq:SKZ}) with respect to $a,u$. 
As is clear from the approximated results, to get an extended brane solution 
we must permit some of the $u_i$'s to be zero. But $Z_1(z)$~(\ref{eq:Z1}) 
is singular as $z\to 0$, so it must be regularized. 
Since the noncompact volume is responsible for this divergence, 
we compactify $X^i$ as $X^i\simeq X^i+
2\pi R^i \, ; \ \ i=n+1,\cdots , 26$. For $T=a$ (namely, all $u_i=0$) the 
action~(\ref{eq:Stach}) is exact and 
\begin{equation}
\tilde{S}(T=a)=K\int\frac{d^{26}x}{(2\pi\ap)^{13}}(a+1)e^{-a}=K(a+1)e^{-a}
\prod_{i=1}^{26}\frac{\sqrt{2\pi}R^i}{\sqrt{\ap}}. \label{eq:Su0}
\end{equation}
This should be equal to the $u\to 0$ limit of the exact action~(\ref{eq:SKZ}), 
\begin{equation}
S=K(a+1)e^{-a}\prod_{i=1}^{26}Z_1(2\ap u_i)\Big|_{u\to 0}. \label{eq:SZu0}
\end{equation}
Hence we set
\[ \lim_{u\to 0}Z_1(2\ap u)=\lim_{u\to 0}\sqrt{2\ap u}e^{2\ap \gamma u}
\Gamma(2\ap u)=\frac{\sqrt{2\pi}R}{\sqrt{\ap}}.\]
Then the exact action for $u_i=0$ $(n+1\le i\le 26)$ becomes 
\begin{equation}
S=K\left(a+1+\sum_{i=1}^n2\ap u_i-\sum_{i=1}^nu_i\frac{\partial}{\partial u_i}
\right)e^{-a}\prod_{i=1}^nZ_1(2\ap u_i)\prod_{j=n+1}^{26}\left(\frac{
\sqrt{2\pi}R^j}{\sqrt{\ap}}\right). \label{eq:sphere}
\end{equation}
Varying it with respect to $a$, we get
\[ \frac{\partial S}{\partial a}\propto \left(a+\sum_{i=1}^n2\ap u_i-
\sum_{i=1}^nu_i\frac{\partial}{\partial u_i}\right)\prod_{i=1}^nZ_1(2\ap u_i)
=0,\]
hence $a$ is determined as a function of $u_i$,
\begin{equation}
a=-2\ap \sum_{i=1}^nu_i+\sum_{i,j=1}^nu_i\frac{\partial}{\partial u_i}\ln Z_1
(2\ap u_j). \label{eq:Lively}
\end{equation}
Substituting it into the action~(\ref{eq:sphere}),
\[ S=K\exp\left[\sum_{j=1}^n\left(\ln Z_1(2\ap u_j)+2\ap u_j-\sum_{i=1}^nu_i
\frac{\partial}{\partial u_i}\ln Z_1(2\ap u_j)\right)\right]\prod_{j=n+1}^{26}
\frac{\sqrt{2\pi}R^j}{\sqrt{\ap}}. \]
This turns out to be a monotonically decreasing function of $u_j$. Therefore 
the action $S$ is easily minimized by taking $u_j\to \infty$, 
the infrared fixed point. From Stirling's formula, we get 
\begin{equation}
\log Z_1(z)=z \log z+\gamma z-z+\log\sqrt{2\pi}+\cO\left(\frac{1}{z}\right). 
\label{eq:Stir}
\end{equation}
The value of $S$ in the limit $u_j\to \infty$ is given by
\begin{equation}
S=K(2\pi)^{n/2}\prod_{j=n+1}^{26}\frac{2\pi R^j}{\sqrt{2\pi\ap}}=\tau_{25}(
2\pi\sqrt{\ap})^n\left(\prod_{j=n+1}^{26}2\pi R^j\right)\equiv \tau_{25-n}
V_{26-n}.
\end{equation}
In this case, the ratio of the tensions is 
\begin{equation}
\frac{\tau_{25-n}}{\tau_{25}}=(2\pi\sqrt{\ap})^n, \label{eq:RandB}
\end{equation}
so we have obtained the exact result. 
\smallskip

Since we have just seen that the known relation between the D-brane tensions 
is exactly reproduced, we now construct the explicit form of the action on a 
lower dimensional D-brane~\cite{11002}. For simplicity, we consider the case 
of a D24-brane. For that purpose, we decompose the constant mode $a$ of the 
tachyon field $T(X)$ into two parts as $a=a_1+\tilde{a}$ and impose on $a_1$ 
the `minimizing' relation
\begin{equation}
a_1=-2\ap u_1+u_1\frac{\partial}{\partial u_1}\ln Z_1(2\ap u_1) \label{eq:aaa}
\end{equation}
as in (\ref{eq:Lively}). Substituting (\ref{eq:aaa}) back into 
(\ref{eq:sphere}) and taking the limit $u_1\to\infty$ to obtain a D24-brane, 
the action becomes 
\begin{eqnarray}
S&=&\lim_{u_1\to\infty}\exp\left[2\ap u_1-u_1\frac{\partial}{\partial u_1}\ln
Z_1(2\ap u_1)+\ln Z_1(2\ap u_1)\right] \label{eq:aab} \\
& &\times K\left(\tilde{a}+1+\sum_{i=2}^n2\ap u_i-\sum_{i=2}^nu_i
\frac{\partial}{\partial u_i}\right)e^{-\tilde{a}}\prod_{i=2}^nZ_1(2\ap u_i)
\prod_{j=n+1}^{26}\left(\frac{\sqrt{2\pi}R^j}{\sqrt{\ap}}\right). \nonumber
\end{eqnarray}
The first line simply gives a factor of $\sqrt{2\pi}$, while the second line 
has the same structure as the original action~(\ref{eq:sphere}) on the 
D25-brane, with $a$ replaced by $\tilde{a}$ and $i=1$ removed. Therefore, 
by using the relation~(\ref{eq:aab}) recursively we conclude that the action 
of the tachyon mode on the tachyonic lump, which is identified with the lower 
dimensional D-brane, takes the same form as the action for the tachyon 
on the D25-brane. 
\medskip

Next we consider the following field redefinition
\[e^{-T/2}=\frac{1}{\sqrt{8\ap}}\phi.\]
Then the derivative-truncated action~(\ref{eq:Stach}) of the tachyon field 
becomes
\begin{equation}
\tilde{S}=\tau_{25}\int d^{26}x\left[\frac{1}{2}\partial_{\mu}\phi
\partial^{\mu}\phi-\frac{\phi^2}{8\ap}\log\frac{\phi^2}{8\ap e}\right],
\label{eq:philogphi}
\end{equation}
which has the kinetic term of the standard form. 
(This model will be treated in Chapter~\ref{ch:others}.) In terms of $\phi$ 
field, the soliton solution takes the form
\[ \phi=\sqrt{8\ap}e^{-a/2}\exp\left(-\frac{u}{2}X^2\right), \]
which is a Gaussian with the size $\sim1/\sqrt{u}$. Since the exact soliton 
solutions were found in the limit $u_j\to\infty$, the widths of the solitons 
are zero, in agreement with those of the conventional D-branes. This result is 
well contrasted with that of the level truncated string field theory, where 
the D-brane has the finite size of order $\sqrt{\ap}$. 

\subsection*{\underline{Degrees of freedom after tachyon condensation}}
When we calculate the partition function in the constant tachyon background 
$T=a$, $a$-dependence is quite simple:
\begin{equation}
Z(a,t)=\int \mathcal{D}X \ e^{-\cS_0-a-\cdots}=e^{-a}\tilde{Z}(t), 
\label{eq:ZN}
\end{equation}
where other couplings are collectively denoted by $t$. Then the spacetime 
action is given by (see the next section)
\begin{eqnarray}
S(T=a,t)&=&K\left(-a\frac{\partial}{\partial a}+\beta^i\frac{\partial}
{\partial t^i}+1\right)e^{-a}\tilde{Z}(t) \nonumber \\
&=&KU(a)\tilde{Z}(t)+Ke^{-a}\beta^i\frac{\partial}{\partial t^i}\tilde{Z}(t),
\label{eq:SUZ}
\end{eqnarray}
where $U(T)$ is the tachyon potential $(T+1)e^{-T}$. 
We can see that the action vanishes as $T$ relaxes to the stable vacuum at 
infinity.
\medskip

As mentioned before, the normalization `$K$' has not yet been determined. 
Now is the time to fix it. In~\cite{GhS} the normalization factor is 
determined by comparing the background independent open string field theory 
action with the cubic string field theory action, but we give another 
argument. To begin with, let's consider the effect of including 
gauge fields in the boundary perturbation. The world-sheet action is
\begin{equation}
\cS=\cS_0+\int\frac{d\theta}{2\pi}T+i\int\frac{d\theta}{2\pi}\partial_{\theta}
X^{\mu}A_{\mu}(X).\label{eq:STA}
\end{equation}
We consider the simple case $A_{\mu}(X)=-\frac{1}{2}F_{\mu\nu}X^{\nu}$ with 
constant $F_{\mu\nu}$. As the equations of motion (Maxwell eqs.) 
$\partial_{\mu}F^{\mu\nu}=0$ are always satisfied, conformal invariance is 
not violated by the perturbation and hence $\beta^{A_{\mu}}=0$. For constant 
$F_{\mu\nu}$, partition function $\tilde{Z}(A_{\mu})$ is calculated 
in~\cite{FT} and the result coincides with the Born-Infeld action (up to 
normalization)
\begin{equation}
\tilde{Z}(A_{\mu})\propto \int d^{26}x\mathcal{L}_{BI}(A_{\mu})=\int d^{26}x 
\sqrt{|\det (g_{\mu\nu}+2\pi\ap F_{\mu\nu})|}.
\end{equation}
Then we conclude that for constant $T,F_{\mu\nu}$, the spacetime 
action~(\ref{eq:SUZ}) with $\beta^i=0$ is given by 
\begin{equation}
S(T,F_{\mu\nu})=K\int\frac{d^{26}x}{(2\pi\ap)^{13}}U(T)\mathcal{L}_{BI}
(A_{\mu}), \label{eq:ZO}
\end{equation}
where we determined an 
overall factor $1/(2\pi\ap)^{13}$ by requiring that for $F_{\mu\nu}=0$ 
$(\mathcal{L}_{BI}=1)$ the action should coincide with the potential term 
in~(\ref{eq:Stach}). On the D25-brane, where $U(T=0)=1$, the fact 
that the action is of the standard Born-Infeld form and that $A_{\mu}$ has 
the standard normalization (as is seen from~(\ref{eq:STA})) implies that the 
coefficient $K/(2\pi\ap)^{13}$ should coincide with the D25-brane 
tension $\tau_{25}$. With this result, 
we complete the previous arguments about the brane annihilation.

From the above argument, 
we have determined the spacetime action for the slowly varying 
tachyon and gauge fields as
\begin{equation}
S(T,F_{\mu\nu})=\tau_{25}\int d^{26}x \ U(T)\sqrt{|\det (g_{\mu\nu}+2\pi\ap 
F_{\mu\nu})|}+\cO(\partial T,\partial F), \label{eq:Born}
\end{equation}
which supports the action proposed in~\cite{NonBPSD}. 
\smallskip

Now let us consider the fluctuations of the tachyon and the gauge field 
around the closed string vacuum. In terms of the $\phi$-field defined 
above~(\ref{eq:philogphi}), the closed string vacuum corresponds to $\phi=0$. 
Since the second derivative of the tachyon potential in~(\ref{eq:philogphi}) 
diverges logarithmically at $\phi=0$, the `tachyon' field $\phi$ acquires 
infinite mass and hence decouples from the spectrum. For the gauge field, it 
was pointed out in~\cite{11009} that the effective spacetime 
action\footnote{This is obtained by modifying~(\ref{eq:Born}) in the 
following way. Replace $F$ with $F+B$, expand the Born-Infeld form to second 
order in fields, and add the kinetic term for the tachyon as 
in~(\ref{eq:Stach}).} for low-lying fields (open string tachyon $T$, $U(1)$ 
gauge field $A_{\mu}$; symmetric 2-tensor $G_{\mu\nu}$, 2-form gauge field 
$B_{\mu\nu}$) in open/closed string theory has some resemblance to that of 
the standard abelian Higgs model, namely the theory of a $U(1)$ gauge field 
interacting with a complex scalar field whose nonvanishing expectation value 
breaks the gauge invariance spontaneously. This analogy provides us with the 
following picture: When we think of the closed string mode $B_{\mu\nu}$ as 
a fixed background, the open string gauge field $A_{\mu}$ becomes singular 
at the closed string vacuum $\phi=0$. However, if $B_{\mu\nu}$ is promoted 
to a dynamical field, $A_{\mu}$ can be absorbed into $B_{\mu\nu}$ as a gauge 
parameter because the string action has the following gauge invariance 
\begin{eqnarray}
B&\longrightarrow& B+d\Lambda \nonumber \\
A&\longrightarrow& A-\Lambda, \label{eq:rice}
\end{eqnarray}
where $\Lambda$ is a 1-form gauge parameter. Furthermore, it was 
conjectured~\cite{11009} that all open string modes except for tachyon could 
be considered as gauge parameters for the corresponding closed string modes. 
To carry out this program, we \textit{must} include closed string modes as 
dynamical variables. In this case, the problem that open string modes become 
singular at the closed string vacuum can be cured. Then it is clear that the 
only remaining dynamical degrees of freedom at the closed string vacuum are 
the closed string modes (recall that the open string tachyon has decoupled 
as well). On the other hand, at the open string vacuum $T=0$ $(\phi=
\sqrt{8\ap})$ the vacuum expectation value $A_{\mu}=0$ of the open string 
gauge field spontaneously breaks the closed string gauge 
invariance~(\ref{eq:rice}) and the massless vector field $A_{\mu}$ arises as 
a `Goldstone boson' since we consider $B_{\mu\nu}$ which has acquired nonzero 
mass due to the stringy Higgs mechanism as a nondynamical background at the 
open string perturbative vacuum. 

\section{Evaluation of the Action}\label{sec:app}
World-sheet action is 
\begin{eqnarray}
\cS&=&\frac{1}{4\pi\ap}\sum_{i=1}^{26}\int_{\Sigma}d^2\sigma\sqrt{g}
g^{\alpha\beta}\partial_{\alpha}X_i\partial_{\beta}X_i+a+\sum_{i=1}^{26}
u_i\int_{\partial\Sigma}\frac{d\theta}{2\pi}X_i^2(\theta) \nonumber \\ 
&\equiv& \sum_{i=1}^{26}\cS_1^i+a,
\end{eqnarray}
where $\Sigma,g$ represent the disk world-sheet and the Euclidean flat 
metric on it, respectively. And for simplicity, 
we also take the spacetime to be 
Euclidean. Boundary conditions derived from this action are
\begin{equation}
n^{\alpha}\partial_{\alpha}X_i+2\ap u_iX_i=0 \quad \mathrm{at} \,\, \partial
\Sigma,
\end{equation}
where $n^{\alpha}$ is a unit vector normal to the disk boundary. Then it is 
clear that $X_i$ obeys Neumann boundary condition if $u_i=0$, and Dirichlet 
if $u_i\to \infty$. The Green's function satisfying these boundary conditions 
is given in~\cite{WiBI2}:
\begin{eqnarray*}
G_{ij}(z,w)=\langle X_i(z)X_j(w)\rangle_{\mathrm{norm.}}&=&-\frac{\ap}{2}
\delta_{ij}\ln |z-w|^2-\frac{\ap}{2}\delta_{ij}\ln |1-z\bar{w}|^2 \\
&+&\frac{\delta_{ij}}{2u_i}-\delta_{ij}\sum_{k=1}^{\infty}\frac{2\ap{}^2u_i}
{k(k+2\ap u_i)}\Bigl((z\bar{w})^k+(\bar{z}w)^k\Bigr),
\end{eqnarray*}
with $z=e^{i\theta},w=e^{i\theta'}\in \partial\Sigma$, and $\langle\cdots
\rangle_{\mathrm{norm.}}$ is the normalized (\textit{i.e.} divided by 
partition function such that $\langle 1\rangle_{\mathrm{norm.}}=1$) 
correlation function.

We need $\langle X_i^2(\theta)\rangle$ below, but since $G(z,w)$ diverges 
as $z\to w$ na\"{\i}vely, we define it by the conformal normal ordering 
(or point splitting)
\begin{equation}
X_i^2(\theta)\equiv \lim_{\epsilon\to 0}\left(X_i(\theta)X_i(\theta+
\epsilon)+\ap\ln |1-e^{i\epsilon}|^2\right). \label{eq:XXreg}
\end{equation}
From this definition, it follows 
\[\langle X_i^2(\theta)\rangle_{\mathrm{norm.}}=\frac{1}{2u_i}-\sum_{k=1}^{
\infty}\frac{4\ap{}^2u_i}{k(k+2\ap u_i)}.\]
Now we calculate the matter partition function on the disk,
\begin{equation}
Z^{\mathrm{m}}=\int \mathcal{D}X\exp \left(-a-\cS_0-\sum_{i=1}^{26}u_i\int
\frac{d\theta}{2\pi}X_i^2(\theta)\right).
\end{equation}
It follows from the definition that
\begin{eqnarray*}
\frac{\partial}{\partial u_i}\ln Z^{\mathrm{m}}&=&\frac{-1}{Z^{\mathrm{m}}}
\int\mathcal{D}X\exp\left[-\cS_0-\int\frac{d\theta}{2\pi}\cV(\theta)
\right]\int\frac{d\theta}{2\pi}X_i^2(\theta) \\ &=&-\int_0^{2\pi}
\frac{d\theta}{2\pi}\langle X_i^2(\theta)\rangle_{\mathrm{norm.}}=-\frac{1}
{2u_i}+\sum_{k=1}^{\infty}\frac{4\ap{}^2u_i}{k(k+2\ap u_i)}.
\end{eqnarray*}
Recalling that the gamma function $\Gamma(z)$ satisfies
\[\frac{d}{dz}\ln \Gamma(z)=-\frac{1}{z}+\sum_{k=1}^{\infty}\frac{z}{k(k+z)}
-\gamma\]
($\gamma\simeq 0.577\ldots $ is the Euler-Mascheroni constant), we get
\begin{equation}
\frac{\partial}{\partial u_i}\ln Z^{\mathrm{m}}(u_i,a)=\frac{\partial}
{\partial u_i}\ln\Gamma(2\ap u_i)+\frac{1}{2u_i}+2\ap\gamma.
\end{equation}
Consequently,
\begin{equation}
Z^{\mathrm{m}}(u_i,a)=e^{-a}\prod_{i=1}^{26}\sqrt{2\ap u_i}e^{2\ap \gamma u_i}
\Gamma(2\ap u_i)\equiv e^{-a}\prod_{i=1}^{26}Z_1(2\ap u_i).
\end{equation}
In the last step, overall normalization (integration constant) 
is absorbed in the definition 
of $a$. Next let us see that the spacetime action can be expressed in terms 
of $Z^{\mathrm{m}}$. As a preliminary, we calculate the BRST transformation 
of the tachyon field, 
\begin{eqnarray*}
\{Q_B,cT(X(0))\}&=&\oint\frac{dz}{2\pi i}j_B(z)cT(X(0)) \\
&=& \oint \frac{dz}{2\pi i}(cT^{\mathrm{m}}(z)+bc\partial c(z))cT(X(0)) \\ 
&=&(1-h)c\partial c(0)T(X(0)),
\end{eqnarray*}
where $h$ is the conformal weight of $T(X)$. Since $T(X)$ has no $\theta$ 
derivative, only its momentum contributes to $h$ and $h=\ap p^2$. Hence
\begin{equation}
\{Q_B,cT(X)\}=c\partial c\left(1+\ap\sum_{i=1}^{26}\frac{\partial^2}
{\partial X_i^2}\right)T(X).
\end{equation}
Substituting $T=a+\sum u_iX_i^2$, and denoting $cT=\cO$,
\begin{equation}
\{Q_B,\cO(X)\}=c\partial c\left(a+\sum_{i=1}^{26}u_i(X_i^2+2\ap)\right).
\end{equation}
Using this expression, the defining equation~(\ref{eq:dS}) of the spacetime 
action can be written in the following way, 
\begin{eqnarray}
dS&=&\frac{K}{2}\int\frac{d\theta}{2\pi}\frac{d\theta'}{2\pi}\Biggl\langle
c(\theta)\left(da+\sum_idu_iX_i^2(\theta)\right)c\partial c(\theta') \nonumber 
\\ & & \times \left(a+\sum_iu_i(X_i^2(\theta')+2\ap)\right)\Biggr\rangle_{\cV}
\nonumber \\ &=&\frac{K}{2}\int\frac{d\theta d\theta'}{(2\pi)^2}
\langle c(\theta)c\partial c(\theta')\rangle_{\mathrm{ghost}} \\ & & \times 
\left\langle\left(da+\sum du_iX_i^2(\theta)\right)\left(a+\sum u_i
(X_i^2(\theta')+2\ap)\right)\right\rangle_{\mathrm{matter},\cV}.\nonumber
\end{eqnarray}
All these correlators can be evaluated as follows:
\[ \langle c(\theta)c\partial c(\theta')\rangle_{\mathrm{ghost}}=2(\cos (
\theta-\theta')-1), \]
\[ \int\frac{d\theta}{2\pi}\langle X_i^2(\theta)
\rangle_{\cV}=-\frac{\partial Z^{\mathrm{m}}}{\partial u_i}, \]
\[\int\frac{d\theta d\theta'}{(2\pi)^2}\langle X_i^2(\theta)X_j^2(\theta')
\rangle_{\cV}=\frac{\partial^2Z^{\mathrm{m}}}{\partial u_i\partial u_j}, \]
\[\int\frac{d\theta d\theta'}{(2\pi)^2}\cos (\theta-\theta')\langle X_i^2
(\theta)X_i^2(\theta')\rangle_{\cV}=\frac{4}{u_i}.\]
Detailed calculations are found in~\cite{WiBI2}. With the above results, 
we finally obtain 
\begin{eqnarray}
dS&=&Kd\left(\sum_i2\ap u_i-\sum_i u_i\frac{\partial}{\partial u_i}+(1+a)
\right)Z^{\mathrm{m}} \nonumber \\ &=& Kd\left(-\sum_iu_i\frac{\partial}
{\partial u_i}-(a+\sum 2\ap u_i)\frac{\partial}{\partial a}+1\right)
Z^{\mathrm{m}}.\label{eq:dSdZ}
\end{eqnarray}
(Remember that $Z^{\mathrm{m}}(u_i,a)$ depends on $a$ only through a simple 
overall factor $e^{-a}$.) Furthermore, if $\cV$ does not contain any ghost, 
the spacetime action was perturbatively shown to be given by~\cite{Sh1,Sh2} 
\begin{equation}
S=K\left(\sum_i\beta^i\frac{\partial}{\partial t^i}+1\right)Z^{\mathrm{m}},
\label{eq:Sbeta}
\end{equation}
where $\beta^i$ is the $\beta$-function for the coupling $t^i$. 
The result~(\ref{eq:dSdZ}) is a special case of~(\ref{eq:Sbeta}), where we 
take $t^i=(u_i,a)$ and $\beta^{u_i}=-u_i , \beta^a=-a-\sum_i2\ap u_i$. 
The additive normalization of the spacetime action 
$S$ cannot be determined by the definition~(\ref{eq:dS}), so we fixed it by 
requiring that $S$ be proportional to $Z^{\mathrm{m}}$ on-shell ($\beta^i=0$).
\medskip

Finally, we refer to the relation between the spacetime action and so-called 
boundary entropy. The \textit{boundary entropy} $g$~\cite{Affleck,Ludwig} 
is defined in terms of the boundary state $|B\rangle$ in the perturbed theory 
as $g=\langle 0|B\rangle$ and has the following properties:
\begin{itemize}
	\item The boundary entropy decreases along the renormalization group flow 
	to the infrared, \textit{i.e.} 
	\[ \frac{dg}{dx}<0. \]
	This is the postulated `$g$-theorem'.
	\item The boundary entropy coincides with the disk partition function 
	at fixed points of the boundary renormalization group.
	\item The boundary entropy is stationary at the fixed points.
\end{itemize}
In fact, the spacetime action defined in~(\ref{eq:dS}) is equipped with 
the above properties. For the first, we have 
\[ x\frac{dS}{dx}=x\frac{dt^i}{dx}\frac{\partial S}{\partial t^i}=
-\beta^i(t)\cdot \beta^jG_{ij}(t)\le 0 \]
because the metric $G_{ij}(t)$ is positive definite in a unitary theory. And 
we have assumed that the action does not explicitly depend on the scale, 
\textit{i.e.} $\partial S/\partial x=0$. We can see that the second property 
is true of the spacetime action if we set $\beta^i=0$ in~(\ref{eq:Sbeta}) 
at the fixed point. And the third can be verified by looking at the 
expression~(\ref{eq:kurikomi}). Hence, we may identify the boundary string 
field theory action with the boundary entropy, at least in this case. This 
correspondence was further investigated in~\cite{12150} using the integrable 
boundary sine-Gordon model.

\section{Application to the Superstring Case}
Though we have focused on the bosonic string case so far, the background 
independent open string field theory can be extended to include the 
superstring~\cite{KMM2}.
We consider the open superstring theory on non-BPS D-branes below.
The world-sheet action is now 
\begin{eqnarray}
\cS &=& \cS_0+\cS_{\partial\Sigma}, \nonumber \\
\cS_0&=&\frac{1}{2\pi\ap}\int d^2z\biggl(\partial X^{\mu}\overline{\partial}
X_{\mu}+\frac{\ap}{2}\left(\psi^{\mu}\overline{\partial}\psi_{\mu}+
\widetilde{\psi}^{\mu}\partial\widetilde{\psi}_{\mu}\right)\biggr), 
\label{eq:ZA} \\
e^{-\cS_{\partial\Sigma}}&=&\mathrm{Tr}_{\mbox{\scriptsize{Chan-Paton}}}P\exp
\left[\int_0^{2\pi}\frac{d\tau}{2\pi}\int d\theta(\Gamma D\Gamma+
T(\mathbf{X}^{\mu})\Gamma)\right]. \nonumber 
\end{eqnarray}
In the boundary interaction term, $(\tau,\theta)$ are the superspace 
coordinates of the disk boundary. There are two superfields $\displaystyle 
\sqrt{\frac{2}{\ap}}\mathbf{X}^{\mu}=\sqrt{\frac{2}{\ap}}X^{\mu}+\theta
\psi^{\mu}$ and $\Gamma=\eta+\theta F$ on the boundary, where $\Gamma$ is an 
anticommuting quantum mechanical degree of freedom~\cite{HKM}. $D=
\partial_{\theta}+\theta\partial_{\tau}$ is the superderivative, and $P$ 
represents the path-ordering. The boundary conditions for the fermions 
$\psi^{\mu},\eta$ are taken to be antiperiodic around the circle ($\psi^{\mu}
(\tau+2\pi)=-\psi^{\mu}(\tau)$) to consider the Neveu-Schwarz sector. 
Decomposing the superfields into the component fields and carrying out the 
$\theta$-integration in~(\ref{eq:ZA}), we can write the boundary action as
\begin{eqnarray}
e^{-\cS_{\partial\Sigma}}&=&\mathrm{Tr_{CP}}P\exp\left[\int\frac{d\tau}{2\pi}
\left(\partial_{\tau}\eta\ \eta+F^2+T(X)F+\sqrt{\frac{\ap}{2}}(\psi^{\mu}
\partial_{\mu}T(X))\eta\right)\right] \label{eq:ZB} \\
&\longrightarrow& \mathrm{Tr_{CP}}P\exp\left[-\frac{1}{4}\int
\frac{d\tau}{2\pi}\left(\frac{\ap}{2}(\psi^{\mu}\partial_{\mu}T)
\partial_{\tau}^{-1}(\psi^{\nu}\partial_{\nu}T)+T(X)^2\right)\right].
\label{eq:ZC}
\end{eqnarray}
In the second line, we have integrated out the `auxiliary' fields $\eta$ and 
$F$ using equations of motion derived from~(\ref{eq:ZB}). The operator 
$\partial_{\tau}^{-1}$ is, more precisely, defined by 
\begin{equation}
\partial_{\tau}^{-1}f(\tau)=\frac{1}{2}\int d\tau^{\prime}\ \epsilon(\tau-
\tau^{\prime})f(\tau^{\prime}), \label{eq:ZD}
\end{equation}
where
\[ \epsilon(\tau)=\left\{
	\begin{array}{ccc}
	+1 & \mathrm{for} & \tau>0 \\
	-1 & \mathrm{for} & \tau<0
	\end{array}
\right. \]
so that $\partial_{\tau}\epsilon(\tau-\tau^{\prime})=2\delta(\tau-
\tau^{\prime})$. Then the equation~(\ref{eq:ZD}) correctly gives 
$\partial_{\tau}(\partial_{\tau}^{-1}f(\tau))=f(\tau)$.
\medskip

To discuss the tachyon condensation, we determine the explicit profile of the 
tachyon field $T(X^{\mu})$ as in the bosonic case~(\ref{eq:Tau}). When 
there are no non-trivial Chan-Paton factors, we take it here to be of the form
\[ T(X)=a+u_{\mu}X^{\mu}. \]
In this case, the world-sheet field theory remains free (\textit{i.e.} 
quadratic) even after including the boundary perturbation~(\ref{eq:ZC}). But
as opposed to the bosonic case~(\ref{eq:Tau}), by exploiting the 
translational and Lorentz invariance we can always rearrange the above 
profile into the following form,
\begin{equation}
T(X)=uX, \label{eq:ZE}
\end{equation}
where $X$ is the coordinate of a single direction, say, $X^1$. Then the 
matter partition function on the disk is given by
\begin{equation}
Z^{\mathrm{m}}(u)=\int\cD X\cD\psi\exp\left[-\cS_0-\frac{u^2}{4}\int_0^{2\pi}
\frac{d\tau}{2\pi}\left(X^2+\frac{\ap}{2}\psi\partial_{\tau}^{-1}\psi\right)
\right]. \label{eq:ZF}
\end{equation}
As in the bosonic case, the partition function is evaluated as 
follows~\cite{KMM2}: By differentiating with respect to $u^2$, we find
\begin{equation}
\frac{\partial}{\partial (u^2)}\ln Z^{\mathrm{m}}(u)=-\frac{1}{4}\int_0^{2\pi}
\frac{d\tau}{2\pi}\left\langle X^2+\frac{\ap}{2}\psi\partial_{\tau}^{-1}\psi
\right\rangle_{\mathrm{norm.}}. \label{eq:ZG}
\end{equation}
The 2-point function in the right hand side can be calculated by regularizing 
as 
\begin{eqnarray}
\left\langle X^2+\frac{\ap}{2}\psi\partial_{\tau}^{-1}\psi\right\rangle&=&
\lim_{\epsilon\to 0}\left\langle X(\tau)X(\tau+\epsilon)+\frac{\ap}{2}
\psi(\tau)\partial_{\tau}^{-1}\psi(\tau+\epsilon)\right\rangle \nonumber \\
&=&-\frac{2}{u}\lim_{\epsilon\to 0}\int d\theta\ \langle\mathbf{X}(\tau)
\Gamma(\tau+\epsilon)\rangle\bigg|_{\eta,F}, \label{eq:ZH}
\end{eqnarray}
where $|_{\eta,F}$ means that the equation holds only after integrating out 
the auxiliary fields $\eta,F$ using~(\ref{eq:ZB}). This expression makes it 
manifest that the regularization preserves world-sheet supersymmetry. And we 
do not need any conformal normal ordering as in~(\ref{eq:XXreg}) due to the 
supersymmetric cancellation. The detailed calculations of the 
correlator~(\ref{eq:ZH}) were shown in~\cite{KMM2,10218}: the result is
\begin{equation}
\left\langle X^2+\frac{\ap}{2}\psi\partial_{\tau}^{-1}\psi\right\rangle=-4\ap
\ln 2+\left[\frac{\ap}{y}-4\ap\sum_{k=1}^{\infty}\frac{y}{k(k+y)}+4\ap
\sum_{k=1}^{\infty}\frac{2y}{k(k+2y)}\right]_{y=\frac{\ap}{2}u^2}.
\label{eq:ZI}
\end{equation}
Combining the identity 
\[ \sum_{k=1}^{\infty}\frac{z}{k(k+z)}=\frac{d}{dz}\ln \Gamma(z)+\frac{1}{z}
+\gamma, \]
the equation~(\ref{eq:ZG}) can be expressed as 
\begin{eqnarray*}
\frac{\partial}{\partial (u^2)}\ln Z^{\mathrm{m}}(u)&=&\ap\ln 2+\frac{\ap}{4y}
+\ap\frac{d}{dy}\ln\Gamma(y)-\frac{\ap}{2}\frac{d}{dy}\ln\Gamma(2y)\bigg|_{y=
\frac{\ap}{2}u^2} \\ &=&\frac{\partial}{\partial (u^2)}\left[\ln\frac{
\Gamma(y)^2\exp(2y\ln 2)\sqrt{y}}{\Gamma(2y)}\right]_{y=\frac{\ap}{2}u^2}.
\end{eqnarray*}
Using the definition (\ref{eq:Z1}) $Z_1(y)=\sqrt{y}e^{\gamma y}\Gamma(y)$, 
the above equation can be integrated to become
\begin{equation}
Z^{\mathrm{m}}(u)=Ce^{\ap u^2\ln 2}\frac{Z_1(\ap u^2/2)^2}{Z_1(\ap u^2)},
\label{eq:ZJ}
\end{equation}
where $C$ is a positive integration constant. The partition 
function~(\ref{eq:ZJ}) is a monotonically decreasing function of $u$, and the 
asymptotic value as $u\to\infty$ is found to be
\begin{equation}
\lim_{u\to\infty}Z^{\mathrm{m}}(u)=C\sqrt{2\pi} \label{eq:ZK}
\end{equation}
by using the Stirling's formula (\ref{eq:Stir}).
\medskip

Now we relate the partition function to the spacetime action, as 
in~(\ref{eq:SKZ}) or (\ref{eq:Sbeta}). In bosonic string case, the classical 
spacetime action~(\ref{eq:dS}) was defined by applying the Batalin-Vilkovisky 
formalism to the space of world-sheet field theories~\cite{WiBI1}. And it was 
shown~\cite{Sh1,Sh2} that the spacetime action can be expressed in terms of 
the disk partition function and $\beta$-functions as in~(\ref{eq:Sbeta}),
at least up to first few orders in perturbation theory.  
Then we found that the spacetime action had the properties which were 
required to be regarded as a generalization of the boundary entropy. In 
superstring case, however, even the definition of the \textit{off-shell} 
spacetime action has not yet been discovered. But the on-shell effective 
action, which is constructed from the $S$-matrices calculated in the first 
quantized superstring theory, turned out to coincide with the disk partition 
function. From this fact, it has been proposed in~\cite{Tsey,KMM2} 
that the off-shell spacetime action, if properly defined, should also be 
equal (up to normalization) to the disk partition function itself, namely,
\begin{equation}
S(t)=K\ Z^{\mathrm{m}}(t), \label{eq:ZL}
\end{equation}
where we again denote the couplings collectively by $t$. In fact, the 
partition function itself has some properties in common with the boundary 
entropy in the superstring case. For example, it is stationary at the fixed 
points of the renormalization group in general~\cite{KMM2}. And in the 
special case of~(\ref{eq:ZE}), we have shown that the partition 
function~(\ref{eq:ZJ}) monotonically decreases along the renormalization 
group flow to infrared ($u\to\infty$). Anyway, we will use~(\ref{eq:ZL}) 
to discuss the tachyon condensation, hoping that the results support the 
conjecture~(\ref{eq:ZL}).

\section{Tachyon Condensation on a Non-BPS D-Brane}\label{sec:aoitori}
In determining the tachyon profile to be $T(X)=uX$ (\ref{eq:ZE}), we assumed 
that the Chan-Paton space is trivial. That is, we are now considering a 
single non-BPS D-brane. We will discuss the generalization to the system of 
plural non-BPS D-branes in the next section. Here, we take the single D-brane 
to be a non-BPS D9-brane in Type IIA theory.
\medskip

To obtain the (exact) tachyon potential, we must go back to~(\ref{eq:ZC}) and 
set $T(X)=a=\mathrm{constant}$. Then the matter partition function becomes
\[ Z^{\mathrm{m}}(T=a)=\int\cD X\cD\psi\ \exp\left(-\cS_0-\frac{a^2}{4}\right)
=\mathcal{Z}e^{-\frac{a^2}{4}}, \]
where $\displaystyle \mathcal{Z}=\int\cD X\cD\psi \ e^{-\cS_0}$ is considered 
to be an unimportant constant because it does not depend on any coupling. 
Using~(\ref{eq:ZL}), we get the spacetime action as
\begin{equation}
S(T=a)=\tilde{K}\tilde{\tau}_9V_{10}e^{-\frac{T^2}{4}}, \label{eq:ZM}
\end{equation}
where we explicitly write the ten-dimensional volume factor $V_{10}=\int 
d^{10}\!x$ and the tension $\tilde{\tau}_9$ of a non-BPS D9-brane because 
the action should be proportional to the product of these two factors. 
And $\tilde{K}$ is a combination of the normalization factors $K,\mathcal{Z}$ 
and $V_{10},\tilde{\tau}_9$ above. But by repeating the argument given 
between (\ref{eq:ZN}) and (\ref{eq:ZO}) in this case, we eventually find 
$\tilde{K}=1$. Then, the energy density at the maximum $T=0$ (representing 
the D9-brane) of the tachyon potential is exactly equal to the non-BPS 
D9-brane tension $\tilde{\tau}_9$, while that of the minimum $T=\pm\infty$ 
(closed string vacuum) is zero. Hence, we can conclude that the difference 
in the energy density between two vacua $T=0$ and $T=\pm\infty$ precisely 
coincides with the tension of an unstable non-BPS 
D-brane, just like the bosonic case. 
Of course, the fact that the non-BPS D-brane is `nine'-brane is not 
essential. We will arrive at the same conclusion 
in the case of a non-BPS D-brane of 
arbitrary dimensionality. In addition, the tachyon potential is symmetric 
under $T\to -T$ and is bounded from below, as is expected for superstring. 
\medskip

In order to describe the lower dimensional D-brane as a soliton on the 
non-BPS D9-brane, we now consider the kinetic term for the tachyon field 
$T(X)$. We begin with the two-derivative approximation. Using the following 
relation
\[ Z_1(y)=\sqrt{y}e^{\gamma y}\Gamma(y)\simeq\sqrt{y}(1+\gamma y)\left(
\frac{1}{y}-\gamma\right)=\frac{1}{\sqrt{y}}(1+\cO(y^2)) \]
for small $y$, we can expand the spacetime action~(\ref{eq:ZL}) 
around $u=0$ as 
\begin{eqnarray}
\tilde{S}&=&(KC)e^{2y\ln 2}\frac{Z_1(y)^2}{Z_1(2y)}\Bigg|_{y=\frac{\ap}{2}
u^2}\simeq KC(1+2y\ln 2)\frac{\sqrt{2y}}{y}\Bigg|_{y=\frac{\ap}{2}u^2}
\nonumber \\ &=& KC\left(\frac{2}{\sqrt{\ap u^2}}+(2\ln 2)\sqrt{\ap u^2}
\right), \label{eq:ZP}
\end{eqnarray}
where the tilde on $\tilde{S}$ indicates that this action is an 
approximate form for small $u$.
From this expression, we deduce the action truncated up to two 
derivatives to be
\begin{equation}
\tilde{S}(T)=\tilde{\tau}_9\int d^{10}\!x\left((\ap\ln 2)e^{-\frac{T^2}{4}}
\partial_{\mu}T\partial^{\mu}T+e^{-\frac{T^2}{4}}\right), \label{eq:ZQ}
\end{equation}
because (1) it reproduces the exact tachyon potential~(\ref{eq:ZM}) 
for constant $T$ with the correct normalization factor ($\tilde{K}=1$), and 
(2) the value of the action~(\ref{eq:ZQ}) evaluated for the 
tachyon profile $T=uX$ becomes 
\begin{equation}
\tilde{S}(uX)=(\tilde{\tau}_9V_9\sqrt{\ap\pi})\left(\frac{2}{\sqrt{\ap u^2}}
+(2\ln 2)\sqrt{\ap u^2}\right), \label{eq:ZQa}
\end{equation}
so that the relative normalization of~(\ref{eq:ZP}) is realized. Now that we 
have determined the form of the spacetime action 
up to two derivatives, we can construct a 
soliton solution in this level of approximation. The equation of motion from 
the action~(\ref{eq:ZQ}) is 
\begin{equation}
(4\ap \ln 2)\partial_{\mu}\partial^{\mu}T-(\ap\ln 2)T\partial_{\mu}T
\partial^{\mu}T+T=0. \label{eq:ZR}
\end{equation}
The `kink' ansatz $T(X)=uX$ solves~(\ref{eq:ZR}) if $u=1/\sqrt{\ap\ln 2}$. 
In this case, the energy density $\cT_8$ of the kink solution is obtained by 
substituting $u=1/\sqrt{\ap\ln 2}$ into~(\ref{eq:ZQa}) and it becomes 
\[ \tilde{S}=(\tilde{\tau}_9V_9\sqrt{\ap\pi})4\sqrt{\ln 2}\equiv \cT_8V_9, \]
or 
\begin{equation}
\cT_8=4\sqrt{\frac{\ln 2}{2\pi}}\cdot 2\pi\sqrt{\ap}\frac{\tilde{\tau}_9}
{\sqrt{2}}\simeq 1.33\cdot 2\pi\sqrt{\ap}\frac{\tilde{\tau}_9}{\sqrt{2}}.
\label{eq:ZS}
\end{equation}
We arranged it in this way because it is conjectured that the tachyonic kink 
configuration on a non-BPS D9-brane represents a BPS D8-brane in Type IIA 
theory, and the BPS D8-brane tension is given by 
$\displaystyle \tau_8=2\pi\sqrt{\ap}
\frac{\tilde{\tau}_9}{\sqrt{2}}$. Since this result was obtained from the 
derivative-truncated action~(\ref{eq:ZQ}), the relation~(\ref{eq:ZS}) is 
not exact, as expected.
\smallskip

Then we turn to the exact treatment. Since the theory remains free under the 
particular boundary perturbation $T(X)=uX$, no new couplings are generated 
throughout the renormalization group flow. This means that the exact tachyon 
profile can be achieved by minimizing the exact action~(\ref{eq:ZL}) together 
with~(\ref{eq:ZJ}) which was exactly obtained for $T=uX$. In fact, we have 
already seen that the partition function $Z^{\mathrm{m}}(u)$ is monotonically 
decreasing and the limiting value is $C\sqrt{2\pi}$ (\ref{eq:ZK}). Thus
\begin{equation}
\mathrm{min}\ S=K\ \mathrm{min}Z^{\mathrm{m}}(u)=\sqrt{2\pi}KC. 
\label{eq:ZT}
\end{equation}
And the normalization constant $KC$ is immediately determined by comparing 
(\ref{eq:ZP}) with (\ref{eq:ZQa}). Therefore the final result is 
\begin{equation}
\mathrm{min}\ S=S(u\to\infty)=2\pi\sqrt{\ap}\frac{\tilde{\tau}_9}{\sqrt{2}}
V_9=\tau_8V_9. \label{eq:ZU}
\end{equation}
This shows that the tension of the kink solution $T(X)=\lim_{u\to\infty}uX$ 
precisely agrees with the tension of the BPS D8-brane. After an appropriate 
change of variable, we can see that the width of the kink goes to zero 
as $u\to\infty$. 
\smallskip

Finally, let us see what happens to the spacetime action as the tachyon 
condenses. From~(\ref{eq:ZQ}), it is natural to guess that the full spacetime 
action for the tachyon field can be written in the form 
\begin{equation}
S(T)=\tilde{\tau}_9\int d^{10}x\ U(T)\biggl(1+(\ap\ln 2)\partial_{\mu}T
\partial^{\mu}T+\ldots\biggr), \label{eq:TD}
\end{equation}
where we defined the tachyon potential 
\[ U(T)=e^{-\frac{T^2}{4}}, \]
and $\ldots$ represents the higher derivative terms. Note that 
the Lagrangian density is proportional to the tachyon potential itself. Since 
$U(T)$ vanishes at the closed string vacuum $T=\pm\infty$, the action for the 
tachyon field also identically vanishes there. We hope that the same is true 
for the complete spacetime action 
including all modes, and that there remain no 
physical excitations of the open string modes around the closed string 
vacuum. 

\section{Higher Codimension Branes}
In this section, we consider the possibility of constructing D-branes of 
higher codimension. In doing so, we introduce the Chan-Paton matrices 
$\{\gamma_i\}_{i=1}^n$ and take the tachyon field to be 
\begin{equation}
T(X)=\sum_{i=1}^nu_iX^i\gamma_i. \label{eq:ZV}
\end{equation}
This profile will turn out to represent a codimension $n$ D-brane in Type IIA 
theory. We require $\gamma_i$'s to be Hermitian and to form a Clifford 
algebra\footnote{We are taking the target space to be Euclidean for 
simplicity.}
\[ \{\gamma_i,\gamma_j\}=2\delta_{ij}. \]
Hence, $\gamma_i$ is a $2^{[n/2]}\times 2^{[n/2]}$ matrix so that the 
original configuration of the D-branes is the system of $2^{[n/2]}$ 
coincident space-filling non-BPS D9-branes of Type IIA theory. Here, 
\[ \left[\frac{n}{2}\right]=\left\{
	\begin{array}{cccc}
	\frac{n}{2} & \mathrm{for} & n & \mathrm{even} \\
	\frac{n-1}{2} & \mathrm{for} & n & \mathrm{odd}
	\end{array}
\right. . \]
Using~(\ref{eq:ZC}), the partition function is now
\begin{equation}
Z^{\mathrm{m}}=\int\cD X\cD\psi\ e^{-\cS_0}\mathrm{Tr_{CP}}P\exp\left[
-\frac{1}{4}\int\frac{d\tau}{2\pi}\left(T(X)^2+\frac{\ap}{2}(\psi^{\mu}
\partial_{\mu}T)\partial_{\tau}^{-1}(\psi^{\nu}\partial_{\nu}T)\right)
\right]. \label{eq:ZW}
\end{equation}
We substitute the tachyon profile~(\ref{eq:ZV}) into (\ref{eq:ZW}), and 
simplify it using the Clifford algebra. The result is
\begin{equation}
Z^{\mathrm{m}}(u_i)=\int\cD X\cD\psi\ e^{-\cS_0}\mathrm{Tr_{CP}}P\exp
\left[-\frac{1}{4}\int\frac{d\tau}{2\pi}\sum_{i=1}^nu_i^2\left(X_i^2+
\frac{\ap}{2}\psi^i\partial_{\tau}^{-1}\psi^i\right)\right]. \label{eq:ZX}
\end{equation}
Note that the Chan-Paton structure has become a simple overall factor 
$\mathrm{Tr_{CP}}I=2^{[n/2]}$. Since the form of the partition 
function~(\ref{eq:ZX}) is quite similar to the previous one~(\ref{eq:ZF}), 
we can proceed in the same way. Differentiating with respect to 
$u_i^2$, we find
\[ \frac{\partial}{\partial (u_i^2)}\ln Z^{\mathrm{m}}(u)=-\frac{1}{4}
\int\frac{d\tau}{2\pi}\left\langle X_i^2+\frac{\ap}{2}\psi^i
\partial_{\tau}^{-1}\psi^i\right\rangle_{\mathrm{norm.}}, \]
where the correlator is given by~(\ref{eq:ZI}). It is integrated to give 
\begin{equation}
Z^{\mathrm{m}}(u_i)=2^{[n/2]}C_n\exp\left((\ap\ln 2)\sum_{i=1}^nu_i^2\right)
\prod_{i=1}^n\frac{Z_1(\ap u_i^2/2)^2}{Z_1(\ap u_i^2)}, \label{eq:ZY}
\end{equation}
where $C_n$ is an integration constant. We again take the spacetime action 
to be 
\begin{equation}
S=K\ Z^{\mathrm{m}}. \label{eq:ZZ}
\end{equation}
For small $u_i$'s, the action is approximately given by
\begin{equation}
\tilde{S}\simeq 2^{[n/2]}KC_n\frac{2^n}{\prod_{i=1}^n\sqrt{\ap u_i^2}}
\left(1+\ap\ln 2\sum_{i=1}^nu_i^2\right). \label{eq:YA}
\end{equation}
The spacetime action up to two derivatives is now 
\[ \tilde{S}(T)=\tilde{\tau}_9\mathrm{Tr_{CP}}\int d^{10}\!x\left((\ap\ln 2)
e^{-\frac{T^2}{4}}\partial_{\mu}T\partial^{\mu}T+e^{-\frac{T^2}{4}}\right) \]
because for constant $T=aI$ the value of $\tilde{S}$ is $2^{[n/2]}
\tilde{\tau}_9V_{10}e^{-a^2/4}$, which is suitable for $2^{[n/2]}$ non-BPS 
D9-branes, and for $T=\sum u_iX^i\gamma_i$ 
\begin{eqnarray*}
\tilde{S}(T=\sum u_iX^i\gamma_i)&=&2^{[n/2]}\tilde{\tau}_9V_{10-n}\left(
1+\ap\ln 2\sum_{i=1}^nu_i^2\right)\prod_{i=1}^n\left[\int dx_ie^{-\frac{1}{4}
u_i^2x_i^2}\right] \\ &=&2^{[n/2]}\tilde{\tau}_9V_{10-n}(\pi\ap)^{\frac{n}{2}}
\frac{2^n}{\prod_{i=1}^n\sqrt{\ap u_i^2}}\left(1+\ap\ln 2\sum_{i=1}^nu_i^2
\right). 
\end{eqnarray*}
Comparing it with (\ref{eq:YA}), we can fix the normalization constant as 
\[ KC_n=\tilde{\tau}_9V_{10-n}(\pi\ap)^{n/2}. \]
Then the exact spacetime action is given by
\begin{equation}
S=2^{[n/2]}\tilde{\tau}_9V_{10-n}(\pi\ap)^{n/2}\exp\left(\ap\ln 2\sum_{i=1}^n
u_i^2\right)\prod_{i=1}^n\frac{Z_1(\ap u_i^2/2)^2}{Z_1(\ap u_i^2)}. 
\label{eq:YB}
\end{equation}
Using (\ref{eq:ZK}), we can easily minimize the above action and its value is 
\begin{equation}
\mathrm{min}\ S=(2\pi\sqrt{\ap})^n\frac{\tilde{\tau}_9}{2^{\frac{n}{2}-[
\frac{n}{2}]}}V_{10-n}=\left\{
	\begin{array}{lccc}
	(2\pi\sqrt{\ap})^n\tilde{\tau}_9V_{10-n} & \mathrm{for} & n & 
	\mathrm{even} \\
	(2\pi\sqrt{\ap})^n(\tilde{\tau}_9/\sqrt{2})V_{10-n} & \mathrm{for} 
	& n & \mathrm{odd} 
	\end{array}
\right. . \label{eq:YC}
\end{equation}
These are just the exact known values: When $n$ is even, the tension of the 
tachyonic soliton represented by the profile~(\ref{eq:ZV}) in the 
limit $u_i\to\infty$ has the dimension of $(9-n)$-brane, and it precisely 
agrees with the tension of a non-BPS D($9-n$)-brane. When $n$ is odd, the 
tension of the tachyonic soliton coincides with the tension of a BPS 
D($9-n$)-brane because of an additional factor of $\frac{1}{\sqrt{2}}$. 

\section{Comments on Some Developments}
In the previous three sections, we focused on the tachyon condensation on 
non-BPS D-branes. However, its generalization to the brane-antibrane system 
was discussed in the context of boundary string field theory in recent 
papers~\cite{12198,TTU,mahoh}. In this case, 
the tachyon field $T(\mathbf{X})$ 
becomes complex-valued and the tachyon potential takes the form $e^{-T
\overline{T}/4}$, as expected. For a D$p$-$\overline{\mathrm{D}p}$ pair, 
a kink-like configuration of the tachyon leads to a non-BPS D($p-1$)-brane, 
whereas a vortex-type configuration leaves a BPS D($p-2$)-brane behind. 
In both cases the tension of the resulting lower dimensional D-brane is 
reproduced exactly. Moreover, the authors found the D-brane world-volume action 
of the Dirac-Born-Infeld type including the gauge fields and the 
Ramond-Ramond couplings\footnote{Couplings between a closed string RR field and 
open string tachyons had been calculated in \cite{Billo} and \cite{Wilk} 
for non-BPS D-branes and 
brane-antibrane systems, respectively, in the conformal field theory approach.} 
for both non-BPS D-branes and brane-antibrane system. 
\medskip

In connection with the noncommutative geometry~\cite{SW}, the spacetime 
action of background independent open string field theory was analysed in 
the presence of a constant $B$-field background \cite{10021,10028,11108}. 
In bosonic string, the world-sheet action is now modified as 
\begin{equation}
\cS_0=\frac{1}{2\pi\ap}\int_{\Sigma} d^2z\ g_{\mu\nu}\partial X^{\mu}
\overline{\partial}X^{\nu}-i\int_{\Sigma} B, \label{eq:YE}
\end{equation}
where
\[ B=\frac{1}{2}B_{\mu\nu}dX^{\mu}\wedge dX^{\nu}=\frac{1}{2}B_{\mu\nu}
\frac{\partial X^{\mu}}{\partial\sigma^a}
\frac{\partial X^{\nu}}{\partial\sigma^b}d\sigma^a\wedge d\sigma^b. \]
Since the world-sheet field theory remains free for a constant $B_{\mu\nu}$, 
we can exactly compute the disk partition function $Z^{\mathrm{m}}(a,u)$ and 
hence the spacetime action $S$ for $T(X)=a+u_{\mu\nu}X^{\mu}X^{\nu}$. The 
result is given~\cite{10028} by
\begin{eqnarray}
Z^{\mathrm{m}}(a,u)&=&\exp\left[-a+\gamma\mathrm{Tr}\left\{\left(\frac{1}
{g+2\pi\ap B}\right)_{\mathrm{sym}}2\ap u\right\}\right] \label{eq:YF} \\
& &\times \sqrt{\det\left\{\Gamma\left(\left(\frac{1}{g+2\pi\ap B}
\right)_{\mathrm{sym}}2\ap u\right)\Gamma(2\ap uE_-)2\ap uE_-\right\}},
\nonumber
\end{eqnarray}
where the subscript of $A_{\mathrm{sym}}$ means that we should take the 
symmetric part of the matrix $A$, and we defined 
\[ E_-=G^{-1}-\frac{\theta}{2\pi\ap} \quad \mathrm{with} \quad 
G^{-1}=\left(\frac{1}{g+2\pi\ap B}\right)_{\mathrm{sym}}\> , \quad 
\theta=2\pi\ap\left(\frac{1}{g+2\pi\ap B}\right)_{\mathrm{antisym}}. \]
Trace and determinant are taken over the spacetime indices $\mu,\nu$.
Then 
\begin{equation}
S(a,u)=K\left[\mathrm{Tr}\left(G^{-1}\cdot 2\ap u\right)-a\frac{\partial}
{\partial a}-\mathrm{Tr}\left(u\frac{\partial}{\partial u}\right)+1\right]
Z^{\mathrm{m}}(a,u). \label{eq:YG}
\end{equation}
Further, by expanding the above exact action for small $u$, we find the 
effective action for the tachyon field $T$ to be
\begin{equation}
S=\tau_{25}\int d^{26}\!x \sqrt{|\det(g+2\pi\ap B)|}e^{-T}(T+1+\ap G^{\mu\nu}
\partial_{\mu}T\partial_{\nu}T+\ldots ). \label{eq:YH}
\end{equation}
As will be explained in section \ref{sec:NC}, in the large noncommutativity 
limit $\theta\to\infty$ the kinetic term and still higher derivative terms 
can be ignored as compared to the potential term. 
Then the effective action for 
the tachyon field \textit{exactly} becomes
\begin{equation}
S=\tau_{25}\int d^{26}\!x\ \mathrm{Pf}(2\pi\ap\theta^{-1})(T(x)+1)*e^{-T(x)},
\label{eq:YI}
\end{equation}
where all the products among the fields are taken by the star product
\[ f*g(x)\equiv \exp\left(\frac{i}{2}\theta^{\mu\nu}\frac{\partial}{\partial
x^{\mu}}\frac{\partial}{\partial y^{\nu}}\right)f(x)g(y)\bigg|_{y=x}, \]
and the Pfaffian is defined for an antisymmetric matrix $X$ as\footnote{The 
sign of Pf$X$ is not determined by this equation only. It is fixed by 
requiring that Pf$X=+\sqrt{\det X}$ if $X$ is self-dual. For more details, 
consult suitable mathematics reference books.}
\[ \det X=(\mathrm{Pf}X)^2. \]
\smallskip

In~\cite{11108}, these results were extended to the supersymmetric case and 
in particular the effective tachyon action was found to be of the form
\begin{equation}
S=\tilde{\tau}_9\int d^{10}\!x \sqrt{|\det(g+2\pi\ap B)|}e^{-\frac{T^2}{4}}
\Bigl(1+(\ap\ln 2)G^{\mu\nu}\partial_{\mu}T\partial_{\nu}T+\ldots \Bigr).
\label{eq:YJ}
\end{equation}

As an application of this formalism, the disk partition function was 
evaluated for the D0-D2 system of Type IIA theory in the presence of a large 
Neveu-Schwarz $B$-field background~\cite{12089}. Since the configuration 
where a D0-brane coexists with a D2-brane does not saturate the BPS bound, 
tachyonic 
modes appear on the fundamental strings stretched between these branes. 
The unstable stationary point of the tachyon potential under consideration 
corresponds to the system of the D-particle 
plus D-membrane, while the stable minimum represents the D0-D2 bound state, 
whose energy is truly lower than that of the unbound state. 
As the mass defect 
\[ \Delta M=M_{\mbox{\scriptsize{D0-D2}}}-(M_{\mbox{\scriptsize{D0}}}+
M_{\mbox{\scriptsize{D2}}}) \]
is known in the first-quantized string theory, we can compare it with the 
result from background independent open superstring field theory. Though 
the partition function was evaluated perturbatively by expanding it in 
powers of the boundary interactions, namely
\begin{eqnarray*}
Z^{\mathrm{m}}&=&\int\cD X\cD\psi\ e^{-\cS_0-\cS_{\partial\Sigma}} \\
&=&\int\cD X\cD\psi\ e^{-\cS_0}-\int\cD X\cD\psi\ \cS_{\partial\Sigma}
e^{-\cS_0}+\frac{1}{2!}\int\cD X\cD\psi\ \cS_{\partial\Sigma}^2
e^{-\cS_0}+\cdots,
\end{eqnarray*}
the resulting value of the mass defect was found to agree exactly with the 
expected one in the large $B$ limit~\cite{12089}. Considering that all the 
results obtained so far from background independent open superstring field 
theory coincide with the expected ones, the conjecture~(\ref{eq:ZL}) that 
the spacetime action of this theory is given by the disk partition function 
itself seems to be plausible. 

\chapter{Some Related Topics}\label{ch:others}

\section{Noncommutative Solitons}\label{sec:NC}
It has been shown that introducing large noncommutativety into spacetime 
simplifies the description of D-branes as tachyonic solitons. In this 
section, we outline the construction of soliton solutions and their 
applications to D-brane physics. 

\subsection{Solitons in noncommutative scalar field theory}
We consider the theory of a single scalar field $\phi$ in $(2+1)$-dimensional 
spacetime whose coordinates are $x^{\mu}=(x^0,x^1,x^2)$. We introduce the 
spacetime noncommutativity as 
\begin{equation}
[x^{\mu},x^{\nu}]=i\theta^{\mu\nu}\quad \mathrm{with} \quad \theta^{\mu\nu}=
\left(
	\begin{array}{ccc}
	0 & 0 & 0 \\ 0 & 0 & \theta \\ 0 & -\theta & 0
	\end{array}
\right). \label{eq:NS}
\end{equation}
In this noncommutative spacetime, arbitrary two functions (fields) $f,g$ of 
$x^{\mu}$ are multiplied using the following Moyal $*$-product
\begin{eqnarray}
f*g(x)&=&\exp\left(\frac{i}{2}\theta^{\mu\nu}\frac{\partial}{\partial x^{\mu}}
\frac{\partial}{\partial y^{\nu}}\right)f(x)g(y)\bigg|_{y=x} \label{eq:NT} \\
&=&\exp\left(\frac{i}{2}\theta\left(\frac{\partial}{\partial x^1}
\frac{\partial}{\partial y^2}-\frac{\partial}{\partial x^2}\frac{\partial}
{\partial y^1}\right)\right)f(x)g(y)\Bigg|_{y=x}. \nonumber 
\end{eqnarray}
We will discuss the field theory action 
\begin{equation}
S=\frac{1}{g^2}\int d^3x\left(-\frac{1}{2}\partial_{\mu}\phi\partial^{\mu}\phi
-V(\phi)\right). \label{eq:NU}
\end{equation}
We allow the potential $V(\phi)$ to have the generic polynomial form 
\begin{equation}
V(\phi)=\frac{1}{2}m^2\phi^2+\sum_{j=3}^r\frac{b_j}{j}\overbrace{\phi*\cdots
*\phi}^j, \label{eq:NV}
\end{equation}
noting that there is no linear term. In writing (\ref{eq:NU}) and 
(\ref{eq:NV}), we used the fact that for arbitrary $f,g$ 
\begin{equation}
\int d^3x\ f*g=\int d^3x\ fg \label{eq:NVb}
\end{equation}
holds because of the antisymmetry of $\theta^{\mu\nu}$. Here we perform the 
rescaling $x^{\mu}\longrightarrow\sqrt{\theta/\ap}x^{\mu}$. Then the action 
can be rewritten as 
\begin{equation}
S=\sqrt{\frac{\theta}{\ap}}\frac{1}{g^2}\int d^3x\left(-\frac{1}{2}
\partial_{\mu}\phi\partial^{\mu}\phi-\frac{\theta}{\ap}V(\phi)\right). 
\label{eq:NVa}
\end{equation}
From the above expression, it is clear that in the limit $\theta/\ap\to\infty$
of large noncommutativity \textit{the kinetic term can be ignored} as 
compared to the potential energy term. Hence in finding a soliton solution it 
suffices to extremize the potential, 
\[ \frac{\partial V}{\partial\phi}=0. \]
In a commutative theory, this equation has only constant solutions 
$\Lambda\equiv\{\phi=\lambda_i\}$. 
In the noncommutative theory, however, it can support soliton solutions 
which non-trivially depend on $x^{\mu}$. Concretely, assume that 
we can construct a function $\phi_0(x)$ which squares to itself under the 
$*$-product, \textit{i.e.}
\begin{equation}
\phi_0 *\phi_0(x)=\phi_0(x). \label{eq:NW}
\end{equation}
Then, using the explicit form (\ref{eq:NV}) of the potential, we find 
\begin{eqnarray*}
\frac{\partial V}{\partial\phi}(\lambda_i\phi_0)&=&m^2(\lambda_i\phi_0)+
\sum_{j=3}^rb_j\lambda_i^{j-1}\overbrace{\phi_0*\cdots *\phi_0}^{j-1} \\
&=&\left(m^2\lambda_i+\sum_{j=3}^rb_j\lambda_i^{j-1}\right)\phi_0=
\frac{\partial V}{\partial\phi}(\lambda_i)\cdot\phi_0, 
\end{eqnarray*}
so it vanishes if $\lambda_i\in\Lambda$. Therefore we can obtain solitonic 
solutions if we find a function $\phi_0$ which satisfies the 
idempotency~(\ref{eq:NW}) and is spatially localized. To do so, it is 
convenient to make use of the correspondence between fields on noncommutative 
space and operators in one-particle Hilbert space, which will be explained 
below. 

Given a function $f(\bar{x}^1,\bar{x}^2)$ on the commutative space, we can 
promote it to the noncommutative version by replacing the commutative 
coordinates $\bar{x}^i$ with the noncommutative ones $x^i$ which 
obey~(\ref{eq:NS}). But a problem arises here. Writing $\widehat{q}=x^1/
\sqrt{\theta}$ and $\widehat{p}=x^2/\sqrt{\theta}$, the commutation 
relation~(\ref{eq:NS}) becomes 
\[ [\widehat{q},\widehat{p}]=i, \]
which looks like the canonical commutation relation in quantum mechanics. 
So just as in the quantum mechanical case we must pay attention to the 
ordering of the noncommutative coordinates. Here we employ the Weyl- or 
symmetric-ordering prescription. This can be achieved by the following 
processes: We Fourier-transform the given function $f(\bar{x}^1,\bar{x}^2)$ 
to the momentum space as 
\begin{equation}
\tilde{f}(k_1,k_2)=\int d^2\bar{x}\ f(\bar{x}^1,\bar{x}^2)e^{ik_1\bar{x}^1
+ik_2\bar{x}^2}, \label{eq:OA}
\end{equation}
and then inverse-Fourier-transform back to the \textit{noncommutative} 
coordinate space by using the noncommutative coordinates $x^i$ as 
\begin{equation}
\widehat{f}(x^1,x^2)=\int\frac{d^2k}{(2\pi)^2}\tilde{f}(k_1,k_2)
e^{-ik_1x^1-ik_2x^2}. \label{eq:OB}
\end{equation}
Through the correspondence $(x^1,x^2)=\sqrt{\theta}(\widehat{q},
\widehat{p})$, we can regard the above Weyl-ordered function $\widehat{f}
(x^1,x^2)$ on the noncommutative space as the quantum mechanical 
\textit{operator} acting on a Hilbert space. Next we consider the product of 
two such operators $\widehat{f}(x)$ and $\widehat{g}(x)$. Using the 
Baker-Campbell-Hausdorff formula it becomes 
\begin{eqnarray}
\widehat{f}\cdot\widehat{g}(x)&=&\int\frac{d^2k\ d^2l}{(2\pi)^4}
\tilde{f}(k)\tilde{g}(l)e^{-ik_1x^1-ik_2x^2}e^{-il_1x^1-il_2x^2} \nonumber \\
&=&\int\frac{d^2k\ d^2l}{(2\pi)^4}\tilde{f}(k)\tilde{g}(l)e^{-\frac{i}{2}
\theta(k_1l_2-k_2l_1)}e^{-i(k_1+l_1)x^1-i(k_2+l_2)x^2} \nonumber \\ &=&\left[
\exp\left\{\frac{i}{2}\theta\left(\frac{\partial}{\partial \bar{x}^1}
\frac{\partial}{\partial\bar{y}^2}-\frac{\partial}{\partial\bar{x}^2}
\frac{\partial}{\partial\bar{y}^1}\right)\right\}f(\bar{x})g(\bar{y})
\Bigg|_{\bar{x}=\bar{y}=x}\right]_W \nonumber \\ &=&[\widehat{f*g}(x)]_W,
\label{eq:OC}
\end{eqnarray}
where $[\cO]_W$ indicates that we should take the Weyl-ordering of $\cO$, and 
$f*g$ in the final line is the $*$-product defined in~(\ref{eq:NT}). And in 
the third line we used the expression~(\ref{eq:OA}) for $\tilde{f}(k)$ and 
$\tilde{g}(l)$. To sum up, the product of two Weyl-ordered operators 
$\widehat{f},\widehat{g}$ is given by the operator version of $f*g$. 
This is why we adopt the Weyl ordering in this context. 

Now that we have found an operator $\widehat{f}$ for each ordinary function 
$f$, we then construct a Hilbert space on which the operators act. By taking 
the following linear combinations
\begin{equation}
a=\frac{x^1+ix^2}{\sqrt{2\theta}}\ , \quad a^{\dagger}=\frac{x^1-ix^2}{
\sqrt{2\theta}}\ ,  \label{eq:NX}
\end{equation}
the noncommutative algebra~(\ref{eq:NS}) can be written as 
\[ [a,a^{\dagger}]=1, \]
from which we interpret $a^{\dagger},a$ as creation and annihilation 
operators respectively in the one-particle quantum mechanical Hilbert space 
$\cH$. We take a basis of states in $\cH$ to be the number eigenstates $\{|n
\rangle \ ;\ n=0,1,2,\cdots\}$, on which $a,a^{\dagger}$ act as 
\[ a|n\rangle=\sqrt{n}|n-1\rangle\ , \quad a^{\dagger}|n\rangle=\sqrt{n+1}
|n+1\rangle \ , \quad a^{\dagger}a|n\rangle=n|n\rangle. \]
A basis of \textit{operators} 
which act in $\cH$ is formed by $\{|m\rangle\langle n|\ ;
\ m,n\ge 0\}$. In terms of $a$ and $a^{\dagger}$, each basis operator can be 
expressed as 
\begin{equation}
|m\rangle\langle n|=:\frac{a^{\dagger m}}{\sqrt{m!}}e^{-a^{\dagger}a}
\frac{a^n}{\sqrt{n!}}:\ . \label{eq:NY}
\end{equation}
Since $2\theta a^{\dagger}a=(x^1)^2+(x^2)^2-\theta$, any radially symmetric 
function in $(x^1,x^2)$-space should correspond to some operator-valued 
function of $a^{\dagger}a$. So we can restrict ourselves to operators of the 
diagonal form
\begin{equation}
\cO=\sum_{n=0}^{\infty}a_n|n\rangle\langle n| \label{eq:NZ}
\end{equation}
in searching for radially symmetric solutions. It was shown in~\cite{GMS,DMR} 
that the operator $|n\rangle\langle n|$ corresponds to the function 
(field)
\begin{equation}
\phi_n(r^2)=2(-1)^ne^{-\frac{r^2}{\theta}}L_n\left(\frac{2r^2}{\theta}\right),
\label{eq:OD}
\end{equation}
where $r^2=(\bar{x}^1)^2+(\bar{x}^2)^2$ and $L_n(x)$ is the $n$-th Laguerre 
polynomial. These functions are suitable for the `soliton' solutions 
because they are localized near $r=0$ due to the Gaussian damping factor. 
\medskip

In the operator language, the idempotency condition~(\ref{eq:NW}) is 
translated into
\begin{equation}
\widehat{\phi *\phi}=\widehat{\phi} \quad \longrightarrow \quad 
\widehat{\phi}^2=\widehat{\phi} \label{eq:OE}
\end{equation}
as explained in (\ref{eq:OC}). In other words, we can obtain non-trivial 
solutions to equation of motion by finding projection operators in $\cH$. 
We immediately see that the operators $P_n=|n\rangle\langle n|$ play such a 
role. Since $P_n$'s are mutually orthogonal, the general solution to the 
equation of motion $\partial V/\partial\phi=0$ is given by
\begin{equation}
\widehat{\phi}=\sum_{n=0}^{\infty}\lambda_nP_n, \label{eq:OF}
\end{equation}
where $\lambda_n\in\Lambda$. To show that this solves the equation of motion, 
one should note that 
\[\left(\sum_{n=0}^{\infty}\lambda_nP_n\right)^m=\sum_{n=0}^{\infty}
\lambda_n^mP_n \quad \mbox{so that} \quad \frac{\partial V}{\partial\phi}
\left(\sum_n\lambda_nP_n\right)=\sum_nP_n\frac{\partial V}{\partial\phi}
(\lambda_n)=0. \]
Mapped to the coordinate space, the solution~(\ref{eq:OF}) corresponds to 
\begin{equation}
\phi(\bar{x})=\sum_{n=0}^{\infty}\lambda_n\phi_n(r^2), \label{eq:OFf}
\end{equation}
where $\phi_n(r^2)$ is given in~(\ref{eq:OD}). This represents a soliton 
solution we have been looking for.
\smallskip

In operator form, the integration over $(x^1,x^2)$-space is replaced by the 
trace over $\cH$ as 
\begin{equation}
\int d^2x\ f(x^1,x^2)=2\pi\theta\mathrm{Tr}_{\cH}\widehat{f}
(x^1,x^2). \label{eq:OG}
\end{equation}
The orthonormality of the Laguerre polynomials is reflected as the mutual 
orthogonality of $P_n$'s, while the normalization is determined by
\[ \int d^2x\ 2e^{-r^2/\theta}=4\pi\int_0^{\infty}r\ dr\ e^{-r^2/
\theta}=2\pi\theta \quad \mathrm{and} \quad \mathrm{Tr}_{\cH}|0\rangle\langle
0|=1. \]
After neglecting the kinetic term, the action~(\ref{eq:NU}) can be rewritten 
as 
\begin{equation}
S=-\frac{2\pi\theta}{g^2}\int dt\ \mathrm{Tr}_{\cH}V(\widehat{\phi}). 
\label{eq:OH}
\end{equation}
This action is invariant under the unitary transformation
\begin{equation}
\widehat{\phi}\longrightarrow U\widehat{\phi}U^{\dagger}, \label{eq:OHm}
\end{equation}
thus it possesses a global `$U(\infty)$' symmetry. Using this symmetry 
transformation, a general non-radially symmetric solution $\phi(\bar{x})$ is 
unitarily equivalent to some radially symmetrical one, because any Hermitian 
operator corresponding to the real field $\phi$ can always be diagonalized 
by a suitable unitary transformation. Therefore it is sufficient to examine 
the radially symmetric solutions~(\ref{eq:OF}) or (\ref{eq:OFf}).
\smallskip

The energy of a static soliton of the form $\widehat{\phi}=\sum_{n=0}^{\infty}
\lambda_nP_n$ is given by 
\begin{equation}
E=-\frac{S}{\int dt}=\frac{2\pi\theta}{g^2}\mathrm{Tr}_{\cH}\left(
\sum_{n=0}^{\infty}P_nV(\lambda_n)\right)=\frac{2\pi\theta}{g^2}
\sum_{n=0}^{\infty}V(\lambda_n). \label{eq:OI}
\end{equation}
So the soliton solution is stable iff \textit{all} $\lambda_n$'s correspond 
to local minima of the potential $V(\phi)$, rather than maxima. Surprisingly, 
the energy of the soliton solution depends only on the value of the potential 
at extrema $\phi=\lambda_n$. In this sense, the energy of the soliton is 
insensitive to the detailed shape of the potential. 
\bigskip

So far, we have discussed the soliton solutions only in the limit $\theta\to
\infty$,\footnote{For the sake of convenience, we set $\ap=1$ in the 
remainder of this subsection.} where we can completely ignore the kinetic 
term.\footnote{Precisely, the derivatives in the commutative directions, 
$x^0=t$ in this case, cannot be dropped because the rescaling $x\to
\sqrt{\theta}x$ does not make sense in such directions. However, since we are 
considering static solitons, $t$-derivative term does not contribute to the 
action anyway.} But the behavior of the solitons at finite $\theta$ has also 
been argued in~\cite{GMS,Zhou,Nest}. In the case of finite $\theta$, the 
solution $\lambda P_n=\lambda |n\rangle\langle n|$ for $n>0$ is energetically 
unstable and can decay to $\lambda |0\rangle\langle 0|$ due to the 
contribution from the kinetic energy ($L_n(x)$ for large $n$ intensively 
fluctuates near $x=0$). More drastically, it happens that some solitons 
even cease to exist as $\theta$ is decreased. When the highest power $r$ of 
the polynomial potential is odd, the potential is not bounded below. As 
mentioned in section~\ref{sec:kink}, the Derrick's theorem is not applicable 
to this case. So there exist solitonic solutions for any nonzero value of 
$\theta$ as well as for the commutative case $\theta=0$. However, if the 
potential is bounded below, it depends on the value of $\theta$ whether the 
solitons exist or not. It was shown that (spherically symmetric) solutions 
exist for a sufficiently large value of $\theta$ just as in the $\theta\to
\infty$ case. But there is a critical value $\theta_c$ below which the 
soliton solutions are absent~\cite{Zhou,Nest}. The reason why the solitons 
of codimension 2 
exist for large $\theta$ in spite of the Derrick's theorem is explained as 
follows: The Derrick's scaling argument tells us 
that the energy of any soliton can always 
be lowered by shrinking to zero size if the value of codimension is 
more than 1. But in the noncommutative field theory, sharply peaked field 
configurations cost high energy so that the solitons are stabilized by 
spreading out. Thus, if the noncommutativity is strong enough, the solitons 
can exist in finite size. 

\subsection{D-branes as noncommutative solitons}
In this subsection, we construct soliton solutions in the noncommutative 
effective tachyon field theory and then identify them with D-branes both in 
bosonic string and in superstring theory. First of all, we introduce the 
noncommutativity by turning on a constant (Neveu-Schwarz) $B$-field 
background. As explained in~\cite{SW}, the effect of $B$-field is 
incorporated by replacing the closed string metric $g_{\mu\nu}$ with the 
effective open string metric 
\[ G^{\mu\nu}=\left(\frac{1}{g+2\pi\ap B}\right)^{\mu\nu}_{\mathrm{sym}}
=\left(\frac{1}{g+2\pi\ap B}g\frac{1}{g-2\pi\ap B}\right)^{\mu\nu} \]
or 
\begin{equation}
G_{\mu\nu}=g_{\mu\nu}-(2\pi\ap)^2(Bg^{-1}B)_{\mu\nu}, \label{eq:OJ}
\end{equation}
replacing the closed string coupling constant $g_s$ with the effective 
coupling
\begin{equation}
G_s=g_s\sqrt{\frac{\det G}{\det(g+2\pi\ap B)}}, \label{eq:OK}
\end{equation}
and replacing ordinary products among fields by the $*$-product~(\ref{eq:NT}) 
with
\begin{equation}
\theta^{\mu\nu}=2\pi\ap\left(\frac{1}{g+2\pi\ap B}\right)^{\mu\nu}_{
\mathrm{antisym}}=-(2\pi\ap)^2\left(\frac{1}{g+2\pi\ap B}B\frac{1}{
g-2\pi\ap B}\right)^{\mu\nu}. \label{eq:OL}
\end{equation}
Here we want to take a limit where the value of the dimensionless 
noncommutativity parameter $\theta/\ap$ goes to infinity while the open 
string metric $G_{\mu\nu}$ remains finite. This can be achieved by the 
following scaling 
\begin{equation}
\ap B\sim \epsilon\ , \quad g\sim\epsilon^2\quad \longrightarrow 
\quad \theta/\ap\sim\epsilon^{-1} \label{eq:OM}
\end{equation}
with $\epsilon\to 0$. Note that it does not require taking the low energy 
limit $\ap\to 0$. 

\subsubsection*{\underline{bosonic D-branes}}
\begin{figure}[htbp]
	\begin{center}
	\includegraphics{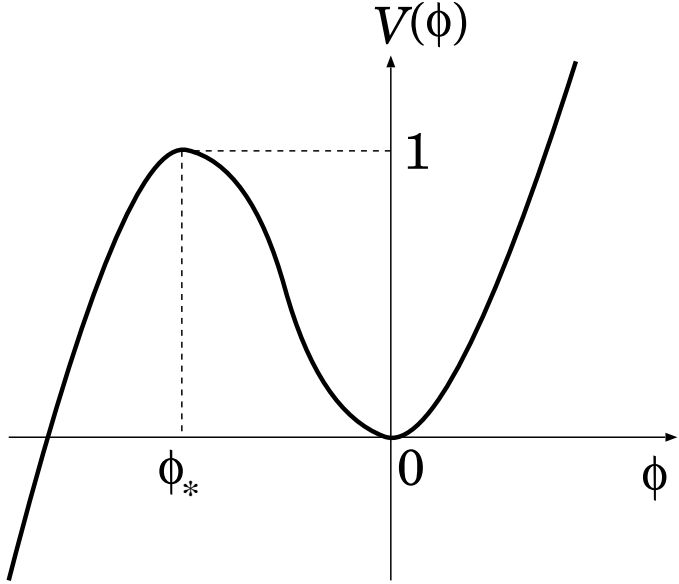}
	\end{center}
	\caption{The effective tachyon potential in bosonic string theory.}
	\label{fig:BF}
\end{figure}

Now we consider the effective action for the tachyon field $\phi$ on a 
bosonic D25-brane~\cite{HKLM} 
\begin{equation}
S=\tau_{25}\int d^{26}x\sqrt{|\det g|}\left(-\frac{1}{2}f(\phi)g^{\mu\nu}
\partial_{\mu}\phi\partial_{\nu}\phi+\ldots -V(\phi)\right), \label{eq:ON}
\end{equation}
where $\ldots$ indicates higher derivative terms, and the potential $V(\phi)$ 
is arranged such that a local minimum (closed string vacuum) is at $\phi=0$ 
and the value of the potential there is $V(0)=0$. Since we are thinking of 
the action~(\ref{eq:ON}) as being obtained by integrating out all fields 
other than the tachyon in the full (\textit{i.e.} not level truncated) cubic 
string field theory action, the maximum value of the `exact' potential $V$ is 
considered to be precisely $V(\phi=\phi_*)=1$, in which case the difference 
in the energy density between two vacua $\phi=0$ and $\phi=\phi_*$ is equal 
to the tension of the D25-brane. In addition, the unknown function $f$ is 
conjectured to take the following values at these vacua,
\begin{equation}
f(\phi_*)=1 \ , \quad f(0)=0 \ . \label{eq:ONa}
\end{equation}
Though we have no more information about the potential $V(\phi)$, it is 
sufficient to know the minimum and maximum values of $V$ because the 
properties of the noncommutative soliton in the $\theta/\ap\to\infty$ limit 
are insensitive to the detailed form of the potential, as we saw in the 
preceding subsection. The form of the potential is schematically illustrated 
in Figure~\ref{fig:BF}.
Note that the potential has no other extrema. Although the potential $V(\phi)=
(\phi+1)e^{-\phi}$ obtained from background independent open string field 
theory will do, we must perform a field redefinition such that the minimum of 
the potential occurs at finite $\phi$ in order to construct noncommutative 
soliton solutions, so we do not consider it here. 
\smallskip

Now we turn on the $B$-field. According to the prescription mentioned 
earlier, the action~(\ref{eq:ON}) is changed to
\begin{equation}
S=\tau_{25}\frac{g_s}{G_s}\int d^{26}\!x\sqrt{|\det G|}*\left(-\frac{1}{2}
f(\phi)*G^{\mu\nu}*\partial_{\mu}\phi *\partial_{\nu}\phi+\ldots -V(\phi)
\right), \label{eq:OO}
\end{equation}
where the precise ordering among the fields does not matter in the 
following analysis. 
The factor $g_s/G_s$ arose because the tension $\tau_{25}$ is proportional to 
$g_s^{-1}$. As the simplest case, we take the $B$-field to be 
\begin{equation}
B_{12}=-B_{21}=B<0\ , \quad B_{\mu\nu}=0 \quad \mbox{for any other }\mu, \nu.
\label{eq:OOa}
\end{equation}
In this case, $x^1$- and $x^2$-directions become noncommutative ($\theta^{12}
\neq 0$) whereas the remaining ($23+1$)-dimensional subspace is kept 
commutative, so that we can make direct use of the noncommutative soliton 
solutions we constructed in the last subsection. In fact, the soliton 
solution~(\ref{eq:OFf}) correctly asymptotes to the closed string vacuum 
$\phi=0$ at infinity in the $(x^1,x^2)$-plane. In the large noncommutativity 
limit~(\ref{eq:OM}), the derivatives in noncommutative directions can be 
dropped for the same reason as in~(\ref{eq:NVa}). If we consider soliton 
solutions which do not depend on the commutative coordinates, the kinetic 
term in~(\ref{eq:OO}) gives no contribution so that the total action is 
written as 
\begin{equation}
S(\widehat{\phi})=-\tau_{25}\sqrt{\Bigg|\frac{\det(g+2\pi\ap B)}{\det G}
\Bigg|}\int d^{24}\!x\sqrt{|\det G|}2\pi\theta\mathrm{Tr}_{\cH}
V(\widehat{\phi}(x^1,x^2)), \label{eq:OP}
\end{equation}
where we used (\ref{eq:OK}), (\ref{eq:OG}). Since extrema of the potential 
are at $\Lambda=\{0,\phi_*\}$, we can take the simplest soliton solution 
to be 
\begin{equation}
\widehat{\phi}=\phi_*|0\rangle\langle 0|\ \sim\ 2\phi_*e^{-r^2/\theta}, 
\label{eq:OQ}
\end{equation}
where $\theta=\theta^{12}=-1/B\ (>0)$ in the limit~(\ref{eq:OM}). 
Substituting it into the action~(\ref{eq:OP}), we find 
\begin{equation}
S(\phi_*|0\rangle\langle 0|)=-\tau_{25}2\pi\theta V(\phi_*)\int d^{24}\!x 
\sqrt{|\det g_{24}|}\sqrt{g^2+(2\pi\ap B)^2}, \label{eq:OR}
\end{equation}
where we divided 26-dimensional matrix $g+2\pi\ap B$ into 
\[ g+2\pi\ap B=\left(
	\begin{array}{c|cc}
	g_{24} & 0 & 0 \\ \hline 0 & g_{11}=g & 2\pi\ap B \\ 0 & -2\pi\ap B & 
	g_{22}=g
	\end{array}
\right) \]
($g_{24}$ is the metric in the 24-dimensional commutative subspace). Since 
$g^2$ is negligible as compared to $(2\pi\ap B)^2$ in the 
limit~(\ref{eq:OM}), the last factor in~(\ref{eq:OR}) gives $2\pi\ap |B| =
2\pi\ap/\theta$. Using the fact that $V(\phi_*)=1$, we finally obtain 
\begin{equation}
S(\phi_*|0\rangle\langle 0|)=-(2\pi\sqrt{\ap})^2\tau_{25}\int d^{24}\!x
\sqrt{|\det g_{24}|}. \label{eq:OS}
\end{equation}
Therefore, the tension of the soliton solution~(\ref{eq:OQ}) precisely 
agrees with the D23-brane tension. To get more evidence that the codimension 
2 soliton~(\ref{eq:OQ}) can be identified with a D23-brane, let us see the 
fluctuation of the tachyon field around the soliton background~(\ref{eq:OQ}). 
We begin with noting that the $U(\infty)$ symmetry~(\ref{eq:OHm}) is 
spontaneously broken to $U(1)\times U(\infty -1)$ by the soliton. This is 
because the 00-component of the `matrix' $\widehat{\phi}=\sum\phi_{mn}|m
\rangle\langle n|$ has a nonvanishing expectation value $\phi_{00}=\phi_*$. 
Though the general fluctuation can be written as 
\begin{equation}
\widehat{\phi}=\overline{\phi}+\widehat{\delta\phi}\equiv\phi_*|0\rangle
\langle 0|+\sum_{m,n=0}^{\infty}\delta\phi_{mn}(x^a)|m\rangle\langle n|,
\label{eq:OT}
\end{equation}
where, and from now on, the indices $a,b$ on $x$ denote the commutative 
directions ($a=0,3,4,\cdots,25$ in this case), the $U(\infty -1)$ 
symmetry transformation can be used to set 
\[ \delta\phi_{0m}=\delta\phi_{m0}=0 \quad \mathrm{for}\ m\ge 1. \]
In order for $\widehat{\phi}$ to be real, the matrix $(\delta\phi)_{mn}$ must 
be Hermitian. Expanding the potential $V(\widehat{\phi})$ to second order in 
$\delta\phi$ and using the fact that $V(\phi)$ is stationary at 
$\phi=\phi_*$, we have\footnote{To be precise, we should take into account 
the ordering of the operators. But it is not important for us.}
\begin{eqnarray*}
V(\widehat{\phi})&=&V(\overline{\phi})+\frac{1}{2}V^{\prime\prime}
(\overline{\phi})\left(\widehat{\delta\phi}\right)^2 \\ &=& V(\phi_*)|0
\rangle\langle 0|+\frac{1}{2}\Bigl(V^{\prime\prime}(\phi_*)|0\rangle\langle 
0|+V^{\prime\prime}(0)(1-|0\rangle\langle 0|)\Bigr) \\ & &\times \left(
\delta\phi_{00}^2|0\rangle\langle 0|+\sum_{m,n=1}^{\infty}(\delta\phi^2)_{mn}
|m\rangle\langle n|\right) \\ &=&\left(V(\phi_*)+\frac{1}{2}V^{\prime\prime}
(\phi_*)\delta\phi_{00}^2\right)|0\rangle\langle 0|+\frac{1}{2}
V^{\prime\prime}(0)\sum_{n,m=1}^{\infty}(\delta\phi^2)_{mn}|m\rangle\langle
n|,
\end{eqnarray*}
where $V^{\prime\prime}(0)$ term appears because the constant term in 
$V^{\prime\prime}(\overline{\phi})$ is not proportional to $|0\rangle\langle
0|$, and we used $\langle m|n\rangle =\delta_{mn}$. Putting these into the 
D25-brane action~(\ref{eq:OO}), the quadratic action for the fluctuation 
becomes
\begin{eqnarray*}
S_{\mathrm{quad}}&=&(2\pi\sqrt{\ap})^2\tau_{25}\int d^{24}\!x\sqrt{|\det 
g_{24}|}\mathrm{Tr}_{\cH}\left(\frac{1}{2}f(\widehat{\phi})
g_{24}^{ab}\partial_a
\widehat{\phi}\partial_b\widehat{\phi}-V(\widehat{\phi})\right) \\
&=&\tau_{23}\int d^{24}\!x\Biggl[-\frac{1}{2}f(\phi_*)(\partial_a\delta
\phi_{00})^2-V(\phi_*)-\frac{1}{2}V^{\prime\prime}(\phi_*)(\delta\phi_{00})^2
\\ & &{}-\frac{1}{2}f(0)\sum_{m,n=1}^{\infty}\partial^a\delta\phi_{mn}
\partial_a\delta\phi_{nm}-\frac{1}{2}V^{\prime\prime}(0)\sum_{m,n=1}^{\infty}
\delta\phi_{mn}\delta\phi_{nm}\Biggr], 
\end{eqnarray*}
where we set $(g_{24})_{ab}=\eta_{ab}=\mathrm{diag}(-1,+1,\cdots,+1)$. As 
$f(0)=0$ from~(\ref{eq:ONa}), the kinetic terms for $\delta\phi_{nm}$ with 
nonzero $m,n$ vanish. It means that these modes are not physical, so we drop 
them. Substituting $f(\phi_*)=V(\phi_*)=1$ and writing $\delta\phi_{00}=
\varphi$, we obtain
\begin{equation}
S_{\mathrm{quad}}=-\tau_{23}V_{24}+\tau_{23}\int d^{24}\!x \left(-\frac{1}{2}
\partial^a\varphi\partial_a\varphi-\frac{1}{2}V^{\prime\prime}(\phi_*)
\varphi^2\right). \label{eq:OU}
\end{equation}
Aside from the constant D23-brane mass term, the above action correctly 
describes the tachyon on a D23-brane with the same mass $m^2=V^{\prime\prime}
(\phi_*)=-1/\ap$ as that of the original tachyon on the D25-brane. The 
tachyonic instability arises on the soliton (D23-brane) world-volume because 
$\phi_*\in\Lambda$ corresponds to the \textit{maximum} of the potential 
$V(\phi)$ rather than a minimum. Furthermore, the fluctuation spectrum of the 
noncommutative $U(1)$\footnote{This $U(1)$ gauge symmetry is in fact the 
gauged $U(\infty)$ symmetry.} gauge field on the D25-brane around the soliton 
solution~(\ref{eq:OQ}) was studied in~\cite{HKLM}. It was shown there that 
the coupled gauge-tachyon field fluctuations on the D25-brane provide 
for the soliton world-volume the correct action which describes the tachyon, 
gauge field and transverse scalars representing the translation modes of 
the soliton, though we do not show it here. However, there also remain 
infinitely many massive gauge fields associated with the generators of the 
gauge symmetry spontaneously broken by the soliton background, namely 
$0m$- and $m0$-components. They acquired mass via the usual Higgs mechanism. 
It was conjectured that such unwanted massive fields were removed from the 
physical spectrum by the mechanism explained in subsection~\ref{sub:fluc}. 
Anyway, it seems very plausible that the codimension 2 noncommutative soliton 
solution on a D25-brane is identified with a D23-brane. 
\medskip

It is clear from the above arguments that the soliton tension and its 
fluctuation spectrum are completely unchanged even if the soliton $\phi_*
|0\rangle\langle 0|$~(\ref{eq:OQ}) is replaced by $\phi_*|n\rangle\langle n|$ 
for any $n$, as long as we are taking the limit $\theta/\ap\to\infty$. 
So each $\phi_*|n\rangle
\langle n|$ equivalently represents a D23-brane, though the 
profile~(\ref{eq:OD}) is different. Well, let us consider the superposition 
of $k$ such solitons, 
\begin{equation}
\widehat{\phi}_k=\phi_*\sum_{n=0}^{k-1}|n\rangle\langle n|, \label{eq:OV}
\end{equation}
which is termed a level $k$ solution in~\cite{GMS}. Mutual orthogonality of 
the projection operators guarantees that the tension of the level $k$ soliton 
is given by $k\tau_{23}$, which suggests that the level $k$ soliton 
corresponds to the system of $k$ coincident D23-branes. In fact, the study of 
the gauge fluctuation shows that the gauge symmetry is properly enhanced to 
$U(k)$ on the soliton world-volume~\cite{HKLM}: the soliton 
background~(\ref{eq:OV}) breaks the $U(\infty)$ symmetry to $U(k)\times U(
\infty -k)$. Note that all these conclusions are true only in the limit 
$\theta/\ap\to\infty$, because $\phi_*|n\rangle\langle n| \ (n>0)$ has 
higher energy than $\phi_*|0\rangle\langle 0|$ in the case of finite 
noncommutativity.

Here we consider what happens to the soliton solution~(\ref{eq:OV}) if we 
take the limit $k\to\infty$. Since the set 
$\{|n\rangle\langle n|\}_{n=0}^{\infty}$ of the projection operators is 
complete in the sense that $\sum_{n=0}^{\infty}|n\rangle\langle n|=I$ (unit 
operator), the level $\infty$ solution becomes $\widehat{\phi}_{\infty}=
\phi_*I=$constant. It represents nothing but the original D25-brane. From 
this fact, it may be that a bosonic D-brane is constructed out of infinitely 
many lower dimensional D-branes.
\bigskip

\begin{figure}[htbp]
	\begin{center}
	\includegraphics{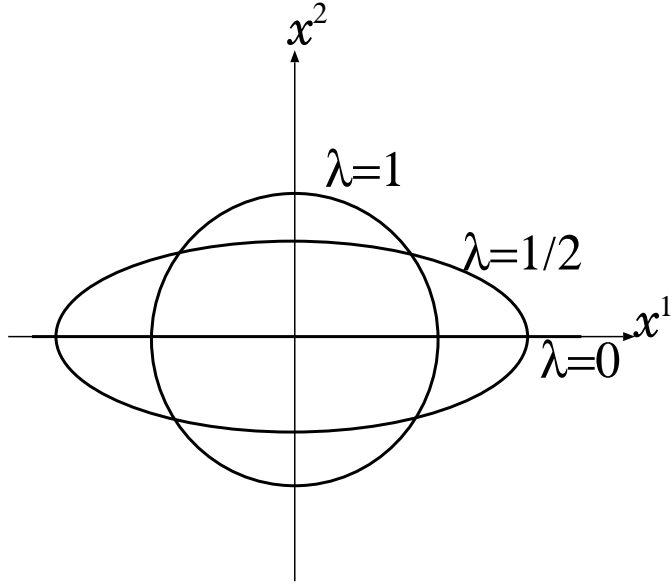}
	\end{center}
	\caption{Deformed soliton.}
	\label{fig:BG}
\end{figure}

Thus far, we have considered only codimension 2 solitons on a D25-brane. One 
can easily generalize them to solitons of arbitrary even codimension, 
say $2q$, by turning on the following $B$-field
\begin{equation}
B=\left. \left(
	\begin{array}{ccccccc}
	0 & B & & & & & \\ -B & 0 & & & & & \\ & & 0 & B & & & \\ & & -B & 0 & & 
	& \\ & & & & \ddots & & \\& & & & & 0 & B \\ & & & & & -B & 0
	\end{array}
\right)\quad\right\}\scriptstyle{2q}. \label{eq:OVa}
\end{equation}
The resulting soliton should be identified with a D($25-2q$)-brane. However, 
as is clear from construction, odd codimension solitons cannot be built 
straightforwardly. It was proposed in~\cite{Rey} that odd codimension 
solitons are constructed by employing the set $\{e^{ipx}\}$ of plain wave 
states as a basis of the Hilbert space $\cH$, instead of the harmonic 
oscillator basis $\{|n\rangle\}$. Equivalently, they are obtained by 
deforming the solitons that have already been constructed before. To see 
this, note that the $U(\infty)$ symmetry group includes an $SL(2,\aaru)$ 
subgroup whose element acts on $(x^1,x^2)$ as
\begin{equation}
\left(
	\begin{array}{c}
	x^1 \\ x^2
	\end{array}
\right)\longrightarrow \left(
	\begin{array}{cc}
	a & b \\ c & d
	\end{array}
\right)\left(
	\begin{array}{c}
	x^1 \\ x^2
	\end{array}
\right) \quad \mathrm{with} \quad ad-bc=1. \label{eq.OW}
\end{equation}
Let us consider the action
\begin{equation}
S=-\frac{1}{g^2}\int dt\ d^2x\ V(\phi(x)) \label{eq:OX}
\end{equation}
which is obtained after neglecting the kinetic term. Under an $SL(2,\aaru)$ 
transformation, the integration measure is invariant:
\[ dx^1\wedge dx^2\longrightarrow (a\ dx^1+b\ dx^2)\wedge(c\ dx^1+d\ dx^2)=
(ad-bc)dx^1\wedge dx^2=dx^1\wedge dx^2. \]
In addition, the $*$-product entering implicitly in~(\ref{eq:OX}) is also 
invariant, which can be verified similarly. Hence, the action~(\ref{eq:OX}) 
is invariant under an arbitrary $SL(2,\aaru)$ transformation. As a special 
case, we take 
\begin{equation}
L=\left(
	\begin{array}{cc}
	\lambda & 0 \\ 0 & 1/\lambda 
	\end{array}
\right)\in SL(2,\aaru). \label{eq:OY}
\end{equation}
Under this transformation, the $n$-th soliton solution~(\ref{eq:OD}) is 
deformed into 
\begin{equation}
\phi_n(\tilde{r}^2)=2(-1)^ne^{-\frac{\tilde{r}^2}{\theta}}L_n\left(
\frac{2\tilde{r}^2}{\theta}\right) \quad \mathrm{where} \quad \tilde{r}^2
=\lambda^2(x^1)^2+\frac{1}{\lambda^2}(x^2)^2. \label{eq:OZ}
\end{equation}
Though the solution is radially symmetric when $\lambda =1$, it gets squeezed 
as in Figure~\ref{fig:BG} for $\lambda\neq 1$. 
In the limit $\lambda\to 0$, the exponential factor becomes
\begin{equation}
\exp\left[-\lambda^2\frac{(x^1)^2}{\theta}-\frac{1}{\lambda^2}\frac{(x^2)^2}
{\theta}\right]\sim\frac{\pi\theta}{2\pi R_1}\delta(x^2), \label{eq:PA}
\end{equation}
where we used 
\[\lim_{a\to 0}\frac{1}{\sqrt{\pi a}}e^{-x^2/a}=\delta(x) \]
and regularized $\int_{-\infty}^{\infty}dx^1$ by compactifying $x^1$ on a 
circle of radius $R_1$ as 
\[ \lim_{\lambda\to 0}\int e^{-(x^1)^2\lambda^2/\theta}dx^1=\lim_{\lambda\to 
0}\frac{\sqrt{\pi\theta}}{\lambda}\equiv\int_0^{2\pi R_1}dx^1=2\pi R_1. \]
Then the soliton profile (\ref{eq:PA}) looks sharply localized in 
$x^2$-direction while it is homogeneously extended in the (compactified) 
$x^1$-direction. If we take the remaining 24 directions to be commutative, 
we have got a 24-brane geometry. The value of the action calculated on the 
24-brane is 
\begin{eqnarray}
S(\phi_*\phi_0(\tilde{r}^2))&=&-\tau_{25}2\pi\ap |B|\int d^{24}\!x\ dx^1\ dx^2
\sqrt{|\det g_{24}|}V(\phi_*)2\frac{\pi\theta}{2\pi R_1}\delta(x^2) \nonumber 
\\ &=& -(2\pi\sqrt{\ap}\tau_{25})\int d^{24}\!x\ dx^1\sqrt{|\det g_{24}|}
\frac{\sqrt{\ap}}{R_1}. \label{eq:PB}
\end{eqnarray}
The prefactor of (\ref{eq:PB}) precisely agrees with the D24-brane tension. 
On the other hand, the last factor $\sqrt{\ap}/R_1$ represents the 
\textit{T-dualized} dimensionless radius of $x^1$-direction, which 
is equal to $\sqrt{g_{11}^{\prime}}$ ($g_{11}^{\prime}$ is the 
(1,1)-component of the metric after $T$-dualizing). After all, we have 
\[ S=-\tau_{24}\int d^{25}\!x \sqrt{|\det g_{25}^{\prime}|}, \]
which is the desired expression for a D24-brane, though the meaning of 
$T$-duality is not so clear. Now that we have obtained a codimension 1 
soliton (D24-brane), we can build a soliton of any odd codimension by 
turning on a $B$-field~(\ref{eq:OVa}) for the D24-brane. Finally, we would 
like to warn that the $SL(2,\aaru)$ subgroup, hence the $U(\infty)$ group, 
is \textit{not} a symmetry of the kinetic term in the action~(\ref{eq:NU}). 
Thus the $U(\infty)$ is only an approximate global symmetry at finite 
noncommutativity, though it becomes exact in the limit $\theta/\ap\to\infty$. 

\subsubsection*{\underline{D-branes in Type II superstring theory}}
\begin{figure}[tbhp]
	\begin{center}
	\includegraphics{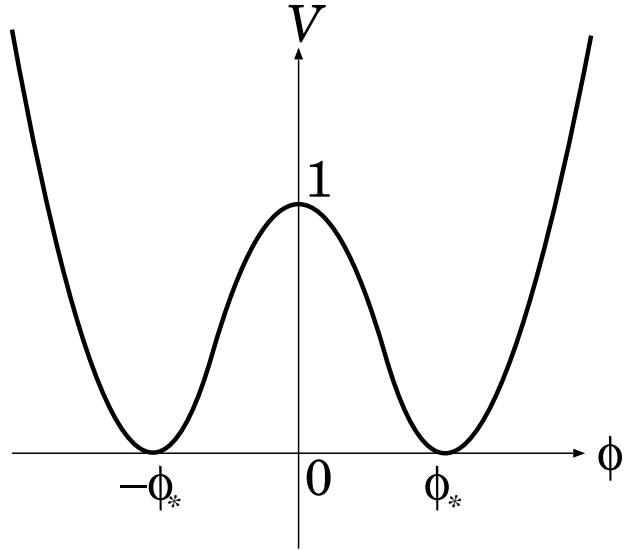}
	\end{center}
	\caption{The effective tachyon potential on a non-BPS D-brane.}
	\label{fig:BH}
\end{figure}

The description of D-branes as noncommutative solitons can be extended to the 
superstring case. We will consider a single non-BPS D9-brane where the 
tachyon on the brane world-volume is described by a real field $\phi$. The 
effective field theory action for the tachyon is given by 
\begin{eqnarray}
S&=&\tilde{\tau}_9\frac{g_s}{G_s}\int d^{10}\!x\sqrt{|\det G|}*\left(-
\frac{1}{2}f(\phi)*G^{\mu\nu}*\partial_{\mu}\phi*\partial_{\nu}\phi+\ldots 
-V(\phi)\right) \label{eq:PC} \\
&\mathop{\longrightarrow}\limits^{\theta/\ap\to\infty}&-\tilde{\tau}_9
(2\pi)^2\ap\int d^8x\sqrt{|\det g_8|}\mathrm{Tr}_{\cH}\left(\frac{1}{2}
f(\widehat{\phi})\partial^a\widehat{\phi}\partial_a\widehat{\phi}+
V\left(\widehat{\phi}(x^1,x^2)\right)\right), \label{eq:PD}
\end{eqnarray}
where we have already turned on a constant $B$-field in the 
$(x^1,x^2)$-directions just as in~(\ref{eq:OOa}). 
We basically follow the notation defined in the 
bosonic case. The potential $V(\phi)$ now has the double-well form as 
illustrated in Figure~\ref{fig:BH}.
That is to say, the potential has two degenerate minima (closed string 
vacuum) at $\phi=\pm\phi_*$ and the value there is $V(\pm\phi_*)=0$, while 
the maximum is at $\phi=0$ and $V(0)=1$, which represents a non-BPS D9-brane 
with the correct tension. The set of extrema is given by $\Lambda=\{0,\pm
\phi_*\}$. For the soliton solution to have the correct asymptotic behavior 
(finite energy density), we choose it to be 
\begin{equation}
\widehat{\phi}(x^1,x^2)=\overline{\phi}\equiv \phi_*(1-|0\rangle\langle 0|)
\sim \phi_*(1-2e^{-r^2/\theta}). \label{eq:PE}
\end{equation}
In fact, $\displaystyle \lim_{r\to\infty}V(\overline{\phi})=V(\phi_*)=0$. 
$\overline{\phi}$ can be rewritten as 
\begin{equation}
\overline{\phi}=0|0\rangle\langle 0|+\sum_{n=1}^{\infty}\phi_*|n\rangle
\langle n|\ \quad \longleftarrow \sum_{n=0}^{\infty}\lambda_n|n\rangle\langle 
n|\mbox{\ \  in (\ref{eq:OF})} \label{eq:PF}
\end{equation}
because of the completeness of $\{|n\rangle\langle n|\}_{n=0}^{\infty}$. 
Since $\lambda_0$ corresponds to the maximum $\phi=0$, the soliton 
solution~(\ref{eq:PE}) is unstable. Together with the fact that the soliton 
is localized in the 2-dimensional space $(x^1,x^2)$, one may guess that the 
soliton represents a non-BPS D7-brane in Type IIA. To verify it, let us 
compute the tension of the soliton. As the potential necessarily contains 
a constant term $V(0)=1$, we proceed as 
\begin{eqnarray*}
V(\overline{\phi})&=&V(0)+\Bigl\{V(\phi_*(1-|0\rangle\langle 0|))-V(0)\Bigr\}
=1+(V(\phi_*)-1)(1-|0\rangle\langle 0|) \\ &=&|0\rangle\langle 0|,
\end{eqnarray*}
where we used $V(\phi_*)=0$ and that $1-|0\rangle\langle 0|$ is also 
idempotent. Dropping the $\partial_a$ term, the action~(\ref{eq:PD}) becomes 
\[ S=-(2\pi\sqrt{\ap})^2\tilde{\tau}_9\int d^8x\sqrt{|\det g_8|}, \]
so the tension of the soliton 
precisely agrees with that of a non-BPS D7-brane. 
As in the bosonic case, other soliton solutions $\overline{\phi}=\phi_*(1-
|n\rangle\langle n|)$ represent the same object in the limit $\theta/\ap\to
\infty$. But a strangely looking phenomenon happens when we calculate the 
Ramond-Ramond charge density of the soliton. A non-BPS D$p$-brane has the 
Chern-Simons coupling 
\begin{equation}
S_{CS}=\frac{1}{2\phi_*}\int_{\widetilde{\mathrm{D}p}}d\phi\wedge C_p, 
\label{eq:PG}
\end{equation}
where $C_p$ is the Ramond-Ramond $p$-form. This coupling is justified in 
the following way: If the tachyon $\phi$ develops a kink profile which 
interpolates between two stable vacua $\pm\phi_*$ along the $x^p$-direction, 
\[ \phi(x^p)=\left\{
	\begin{array}{ccc}
	-\phi_* & \mathrm{for} & x^p\to -\infty \\
	\phi_*  & \mathrm{for} & x^p\to +\infty
	\end{array}
\right. , \]
then (\ref{eq:PG}) becomes 
\[ S_{CS}=\frac{1}{2\phi_*}\int_{\widetilde{\mathrm{D}p}}dx^p\frac{d\phi}
{dx^p}\wedge C_p=\int_{\mathrm{D}(p-1)}C_p, \]
that is to say, the coupling (\ref{eq:PG}) brings about one unit of 
D($p-1$)-brane charge on the tachyonic kink (to be identified with a 
D($p-1$)-brane), as desired. Now we consider the above soliton 
solution~(\ref{eq:PE}) 
\[ \overline{\phi}(x^1,x^2)=\phi_*(1-2e^{-r^2/\theta}). \]
Substituting it into the Chern-Simons coupling~(\ref{eq:PG}) for a non-BPS 
D9-brane, we have 
\begin{eqnarray}
S_{CS}&=&\frac{1}{2\phi_*}\int_{\widetilde{\mathrm{D}9}}dr\frac{d}{dr}(-2
\phi_*e^{-r^2/\theta})\wedge C_9=-\Bigl[e^{-r^2/\theta}\Bigr]_{r=0}^{\infty}
\int_{\mathrm{D}8}C_9 \nonumber \\
&=&\int_{\mathrm{D}8}C_9, \label{eq:PH}
\end{eqnarray}
which shows that the soliton solution \textit{locally} carries one unit of 
D8-brane charge. It must be local because the solitonic configuration is 
not topological so that it does not have any \textit{global} Ramond-Ramond 
charge. From these results, the noncommutative soliton solution~(\ref{eq:PE}) 
on a non-BPS D9-brane is interpreted as a D8-brane which winds along the 
angular coordinate $\vartheta$ in the $(x^1,x^2)$-plane and is 
\textit{smeared} over the radial coordinate $r$, whose distribution function 
in the radial direction is given by $2e^{-r^2/\theta}$~\cite{DMR}. Since we 
found that the D-brane charge~(\ref{eq:PH}) depends on the value of $2e^{-r^2
/\theta}$ only at $r=0$, the $n$-th soliton solution 
\[ \phi_n(r^2)=2(-1)^ne^{-r^2/\theta}L_n\left(\frac{2r^2}{\theta}\right) \quad
\longrightarrow \quad \phi_n(0)=2(-1)^n \]
locally carries D8-brane charge $(-1)^n$. Nevertheless, by carrying out the 
integral (trace) over the noncommutative directions in~(\ref{eq:PD}), we 
obtain the 7-brane action. In this 7-brane picture, 
the issues such as tachyon-gauge 
fluctuations and multiple D-branes with nonabelian gauge symmetry can be 
analyzed in the same way as in the bosonic case, and one finds 
similar conclusions which are suitable for non-BPS D7-branes. 
\medskip

In addition to the unstable soliton solutions $\overline{\phi}=\phi_*(1-|n
\rangle\langle n|)$, we find the configurations 
\begin{equation}
\widehat{\phi}=\underline{\phi}_n=\phi_*(1-2|n\rangle\langle n|) \label{eq:PI}
\end{equation}
to be \textit{solutions} to the equation of motion. Focusing on $n=0$, it can 
be written as
\begin{equation}
\underline{\phi}_0=-\phi_*|0\rangle\langle 0|+\sum_{m=1}^{\infty}\phi_*|m 
\rangle\langle m|. \label{eq:PJ}
\end{equation}
Since all of the coefficients correspond to the minima of the potential, 
namely either $\phi_*$ or $-\phi_*$, this solution is stable. To see it 
more closely, let us calculate the tension of the solution~(\ref{eq:PI}). 
The following property can easily be verified:
\begin{equation}
(1-2|n\rangle\langle n|)^k=\left\{
	\begin{array}{lccc}
	I & \mathrm{for} & k & \mathrm{even} \\
	1-2|n\rangle\langle n| & \mathrm{for} & k & \mathrm{odd}
	\end{array}
\right. , \label{eq:PK}
\end{equation}
where $I$ denotes the unit operator. Considering that the potential $V(\phi)$ 
is invariant under the reflection $\phi\to -\phi$ (\textit{i.e.} $V$ is an 
even function of $\phi$), the potential for $\phi=\underline{\phi}_n$ becomes
\[ V(\underline{\phi}_n)=1+\frac{V^{\prime\prime}(0)}{2}\phi_*^2(1-2|n\rangle
\langle n|)^2+\ldots =V(\phi_*)=0. \]
It is surprising that the seemingly non-trivial configurations 
$\underline{\phi}_n$ are degenerate with the closed string vacuum. However, 
it was pointed out in~\cite{8064} that these `tensionless branes' are in fact 
gauge equivalent to the closed string vacuum under a discrete gauge symmetry. 
So we should not consider the solutions~(\ref{eq:PI}) as representing 
D-branes of a new kind.
\medskip

As opposed to the bosonic case, odd codimension branes are more naturally 
constructed by combining the noncommutative solitons of even codimension 
with the kink configuration along one of the commutative directions. We 
consider the following soliton solution~\cite{8064}
\begin{equation}
\widehat{\phi}(x^1,x^2;x^3)=\phi_*\biggl\{1-\Bigl(1-\cK(x^3)\Bigr)\cP-
\Bigl(1-\overline{\cK}(x^3)\Bigr)\overline{\cP}\biggr\}, \label{eq:PL}
\end{equation}
where $\cK,\overline{\cK}$ are a kink and an anti-kink configuration, 
respectively, along the commuting direction $x^3$, more precisely 
\begin{equation}
\cK(x^3)=\left\{
	\begin{array}{cc}
	+1 & x^3\to +\infty \\ -1 & x^3\to -\infty
	\end{array}
\right. , \quad \overline{\cK}(x^3)=\left\{
	\begin{array}{cc}
	-1 & x^3\to +\infty \\ +1 & x^3\to -\infty
	\end{array}
\right. . \label{eq:PLa}
\end{equation}
$\cP$ is a projection operator onto an $n$-dimensional subspace of the full 
Hilbert space, which corresponds to a solitonic configuration in the 
noncommutative $(x^1,x^2)$-space. And $\overline{\cP}$ is another projection 
operator onto an $\overline{n}$-dimensional subspace. We require $\cP,
\overline{\cP}$ to be mutually orthogonal. For example, we can take them to be
\begin{equation}
\cP=\sum_{m=0}^{n-1}|m\rangle\langle m|\sim\sum_{m=0}^{n-1}\phi_m(r^2), \quad 
\overline{\cP}=\sum_{m=n}^{n+\overline{n}-1}|m\rangle\langle m|. \label{eq:PM}
\end{equation}
Writing $r^2=(x^1)^2+(x^2)^2$ as usual, we see from (\ref{eq:PL}) that 
\begin{equation}
\lim_{r\to\infty}\widehat{\phi}(r;x^3)=\phi_* \label{eq:PN}
\end{equation}
because the soliton profiles corresponding to the projection 
operators~(\ref{eq:PM}) vanish as $r\to\infty$. Moreover, 
\begin{equation}
\widehat{\phi}(r;x^3)=\left\{
	\begin{array}{cc}
	\phi_*(1-2\overline{\cP}) & x^3 \to +\infty \\
	\phi_*(1-2\cP) & x^3\to -\infty
	\end{array}
\right. \label{eq:PO}
\end{equation}
holds due to the properties (\ref{eq:PLa}) of $\cK,\overline{\cK}$. Since the 
asymptotic configurations~(\ref{eq:PO}) are actually gauge equivalent to the 
closed string vacuum as we saw earlier, (\ref{eq:PN}) and (\ref{eq:PO}) show 
that the soliton~(\ref{eq:PL}) is localized near the origin in the 
$(x^1,x^2,x^3)$-space. If we take the original system to be a non-BPS 
D9-brane of Type IIA theory, the soliton solution~(\ref{eq:PL}) should 
represent some 6-brane. Indeed, this solution corresponds to the system of 
coincident $n$ D6-branes and $\overline{n}$ anti-D6-branes. It can be seen 
from the consideration that in the solution~(\ref{eq:PL}) $n$ kinks and 
$\overline{n}$ anti-kinks are put on the $(n+\overline{n})$ non-BPS D7-branes 
represented by $\widehat{\phi}=\phi_*(1-\cP-\overline{\cP})$, though we 
cannot compute the tension of the soliton without knowing the precise 
forms of the kink profiles $\cK,\overline{\cK}$ and the potential $V(\phi)$.
\smallskip

On the other hand, we can also construct odd codimension solitons as the 
singular limit of the squeezing deformation. The resulting 8-brane on a 
non-BPS D9-brane in Type IIA theory was interpreted as a non-BPS D8-brane 
in the \textit{T-dualized} (\textit{i.e.} Type IIB) theory~\cite{Rey}, 
because
\begin{itemize}
\item the codimension 1 soliton has no global Ramond-Ramond charge, 
\item the tension of the soliton is given by $2\pi\sqrt{\ap}\tilde{\tau}_9$ 
in the $T$-dualized geometry as in the bosonic case, 
\item there remains a tachyon on the world-volume of the deformed soliton 
just like the original 
codimension 2 soliton (the former is obtained simply by deforming the latter),
\item the fluctuation of the above tachyon leads to the Chern-Simons coupling 
for a non-BPS D8-brane.
\end{itemize}
Thus, it seems difficult to obtain BPS D-branes in this method. 

\subsubsection*{\underline{gauge field configurations}}
So far, we have assumed that only the tachyon field $\phi$ develops a 
non-trivial solitonic configuration in the infinite noncommutativity. 
However, it was proposed in~\cite{7217,7226} that the noncommutative gauge 
field takes some non-trivial configurations as well. Here we will refer to 
them. 
\medskip

In order to neglect the kinetic term in the action completely, we had to work 
in the limit of infinite noncommutativity, $\theta/\ap\to\infty$. In such a 
case, the equation of motion was simplified to $\partial V/\partial\phi =0$, 
and by finding a function $\overline{\phi}$ which satisfies 
\[ \overline{\phi}*\overline{\phi}(x)=\overline{\phi}(x) \]
we obtained a solitonic solution $\phi_*\overline{\phi}(x)$ which depended on 
noncommutative coordinates. But once we tried to keep the noncommutativity 
parameter $\theta/\ap$ finite, we were not able to acquire such great 
simplifications. The idea of~\cite{7217} to overcome this problem is to 
consider the gauge field background as well as the tachyonic soliton 
background in such a way that the contribution from the kinetic term for the 
tachyon should precisely be canceled by the gauge field, instead of ignoring 
it by taking the limit $\theta/\ap\to\infty$. The key observation is the 
following one: After introducing the noncommutative $U(1)$ gauge field 
$A_{\mu}$, the ordinary derivative $\partial_{\mu}\phi$ of the tachyon field 
in the adjoint representation is replaced by the covariant derivative 
\[ \nabla_{\mu}\phi=\partial_{\mu}\phi-i[A_{\mu},\phi]. \]
In the noncommutative theory, the ordinary derivative is also rewritten as a 
commutator by utilizing the $*$-product as 
\[ \partial_{\mu}\phi=-i[\theta^{-1}_{\mu\nu}x^{\nu},\phi]. \]
Thus we have
\begin{equation}
\nabla_{\mu}\phi=-i[A_{\mu}+\theta^{-1}_{\mu\nu}x^{\nu},\phi]\equiv -i[
C_{\mu},\phi]. \label{eq:PP}
\end{equation}
Similarly, the field strength for $A_{\mu}$ can be rewritten in terms of 
$C_{\mu}$. To satisfy the equation of motion for $\phi$, it is sufficient 
(in fact, too strong) that both 
\begin{equation}
\frac{\partial V}{\partial\phi}=0 \quad \mathrm{and} \quad [C_{\mu},\phi]=
i\partial_{\mu}\phi+[A_{\mu},\phi]=0 \label{eq:PQ}
\end{equation}
hold. It is important that it suffices to consider the equation $\partial V/
\partial\phi =0$ so that we can avoid explicitly solving the differential 
equation $\partial^2 \phi=\partial
V/\partial\phi$ even for \textit{finite} value of $\theta$. The second 
equation in~(\ref{eq:PQ}) requires that the non-trivial gauge field 
configuration should cancel the contribution from the derivative term 
$\partial_{\mu}\phi$ which would be absent in the case of large 
noncommutativity limit. 

Similarly, it was proposed~\cite{7226} that in the true closed string 
vacuum (called `nothing' state there) the classical configuration of tachyon 
$\phi$ and gauge field $A_{\mu}$ is not $\phi=\phi_*,\ A_{\mu}=0$ but
\begin{equation}
\phi=\phi_* \quad \mathrm{and} \quad C_{\mu}=A_{\mu}+\theta^{-1}_{\mu\nu}
x^{\nu}=0, \label{eq:PR}
\end{equation}
where $\phi_*$ represents the minimum of the potential. Since $C_{\mu}$ 
transforms homogeneously under the noncommutative $U(1)$ gauge transformation 
as $\delta C_{\mu}=i[\lambda,C_{\mu}]$, $C_{\mu}=0$ is completely gauge 
invariant. In contrast, the configuration $A_{\mu}=0$ (which is true of a 
single D25-brane) is not gauge invariant as $\delta A_{\mu}=\partial_{\mu}
\lambda$. In this sense, the $U(\infty)$ symmetry we saw in~(\ref{eq:OHm}) 
is fully restored for the `maximally symmetric' configuration~(\ref{eq:PR}). 
As a result, we can see there are no propagating open string degrees of 
freedom around the nothing state in the following way. The gauge invariance 
guarantees that the explicit derivatives always enter the action in the form 
of covariant derivatives $\nabla_{\mu}\varphi=-i[C_{\mu},\varphi]$. Even 
though we consider the fluctuation $\delta\varphi$ of the field $\varphi$, no 
derivative terms of $\delta\varphi$ will remain around the background 
$C_{\mu}=0$. In a similar way, the derivatives do not accompany the gauge 
fluctuation $\delta A_{\mu}$. Although the derivatives are also provided by 
the $*$-product, it does not give the standard kinetic term (quadratic form) 
because $\int dx\ A*B$ reduces to $\int dx\ AB$ as in~(\ref{eq:NVb}). The 
absence of the standard kinetic terms for all fields suggests that open 
string modes cannot propagate in the nothing state~(\ref{eq:PR}) at all. 
We now consider the solitonic field configuration 
\begin{eqnarray}
\phi&=&\phi_{\mathrm{max}}\sum_{n=0}^{N_1-1}|n\rangle\langle n|+\phi_*
\sum_{n=N_1}^{\infty}|n\rangle\langle n|=\phi_*+(\phi_{\mathrm{max}}
-\phi_*)\sum_{n=0}^{N_1-1}|n\rangle\langle n|, \nonumber \\
C_i&=&\cP_{N_2}a_i^{\dagger} \quad \mathrm{for}\ \ i=2,\cdots,13,
\label{eq:PS} \\
C_1&=&C_2=0, \nonumber 
\end{eqnarray}
where $\phi_{\mathrm{max}}$ corresponds to the maximum of the potential, 
$\cP_{N_2}$ is some projection operator onto an $N_2$-dimensional subspace 
in the Hilbert space on which $\displaystyle a_1=\frac{x^1+ix^2}{\sqrt{2
\theta^{12}}},\ a_1^{\dagger}=\frac{x^1-ix^2}{\sqrt{2\theta^{12}}}$ act. 
And we defined $\displaystyle a_i^{\dagger}=\frac{x^{2i-1}-ix^{2i}}{
\sqrt{2\theta^{2i-1,2i}}}$, assuming that the $B$-field (and hence $\theta$) 
here is of maximal rank. If we take 
\begin{equation}
\cP_{N_2}=\sum_{n=0}^{N_1-1}|n\rangle\langle n|, \label{eq:PT}
\end{equation}
then one can verify that the solitonic configuration (\ref{eq:PS}) localized 
in the ($x^1,x^2$)-directions represents $N_1$ coincident D23-branes with
correct tension and fluctuation spectrum~\cite{7226}. However, some choices 
of $\cP_{N_2}$ do not have physical interpretation though they \textit{are} 
solutions to the equation of motion. This poses a problem. 

To resolve it, it was argued~\cite{9038} that at the `closed string vacuum' 
defined to be $\phi=\phi_*$, the original choice of field variables becomes 
singular and that the seemingly different values of fields, for example 
$A_{\mu}=0$ and $C_{\mu}=0$, in fact correspond to the same point in the 
configuration space. This situation is well described by the following 
mechanical analogy. In the three dimensional space, let us consider 
a particle 
moving under the influence of the spherically symmetric potential which has 
a minimum at the origin $r=0$. When we use the polar coordinates $(r,
\vartheta,\varphi)$, it seems that the ground states are infinitely 
degenerate because the value of the potential at $r=0$ is invariant under 
the shift of variables $\vartheta,\varphi$. Although the energy is actually 
degenerate for $r>0$, at the special point $r=0$ the different values of 
$(\vartheta,\varphi)$ all represent a single point `$r=0$' so that the ground 
state should be unique. By replacing $(r,\vartheta,\varphi)$ and $r=0$ with 
$(\phi-\phi_*,A_{\mu},\mbox{other massless fields } \chi^i)$ and $\phi=
\phi_*$ respectively, one can easily expect that at $\phi=\phi_*$ different 
values of $A_{\mu}$ and $\chi^i$ correspond to the same configuration, namely 
the unique closed string vacuum. 

To be concrete, we would like to write down the explicit form of the 
effective action and the field redefinition. Though we have incorporated the 
effect of the constant $B$-field background using the formulae 
(\ref{eq:OJ})--(\ref{eq:OL}), there exist many different descriptions labeled 
by an antisymmetric tensor $\Phi_{\mu\nu}$, as explained in~\cite{SW}. 
Given a value of $\Phi$, we obtain $\theta^{\mu\nu},\ G_{\mu\nu}$ and $G_s$ by
\begin{eqnarray}
\frac{1}{G+2\pi\ap\Phi}&=&-\frac{\theta}{2\pi\ap}+\frac{1}{g+2\pi\ap B}, 
\label{eq:PU} \\ G_s&=&g_s\sqrt{\frac{\det(G+2\pi\ap\Phi)}{\det(g+2\pi\ap B)}}
. \nonumber 
\end{eqnarray}
This general formula contains the previous one (\ref{eq:OJ})--(\ref{eq:OL}) 
as a special case $\Phi=0$. But here, we employ another choice $\Phi=-B$, 
in which case we find 
\begin{eqnarray}
\theta^{\mu\nu}&=&(B^{-1})^{\mu\nu}, \nonumber \\
G_{\mu\nu}&=&-(2\pi\ap)^2(Bg^{-1}B)_{\mu\nu}, \label{eq:PV} \\
G_s&=&g_s\sqrt{\det(2\pi\ap Bg^{-1})}. \nonumber 
\end{eqnarray}
We define the shifted gauge field $C_{\mu}$ to be 
\[ C_{\mu}=A_{\mu}+\theta^{-1}_{\mu\nu}x^{\nu} \] 
as in (\ref{eq:PP}), and further define the new `background independent' 
variables $X^{\mu}$ as~\cite{Seib}
\begin{equation}
X^{\mu}=\theta^{\mu\nu}C_{\nu}=x^{\mu}+\theta^{\mu\nu}A_{\nu}. \label{eq:PW}
\end{equation}
In terms of $C_{\mu}$ or $X^{\mu}$, the field strength for the original gauge 
field $A_{\mu}$ can be rewritten as 
\begin{eqnarray*}
F_{\mu\nu}&=&\partial_{\mu}A_{\nu}-\partial_{\nu}A_{\mu}-i[A_{\mu},A_{\nu}] \\
&=&-i[C_{\mu},C_{\nu}]+\theta^{-1}_{\mu\nu}=iB_{\mu\rho}[X^{\rho},X^{\sigma}]
B_{\sigma\nu}+B_{\mu\nu},
\end{eqnarray*}
where $[A,B]=A*B-B*A$ and the noncommutativity parameter involved in the 
$*$-product is $\theta$ given in~(\ref{eq:PU}). Using these variables, the 
effective action for the tachyon and gauge field is given 
by~\cite{7226,Seib,9038}
\begin{eqnarray}
S&=&\frac{1}{g_s(2\pi)^{12}\ap{}^{13}}\mathrm{Tr}\biggl[V(\phi)\sqrt{\det(
\delta_{\mu}^{\nu}-2\pi\ap ig_{\mu\rho}[X^{\rho},X^{\nu}])} \nonumber \\
& &+\ap f(\phi)g_{\mu\nu}[X^{\mu},\phi][\phi,X^{\nu}]+\cdots\biggr], 
\label{eq:PX}
\end{eqnarray}
where $V(\phi)$ is the tachyon potential which vanishes at $\phi=\phi_*$, 
$f(\phi)$ is some unknown function, and $\cdots$ represents higher derivative 
terms. Since $V(\phi_*)=0$, the kinetic term for $C_{\mu}$ field 
(equivalently $X^{\mu}$) also vanishes at $\phi=\phi_*$. Therefore the field 
$C_{\mu}$ is not a good coordinate around $\phi=\phi_*$, just like the 
angular coordinates $\vartheta,\varphi$ at $r=0$. We must then perform a 
field redefinition 
\begin{equation}
\tilde{C}_{\mu}=\tilde{C}_{\mu}(\phi,C_{\nu})\ , \quad \tilde{\phi}=
\tilde{\phi}(\phi,C_{\nu}) \label{eq:PY}
\end{equation}
such that the new variables satisfy
\begin{equation}
\tilde{C}_{\mu}(\phi_*,\forall C_{\nu})=0\ , \quad \tilde{\phi}(\phi_*,
\forall C_{\nu})=\tilde{\phi}_*=\mathrm{const.}, \label{eq:PZ}
\end{equation}
where $\tilde{\phi}_*$ can be set to zero by suitably defining $\tilde{\phi}$.
That is to say, the newly defined field variables vanish at $\phi=\phi_*$ 
\textit{irrespective of} the value of the original field $C_{\mu}$. Hence, 
any finite $C_{\mu}$ corresponds to the same point $(\tilde{C}_{\mu}=0,
\tilde{\phi}=\tilde{\phi}_*)$ in the configuration space if $\phi=\phi_*$, 
so that the apparent degeneracy caused by the different values of 
expectation values of $A_{\mu}$ is removed. Now we will apply it to the case 
of the soliton solution~(\ref{eq:PS}). Though the soliton solution is of 
course different from the vacuum $\phi=\phi_*$, restricting ourselves to the 
subspace spanned by $\{|n\rangle\}_{n=N_1}^{\infty}$ the tachyon 
configuration effectively becomes $\phi=\phi_*I_{\infty-N_1}$. In that case, 
any value of $C_i$ restricted to the above subspace corresponds to the same 
configuration, as discussed above. For clarity, we rewrite (\ref{eq:PS}) in 
a matrix form as 
\begin{equation}
\phi=\left(
	\begin{array}{cc}
	\phi_{\mathrm{max}}I_{N_1} & 0 \\ 0 & \phi_*I_{\infty-N_1}
	\end{array}
\right)\ , \quad C_i=\left(
	\begin{array}{cc}
	S_i & 0 \\ 0 & V_i
	\end{array}
\right). \label{eq:QA}
\end{equation}
Then we can restate that different values of $V_i$ do not lead to different 
configurations. So we have solved the problem that there is unwanted 
degeneracy of solutions arising from different values of $V_i\sim
\sum_{n=N_1}^{\infty}a_n|n\rangle\langle n|$. On the other hand, different 
values of $S_i$ \textit{really} correspond to different background gauge 
field configurations on the solitonic D-brane represented by $\phi$ 
in~(\ref{eq:QA}). 
\bigskip

The authors of \cite{10060} invented still another method of constructing 
exact noncommutative solitons for any value of noncommutativity. In the 
operator representation of the noncommutative functions, the action 
generally takes the form 
\begin{equation}
S=\mathrm{Tr}\sum\prod\cO_i, \label{eq:QC}
\end{equation}
where we assume that all $\delta S/\delta\cO_i$ do not contain constant 
terms, and that every operator $\cO_i$ transforms as $\cO_i\longrightarrow
U\cO_iU^{\dagger}$ under a $U(\infty)$ transformation $U$. If $U$ satisfies 
\begin{equation}
U^{\dagger}U=I, \label{eq:QD}
\end{equation}
then it follows from (\ref{eq:QC}) that 
\begin{equation}
\frac{\delta S}{\delta\cO_i}\mathop{\longrightarrow}\limits^UU\frac{\delta S}{
\delta\cO_i}U^{\dagger} \label{eq:QE}
\end{equation}
irrespective of $UU^{\dagger}$. The equation~(\ref{eq:QE}) means that the 
transformation $U$ takes solutions of equations of motion $\delta S/\delta
\cO_i=0$ to solutions. If $U$ also satisfies $UU^{\dagger}=I$, it simply 
corresponds to a gauge transformation. But if $UU^{\dagger}\neq I$, the new 
solution $\widetilde{\cO}_i^{\prime}=U\widetilde{\cO}_iU^{\dagger}$ is 
inequivalent to the original solution $\widetilde{\cO}_i$ because such a 
transformation $U$ does not leave the action and the energy 
invariant. So we can use $U$ to 
generate a new solution from a known one. By multiplying~(\ref{eq:QD}) with 
$U,U^{\dagger}$ from left and right respectively, we obtain
\[ \left(UU^{\dagger}\right)\left(UU^{\dagger}\right)=UU^{\dagger}. \]
It requires $UU^{\dagger}$ to be a projection operator, 
\begin{equation}
UU^{\dagger}=P. \label{eq:QF}
\end{equation}
Now we construct an example of $U$. In the Hilbert space $\cH$ constructed 
before, let us consider the raising shift operator 
\begin{equation}
S=\sum_{k=0}^{\infty}|k+1\rangle\langle k|. \label{eq:QG}
\end{equation}
Since it satisfies
\begin{equation}
S^{n\dagger}S^n=I\ , \quad S^nS^{n\dagger}=I-\sum_{k=0}^{n-1}|k\rangle
\langle k|\equiv I-P_n, \label{eq:QH}
\end{equation}
$U=S^n$ has the desired properties (\ref{eq:QD}), (\ref{eq:QF}) to be a 
`solution generating transformation' for any $n$. 
\smallskip

\begin{figure}[tbhp]
	\begin{center}
	\includegraphics{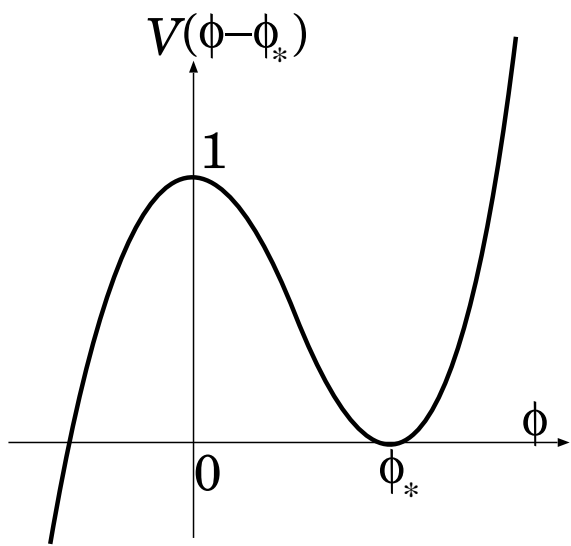}
	\end{center}
	\caption{The tachyon potential $V(\phi-\phi_*)$.}
	\label{fig:BI}
\end{figure}

Next we consider the effective action
\begin{eqnarray}
S&=&\tau_{25}\int d^{26}\!x \sqrt{|\det g|}\biggl[-\frac{1}{2}f(\phi-\phi_*)
g^{\mu\nu}\partial_{\mu}\phi\partial_{\nu}\phi+\cdots -V(\phi-\phi_*) 
\nonumber \\ & &{}-\frac{1}{4}h(\phi-\phi_*)F^{\mu\nu}F_{\mu\nu}+\cdots 
\biggr] \label{eq:QI}
\end{eqnarray}
for the $U(1)$-neutral tachyon and gauge field 
in bosonic string theory. The potential 
$V(\phi-\phi_*)$ has a local minimum at $\phi=\phi_*$ and a local maximum 
at $\phi=0$, and the values there are $V(0)=0$ and $V(-\phi_*)=1$, as 
shown in Figure~\ref{fig:BI}. 
We assume that the function $h(\phi-\phi_*)$ is obtained by expanding the 
product of $V(\phi-\phi_*)$ and the Born-Infeld action in powers of 
$F_{\mu\nu}$, so that $h(\phi-\phi_*)$ vanishes at $\phi=\phi_*$: $h(0)=0$. 
Turning on a $B$-field in the $(x^1,x^2)$-directions, the action becomes
\begin{eqnarray}
S&=&\tau_{25}\frac{g_s}{G_s}\int d^{24}\!x \sqrt{|\det G|}2\pi\theta
\mathrm{Tr}\biggl[-\frac{1}{2}f(\phi-\phi_*)G^{\mu\nu}D_{\mu}\phi D_{\nu}\phi
+\cdots \nonumber \\ & &{}-V(\phi-\phi_*)-\frac{1}{4}h(\phi-\phi_*)
(F+\Phi)^{\mu\nu}(F+\Phi)_{\mu\nu}+\cdots \biggr], \label{eq:QJ}
\end{eqnarray}
where $\Phi$ is the antisymmetric tensor appearing in~(\ref{eq:PU}). 
Here we again 
choose $\Phi=-B$, so that $G^{\mu\nu},G_s,\theta^{\mu\nu}$ are given 
by~(\ref{eq:PV}) for $B_{12}=B\ (<0), \ g_{\mu\nu}=\eta_{\mu\nu}$. 
Explicitly, 
\begin{eqnarray}
\theta&\equiv&\theta^{12}=-1/B=1/|B|, \nonumber \\
G_{\mu\nu}&=&\mathrm{diag}(-1,(2\pi\ap B)^2,(2\pi\ap B)^2,+1,\cdots,+1),
\label{eq:QK} \\ G_s&=&2\pi\ap |B|g_s. \nonumber 
\end{eqnarray}
And we have 
\[ D_{\mu}\phi=\partial_{\mu}\phi-i[A_{\mu},\phi]=-i[C_{\mu},\phi] 
\quad \mathrm{with} \quad C_{\mu}=A_{\mu}+B_{\mu\nu}x^{\nu}, \]
and 
\[ F_{\mu\nu}+\Phi_{\mu\nu}=(-i[C_{\mu},C_{\nu}]+B_{\mu\nu})+(-B_{\mu\nu})
=-i[C_{\mu},C_{\nu}]. \]
The solution which represents the closed string vacuum is given by 
\begin{eqnarray}
\phi=\phi_*, \quad A_1&=&A_2=0 \quad (C_1=Bx^2,\ C_2=-Bx^1), \nonumber \\ 
A_{\mu}&=&0 \quad \mathrm{for} \quad \mu=0,3,\cdots,25. \label{eq:QL}
\end{eqnarray}
Now we act on the above simple solution with the solution generating 
transformation $U=S^n$. Then 
\begin{equation}
\left(
	\begin{array}{c}
	\phi_* \\ C_1 \\ C_2 \\ A_{\mu}=0
	\end{array}
\right)\ \ \mathop{\longrightarrow}\limits^{S^n}\quad S^n\left(
	\begin{array}{c}
	\phi_* \\ C_1 \\ C_2 \\ 0
	\end{array}
\right) S^{n\dagger}=\left(
	\begin{array}{c}
	\phi_*(I-P_n) \\ S^nC_1S^{n\dagger} \\ S^nC_2S^{n\dagger} \\ 0
	\end{array}
\right). \label{eq:QM}
\end{equation}
From the general argument above, it is guaranteed that the transformed field 
configuration~(\ref{eq:QM}) is a \textit{solution} to the equations of motion 
derived from the action~(\ref{eq:QJ}) that is \textit{not equivalent} to 
the original solution~(\ref{eq:QL}) due to the properties~(\ref{eq:QH}) of 
$S^n$. The kinetic term for $\phi$ in the action~(\ref{eq:QJ}) vanishes on 
the solution~(\ref{eq:QM}) because $D_{\mu}\phi=-i[C_{\mu},\phi]$ vanishes 
for constant $\phi=\phi_*$ at the closed string vacuum~(\ref{eq:QL}) and 
the covariant derivative of $\phi$ homogeneously transforms under the 
solution generating transformation as 
\[ [C_{\mu},\phi_*]=0 \quad \mathop{\longrightarrow}\limits^{S^n} \quad 
S^n[C_{\mu},\phi_*]S^{n\dagger}=0. \]
For the same reason, higher derivative terms (represented by $\cdots$) for 
$\phi$ also vanish. For higher derivative terms of $F_{\mu\nu}$, each term 
includes at least one factor of $D_{\rho}(F_{\mu\nu}+\Phi_{\mu\nu})=-[
C_{\rho},[C_{\mu},C_{\nu}]]$, which vanishes on the solution~(\ref{eq:QM}) 
again because $[C_{\mu},C_{\nu}]\sim B^2[x,x]=$constant. Furthermore, we can 
show that the $h\cdot (F+\Phi)^2$ term vanishes as well. After the 
transformation, $h(\phi-\phi_*)$ becomes
\[ h(\phi_*(I-P_n)-\phi_*)=h(-\phi_*P_n)=h(-\phi_*)P_n \]
because the function $h(\phi-\phi_*)$ does not contain a constant term 
$(h(0)=0)$. On the other hand, $(F+\Phi)^2$ becomes 
\begin{eqnarray*}
(F+\Phi)^2=-[C^{\mu},C^{\nu}][C_{\mu},C_{\nu}]\  
&\mathop{\longrightarrow}\limits^{S^n}& \ -S^n[B^{\mu}{}_{\alpha}
x^{\alpha}, 
B^{\nu}{}_{\beta}x^{\beta}][B_{\mu\lambda}x^{\lambda},B_{\nu\rho}x^{\rho}]
S^{n\dagger} \\ &=& B_{\mu\nu}B^{\mu\nu}S^nS^{n\dagger}=B^2(I-P_n).
\end{eqnarray*}
So the product vanishes as 
\[h(\phi-\phi_*)(F+\Phi)^2=h(-\phi_*)B^2\ P_n(I-P_n)=0 \]
because of the property $P_n^2=P_n$ of the projection operator $P_n$. 
Collecting all these results, the value of the action~(\ref{eq:QJ}) 
calculated on the solution~(\ref{eq:QM}) is given by 
\begin{eqnarray*}
S&=&-\frac{2\pi\theta\tau_{25}}{2\pi\ap |B|}\int d^{24}\!x (2\pi\ap B)^2 
\sqrt{|\det g_{24}|}\mathrm{Tr}V(\phi_*(I-P_n)-\phi_*) \\ &=&
-(2\pi\sqrt{\ap})^2\tau_{25}\int d^{24}\!x\sqrt{|\det g_{24}|}V(-\phi_*)
\mathrm{Tr}P_n \\ &=&-(2\pi\sqrt{\ap})^2n\tau_{25}\int d^{24}\!x 
\sqrt{|\det g_{24}|}, 
\end{eqnarray*}
where we used $V(0)=0$. So we have found that the tension of the soliton 
solution~(\ref{eq:QM}) exactly agrees with the tension of $n$ D23-branes, 
even without taking the special value of $\theta$. Note that we succeeded 
in obtaining the exact result for any value of $\theta$ because the solution 
generating transformation preserves the property that the covariant 
derivatives vanish 
so that we only need to consider the tachyon potential term $V(\phi-\phi_*)$, 
as in the case of $\theta\to\infty$ scalar field theory.
\smallskip

This `solution generating technique' can also be applied to the superstring 
case. By replacing the bosonic string tachyon potential $V(\phi-\phi_*)$ 
in~(\ref{eq:QI}) with a suitable potential of the double-well form, we can 
construct exact soliton solutions on a non-BPS D-brane in the same way. 
Furthermore, we can discuss the vortex solutions on the 
D$\overline{\mathrm{D}}$ system~\cite{10060}, where we take the tachyon field 
$\phi$ in the bifundamental representation of the noncommutative gauge group 
$U(1)\times U(1)$. For detailed discussions about the 
effective action for the brane-antibrane system, see~\cite{12198,TTU}. 
And the intersecting brane configurations were constructed in~\cite{0101125}. 

\subsubsection*{\underline{fundamental strings as noncommutative solitons}}
It was proposed that in addition to the lower dimensional D-branes, 
macroscopic fundamental closed strings\footnote{Fundamental closed strings 
are constructed also in the commutative string field theory; 
see~\cite{Conf,9061,10240,12081,0101213}.} 
are also constructed as noncommutative 
solitons~\cite{HKLM,Finn}: We will find electric flux tubes in the form of 
solitonic solutions in open string field theory, even at the closed string 
vacuum. Since we believe that the open string degrees of freedom are frozen 
out after tachyon condensation, such solitonic excitations may be interpreted 
as the fundamental closed strings. In fact, we can show that the electric 
flux tube has the same tension and fluctuation spectrum as that of the 
fundamental string. Though we will consider bosonic strings for definiteness, 
the same results should hold true for superstrings if we could take the 
world-sheet supersymmetry into account. 
\medskip

We begin with the following Born-Infeld action 
\begin{equation}
S=-\tau_{25}\int d^{26}\!x\ V(\phi)\sqrt{-\det (g_{\mu\nu}+2\pi\ap 
F_{\mu\nu})}, \label{eq:QN}
\end{equation}
where the tachyon potential $V(\phi)$ is exactly the same as the previous 
one in~(\ref{eq:ON}), namely 
\[ \left.
	\begin{array}{cll}
	\mbox{a local minimum} & \phi=0 & V(0)=0 \\
	\mbox{a local maximum} & \phi=\phi_* & V(\phi_*)=1 
	\end{array}
\right. \]
and the shape is illustrated in Figure~\ref{fig:BF}. The Born-Infeld action 
is correct only when we can ignore countless higher derivative terms, 
including the kinetic term for the tachyon. To justify it, we again turn on 
a constant $B$-field, but in this case in 24 directions $(x^2,\cdots,
x^{25})$~\cite{HKLM,Finn}. By taking the large noncommutativity 
limit~(\ref{eq:OM}), we can neglect all derivatives along the noncommutative 
directions for the same reason as before. As long as we are considering the 
field configurations which do not depend on $(x^0,x^1)=(t,x)$, we regard the 
Born-Infeld action~(\ref{eq:QN}) as \textit{exact}. To construct flux tubes 
extended in the $x^1$-direction, we only need to keep one component 
\[ F_{01}=-F_{10}=\partial_0A_1\equiv \dot{A}_1 \]
of the field strength tensor. Putting it into~(\ref{eq:QN}), the action, 
after introducing $B$-field and taking the limit, becomes
\begin{eqnarray*}
S&=&-\tau_{25}\frac{g_s}{G_s}\int d^{26}\!x\ V(\phi)\sqrt{-\det(G_{\mu\nu}
+2\pi\ap F_{\mu\nu})} \\ &\longrightarrow& -(2\pi\sqrt{\ap})^{24}\tau_{25}
\int\frac{d^{26}x}{(2\pi\theta)^{12}}V(\phi)\sqrt{1-(2\pi\ap \dot{A}_1)^2},
\end{eqnarray*}
and we define the Lagrangian to be 
\begin{equation}
L=-\tau_1\int\frac{d^{25}x}{(2\pi\theta)^{12}}V(\phi)\sqrt{1-(
2\pi\ap\dot{A}_1)^2}, \label{eq:QO}
\end{equation}
where $\tau_1$ is the D-string tension. The canonically conjugate momentum 
to the gauge field component $A_1$ is given by 
\begin{equation}
\mathcal{E}(t,\vec{x})=(2\pi\theta)^{12}\frac{\delta L(t)}{\delta\dot{A}_1
(t,\vec{x})}=\frac{\tau_1V(\phi)(2\pi\ap)^2\dot{A}_1}{\sqrt{1-(2\pi\ap
\dot{A}_1)^2}}. \label{eq:QP}
\end{equation}
The electric charge density of the solitonic 1-brane we are seeking is 
computed by integrating the electric flux $\mathcal{E}$ over the 24 
transverse coordinates as 
\begin{equation}
p\equiv \int\frac{d^{24}x}{(2\pi\theta)^{12}}\mathcal{E}(x^0,x^1,x^i). 
\label{eq:QR}
\end{equation}
In string theory context, the above charge density is quantized in the 
integer unit and is interpreted as the total number of fundamental strings, 
as we will see below. Moving to the canonical formalism, the Hamiltonian is 
\begin{eqnarray}
H&=&\int\frac{d^{25}x}{(2\pi\theta)^{12}}\mathcal{E}\dot{A}_1-L+\lambda\left(
\int\frac{d^{24}x}{(2\pi\theta)^{12}}\mathcal{E}-p\right) \nonumber \\
&=&\int\frac{d^{25}x}{(2\pi\theta)^{12}}\left[\sqrt{\tau_1^2V(\phi)^2+
\frac{\mathcal{E}^2}{(2\pi\ap)^2}}+\lambda\partial_1\mathcal{E}\right]
-\lambda p, \label{eq:QS}
\end{eqnarray}
where $\lambda$ is a Lagrange multiplier which forces the flux conservation. 
The equation of motion for $\phi$, $\delta H/\delta\phi\propto V^{\prime}
(\phi)=0$, can be solved by finding the functions satisfying 
$\phi *\phi=\phi$, or by the projection operators $\widehat{\phi}^2=
\widehat{\phi}$ as before. The second equation of motion $\delta H/\delta
\mathcal{E}=0$ with respect to $\mathcal{E}$ is, as far as we are not 
interested in the value of $\lambda$, always satisfied by taking 
$\mathcal{E}$ to contain the same projector as $\phi$. 

The simplest solution is given by 
\begin{equation}
\phi=0 \ , \quad \mathcal{E}=\frac{p}{q}P_{(q)}(x^2,x^3;\cdots;x^{24},x^{25}),
\label{eq:QT}
\end{equation}
where $P_{(q)}$ is a level $q$ projector in the operator language to ensure 
the condition~(\ref{eq:QR}). In detail, a basis of operators acting in the 
Hilbert space corresponding to the 24 dimensional noncommutative space is 
given by 
\[ \biggl\{ |m_1\rangle\langle n_1|\otimes\cdots\otimes |m_{12}\rangle\langle
n_{12}| \ ; \ m_i,n_i\ge 0\biggr\}, \]
where each $|m\rangle\langle n|$ is of the form~(\ref{eq:NY}). A level $q$ 
projector is the sum of $q$ mutually orthogonal projectors, 
\[ P_{(q)}=\sum_{\{n\}}^q|n_1\rangle\langle n_1|\otimes\cdots\otimes |n_{12}
\rangle\langle n_{12}| \]
and tracing over it gives $q$. And the mapping between the operators and the 
noncommutative functions is given by naturally extending the previous one to 
the higher dimensional case, for example, 
\begin{eqnarray}
\int \frac{d^{24}x}{(2\pi\theta)^{12}}\quad &\longleftrightarrow& \quad 
\mathrm{Tr}_{\cH\otimes\cdots\otimes\cH}, \label{eq:QU} \\
|0\rangle\langle 0|\otimes\cdots\otimes |0\rangle\langle 0| 
&\longleftrightarrow& 2e^{-[(x^2)^2+(x^3)^2]/\theta}\times\ldots\times 2
e^{-[(x^{24})^2+(x^{25})^2]/\theta}=2^{12}e^{-r^2/\theta} \nonumber \\
& &\mathrm{where} \quad r^2=\sum_{i=2}^{25}(x^i)^2, \nonumber 
\end{eqnarray}
and so on. Now look at the solution~(\ref{eq:QT}). Since the tachyon field 
$\phi$ vanishes everywhere, the original D25-brane has totally disappeared, 
resulting in the closed string vacuum. The electric flux $\mathcal{E}$ is 
localized in the transverse 24 dimensional space, so that the flux 
tube~(\ref{eq:QT}) represents a 1-brane. The energy of the solution can be 
computed by substituting $\mathcal{E}=(p/q)P_{(q)}$ and $V(0)=0$ into the 
Hamiltonian~(\ref{eq:QS}) as 
\[ H=\int dx^1\mathrm{Tr}_{\cH\otimes\cdots\otimes\cH}\left(\frac{|p|/q}{
2\pi\ap}P_{(q)}\right)=\frac{|p|}{2\pi\ap}\int dx^1\ , \]
hence the tension of the soliton~(\ref{eq:QT}) coincides with $|p|$ times 
the tension of the fundamental string. Furthermore, it was shown that the 
effective action describing the fluctuations around a single (\textit{i.e.} 
$p=1$) string-like flux tube~(\ref{eq:QT}) becomes the Nambu-Goto action 
in the realm of long wavelength $\sqrt{\ap}\partial_1F_{\mu\nu}\ll 
F_{\mu\nu}$~\cite{HKLM}. All these results suggest that the flux tube 
solution~(\ref{eq:QT}) should be identified with $p$ coincident fundamental 
closed strings on the closed string vacuum. 

The second solution we consider is 
\begin{equation}
\phi=\phi_*P_{(q)}\ , \quad \mathcal{E}=\frac{p}{q}P_{(q)}, \label{eq:QV}
\end{equation}
whose energy is found to be 
\begin{eqnarray}
H&=&\int dx^1\mathrm{Tr}_{\cH\otimes\cdots\otimes\cH}\sqrt{\tau_1^2V(\phi_*)^2
+\frac{p^2}{(2\pi\ap)^2q^2}}P_{(q)} \nonumber \\
&=&\frac{1}{2\pi\ap}\sqrt{p^2+\frac{q^2}{g_s^2}}\int dx^1, \label{eq:QW}
\end{eqnarray}
where we rewrote the D-string tension $\tau_1$ in terms of the closed string 
coupling $g_s$ as $\tau_1=1/2\pi\ap g_s$. The tension of~(\ref{eq:QW}) 
exactly agrees with that of the $(p,q)$-string. In fact, the 
solution~(\ref{eq:QV}) looks like the superposition of $p$ fundamental 
strings constructed as the flux tube and $q$ coincident D-strings 
(codimension 24 D-branes). 
\medskip

Even though the `fundamental closed string' constructed as the electric flux 
tube is a solitonic object in open string field theory, it can be quantized 
because the most fundamental degrees of freedom, open strings, are absent at 
the closed string vacuum so that the solitonic closed string is the lightest 
excitation there. Since the action for fluctuations of the string takes the 
Nambu-Goto form at long wavelength, the quantization of the closed string 
seems to reproduce the usual closed string theory, at least within the range 
of validity. 

\subsection{Noncommutativity in string field theory}
Thus far, we have considered the noncommutativity in the standard 
scalar-gauge field theory or in the effective field theory obtained by 
integrating out the massive fields in open string 
field theory. In this subsection, we will take a glance at the 
noncommutativity in the full string field theory context. 

In bosonic string theory, a vertex operator which corresponds to some open 
string state in a one-to-one manner generally takes the form 
\begin{equation}
\cV(\tau)=\ :\left(\prod\partial^nX\cdot\partial^mb\cdot\partial^{\ell}c
\right)e^{ip\cdot X}(\tau):\ , \label{eq:HapA}
\end{equation}
where $\tau$ parametrizes the boundary of the open string world-sheet. We 
denote by $\cB$ the full associative algebra constructed by vertex operators 
of the form~(\ref{eq:HapA}) and the multiplication law among them. Now we 
consider decomposing $\cB$ into two parts as $\cB=\cA_0\otimes\cC$ such that 
$\cA_0$ consists of those vertex operators 
which do not depend on the center of mass 
coordinate $x^i$ (\textit{i.e.} out of $\partial^nX$ with $n\ge\mathit{1}$ 
and $\partial^mb,\ \partial^mc$ with $m\ge 0$) while $\cC$ contains only 
$e^{ipX}$. Although the elements of $\cA_0$ are closed under OPEs, those of 
$\cC$ are not because we have 
\begin{equation}
e^{ipX(\tau)}e^{iqX(\tau^{\prime})}\sim \exp\left(\frac{p\cdot q}{2}\ln
(\tau-\tau^{\prime})^2\right)e^{i(p+q)X(\tau^{\prime})}\Bigl(1+i(\tau-
\tau^{\prime})p\cdot\partial X(\tau^{\prime})+\cdots \Bigr). 
\label{eq:HapB}
\end{equation}
It was shown in~\cite{6071}, however, that $\cB$ is actually factorized into 
two \textit{commuting subalgebras} $\cA_0$ and $\cC$ in the limit of large 
noncommutativity. As usual, we introduce noncommutativity by turning 
on a Neveu-Schwarz $B$-field $B=tB_0$, which is now assumed to have maximal 
rank. Then the $XX$ OPE becomes 
\begin{equation}
X^{\mu}(\tau)X^{\nu}(\tau^{\prime})\sim -\ap G^{\mu\nu}\ln(\tau-\tau^{\prime}
)^2+\frac{i}{2}\theta^{\mu\nu}\epsilon(\tau-\tau^{\prime}), 
\label{eq:HapC}
\end{equation}
where we take the upper half-plane representation of the disk world-sheet so 
that $\tau, \tau^{\prime}$ are real, and $\epsilon(x)$ is a step function. 
$G^{\mu\nu},\ \theta^{\mu\nu}$ have already been defined in~(\ref{eq:OJ}) and 
(\ref{eq:OL}). Here we take the limit $t\to\infty$ with $B_0,\ap,g$ (closed 
string metric) held fixed. In this limit, we find 
\begin{equation}
G^{\mu\nu}\to -\frac{1}{(2\pi\ap)^2t^2}(B_0^{-1}gB_0^{-1})^{\mu
\nu}, \qquad \theta^{\mu\nu}\to\frac{1}{t}(B_0^{-1})^{\mu\nu},
\end{equation}
so that the OPE for $Z^{\mu}(\tau)\equiv tX^{\mu}(\tau)$ takes the form 
\begin{eqnarray}
Z^{\mu}(\tau)Z^{\nu}(\tau^{\prime})&\sim&\frac{1}{(2\pi)^2\ap}(B_0^{-1}g
B_0^{-1})^{\mu\nu}\ln(\tau-\tau^{\prime})^2 \nonumber \\
& &{}+\frac{i}{2}t(B_0^{-1})^{\mu\nu}\epsilon(\tau-\tau^{\prime}). 
\label{eq:HapD}
\end{eqnarray}
Since $\partial Z\cdot\partial Z$ OPE 
\begin{equation}
\partial Z^{\mu}(\tau)\partial Z^{\nu}(\tau^{\prime})\sim\frac{(B_0^{-1}g
B_0^{-1})^{\mu\nu}}{2\pi^2\ap}\frac{1}{(\tau-\tau^{\prime})^2}
\end{equation}
is finite in the limit $t\to\infty$, we think of $Z^{\mu}$ 
rather than $X^{\mu}$ as elements of $\cB$. This time, the OPE among 
the elements of $\cC$ becomes 
\begin{eqnarray}
e^{i\frac{p}{\sqrt{t}}Z(\tau)}e^{i\frac{q}{\sqrt{t}}Z(\tau^{\prime})}&\sim&
\exp\left(-\frac{i}{2}p_{\mu}q_{\nu}(B_0^{-1})^{\mu\nu}\right)e^{i
\frac{p+q}{\sqrt{t}}Z(\tau^{\prime})} \label{eq:HapE} \\ & &\times
\left(1+i(\tau-\tau^{\prime})\frac{p_{\mu}}{\sqrt{t}}\partial Z^{\mu}
(\tau^{\prime})+\cdots \right) \nonumber 
\end{eqnarray}
for $\tau\to\tau^{\prime}+0$. We scaled the momentum as $p/\sqrt{t}$ so 
that the prefactor in the right hand side of~(\ref{eq:HapE}) may become 
definite. Notice that the `outside $\cC$' terms $\partial Z^{\mu}$ and 
$\cdots$ in the right hand side of~(\ref{eq:HapE}) all vanish in the limit 
$t\to\infty$. Therefore, we found that $\cC$ also forms a subalgebra of 
$\cB$. Besides, the $(\tau-\tau^{\prime})$-dependent factor (anomalous 
dimension) which was present in~(\ref{eq:HapB}) has dropped in this limit 
as well. As the result of these facts, the center-of-mass algebra $\cC$ 
has been reduced to that of noncommutative functions, in which the 
multiplication among elements is taken by the Moyal $*$-product with the 
noncommutativity parameter $(B_0^{-1})^{\mu\nu}$, as we can see 
from~(\ref{eq:HapE}). Furthermore, the two subalgebras $\cA_0$ and $\cC$ 
mutually commute in the sense that 
\begin{equation}
\partial^nZ^{\mu}(\tau)\cdot e^{i\frac{p}{\sqrt{t}}Z(\tau^{\prime})}\sim
\frac{i(B_0^{-1}gB_0^{-1})^{\mu\nu}p_{\nu}}{2\pi^2\ap}\partial_{\tau}^n
\ln(\tau-\tau^{\prime})\frac{1}{\sqrt{t}}e^{i\frac{p}{\sqrt{t}}Z(
\tau^{\prime})}\mathop{\longrightarrow}\limits^{t\to\infty}0.
\end{equation}
When the $B$-field has less than maximal rank, we classify those elements 
carrying nonzero momenta in the \textit{commutative} directions as $\cA_0$, 
because otherwise $\cC$ would not be closed as in~(\ref{eq:HapB}). 

This factorization $\cB=\cA_0\otimes\cC$ was alternatively shown 
in~\cite{10034} in the language of operator formulation. 
\medskip

Here we outline its application to tachyon condensation. 
For more details, see~\cite{6071,10034}. The information about the tachyon 
potential is contained in $\cA_0$. The string field configuration 
representing the original D25-brane is taken to be $\Phi=0$, while we denote 
by $\Phi_0\in\cA_0$ the closed string vacuum configuration which solves the 
string field equation of motion 
\begin{equation}
Q_B\Phi_0+\Phi_0\star\Phi_0=0, \label{eq:HapF}
\end{equation}
where we used the symbol $\star$ to represent the multiplication (gluing) 
among string fields in order to distinguish it from the Moyal $*$-product. 
The BRST operator $Q_B$ acts only on $\cA_0$ and commutes with any element 
in $\cC$ because $Q_B$ is constructed out of the elements of $\cA_0$ except 
for the term $\ap p^2/t$, which is negligible in the limit $t\to\infty$. 
To find a lump solution, we consider a noncommutative function $\rho\in\cC$ 
which squares to itself under the $*$-product, 
\begin{equation}
\rho *\rho =\rho \label{eq:HapG}
\end{equation}
as in the previous subsections. In terms of the operator language, it 
corresponds, of course, to a projection operator $\widehat{\rho}$ in the 
auxiliary Hilbert space $\cH$. Due to the factorization $\cB=\cA_0\otimes
\cC$, we find that the following string field configuration 
\begin{equation}
\Phi=\Phi_0\otimes\rho\in\cB\qquad (\Phi_0\in\cA_0,\ \rho\in\cC) 
\label{eq:HapH}
\end{equation}
is a solution to the equation of motion, 
\begin{eqnarray*}
Q_B\Phi+\Phi\star\Phi&=&(Q_B\Phi_0)\otimes\rho+(\Phi_0\star\Phi_0)\otimes
(\rho *\rho) \\ &=&(Q_B\Phi_0+\Phi_0\star\Phi_0)\otimes\rho=0
\end{eqnarray*}
with the help of (\ref{eq:HapG}), (\ref{eq:HapF}). Since $\rho$ takes a 
spatially nontrivial solitonic configuration in the noncommutative subspace, 
the solution $\Phi$~(\ref{eq:HapH}) represents a tachyonic lump which is 
to be identified with lower dimensional D-branes. To be more specific, we 
take the rank of the $B$-field to be $2p$ and $\widehat{\sigma}=\mathbf{1}
-\widehat{\rho}$ to be a projection operator onto an $n$-dimensional subspace 
of $\cH$. If we rewrite the codimension $2p$ soliton solution~(\ref{eq:HapH}) 
as 
\begin{equation}
\Phi=0\otimes\widehat{\sigma}+\Phi_0\otimes(\mathbf{1}-\widehat{\sigma})=
\left(
	\begin{array}{cc}
	0\times\mathbf{1}_n & 0 \\
	0 & \Phi_0\times\mathbf{1}_{\infty-n}
	\end{array}
\right), \label{eq:HapI}
\end{equation}
it is clear that the above solution describes $n$ coincident 
D($25-2p$)-branes. The tension of the soliton solution~(\ref{eq:HapI}) can be 
calculated in a similar fashion to~(\ref{eq:OP})--(\ref{eq:OS}) and agrees 
with $n$ times the D($25-2p$)-brane tension if we assume the brane 
annihilation conjecture \[ S(0)-S(\Phi_0)=-\tau_{25}V_{26}. \]

Thus we can construct lower dimensional D-branes as noncommutative solitons 
using the closed string vacuum solution $\Phi_0$. But since the construction 
of the projection operators~(\ref{eq:HapG}) strongly depends on the 
factorization property $\cB=\cA_0\otimes\cC$ which in turn requires taking 
the large noncommutativity limit, it seems to be difficult to build 
noncommutative-solitonic string field configurations in arbitrary finite 
noncommutativity. 

\section{Field Theory Models for Tachyon Condensation}
For the study of tachyon condensation, useful field theory models have been 
constructed and developed by B. Zwiebach and 
J. Minahan~\cite{8227,8231,9246,11226}. These models are built by picking out 
only part of the tachyon sector of the full string field theory. They have 
such simple structure that we can exactly find the lump solutions and the 
fluctuation spectra around them, as we will see below. 

\subsection{Scalar field theory on the lump}\label{sub:Toy}
In section~\ref{sec:HK}, we analysed the scalar field theory for the tachyon. 
Expanding the scalar field $\phi$ on the original D($p+1$)-brane around the 
codimension 1 lump solution $\overline{\phi}(x)$~(\ref{eq:IF}) as 
$\phi(x^M)=\overline{\phi}(x)+\varphi(x,y^{\mu})$, the action for the 
fluctuation field $\varphi$ became
\begin{equation}
S=2\pi^2\ap{}^3\tau_{p+1}\int d^{p+1}\!y\ dx\left[-\frac{1}{2}\partial_{\mu}
\varphi\partial^{\mu}\varphi-\frac{1}{2}\varphi\left(-\frac{\partial^2\varphi}
{\partial x^2}+V^{\prime\prime}(\overline{\phi})\varphi\right)-2\kappa
\varphi^3\right]. \label{eq:LB}
\end{equation}
We are using the following notation as in chapter~\ref{ch:lump}:
\[\overbrace{\underbrace{x^0,x^1,\cdots,x^p}_{\displaystyle{y^{\mu}}}
,x^{p+1}}^{\displaystyle{x^M}}=x. \]
The spectrum on the lump was determined by the eigenvalue equation
\begin{equation}
-\frac{d^2}{dx^2}\xi_m(x)+V^{\prime\prime}(\overline{\phi})\xi_m(x)=M_m^2
\xi_m(x), \label{eq:LA}
\end{equation}
which was the $\ell=3$ case of a series of exactly solvable Schr\"{o}dinger 
potentials. The fluctuation field $\varphi$ was expanded using the 
eigenfunctions $\xi_m(x)$ of~(\ref{eq:LA}) as 
\begin{equation}
\varphi(x,y)=\phi_-(y)\xi_-(x)+\phi_0(y)\xi_0(x)+\phi_+(y)\xi_+(x)+
(\mathrm{continuum}), \label{eq:yui}
\end{equation}
where $\xi_m$'s were given in (\ref{eq:IX})$\sim$(\ref{eq:IZ}). The 
coefficients $\phi_m$ were interpreted as the fields living on the 
world-volume of the lump which was labeled by the coordinates $y^{\mu}$. 
Substituting the above expansion into the action~(\ref{eq:LB}) and carrying 
out the $x$-integral, the action could be written in terms of the lump 
fields $\phi_m$ as~\cite{8227}
\begin{eqnarray}
S&=&2\pi^2\ap{}^3\tau_{p+1}\int d^{p+1}\!y\Biggl[-\frac{1}{2}(\partial_{\mu}
\phi_-)^2+\frac{5}{8\ap}\phi_-^2-\frac{1}{2}(\partial_{\mu}\phi_+)^2
-\frac{3}{8\ap}\phi_+^2 \nonumber \\
& &{}-\frac{525\kappa}{4096}\sqrt{\frac{15}{2\sqrt{\ap}}}\pi\phi_-^3-
\frac{675\kappa}{4096}\sqrt{\frac{3}{2\sqrt{\ap}}}\pi\phi_-^2\phi_+
-\frac{387\kappa}{4096}\sqrt{\frac{15}{2\sqrt{\ap}}}\pi\phi_-\phi_+^2
+\frac{603\kappa}{4096}\sqrt{\frac{3}{2\sqrt{\ap}}}\pi\phi_+^3 \nonumber \\
& &{}-\frac{1}{2}(\partial_{\mu}\phi_0)^2-\left(\frac{225\kappa}{1024}
\sqrt{\frac{15}{2\sqrt{\ap}}}\pi\phi_--\frac{315\kappa}{1024}
\sqrt{\frac{3}{2\sqrt{\ap}}}\pi\phi_+\right)\phi_0^2+(\mathrm{continuum})
\Biggr]. \label{eq:LC}
\end{eqnarray}

Now we will discuss the tachyon condensation on the unstable lump using the 
above action. 
The structure of this action is very similar to that of the level 
truncated action of the 
cubic string field theory on a bosonic D-brane, so we can study the tachyon 
potential in the level expansion scheme in this case as well. 
On the other hand, when we think of 
it as a lump action on a D($p+1$)-brane, we already know the exact lump 
profile as well as the lump tension. By comparing the results from the level 
truncation approximation with the exact ones, we can discuss the convergence 
property and the validity of this approximation scheme in this field theory 
model.
\smallskip

As is clear from the explicit expression~(\ref{eq:LC}), $\phi_0$ enters the 
action only quadratically. This follows from the fact that the associated 
eigenfunction $\xi_0(x)$~(\ref{eq:IY}) is an odd function of $x$. So we can 
consistently set $\phi_0$ to zero when looking for a vacuum. The level of the 
tachyon field $\phi_-$ is defined to be 0 as usual. Since the difference 
between mass squared of $\phi_-$ and that of $\phi_+$ is $\displaystyle 
\left(\frac{3}{4\ap}\right)-\left(-\frac{5}{4\ap}\right)=\frac{2}{\ap}$, 
$\phi_+$ is defined to be at level 2. And we similarly assign level 
$\displaystyle \frac{9+\ap k^2}{4}\ (>2)$ to the continuum state labeled 
by $k$. We begin with the level (0,0) approximation. At this level, the 
action includes only $\phi_-$ and is given by 
\begin{equation}
S_{(0,0)}=2\pi^2\ap{}^3\tau_{p+1}\int d^{p+1}\!y\left[-\frac{1}{2}(
\partial_{\mu}\phi_-)^2+\frac{5}{8\ap}\phi_-^2-\frac{5^2\cdot 7\cdot 3\kappa}
{2^{12}}\sqrt{\frac{15}{2\sqrt{\ap}}}\pi\phi_-^3\right]. \label{eq:LD}
\end{equation}
To study the decay (annihilation) of the lump, we take $\phi_-$ to be a 
constant. Then the equation of motion is solved by
\[ \overline{\phi}_-=\frac{2^{10}}{5\cdot 7\cdot 3^2\pi\kappa\ap}
\sqrt{\frac{2\sqrt{\ap}}{15}} \]
and the value of the action at $\phi_-=\overline{\phi}_-$ is 
\[ S_{(0,0)}(\overline{\phi}_-)=\frac{2^{33}\sqrt{\ap}}{5^2\cdot 7^2\cdot 
3^{13}}\tau_{p+1}V_{p+1}\equiv V_{p+1}\cT_p^{(0,0)}, \]
where $\cT_p^{(0,0)}$ represents the approximate value of the tension of 
the lump at level (0,0). But we have already obtained in~(\ref{eq:IH}) the 
exact lump tension as
\[\cT_p=\frac{2^{14}\pi^2\sqrt{\ap}}{5\cdot 3^8}\tau_{p+1}. \]
So by forming the ratio 
\begin{equation}
r=\frac{\cT_p^{(0,0)}}{\cT_p}=\frac{2^{19}}{5\cdot 7^2\cdot 3^5\pi^2}\simeq 
0.892\ , \label{eq:LE}
\end{equation}
we see that the tachyonic field $\phi_-$ alone reproduces 89\% of the exact 
value. If we also include the massive scalar field $\phi_+$, we can extend 
the above result to level (2,4) and (2,6). By minimizing the multiscalar 
potential~(\ref{eq:LC}) we can analogously obtain the approximate value of 
the lump tension at each level. These results are shown in Table~\ref{tab:Q}.
Since the approximate value of the tension is very close to the exact value 
even at the lowest level, we realize that the contribution from higher level 
continuum state is so small that it is indeed negligible in searching for 
an approximate value. 
\begin{table}[htbp]
	\begin{center}
	\begin{tabular}{|c|c|c|c|c|}
	\hline
	level & $6\kappa\ap{}^{3/4}\overline{\phi}_-$ & $6\kappa\ap{}^{3/4}
	\overline{\phi}_+$ & tension$/2\pi\sqrt{\ap}\tau_{p+1}$ & ratio $r$ 
	\\ \hline
	(0,0) & 2.267 & --- & 0.700 & 0.892 \\ \hline
	(2,4) & 2.416 & $-0.439$ & 0.779 & 0.993 \\ \hline
	(2,6) & 2.405 & $-0.403$ & 0.774 & 0.987 \\ \hline
	exact & 2.420 & $-0.361$ & 0.784 & 1 \\ \hline
	\end{tabular}
	\end{center}
	\caption{Approximate values of expectation values of the fields and the 
	lump tension at each level.}
	\label{tab:Q}
\end{table}
\medskip

\begin{figure}[htbp]
	\begin{center}
	\includegraphics{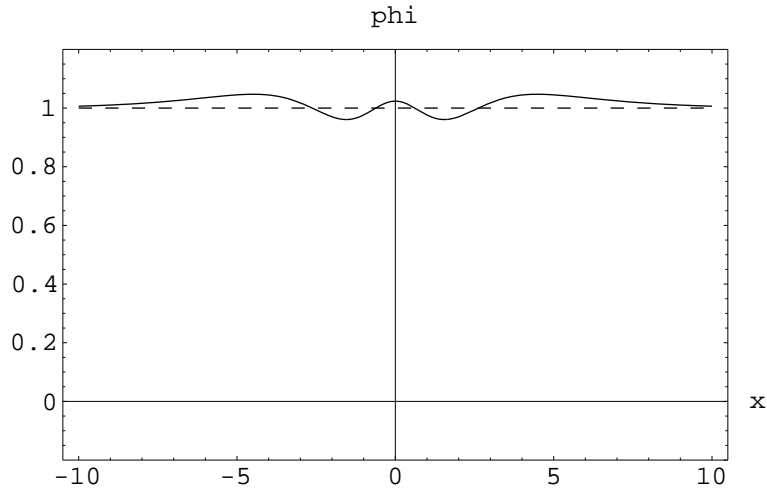}
	\end{center}
	\caption{Plot of the function $6\kappa\ap\phi_{(2,6)}(x)$.}
	\label{fig:BC}
\end{figure}

Thus far, we have seen the decay of the lump which is caused by the 
non-trivial configuration of the scalar fields living on the lump. But seen 
from the point of view of the original D($p+1$)-brane, the whole process of 
the formation of the lump followed by the decay of the lump is identified 
with the annihilation of the D($p+1$)-brane itself. Therefore, letting 
$\overline{\varphi}$ denote the configuration of $\varphi$ representing the 
decay of the lump, the sum of the lump profile $\overline{\phi}(x)$ and the 
fluctuation $\overline{\varphi}(x,y\!\!\!\backslash)$ 
should identically be equal to 
$1/6\kappa\ap$, which corresponds to the total annihilation of the original 
D($p+1$)-brane in $\phi^3$ scalar field theory model described by the 
action~(\ref{eq:IB}). To verify it, we substitute the expectation values 
$\overline{\phi}_+,\overline{\phi}_-$ at level (2,6) into (\ref{eq:yui}) and 
get the following profile
\begin{eqnarray}
\phi_{(2,6)}(x)&\equiv&\overline{\phi}(x)+\overline{\varphi}
(x,y\!\!\!\backslash) \label{eq:LF} \\
&=&\frac{1}{6\kappa\ap}\left[1+0.4936 \sech\frac{x}{2\sqrt{\ap}}-\frac{3}{2}
\sech^2\frac{x}{2\sqrt{\ap}}+1.030\sech^3\frac{x}{2\sqrt{\ap}}\right].
\nonumber
\end{eqnarray}
It is plotted in Figure~\ref{fig:BC}. 
Indeed, the value of $\phi_{(2,6)}(x)$ stays near $1/6\kappa\ap$ over the 
whole range of $x$. Conversely, by imposing the following relation 
\begin{equation}
\frac{1}{6\kappa\ap}=\overline{\phi}(x)+\overline{\phi}_-\xi_-(x)+
\overline{\phi}_+\xi_+(x)+(\mathrm{continuum}), \label{eq:LG}
\end{equation}
we can obtain the exact expectation values for $\overline{\phi}_-$ and 
$\overline{\phi}_+$ as we know the lump profile $\overline{\phi}(x)$ exactly. 
In doing so, we do not need any information about the continuum part because 
the eigenfunctions $\xi_m(x)$ have different eigenvalues so that they are 
orthogonal with each other. Multiplying~(\ref{eq:LG}) by $\xi_-$ or $\xi_+$ 
and integrating over $x$, we get 
\begin{eqnarray}
\overline{\phi}_-&=&\frac{1}{6\kappa\ap}\frac{3}{2}\int_{-\infty}^{\infty}
dx\ \sech^2\frac{x}{2\sqrt{\ap}}\xi_-(x)=\frac{1}{6\kappa\ap{}^{3/4}}
\frac{9\pi}{32}\sqrt{\frac{15}{2}}\simeq\frac{2.420}{6\kappa\ap{}^{3/4}},
\nonumber \\
\overline{\phi}_+&=&\frac{1}{6\kappa\ap}\frac{3}{2}\int_{-\infty}^{\infty}
dx\ \sech^2\frac{x}{2\sqrt{\ap}}\xi_+(x)=\frac{-1}{6\kappa\ap{}^{3/4}}
\frac{3\pi}{32}\sqrt{\frac{3}{2}}\simeq -\frac{0.3607}{6\kappa\ap{}^{3/4}},
\label{eq:LH}
\end{eqnarray}
whose values are already listed in Table~\ref{tab:Q}. Compared with these 
exact values, the approximate values at level (2,6) are rather precise, 
as is expected from the successful result about the tension. 
\medskip

In the full string field theory, there should be no physical excitations 
around the new vacuum which is left after the lump decays. But there is 
no mechanism which undertakes a task of removing the states in this 
field theory model. Instead, 
expanding the lump fields $\phi_{\pm}$ around the expectation 
values~(\ref{eq:LH}), the mass squared of the fields shift. Though the mixing 
term $\phi_-\phi_+$ appears, the mass matrix can be diagonalized with the 
eigenvalues
\[ m_1^2\simeq 1.04 \ , \quad m_2^2\simeq 1.91 \ . \]
Since both are positive, there is no tachyonic instability, so the new 
vacuum~(\ref{eq:LH}) is perturbatively stable in the sense of the 
criterion explained in subsection~\ref{sub:fluc}. 

\subsection{Field theory on the lump in $\ell\to\infty$ model}\label{sub:ell}
In section~\ref{sec:HK} and subsection~\ref{sub:Toy}, we considered the 
$\phi^3$ tachyonic scalar field theory on a bosonic D($p+1$)-brane. 
On it, we constructed a codimension 1 lump solution 
$\overline{\phi}(x)$~(\ref{eq:IF}) exactly. Expanding the scalar field 
$\phi(x^M)$ around the lump solution, the full spectrum of the fluctuation 
field $\varphi$ was determined by the Schr\"{o}dinger equation which was the 
$\ell=3$ case of a series of exactly solvable reflectionless potentials,
\begin{equation}
-\frac{d^2}{du^2}\xi(u)+\Bigl(\ell^2-\ell(\ell+1)\sech^2u\Bigr)\xi(u)
=E(\ell)\xi(u). \label{eq:LI}
\end{equation}
In this subsection, we will generalize the $\ell=3$ model to arbitrary 
integer $\ell$, including the limit $\ell\to\infty$. In particular, we 
will discuss the tachyon condensation in the $\ell\to\infty$ model in detail. 
\medskip

\begin{figure}[htbp]
	\begin{center}
	\includegraphics{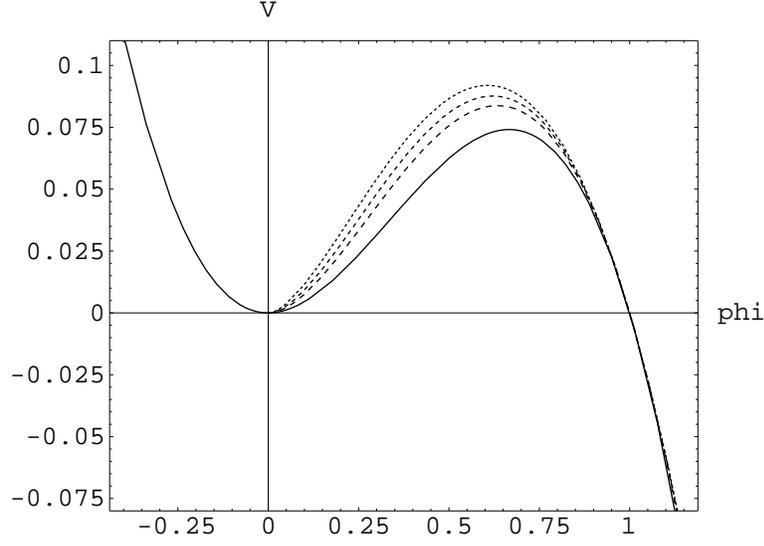}
	\end{center}
	\caption{Plot of $V_{\ell+1}(\phi)$ for $\ell=2$ (solid line), $\ell=5,\ 
	\ell=10,\  \ell=1000$ from below.}
	\label{fig:BD}
\end{figure}

To begin with, we consider the following scalar field theory 
action~\cite{8231} on a D($p+1$)-brane, 
\begin{eqnarray}
S_{\ell+1}&=&\tau_{p+1}\int d^{p+2}\!x\left(-\frac{\ap}{2}\partial_M\phi
\partial^M\phi-V_{\ell+1}(\phi)\right), \label{eq:LJ} \\
V_{\ell+1}(\phi)&=&\frac{\ell}{4}\phi^2\left(1-\phi^{\frac{2}{\ell}}\right).
\label{eq:LK}
\end{eqnarray}
For $\ell=2$ (it was previously called `$\ell=3$'), the potential becomes
\[V_3(\phi)=\frac{1}{2}\phi^2-\frac{1}{2}\phi^3. \]
This potential has a local minimum at $\phi=0$ and a local maximum at $\phi=
\frac{2}{3}$, as illustrated in Figure~\ref{fig:BD}. 
After rescaling and shifting $\phi$, it correctly reproduces the $\phi^3$ 
model considered before. Generally, the potential $V_{\ell+1}(\phi)$ has 
a local minimum at $\phi=0$ and we identify it with the `closed string 
vacuum' in this model. And a local maximum appears at 
\begin{equation}
\phi=\phi_0\equiv \left(\frac{\ell}{\ell+1}\right)^{\frac{\ell}{2}} 
\label{eq:LL}
\end{equation}
with $\displaystyle V_{\ell+1}(\phi_0)=\frac{1}{4}\left(\frac{\ell}{\ell+1}
\right)^{\ell+1}$. Further, we can see $V_{\ell+1}^{\prime\prime}(\phi_0)=-1$ 
holds irrespective of the value of $\ell$. Hence, we regard the maximum as 
the open string perturbative vacuum with a tachyon of 
mass squared $m^2=-1/\ap$ on a D($p+1$)-brane. 

Now we look for a codimension one lump solution $\overline{\phi}(x)$. The 
equation of motion is 
\begin{equation}
\frac{\ap}{2}\left(\frac{d\overline{\phi}}{dx}\right)^2=\frac{\ell}{4}
\overline{\phi}^2(1-\overline{\phi}^{2/\ell}). \label{eq:LM}
\end{equation}
It is solved by
\begin{equation}
\overline{\phi}(x)=\sech^{\ell}\left(\frac{x}{\sqrt{2\ell\ap}}\right). 
\label{eq:LN}
\end{equation}
If we expand $\phi(x^M)=\overline{\phi}(x)+\varphi(x,y^{\mu})$, the 
action~(\ref{eq:LJ}) becomes
\begin{eqnarray}
S_{\ell+1}&=&\tau_{p+1}\int d^{p+1}\!y\ dx\Biggl[\left\{-\frac{\ap}{2}\left(
\frac{d\overline{\phi}}{dx}\right)^2-V_{\ell+1}(\overline{\phi})\right\}-
\frac{\ap}{2}\partial_{\mu}\varphi\partial^{\mu}\varphi \nonumber \\
& &{}-\frac{1}{2}\varphi\left(-\ap\frac{\partial^2\varphi}{\partial x^2}
+V_{\ell+1}^{\prime\prime}(\overline{\phi})\varphi\right)-\sum_{n=3}^{\infty}
\frac{\varphi^n}{n!}V_{\ell+1}^{(n)}(\overline{\phi})\Biggr] \label{eq:LO}
\end{eqnarray}
and 
\[ V_{\ell+1}^{\prime\prime}(\overline{\phi})=\frac{\ell}{2}-\frac{(\ell+1)
(\ell+2)}{2\ell}\sech^2\frac{x}{\sqrt{2\ell\ap}}. \]
So now we should consider the following eigenvalue equation
\[-\ap\frac{d^2}{dx^2}\xi_m(x)+\frac{1}{2\ell}\left(\ell^2-(\ell+1)(\ell+2)
\sech^2\frac{x}{\sqrt{2\ell\ap}}\right)\xi_m(x)=\ap M_m^2\xi_m(x). \]
In terms of $u\equiv x/\sqrt{2\ell\ap}$, it is rewritten as 
\begin{equation}
-\frac{d^2}{du^2}\xi_m(u)+\left[(\ell+1)^2-(\ell+1)(\ell+2)\sech^2u\right]
\xi_m(u)=(2\ell\ap M_m^2+2\ell+1)\xi_m(u), \label{eq:LP}
\end{equation}
which coincides with (\ref{eq:LI}) if $\ell+1$ is replaced by $\ell$. So the 
potential~(\ref{eq:LK}) provides the generalized solvable models for 
arbitrary $\ell$ around the lump solution~(\ref{eq:LN}). Solutions to the 
equation~(\ref{eq:LP}) contain a tachyon of mass squared $m^2=-\frac{1}{\ap}
-\frac{1}{2\ell\ap}$, a massless scalar, $(\ell-1)$ massive scalars and 
a continuum. 
\medskip

We now take the limit $\ell\to\infty$. Then, the potential~(\ref{eq:LK}) 
becomes
\begin{eqnarray}
V_{\infty}(\phi)&=&\frac{\phi^2}{4}\lim_{\ell\to\infty}\ell(1-\phi^{2/\ell})
\mathop{=}\limits^{\frac{1}{\ell}=\lambda}-\frac{\phi^2}{4}\lim_{\lambda\to 0}
\frac{e^{2\lambda\ln\phi}-1}{\lambda-0} \nonumber \\
&=&-\frac{\phi^2}{4}\frac{d}{d\lambda}e^{\lambda\ln\phi^2}\bigg|_{\lambda=0}
=-\frac{1}{4}\phi^2\ln\phi^2. \label{eq:LQ}
\end{eqnarray}
\begin{figure}[htbp]
	\begin{center}
	\includegraphics{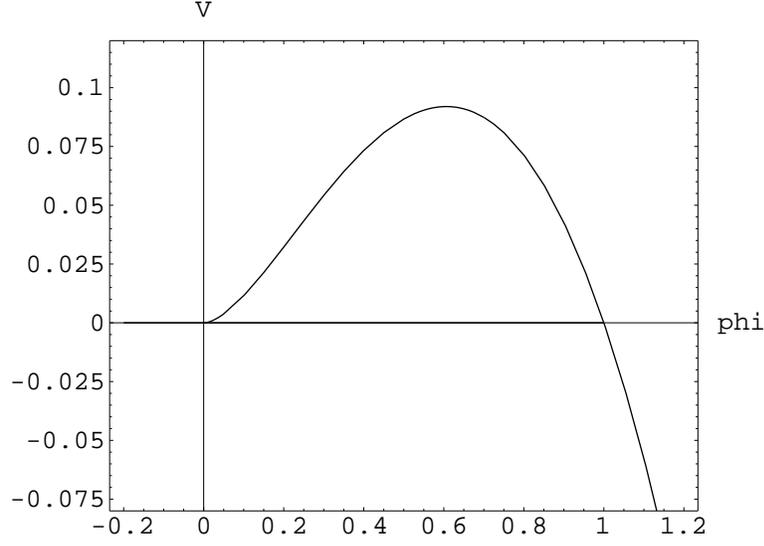}
	\end{center}
	\caption{The potential $V_{\infty}(\phi)=-\frac{1}{4}\phi^2\ln\phi^2$.}
	\label{fig:BE}
\end{figure}
As shown in Figure~\ref{fig:BE}, this potential has a minimum (closed 
string vacuum) at $\phi=0$ with $V_{\infty}(0)=0$ and a maximum (original 
unstable D($p+1$)-brane) at $\phi=\phi_0=e^{-1/2}$ with $V_{\infty}(\phi_0)
=1/4e\simeq 0.092$. Since 
\[ V_{\infty}^{\prime\prime}(\phi)=-\frac{1}{2}(\ln\phi^2+3), \]
the tachyon living on the original D($p+1$)-brane has mass squared $\ap m^2=
V_{\infty}^{\prime\prime}(\phi_0)=-1$ as ever. On the other hand, $\ap m^2
=V_{\infty}^{\prime\prime}$ diverges at the closed string vacuum $\phi=0$ 
so that the scalar field $\phi$ decouples from the spectrum there. Since 
the scalar field theory \textit{model}~(\ref{eq:LJ}) under consideration 
includes only the tachyon field $\phi$, there remain no physical excitations 
at the closed string vacuum, as desired. 

Here we construct a lump solution $\overline{\phi}(x)$. By solving the 
equation of motion
\[ \frac{\ap}{2}\left(\frac{d\overline{\phi}}{dx}\right)^2=V_{\infty}(
\overline{\phi})=-\frac{1}{4}\overline{\phi}^2\ln\overline{\phi}^2 \]
with the boundary conditions 
\[ \lim_{x\to\pm\infty}\overline{\phi}(x)=0\ , \quad \overline{\phi}(x=0)
=1 \quad (V_{\infty}(\phi=1)=0), \]
we find
\[ \frac{x}{\sqrt{\ap}}=-\int_{\overline{\phi}(0)}^{\overline{\phi}(x)}
\frac{d\phi}{\sqrt{-\frac{1}{2}\phi^2\ln\phi^2}}=\int^1_{\overline{\phi}(x)}
d\phi\frac{d\ln\phi}{d\phi}\frac{1}{\sqrt{-\ln\phi}}=
2\sqrt{-\ln\overline{\phi}}, \]
or 
\begin{equation}
\overline{\phi}(x)=e^{-\frac{x^2}{4\ap}}. \label{eq:LR}
\end{equation}
That is, the lump profile is a simple Gaussian. Expanding $\phi(x^M)$ about 
this lump profile, the Schr\"{o}dinger equation for the fluctuation fields 
is found to become 
\begin{equation}
-\ap\frac{d^2}{dx^2}\xi_n(x)+\frac{x^2}{4\ap}\xi_n(x)=\left(\ap M_n^2+
\frac{3}{2}\right)\xi_n(x), \label{eq:LS}
\end{equation}
which is that of the simple harmonic oscillator potential. It is well known 
in quantum mechanics that the $n$-th eigenfunction of~(\ref{eq:LS}) is given 
by the product of the Gaussian factor $e^{-x^2/4\ap}$ and the $n$-th Hermite 
polynomial $H_n(x/\sqrt{2\ap})$, and that it belongs to the eigenvalue 
$n+\frac{1}{2}$. Then the mass spectrum is given by 
\begin{equation}
M_n^2=\frac{n-1}{\ap}\ , \quad n=0,1,2,\ldots \ . \label{eq:LT}
\end{equation}
It includes a tachyonic state of mass squared $m^2=-1/\ap$, a massless state, 
and infinitely many massive states, all of which are discrete. The mass 
spacing between two adjacent states is always equal to $1/\ap$. This spectrum 
reminds us of the spectrum of bosonic string theory: We started by 
considering the scalar field theory model~(\ref{eq:LJ}) with somewhat 
peculiar potential~(\ref{eq:LQ}) on a D($p+1$)-brane, constructed a lump 
solution~(\ref{eq:LR}), and found the spectrum~(\ref{eq:LT}) of the fields 
living on the lump, then the spectrum was similar to that of bosonic string 
theory. Hence this field theory model is very useful in studying tachyon 
condensation in string theory. Looking at the original action
\begin{equation}
S_{\infty}=4e\tau_{p+1}\int d^{p+2}\!x \left(-\frac{\ap}{2}\partial_M\phi 
\partial^M\phi+\frac{1}{4}\phi^2\ln\phi^2\right), \label{eq:LU}
\end{equation}
this field theory action actually coincides with the 
action~(\ref{eq:philogphi}) we encountered in section~\ref{sec:bsft} after 
rescaling $\phi^2\to\phi^2/8\ap e\ , \ \ x\to x/\sqrt{2}$. The 
action~(\ref{eq:philogphi}) was obtained there by truncating the full 
spacetime action for the tachyon up to two derivatives in background 
independent open string field theory (boundary string field theory). 
Though this truncation is an approximation, we can hope that the simple 
field theory model~(\ref{eq:LU}) captures some of the important features 
of tachyon condensation. 
\smallskip

Now we compute the tension $\cT_p$ of the lump~(\ref{eq:LR}). Since the 
action~(\ref{eq:LU}) on the D($p+1$)-brane is normalized such that 
\[ S_{\infty}(\phi=\phi_0)=-\tau_{p+1}V_{p+2}, \]
the lump tension is given by
\begin{equation}
\cT_p=-\frac{S_{\infty}(\overline{\phi})}{V_{p+1}}=4e\ap\tau_{p+1}
\int_{-\infty}^{\infty}dx\left(\frac{d\overline{\phi}}{dx}\right)^2
=\left(\frac{e}{\sqrt{2\pi}}\right)2\pi\sqrt{\ap}\tau_{p+1}. \label{eq:LV}
\end{equation}
We can easily extend the lump solution (\ref{eq:LR}) to higher codimension 
lumps. If we look for a spherically symmetric codimension $n$ lump on a 
D($p+1$)-brane, the equation of motion derived from~(\ref{eq:LU}) is 
\begin{equation}
\ap\frac{\partial^2\overline{\phi}}{\partial r^2}+\ap\frac{n-1}{r}
\frac{\partial\overline{\phi}}{\partial r}=V_{\infty}^{\prime}
(\overline{\phi})=-\frac{1}{2}\overline{\phi}(1+\ln\overline{\phi}^2), 
\label{eq:LW}
\end{equation}
where $r$ is the radial coordinate in the $n$-dimensional transverse space. 
This equation is solved by the following lump profile
\begin{equation}
\overline{\phi}(r)=e^{\frac{n-1}{2}}e^{-\frac{r^2}{4\ap}}. \label{eq:LX}
\end{equation}
The `initial value' $\overline{\phi}(0)$ is an increasing function of $n$, 
which is consistent with our intuition from the mechanical analogy mentioned 
in section~\ref{sec:HK}. The tension of the $(p+2-n)$-dimensional lump 
given by~(\ref{eq:LX}) is calculated as 
\begin{eqnarray}
\cT_{p+1-n}&=&-\frac{S_{\infty}(\overline{\phi})}{V_{p+2-n}}=-4e\tau_{p+1}
\int r^{n-1}dr\ d\Omega_{n-1}\left(-\frac{\ap}{2}\left(\frac{\partial
\overline{\phi}}{\partial r}\right)^2+\frac{1}{4}\overline{\phi}^2\ln
\overline{\phi}^2\right) \nonumber \\
&=&-e\tau_{p+1}\frac{2\pi^{n/2}}{\Gamma(\frac{n}{2})}\int_0^{\infty}dr\ 
r^{n-1}\left(n-1-\frac{r^2}{\ap}\right)e^{n-1}e^{-\frac{r^2}{2\ap}} \nonumber
\\ &=&2\pi^{n/2}e^n\tau_{p+1}\frac{\ap{}^{n/2}}{\Gamma(\frac{n}{2})}\times
\left\{
	\begin{array}{lcc}
	\sqrt{2\pi}(n-2)(n-4)\cdots 3\cdot 1 & \mbox{for odd} & n \\
	(n-2)(n-4)\cdots 4\cdot 2 & \mbox{for even} & n
	\end{array}
\right. \nonumber \\
&=&2^{\frac{n}{2}}e^n\pi^{\frac{n}{2}}\ap{}^{\frac{n}{2}}\tau_{p+1}=
\left\{\left(\frac{e}{\sqrt{2\pi}}\right)2\pi\sqrt{\ap}\right\}^n
\tau_{p+1}, \label{eq:LY}
\end{eqnarray}
where $\displaystyle \int d\Omega_{n-1}=\frac{2\pi^{n/2}}{\Gamma(
\frac{n}{2})}$ is the volume of the $(n-1)$-dimensional sphere $S^{n-1}$. 
The relation~(\ref{eq:LY}) between the lump tensions takes the same form as 
that for the D-brane tensions except for the factor $e/\sqrt{2\pi}\simeq 
1.08$. From the point of view of background independent open string field 
theory, this factor arises because we approximated the spacetime action for 
the tachyon field by ignoring the terms with more than two derivatives. As 
we saw in section~\ref{sec:bsft}\footnote{One may note that the 
formula~(\ref{eq:LY}) differs from the previous result~(\ref{eq:dis}) by a 
factor of $\sqrt{2}^{-n}$. This is attributed to the rescaling $x\to x/
\sqrt{2}$ mentioned below eq.(\ref{eq:LU}).}, if we treated the action 
exactly, we would obtain the exact relation~(\ref{eq:RandB}). The 
Schr\"{o}dinger equation for the fluctuation fields now reads
\[ -\ap \triangle\xi_m(x^i)+\frac{\sum_i(x^i)^2}{4\ap}\xi_m(x^i)=
\left(\ap M_m^2+\frac{n+2}{2}\right)\xi_m(x^i), \]
where $x^i=(x^{p+2-n},\cdots,x^{p+1})$ is the coordinate of the 
$n$-dimensional transverse space and $\displaystyle \triangle =\sum_i\left(
\frac{\partial}{\partial x^i}\right)^2$ is Laplacian in the transverse space. 
Since the potential is that of the $n$-dimensional spherically symmetric 
harmonic oscillator, the eigenfunctions are easily found. The spectrum 
includes a tachyon, $n$ massless scalars, and degenerate massive states. 
\medskip

Now let us consider the decay of the codimension one lump. Since we already 
know that the fluctuation field $\varphi(x,y)$ is expanded using the 
eigenfunctions of the Schr\"{o}dinger equation~(\ref{eq:LS}), the original 
tachyon field $\phi(x^M)$ can be written as 
\begin{equation}
\phi(x^M)=\overline{\phi}(x)+\varphi(x,y)=e^{-\frac{x^2}{4\ap}}+
\sum_{n=0}^{\infty}\eta_n(y)\frac{1}{2^{n/2}\sqrt{n!}}H_n\left(
\frac{x}{\sqrt{2\ap}}\right)e^{-\frac{x^2}{4\ap}}. \label{eq:LZ}
\end{equation}
As we saw in subsection~\ref{sub:Toy}, if $\overline{\varphi}(x,y\!\!\!
\backslash)$ denotes the configuration of $\varphi(x,y)$ which represents 
the decay of the lump, then $\overline{\phi}+\overline{\varphi}$ should be 
equal to 0 (minimum of the potential~(\ref{eq:LQ})) everywhere. But note 
$H_0(x/\sqrt{2\ap})=1$ so that the ground state wavefunction 
$H_0e^{-x^2/4\ap}$ is equal to the lump profile itself. Using the fact that 
the eigenfunctions $\displaystyle \xi_n(x)=\frac{1}{2^{n/2}\sqrt{n!}}H_n\left(
\frac{x}{\sqrt{2\ap}}\right)e^{-\frac{x^2}{4\ap}}$ are mutually orthogonal
\[\int_{-\infty}^{\infty}dx\ \xi_m(x)\xi_n(x)=\sqrt{2\pi\ap}\delta_{mn}, \]
the equation $\overline{\phi}+\overline{\varphi}=0$ determines the 
expectation values of the lump fields $\eta_n(y)$ as
\begin{equation}
1+\overline{\eta}_0=0 \ , \quad \overline{\eta}_n=0 \quad (n\ge 1). 
\label{eq:MA}
\end{equation}
That is, only the tachyonic ground state condenses while all the other modes 
remain zero expectation values. This is why we can draw some exact statements 
from the theory without relying upon the approximations such as level 
truncation. We have already seen it in section~\ref{sec:bsft}: There, we have 
concluded it for the reason that under the `free' perturbation $T(X)=a+uX^2$ 
the spacetime action contains higher modes at least quadratically. Here, we 
have obtained the results~(\ref{eq:MA}) by comparing the lump profile with 
the wavefunctions $\xi_n(x)$. Anyway, this property simplifies the analysis 
of how the spectrum changes as the tachyon condenses. By interpolating 
between the complete lump solution $\overline{\phi}(x)$ and the closed string 
vacuum 0 as 
\begin{equation}
\phi(x^M;\eta_0)=(1+\eta_0)\overline{\phi}(x)+\varphi(x,y), \label{eq:MB}
\end{equation}
where $\eta_0$ runs from 0 (lump) to $-1$ (closed string vacuum), the 
Schr\"{o}dinger equation gets changed into 
\begin{eqnarray*}
& &-\ap\frac{d^2}{dx^2}\xi_n(x)+V_{\infty}^{\prime\prime}\Bigl((1+\eta_0)
\overline{\phi}(x)\Bigr)\xi_n(x) \\
&=&-\ap\frac{d^2}{dx^2}\xi_n(x)+\left[-\frac{1}{2}(3+\ln\overline{\phi}^2)
-\ln(1+\eta_0)\right]\xi_n(x)=\ap M_n^2\xi_n(x).
\end{eqnarray*}
But one can rewrite it as 
\begin{equation}
-\ap\frac{d^2}{dx^2}\xi_n(x)+\frac{x^2}{4\ap}\xi_n(x)=\left(\ap M_n^2+
\frac{3}{2}+\ln(1+\eta_0)\right)\xi_n(x), \label{eq:MC}
\end{equation}
which differs from~(\ref{eq:LS}) only in the right hand side. Since the 
$n$-th eigenvalue is given by $n+\frac{1}{2}$ as ever, the mass squared of 
the $n$-th lump field is now
\[M_n^2=\frac{n-1}{\ap}-\frac{1}{\ap}\ln(1+\eta_0). \]
Hence the mass${}^2$ of every state blows up to $+\infty$ as $\eta_0\to -1$, 
irrespective of $n$. So there remain no physical states at all after the lump 
decays. This conclusion is of course consistent with the previous assertion 
that the mass of the tachyon field $\phi$ diverges when the original 
D($p+1$)-brane directly decays into the closed string vacuum without 
passing through the formation of the lump.
\smallskip

In section~\ref{sec:bsft}, we showed that the action for the tachyon field 
living on the tachyonic lump takes the same form as that of the 
tachyon on the D25-brane in the context of 
background independent open string field theory. Here we see it 
in the field theory model~\cite{11002}. We expand the original tachyon field 
$\phi(x^M)$ about the lump solution $\overline{\phi}(x)$ as in~(\ref{eq:LZ}), 
but we set to zero all modes other than the tachyon field $\eta_0(y)$ living 
on the lump. That is, we consider 
\begin{equation}
\phi(x^M)=\overline{\phi}(x)+\eta_0(y)e^{-x^2/4\ap}=(1+\eta_0(y))
e^{-x^2/4\ap}, \label{eq:MG}
\end{equation}
where we used the fact that the ground state wavefunction coincides with the 
lump profile $e^{-x^2/4\ap}$. Then the action~(\ref{eq:LU}) becomes 
\begin{eqnarray}
S_{\infty}&=&4e\left(\frac{e}{\sqrt{2\pi}}2\pi\sqrt{\ap}\tau_{p+1}\right)\int 
d^{p+1}\!y\left(-\frac{\ap}{2}\frac{(\partial_{\mu}\eta_0(y))^2}{e}+
\frac{1}{4}\frac{(1+\eta_0)^2}{e}\ln\frac{(1+\eta_0)^2}{e}\right) \nonumber \\
&=&4e\cT_p\int d^{p+1}\!y\left(-\frac{\ap}{2}\partial_{\mu}\widehat{\phi}
\partial^{\mu}\widehat{\phi}+\frac{1}{4}\widehat{\phi}^2\ln\widehat{\phi}^2
\right), \label{eq:MH}
\end{eqnarray}
where we defined 
\[ \widehat{\phi}(y)=\frac{1+\eta_0(y)}{\sqrt{e}} \]
and used the expression~(\ref{eq:LV}) for the tension of the lump. The 
action~(\ref{eq:MH}) precisely takes the same form as the original 
action~(\ref{eq:LU}) on the D($p+1$)-brane, except that the dimensionality 
of the brane is lowered by one (accordingly $M\to\mu$) and that the 
D($p+1$)-brane tension $\tau_{p+1}$ is replaced by the lump tension $\cT_p$. 
From this result, it is obvious that the tachyon field $\widehat{\phi}$ 
living on the lump has the same property as the original tachyon field 
$\phi$ on the D$(p+1)$-brane. In particular, they have the common mass 
$m^2=-1/\ap$. These facts strongly support the conjecture that the tachyonic 
lump solution on the original D-brane is identified with a lower dimensional 
D-brane, even if the relation~(\ref{eq:LY}) between the tensions of the lumps 
of various dimensionality slightly differs from the known relation between 
the D-brane tensions. 
\medskip

We consider here the translation of the lump in the $x$-direction by an 
infinitesimal amount $dx$. The shifted lump profile is expanded as 
\begin{eqnarray}
\overline{\phi}(x+dx)&=&\overline{\phi}(x)+\overline{\phi}^{\prime}(x)dx
=e^{-\frac{x^2}{4\ap}}-\frac{x}{2\ap}e^{-\frac{x^2}{4\ap}}dx \nonumber \\
&=&\overline{\phi}(x)-\frac{1}{2\ap}x\overline{\phi}(x)dx=\xi_0(x)-
\frac{1}{2\sqrt{\ap}}\xi_1(x)dx, \label{eq:MD}
\end{eqnarray}
where $\xi_n(x)$ is the $n$-th eigenfunction of the Schr\"{o}dinger 
equation~(\ref{eq:LS}) and $\xi_n\sim H_n(x/\sqrt{2\ap})e^{-x^2/4\ap}$. 
So the expectation value of the scalar field $\eta_1(y)$ living on the lump, 
associated with the wavefunction $\xi_1(x)$, represents infinitesimal 
translation of the lump\footnote{For finite translations, higher scalar 
modes must also be turned on: see section~\ref{sec:marginal}.}. 
\bigskip

Finally, we discuss the possibility of adding gauge fields to the scalar 
field theory model~(\ref{eq:LJ}). We show here three different types of gauge 
couplings. The first~\cite{8231} is inspired by the scalar-vector coupling in 
cubic string field theory. The others~\cite{11226} are of the Born-Infeld type 
and its variant, based on the results from boundary string field theory. Let 
us see them in turn. In cubic string field theory, there is a cubic coupling 
of the form $\phi A_{\mu}A^{\mu}$, as in~(\ref{eq:AM}). Since the potential 
$V_3(\phi)$ is cubic, $\phi$ can be written as $V_3^{\prime\prime}(\phi)-
V_3^{\prime\prime}(0)$ up to normalization, where `0' corresponds to the 
maximum of the potential. Hence, the cubic coupling term $(\phi-\phi_0)A_{\mu}
A^{\mu}$ in the $\ell =2$ model may be generalized to 
\begin{equation}
\biggl(V_{\ell+1}^{\prime\prime}(\phi)-V_{\ell+1}^{\prime\prime}(\phi_0)
\biggr)A_{\mu}A^{\mu} \label{eq:ME}
\end{equation}
in the model~(\ref{eq:LJ}) labeled by an arbitrary integer $\ell$. 
The potential 
has a maximum at $\displaystyle \phi=\phi_0=\left(\frac{\ell}{\ell+1}
\right)^{\ell/2}$ at which~(\ref{eq:ME}) vanishes so that the gauge field 
$A_{\mu}$ is guaranteed to be massless at the `open string perturbative 
vacuum'. Using the explicit formulae~(\ref{eq:LK}) and (\ref{eq:LL}), we find 
\[ V_{\ell+1}^{\prime\prime}(\phi)-V_{\ell+1}^{\prime\prime}(\phi_0)=
\frac{\ell+2}{\ell}\left(\frac{\ell}{2}-\frac{\ell+1}{2}\phi^{\ell/2}\right)
=\frac{\ell+2}{\ell}\frac{V_{\ell+1}^{\prime}(\phi)}{\phi}. \]
Combining this with (\ref{eq:ME}), we propose the gauge fixed action for the 
gauge field $A_M$ to be of the form 
\begin{equation}
S_{\ell+1}^A=\tau_{p+1}\int d^{p+2}\!x\left(-\frac{\ap}{2}\partial_MA_N
\partial^MA^N-\frac{1}{2}\frac{V_{\ell+1}^{\prime}(\phi)}{\phi}A_MA^M+\ldots
\right). \label{eq:MF}
\end{equation}
Note that the coefficient of the interaction term is different from the 
one~(\ref{eq:AM}) obtained in cubic string field theory even when $\ell=2$. 
This is because we are interested in an exactly solvable model, as we will 
see below. For simplicity, we consider the $\ell\to\infty$ model. To see the 
fluctuation spectrum of the gauge field $A_{\mu}(x^N)=\sum_n \cA_{\mu}^n(y)
\zeta_n(x)$ polarized \textit{along} the codimension 1 lump, substitute 
$\phi=\overline{\phi}(x)$ and look at the mass term for $\cA_{\mu}^n$, 
\[ -\frac{1}{2}\sum_{n,m}\zeta_m(x)\left(-\ap\frac{d^2}{dx^2}\zeta_n(x)+
\frac{V_{\infty}^{\prime}(\overline{\phi})}{\overline{\phi}}\zeta_n(x)\right)
\cA_{\mu}^m(y)\cA^{n\mu}(y). \]
The mass${}^2$ of the gauge field $\cA_{\mu}^n$ living on the lump is given 
by the eigenvalue of the following Schr\"{o}dinger equation
\begin{equation}
-\ap\frac{d^2}{dx^2}\zeta_n(x)+\left(\frac{x^2}{4\ap}-\frac{1}{2}\right)
\zeta_n(x)=\ap m_n^2\zeta_n(x). \label{eq:MFa}
\end{equation}
By solving the above equation, we find the eigenvalues 
$\ap m_n^2=n, \ n=0,1,2,\cdots$. 
That is, a \textit{massless} gauge field $\cA_{\mu}^0(y)$ as well as 
discrete massive gauge fields $\cA_{\mu}^n(y)$ lives on the lump 
world-volume. For a finite value of $\ell$, we also have a massless gauge 
field on the lump, as shown in~\cite{8231}. 

We should also consider the gauge field component polarized perpendicular 
to the lump. If there were any massless mode associated with it, such a 
zero mode would represent the collective motion of the lump. But the zero 
mode must not arise from the `bulk' gauge field $A_M$ because the translation 
mode whose wavefunction is the derivative of the lump profile has already 
been provided by the `bulk' tachyonic scalar field $\phi(x^M)$ as we saw 
before. The authors of~\cite{8231} have argued that the massless mode 
perpendicular to the lump is actually removed from the spectrum, but we do 
not show it here because there is not much evidence to justify the term 
that is required to obtain the desired result. 

To sum up, we consider the massless gauge field $A_M$ on the original 
D($p+1$)-brane and assume that the gauge field couples to the tachyonic 
scalar field $\phi$ through the coupling in~(\ref{eq:MF}). If we expand 
$\phi$ about the codimension 1 lump background $\overline{\phi}(x)$, we find 
that a massless gauge field $\cA_{\mu}^0(y)$ is induced on the lump 
world-volume from the original gauge field $A_{\mu}$, 
whereas a massless scalar mode representing the translation of the lump 
arises not from the transverse component $A_{p+1}$ of the bulk gauge field 
but from the bulk scalar field $\phi$.
\medskip

Now let us turn to the second possibility. Performing the field redefinition 
$\phi=\exp (-\frac{1}{4}T)$, the $\ell\to\infty$ action~(\ref{eq:LU}) is 
written as 
\begin{equation}
S_{\infty}=\frac{e\tau_{p+1}}{4}\int d^{p+2}\!x\ \exp\left(-\frac{T}{2}
\right)\left(
-\frac{\ap}{2}\partial_MT\partial^MT-2T\right). \label{eq:MR}
\end{equation}
In this variable, the codimension 1 lump solution is simply given by 
$\displaystyle \overline{T}(x)=\frac{x^2}{\ap}$. Here recall that for 
slowly varying gauge fields the spacetime action in background independent 
open string field theory can be written as the Born-Infeld 
form~(\ref{eq:Born}). By approximating it as 
\[ \sqrt{-\det(\eta_{\mu\nu}+F_{\mu\nu})}=1+\frac{1}{4}F_{\mu\nu}F^{\mu\nu}
+\cdots, \]
we propose the following coupling of the gauge field and tachyon,
\begin{equation}
S=\frac{e\tau_{p+1}}{4}\int d^{p+2}\!x\ e^{-\frac{T}{2}}\left(-\frac{\ap}{2}
\partial_MT\partial^MT-2T-2T\cdot \frac{1}{4}F_{MN}F^{MN}\right). 
\label{eq:MS}
\end{equation}
Expecting that the transverse component $A_{p+1}$ of the gauge field is 
absent in the massless sector, we choose the `axial gauge' $A_{p+1}=0$. 
If we define a new field variable as 
\begin{equation}
B_{\mu}(x,y)=\sqrt{\frac{e\tau_{p+1}}{2}}\sqrt{T}e^{-\frac{T}{4}}A_{\mu} 
\quad \mathrm{and} \quad \tilde{F}_{\mu\nu}=\partial_{\mu}B_{\nu}-
\partial_{\nu}B_{\mu}, \label{eq:MT}
\end{equation}
the $F_{MN}^2$ term in the action becomes
\begin{eqnarray}
S_B&=&-\frac{e\tau_{p+1}}{8}\int d^{p+1}\!y\ dx\ e^{-\frac{\overline{T}}{2}}
\overline{T}\left(2\left(\frac{\partial A_{\mu}}{\partial x}\right)^2+
(\partial_{\mu}A_{\nu}-\partial_{\nu}A_{\mu})^2\right) \nonumber \\
&=&\int d^{p+1}\!y\ dx\left[-\frac{1}{4}\tilde{F}_{\mu\nu}\tilde{F}^{\mu\nu}
-\frac{1}{2\ap}B_{\mu}\left(-\ap\frac{\partial^2}{\partial x^2}+\frac{x^2}
{4\ap}-\frac{3}{2}\right)B^{\mu}\right] \label{eq:MU}
\end{eqnarray}
around the lump solution $\overline{T}=x^2/\ap$. Hence, the fluctuation 
spectrum for $B_{\mu}$ on the lump is determined by the same Schr\"{o}dinger 
equation as that of the tachyon field, (\ref{eq:LS}). This means that the 
lowest eigenstate is tachyonic, the first excited state is massless, 
and so on. Though the existence of tachyonic mode seems problematic, we find 
this mode should be discarded by the following argument. The ground state 
wavefunction $\xi_0(x)$ is a simple Gaussian, which does not vanish at $x=0$. 
From the definition~(\ref{eq:MT}) of $B_{\mu}$, around the lump solution 
$\overline{T}=x^2/\ap$ we have $A_{\mu}\sim B_{\mu}/x$ for small $x$. Then 
$B_{\mu}(x=0)\neq 0$ means that the original gauge field $A_{\mu}$ is 
singular near the lump world-volume. Since the same situation occurs also for 
$n=\mathrm{even}=2m$ wavefunctions $\xi_{2m}(x)$, 
these modes should also be removed from 
the spectrum. As a result, the lowest mode happens to be massless. But 
another (though less troublesome) problem arises: The spectrum consists of 
states of mass levels $\ap m^2=0,2,4,\cdots$ because the states of $\ap m^2=
-1,1,3,\cdots$ have been discarded. Considering that the tachyon has mass 
squared $\ap m^2=-1$, the mass spacing between two nearby states is twice 
as large as the expected one in bosonic string theory. Although we have 
focused on the $\ell\to\infty$ model so far, the gauge coupling of the 
form~(\ref{eq:MS}) is also applicable to finite $\ell$ models. This fact 
and other things are explained in~\cite{11226} more generally than here. 
\smallskip

The third candidate for the gauge coupling resembles the second 
one~(\ref{eq:MS}), but a factor of $2T$ is absent. That is, we consider 
the following coupling 
\begin{equation}
S=-\frac{e\tau_{p+1}}{4}\int d^{p+2}\!x\ e^{-\frac{T}{2}}\frac{1}{4}F_{MN}
F^{MN}, \label{eq:MV}
\end{equation}
where we have omitted the tachyon kinetic and potential terms. This coupling 
is, of course, not derived from the Born-Infeld form as opposed 
to~(\ref{eq:MS}). Further, this coupling does not provide a solvable 
Schr\"{o}dinger problem for the gauge fluctuations in the \textit{finite} 
$\ell$ models. Nevertheless, we will consider the particular 
coupling~(\ref{eq:MV}) because the results obtained 
from it are preferable to the 
previous ones. In any case, we take the tachyon field $T$ to be the lump 
solution $\overline{T}(x)=x^2/\ap$ in the action~(\ref{eq:MV}) and define 
\begin{equation}
B_{\mu}(x,y)=\frac{\sqrt{e\tau_{p+1}}}{2}e^{-\frac{\overline{T}}{4}}A_{\mu}
\quad \mathrm{and} \quad \tilde{F}_{\mu\nu}=\partial_{\mu}B_{\nu}-
\partial_{\nu}B_{\mu}. \label{eq:MW}
\end{equation}
In the axial gauge $A_{p+1}=0$, the action~(\ref{eq:MV}) is rewritten as 
\begin{equation}
S_B=\int d^{p+1}\!y\ dx \left[-\frac{1}{4}\tilde{F}_{\mu\nu}\tilde{F}^{\mu\nu}
-\frac{1}{2\ap}B_{\mu}\left(-\ap\frac{\partial^2}{\partial x^2}+\left(
\frac{x^2}{4\ap}-\frac{1}{2}\right)\right)B^{\mu}\right]. \label{eq:MX}
\end{equation}
Solving the Schr\"{o}dinger equation, we find that the ground state 
wavefunction belongs to the eigenvalue $m^2=0$, hence there is 
no tachyonic mode. 
Moreover, since no modes need to be removed as is seen from the 
definition~(\ref{eq:MW}) of $B_{\mu}$, the mass spacing agrees with 
(the absolute value of) the mass squared of the tachyon, as desired for a 
bosonic string model. Incidentally, the Schr\"{o}dinger equation we are 
considering here  has exactly the same form as that 
in the first type of coupling, (\ref{eq:MFa}). 
\medskip

By imposing the axial gauge condition on the gauge field $A_M$ living on the 
D($p+1$)-brane, the transverse component $A_{p+1}$ is `lost' via 
localization. It is interesting to see that the massless translation mode of 
the $p$-brane lump, which we expect is provided by $A_{p+1}$ in the 
conventional D-brane 
picture, is instead filled in by the first excited massless scalar mode 
belonging to the tachyon tower, \textit{i.e.} the set of infinitely many 
scalar modes on the lump arising from the tachyon on the original 
D($p+1$)-brane. During this process, the number of physical degrees of 
freedom is preserved. It is explained in~\cite{11226} that this pattern also 
continues after including massive gauge and scalar fields as well as 
symmetric rank two tensor. 

\subsection{$\ell\to\infty$ field theory model for superstring tachyon}
We have seen in the preceding subsection that the $\ell\to\infty$ field 
theory model~(\ref{eq:LU}) for the tachyon dynamics constructed 
as the derivative 
truncated action of background independent bosonic open string field theory 
provides a useful description of the tachyonic lump solution and the 
string-like fluctuation spectrum on it. In this subsection, 
we will discuss the similar model representing the tachyon dynamics 
on a non-BPS D-brane of superstring theory~\cite{9246}.
\medskip

Expecting that a suitable field theory model can be obtained from the 
background independent open superstring field theory action~(\ref{eq:ZQ}) up 
to two derivatives, we consider the following action for the tachyon field 
$T(x)$ on a non-BPS D($p+1$)-brane, 
\begin{equation}
S(T)=8\tilde{\tau}_{p+1}\int d^{p+2}\!x\left(-\frac{\ap}{2}e^{-2T^2}\partial_M
T\partial^MT-\frac{1}{8}e^{-2T^2}\right). \label{eq:MI}
\end{equation}
This action is obtained from~(\ref{eq:ZQ}) after rescaling $T\to 2\sqrt{2}T,\ 
x^M\to\sqrt{2\ln 2}x^M$ and $(2\ln 2)^{(p+2)/2}\tilde{\tau}_{p+1}\to
\tilde{\tau}_{p+1}$. The above rescaling was taken for the following reason: 
By expanding the exponential in powers of $T^2$ around the unstable vacuum 
$T=0$ representing the D($p+1$)-brane, we get 
\[ S(T)=-\tilde{\tau}_{p+1}V_{p+2}+8\ap\tilde{\tau}_{p+1}\int d^{p+2}\!x\left(
-\frac{1}{2}\partial_MT\partial^MT+\frac{1}{4\ap}T^2\right)+\ldots. \]
The first constant term correctly reproduces the mass of the non-BPS 
D($p+1$)-brane, while the quadratic terms in $T$ shows that the tachyon field 
$T$ has the correct mass squared $m^2=-1/2\ap$. The exact tachyon potential 
$e^{-2T^2}$ is positive definite and has the stable minima at $T=\pm\infty$. 
To have the standard form for the kinetic term, we further perform a field 
redefinition
\begin{equation}
\phi=\mathrm{erf}(T)\equiv \frac{2}{\sqrt{\pi}}\int_0^Te^{-u^2}du, 
\label{eq:MJ}
\end{equation}
where `erf' is the error function which has the kink-like profile and 
erf$(\infty)=1$. Then the action~(\ref{eq:MI}) becomes
\begin{equation}
S(\phi)=2\pi\tilde{\tau}_{p+1}\int d^{p+2}\!x \left(-\frac{\ap}{2}\partial_M
\phi\partial^M\phi-\frac{1}{2\pi}\exp\left[-2(\mathrm{erf}^{-1}(\phi))^2
\right]\right), \label{eq:MK}
\end{equation}
where erf${}^{-1}(\phi)$ is the inverse function to erf (\ref{eq:MJ}) and 
defined for $-1\le \phi\le 1$ in a single-valued manner. The potential 
\begin{equation}
V(\phi)=\frac{1}{2\pi}\exp\left[-2(\mathrm{erf}^{-1}(\phi))^2\right]
\label{eq:ML}
\end{equation}
has a maximum (original non-BPS D($p+1$)-brane) at $\phi=0$ with 
$V(0)=1/2\pi$ and two minima (closed string vacuum) at $\phi=\pm 1$ with 
$V(\pm 1)=0$. 
\medskip

Now we look for a kink solution $\overline{\phi}(x)$. The equation of motion 
from~(\ref{eq:MK}) is 
\begin{equation}
\frac{\ap}{2}\left(\frac{d\overline{\phi}}{dx}\right)^2=V\left(\overline{\phi}
(x)\right)=\frac{1}{2\pi}\exp\left[-2(\mathrm{erf}^{-1}(\overline{\phi}))^2
\right], \label{eq:MM}
\end{equation}
where we assumed that the solution $\overline{\phi}$ depends on $x=x^{p+1}$ 
(its codimension must be 1 since it is to be a kink). This equation is 
solved by
\begin{equation}
\overline{\phi}(x)=\mathrm{erf}\left(\frac{x}{2\sqrt{\ap}}\right), 
\label{eq:MN}
\end{equation}
which obeys the suitable boundary conditions 
\[ \lim_{x\to\pm\infty}\overline{\phi}(x)=\pm 1 \ , \quad \overline{\phi}(-x)
=-\overline{\phi}(x) \]
for a kink solution with finite energy density. As usual, we expand the 
tachyon field $\phi(x^M)$ around the kink solution $\overline{\phi}(x)$ as 
$\phi(x^M)=\overline{\phi}(x)+\varphi(x,y^{\mu})$. This gives
\begin{eqnarray}
S(\phi)&=&2\pi\tilde{\tau}_{p+1}\int d^{p+1}\!y\ dx \Biggl[\left\{
-\frac{\ap}{2}\left(\frac{d\overline{\phi}}{dx}\right)^2-V(\overline{\phi})
\right\}-\frac{\ap}{2}\partial_{\mu}\varphi\partial^{\mu}\varphi \nonumber \\
& &-\frac{1}{2}\varphi\left(-\ap\frac{\partial^2\varphi}{\partial x^2}+
V^{\prime\prime}(\overline{\phi})\varphi\right)+\cdots\Biggr] \label{eq:MO}
\end{eqnarray}
as in (\ref{eq:LO}). The tension $\cT_p$ of the kink is given by the 
$\varphi$-independent term as
\begin{eqnarray}
\cT_p&=&2\pi\tilde{\tau}_{p+1}\int dx\ 2V(\overline{\phi}(x))=2
\tilde{\tau}_{p+1}\int_{-\infty}^{\infty}dx\ e^{-x^2/2\ap} \nonumber \\
&=&\frac{2}{\sqrt{\pi}}\cdot 2\pi\sqrt{\ap}\frac{\tilde{\tau}_{p+1}}{\sqrt{2}}
\simeq 1.13 \tau_p, \label{eq:MP}
\end{eqnarray}
where $\tau_p$ is the BPS D$p$-brane tension. That is, the tension $\cT_p$ of 
the ($p+1$)-dimensional kink solution approximately reproduced the tension 
$\tau_p$ of the BPS D$p$-brane. This fact supports the conjecture that the 
kink solution on a non-BPS D($p+1$)-brane represents a BPS D$p$-brane. 
Actually, we have already seen the above result in the context of background 
independent open superstring field theory in section~\ref{sec:aoitori}. The 
result~(\ref{eq:ZS}) there of course agrees with~(\ref{eq:MP}) if we perform 
the above mentioned rescaling $\sqrt{2\ln 2}\tilde{\tau}_9\to\tilde{\tau}_9$. 
We turn to the fluctuation spectrum on the kink. From the equation of motion 
for $\overline{\phi}$, we get 
\[ V^{\prime}(\overline{\phi})=\ap\frac{d^2\overline{\phi}}{dx^2} \quad 
\mathop{\longrightarrow}\limits^{d/dx} 
\quad V^{\prime\prime}(\overline{\phi})
\frac{d\overline{\phi}}{dx}=\ap\frac{d^3\overline{\phi}}{dx^3}. \]
Since \[ \frac{d\overline{\phi}}{dx}=\frac{d}{dx}\frac{2}{\sqrt{\pi}}
\int_0^{x/2\sqrt{\ap}}e^{-u^2}du=\frac{1}{\sqrt{\ap\pi}}e^{-\frac{x^2}{4\ap}}
\] and \[ \frac{d^3\overline{\phi}}{dx^3}=\frac{1}{\sqrt{\pi\ap}}
e^{-\frac{x^2}{4\ap}}\left(\frac{x^2}{4\ap{}^2}-\frac{1}{2\ap}\right), \]
we find \[ V^{\prime\prime}\left(\overline{\phi}(x)\right)=\frac{x^2}{4\ap}
-\frac{1}{2}. \]
Expanding the fluctuation field as $\varphi=\sum_{n=0}^{\infty}\phi_n(y)
\xi_n(x)$, the mass squared of the field $\phi_n(y)$ living on the kink 
world-volume is determined by the Schr\"{o}dinger equation 
\begin{equation}
-\ap\frac{d^2}{dx^2}\xi_n(x)+\left(\frac{x^2}{4\ap}-\frac{1}{2}\right)
\xi_n(x)=\ap M^2_n\xi_n(x), \label{eq:MQ}
\end{equation}
and its eigenvalue is given by $M_n^2=n/\ap\ , \quad n=0,1,2,\cdots$. That 
is, the tachyonic mode is absent and the lightest state is massless, and 
discrete massive states are equally spaced. Note that the ground state 
wavefunction \[ \xi_0(x)=(2\pi\ap)^{-\frac{1}{4}}e^{-\frac{x^2}{4\ap}} \] 
is now proportional to the \textit{derivative} of the kink profile 
$\overline{\phi}=\mathrm{erf}(x/2\sqrt{\ap})$, which means that the kink 
field $\phi_0$ associated with $\xi_0(x)$ represents the translation mode of 
the kink. This is consistent with the masslessness of $\phi_0(y)$. 
Furthermore, the absence of the tachyonic mode indicates that the resulting 
kink configuration is stable against the decay. This fact strongly supports 
the identification of the kink with a BPS D$p$-brane.
\medskip

Though we started with the derivative truncated action~(\ref{eq:MI}) which 
had already been obtained from boundary string field theory~\cite{KMM2}, 
we can instead reconstruct the field theory action from a given kink 
profile~\cite{9246}. Now we describe how to do that. First of all, we assume 
that we are given a kink profile
\begin{equation}
\overline{\phi}(x)=\cK(x/\sqrt{\ap}). \label{eq:MY}
\end{equation}
`Kink' means that $\cK$ is a monotonically increasing or monotonically 
decreasing function of $x/\sqrt{\ap}$ so that $d\cK/dx$ never vanishes. 
If $\cK$ solves the equation of motion derived from the action
\begin{equation}
S=\cT\int d^dx\left(-\frac{\ap}{2}\partial_{\mu}\phi\partial^{\mu}\phi
-V(\phi)\right), \label{eq:MZ}
\end{equation}
then the potential $V(\phi)$ satisfies 
\begin{equation}
V(\overline{\phi}(x))=\frac{\ap}{2}\left(\frac{d\overline{\phi}}{dx}\right)^2
=\frac{1}{2}\left(\cK^{\prime}\left(\frac{x}{\sqrt{\ap}}\right)\right)^2.
\label{eq:NA}
\end{equation}
Making use of the function $\cK$, we perform a field redefinition 
\begin{equation}
\phi(x)=\cK(T(x)) \quad \mathrm{or} \quad T(x)=\cK^{-1}(\phi(x)), 
\label{eq:NB}
\end{equation}
where the monotonicity of $\cK$ guarantees the single-valuedness of 
$\cK^{-1}$. In terms of $T$, the kink profile is 
\[ \overline{T}(x)=\cK^{-1}(\overline{\phi}(x))=\cK^{-1}\left(\cK\left(
\frac{x}{\sqrt{\ap}}\right)\right)=\frac{x}{\sqrt{\ap}}, \]
and the potential is rewritten as 
\begin{equation}
V(\overline{\phi}(x))=\frac{1}{2}\left(\cK^{\prime}\left(\frac{x}{\sqrt{\ap}}
\right)\right)^2=\frac{1}{2}\left(\cK^{\prime}(\overline{T})\right)^2. 
\label{eq:NC}
\end{equation}
The kinetic term now becomes 
\begin{equation}
(\partial_{\mu}\phi)^2=\Bigl(\partial_{\mu}\cK(T(x))\Bigr)^2=\left(
\cK^{\prime}(T)\right)^2(\partial_{\mu}T)^2. \label{eq:ND}
\end{equation}
Plugging (\ref{eq:NC}) and (\ref{eq:ND}) into (\ref{eq:MZ}), we eventually 
get the action
\begin{equation}
S=\cT\int d^dx \left(\cK^{\prime}(T)\right)^2\left(-\frac{\ap}{2}\partial_{\mu}
T\partial^{\mu}T-\frac{1}{2}\right). \label{eq:NE}
\end{equation}
If we choose $\cK^{\prime}(T)=e^{-T^2}$ and rescale $x^{\mu}$, we correctly 
recover the action~(\ref{eq:MI}). In fact, the function 
$\displaystyle \cK(\overline{T})=\int^{\overline{T}}e^{-u^2}du\propto
\mathrm{erf}(\overline{T})$ is nothing but 
the kink solution~(\ref{eq:MN}) we found previously. From now on we consider 
the action on a non-BPS D($p+1$)-brane normalized as 
\begin{equation}
S(T)=8\tilde{\tau}_{p+1}\int d^{p+2}\!x (\cK^{\prime}(T))^2\left(-
\frac{\ap}{2}\partial_MT\partial^MT-\frac{1}{8}\right). \label{eq:NF}
\end{equation}
Recalling the previous results, the tachyon $T$ has mass squared $m^2=-
\frac{1}{2\ap}$ and the fluctuation spectrum about the kink solution 
$\overline{T}(x)=\frac{x}{2\sqrt{\ap}}$ contains states of mass${}^2$ 
$M_n^2=n/\ap\ , \quad n=0,1,2,\cdots$, in particular there is no tachyonic 
state on the kink which we want to identify with a BPS D$p$-brane. Note that 
the mass spacing $\Delta m^2=1/\ap$ is twice the magnitude of the mass 
squared of the original tachyon, which agrees with the GSO-projected spectrum 
of the superstring theory. Now we incorporate the gauge field into 
the tachyonic scalar field theory~(\ref{eq:NF}). 
To do so, we again assume the Born-Infeld form
\begin{eqnarray}
\hspace{-1cm}
S(T,A)&=&8\tilde{\tau}_{p+1}\int d^{p+2}\!x (\cK^{\prime}(T))^2\left(-
\frac{\ap}{2}\partial_MT\partial^MT-\frac{1}{8}\right)\sqrt{-\det(\eta_{MN}
+F_{MN})} \nonumber \\ &=& 8\tilde{\tau}_{p+1}\int d^{p+2}\!x (\cK^{\prime}
(T))^2\left(-\frac{\ap}{2}\partial_MT\partial^MT-\frac{1}{8}-\frac{1}{32}
F_{MN}F^{MN}+\cdots\right). \label{eq:NG}
\end{eqnarray}
As in the bosonic case, we choose the axial gauge $A_{p+1}=0$ and define 
a new variable by 
\[ B_{\mu}(x,y)=\cK^{\prime}(\overline{T})A_{\mu} \quad \mathrm{and} \quad 
\tilde{F}_{\mu\nu}=\partial_{\mu}B_{\nu}-\partial_{\nu}B_{\mu}. \] 
Then the $F_{MN}^2$ term becomes 
\begin{eqnarray}
S_B(\overline{T},B)&=&\tilde{\tau}_{p+1}\int d^{p+1}\!y\ dx\Biggl[-\frac{1}{4}
\tilde{F}_{\mu\nu}\tilde{F}^{\mu\nu} \nonumber \\
& &-\frac{1}{2\ap}B^{\mu}\left(-\ap\frac{\partial^2}{\partial x^2}+\frac{\cK^{
\prime\prime\prime}(x/2\sqrt{\ap})}{4\cK^{\prime}(x/2\sqrt{\ap})}\right)
B_{\mu}\Biggr]. \label{eq:NH}
\end{eqnarray}
It can be applied to the finite $\ell$ models by taking 
\[ \cK^{\prime}_{\ell}(x/2\sqrt{\ap})=\sech^{\ell}(x/2\sqrt{\ap}). \]
In this case, the Schr\"{o}dinger potential 
\[ \frac{\cK^{\prime\prime\prime}_{\ell}(x/2\sqrt{\ap})}{4\cK^{\prime}_{\ell}
(x/2\sqrt{\ap})}=\frac{\ell^2}{4}-\frac{\ell(\ell+1)}{4}\sech^2\left(\frac{x}
{2\sqrt{\ap}}\right) \]
is just the $\ell$-th exactly solvable reflectionless 
potential~(\ref{eq:LI}). In the $\ell\to\infty$ model, in which we choose 
$\cK^{\prime}(T)=e^{-T^2}$ as remarked above, we obtain 
\[ \frac{\cK^{\prime\prime\prime}(x/2\sqrt{\ap})}{4\cK^{\prime}
(x/2\sqrt{\ap})}=\frac{x^2}{4\ap}-\frac{1}{2} \]
which is again the harmonic oscillator potential and the Schr\"{o}dinger 
equation precisely agrees with~(\ref{eq:MQ}) for the scalar fluctuations. 
So there are gauge fields with mass squared $\ap M_n^2=n \quad (n=0,1,2,
\cdots)$ on the kink world-volume, the mass spacing being twice the mass 
squared of the original tachyon. Though we partially fixed the gauge by 
imposing the axial gauge condition $A_{p+1}=0$, the action~(\ref{eq:NG}) on 
the D($p+1$)-brane still has a residual gauge invariance under 
\begin{equation}
A_{\mu}(x^M)\longrightarrow A_{\mu}(x^M)+\partial_{\mu}\varepsilon(y) 
\label{eq:NI}
\end{equation}
if $\partial\varepsilon /\partial x^{p+1}=0$ holds (this condition is 
necessary for $F_{p+1,\mu}=\partial A_{\mu}/\partial x^{p+1}$ 
to be invariant). This residual 
gauge symmetry is succeeded to by the \textit{massless} gauge field on the 
kink world-volume. This fact is immediately seen from the 
action~(\ref{eq:NH}). Since the mass term vanishes for the massless gauge 
field, the Lagrangian is simply $-\frac{1}{4}\tilde{F}_{\mu\nu}
\tilde{F}^{\mu\nu}$, which is manifestly invariant under 
\begin{equation}
B_{\mu}(y)\longrightarrow B_{\mu}(y)+\partial_{\mu}\varepsilon(y). 
\label{eq:NJ}
\end{equation}
This can be recast as follows: In terms of the newly defined $B_{\mu}$ field, 
the gauge transformation~(\ref{eq:NI}) has the form 
\begin{equation}
B_{\mu}(x,y)\longrightarrow B_{\mu}(x,y)+\cK^{\prime}(x/2\sqrt{\ap})
\partial_{\mu}\varepsilon(y). \label{eq:NK}
\end{equation}
Now we decompose $B_{\mu}(x,y)$ into the product of the field 
$\mathcal{B}_{\mu}(y)$ living on the kink world-volume and the eigenfunction 
$\xi(x)$ of the Schr\"{o}dinger equation (we concentrate only on the massless 
field). But noticing that the massless ground state wavefunction has the same 
form as the derivative $\cK^{\prime}$ of the kink profile, the decomposition 
becomes \[ B_{\mu}(x,y)=\mathcal{B}_{\mu}(y)\cK^{\prime}(x/2\sqrt{\ap}). \]
Substituting it into (\ref{eq:NK}), we eventually find the gauge 
transformation of the massless gauge field living on the kink to be 
\[ \mathcal{B}_{\mu}(y)\longrightarrow \mathcal{B}_{\mu}(y)+\partial_{\mu}
\varepsilon(y), \]
which exactly coincides with (\ref{eq:NJ}). In other words, due to the 
special property that the ground state wavefunction agrees with the 
derivative of the kink profile, the definition combined with the 
decomposition 
\[ \cK^{\prime}(x/2\sqrt{\ap})A_{\mu}\equiv B_{\mu}(x,y)=\mathcal{B}_{\mu}
(y)\cK^{\prime}(x/2\sqrt{\ap}) \]
leads to $A_{\mu}=\mathcal{B}_{\mu}(y)$. From this expression, we can easily 
see that the massless gauge field $\mathcal{B}_{\mu}$ on the kink inherits 
the gauge transformation property from the bulk gauge field $A_{\mu}$. 
On the other hand, the \textit{massive} gauge fields are gauge-fixed by 
the Lorentz-type gauge condition $\partial_{\mu}\mathcal{B}^{\mu(n)}(y)=0$ 
\cite{11226}, as is seen from the action~(\ref{eq:NH}). 
\bigskip

It was also considered in~\cite{11226} to incorporate the massless fermion 
field into the tachyon field theory model on a non-BPS D9-brane of Type IIA. 
Since there are both GSO(+) and GSO($-$) sectors on a non-BPS D-brane, 
a ten-dimensional massless Majorana fermion (or two Majorana-Weyl fermions 
of opposite chirality) has 16 real degrees of freedom after imposing the 
physical condition (Dirac equation). In contrast, the massless bosonic 
content is the gauge field $A_{\mu}$ transforming as an 8-dimensional vector 
representation under the little group $SO(8)$ of the full Lorentz group 
$SO(9,1)$. If the kink configuration of the tachyon on the non-BPS D9-brane 
is to describe a supersymmetric BPS D8-brane, then the spectrum on the kink 
world-volume must have the Bose-Fermi degeneracy at least at the massless 
level. We will explain below how the number of fermionic degrees of freedom 
is reduced to 8. We begin by considering the following coupling of the 
fermion $\psi$ to the tachyon $T$, 
\begin{equation}
S=-\tilde{\tau}_9\int d^{10}\!x (\cK^{\prime}(T))^2\left[\frac{i\sqrt{\ap}}{2}
\overline{\psi}\Gamma^M\mathop{\partial}\limits^{\leftrightarrow}{}_M\psi+
W(T)\overline{\psi}
\psi\right], \label{eq:NL}
\end{equation}
where $\displaystyle \overline{\psi}=\psi^{\dagger}\Gamma^0, \ 
f\mathop{\partial}\limits^{\leftrightarrow}g\equiv f(\partial g)-(\partial f)
g,\ W(T)=-\frac{\cK^{\prime\prime}(T)}{2\cK^{\prime}(T)}$ and 10-dimensional 
gamma matrices $\Gamma^M$ are constructed from the 9-dimensional gamma 
matrices $\gamma^{\mu}$ as 
\begin{equation}
\Gamma^9=i\left(
	\begin{array}{cc}
	0 & I_{16} \\ I_{16} & 0
	\end{array}
\right) \ , \quad \Gamma^{\mu}=\left(
	\begin{array}{cc}
	\gamma^{\mu} & 0 \\ 0 & -\gamma^{\mu}
	\end{array}
\right) \quad \mu=0,\cdots,8. \label{eq:NM}
\end{equation}
One can easily verify that $\{\Gamma^M,\Gamma^N\}=-2\eta^{MN}$ holds if 
$\{\gamma^{\mu},\gamma^{\nu}\}=-2\eta^{\mu\nu}$, where $\eta^{MN}=
\mathrm{diag}(-1,1,\cdots,1)$. We set $x=x^9, \ y^{\mu}=(x^0,\cdots,x^8)$. 
As before, we perform a field redefinition 
\[ \chi(x,y)=\cK^{\prime}(T)\psi . \]
The action (\ref{eq:NL}) can be rewritten in terms of $\chi$ as
\begin{eqnarray}
S&=&-\tilde{\tau}_9\int d^9y\ dx\left(i\sqrt{\ap}\overline{\chi}\Gamma^M
\partial_M\chi+W(T)\overline{\chi}\chi\right) \label{eq:NN} \\ &=&
-\tilde{\tau}_9\int d^9y\ dx\left[i\sqrt{\ap}\overline{\chi}\Gamma^{\mu}
\partial_{\mu}\chi+\overline{\chi}\left(
	\begin{array}{cc}
	W(T) & -\sqrt{\ap}\partial/\partial x \\ -\sqrt{\ap}\partial/\partial x & 
	W(T)
	\end{array}
\right)\chi\right]. \nonumber 
\end{eqnarray}
To see the fluctuation spectrum of the fermion field around the kink 
background, we put $\displaystyle T=\overline{T}=\frac{x}{2\sqrt{\ap}}$. And 
we decompose the ten-dimensional 32 component Majorana spinor $\chi$ into two 
16 component Majorana-Weyl spinors $\chi_1,\chi_2$ as 
$\displaystyle \chi=\left(
	\begin{array}{c}
	\chi_1 \\ \chi_2
	\end{array}
\right)$. If $\chi$ satisfies the following eigenvalue equation 
\begin{equation}
\left(
	\begin{array}{cc}
	W(\overline{T}) & -\sqrt{\ap}\partial/\partial x \\ -\sqrt{\ap}\partial/
	\partial x & W(\overline{T})
	\end{array}
\right)\left(
	\begin{array}{c}
	\chi_1 \\ \chi_2
	\end{array}
\right)=\sqrt{\ap}m\left(
	\begin{array}{c}
	\chi_1 \\ -\chi_2
	\end{array}
\right), \label{eq:NO}
\end{equation}
the action (\ref{eq:NN}) becomes 
\[ S=-\tilde{\tau}_9\int d^9y\ dx\sum_{i=1}^2\sqrt{\ap}\Bigl(i
\overline{\chi}_i\gamma^{\mu}\partial_{\mu}\chi_i+m\overline{\chi}_i\chi_i
\Bigr),\]
where $\overline{\chi}_i=\chi^{\dagger}_i\gamma^0$. Hence the $-$ sign on the 
right hand side of~(\ref{eq:NO}) is needed. If we take the linear 
combinations $\chi_{\pm}=\chi_1\pm\chi_2$, two equations~(\ref{eq:NO}) are 
written as 
\[ \left\{
	\begin{array}{l}
	\ap{}^{-1/2}W(\overline{T})\chi_+-\partial\chi_+/\partial x=m\chi_- \\
	\ap{}^{-1/2}W(\overline{T})\chi_-+\partial\chi_-/\partial x=m\chi_+
	\end{array}
\right. . \]
Combining these two equations, we get the Schr\"{o}dinger equations
\begin{equation}
\left(-\ap\frac{\partial^2}{\partial x^2}+W^2\left(\frac{x}{2\sqrt{\ap}}
\right)\pm\frac{1}{2}W^{\prime}\left(\frac{x}{2\sqrt{\ap}}\right)-\ap m^2
\right)\chi_{\pm}=0. \label{eq:NP}
\end{equation}
Though it can be applied to the finite $\ell$ models, we consider the $\ell\to
\infty$ case only. Using the kink profile $\cK^{\prime}(T)=e^{-T^2}$, we can 
compute 
\begin{eqnarray*}
W(\overline{T})&=&-\frac{\cK^{\prime\prime}(\overline{T})}{2\cK^{\prime}
(\overline{T})}=\overline{T}=\frac{x}{2\sqrt{\ap}}, \\
W^{\prime}(\overline{T})&=&-\frac{\cK^{\prime\prime\prime}(\overline{T})}
{2\cK^{\prime}(\overline{T})}+\frac{1}{2}\left(\frac{\cK^{\prime\prime}
(\overline{T})}{\cK^{\prime}(\overline{T})}\right)^2=1.
\end{eqnarray*}
Focusing on the single mode $\chi_{\pm}(x,y)=\zeta_{\pm}(y)\xi_{\pm}(x)$, 
the equations~(\ref{eq:NP}) now read
\begin{equation}
\left[-\ap\frac{d^2}{dx^2}+\left(\frac{x^2}{4\ap}\pm\frac{1}{2}\right)\right]
\xi_{\pm}(x)=\ap m^2\xi_{\pm}(x). \label{eq:NQ}
\end{equation}
For the $\xi_-$ modes, mass eigenvalues are $\ap m^2=n=0,1,2,\cdots$ so that 
the ground state is massless, as usual. However, since the harmonic oscillator
potential is shifted by one unit for $\xi_+$ modes, the resulting mass 
eigenvalues are also shifted to $\ap m^2=n+1=1,2,3,\cdots$. Therefore, no 
massless fields are induced by the localization of $\chi_+$ spinor on the 
kink world-volume. After all, the number of physical massless 
fermionic degrees of 
freedom gets reduced to 8 (half of original 16) on the kink. For the sake 
of confirmation, let us count the number of bosonic degrees of freedom as 
well. The original gauge field $\mathbf{8}$ of $SO(8)$ on the non-BPS 
D9-brane gives rise to the massless gauge field transforming as $\mathbf{7}$ 
under $SO(7)$ after being localized on the kink world-volume, as discussed 
before. The `lost' eighth component, which represents the translation mode 
of the kink, is provided by the lowest massless scalar mode arising from the 
localization of the original tachyon field $T$. In this way, we have seen 
that the Bose-Fermi degeneracy is achieved on the kink world-volume by 
considering the couplings~(\ref{eq:NG}), (\ref{eq:NL}) of the gauge field and 
fermion to the tachyon, when restricting ourselves to the massless sector. 
Although this result is necessary to identify the kink with the 
supersymmetric BPS D8-brane, it is not sufficient at all. In particular, 
in order to see whether the Bose-Fermi degeneracy is extended to all the 
massive states we must consider the full string field theory on the 
non-BPS D9-brane. 
\bigskip

We close this section by giving a brief comment on the extension of the 
tachyon field theory model to include higher derivative terms~\cite{11226}. 
As one possibility, the following `modified Born-Infeld' 
action~\cite{3122,pyo}
\begin{eqnarray}
S&=&-\tilde{\tau}_{p+1}\int d^{p+2}\!x\ U(T)\sqrt{-\det(\eta_{MN}+F_{MN}+8\ap
\partial_MT\partial_NT)}, \label{eq:NR} \\
U(T)&=&e^{-2T^2}, \nonumber
\end{eqnarray}
was considered. It was shown that this action has the following pleasant 
features:
\begin{itemize}
\item The kink-like profile $T(x)=ux$ of the tachyon field solves the 
equation of motion derived from~(\ref{eq:NR}) if the value of $u$ is pushed 
to infinity, just like the exact solution in boundary string field theory. 
\item The open string tachyon on the original non-BPS D($p+1$)-brane has 
mass \\ $m^2=-1/2\ap$. 
\item When expanding the tachyon field about the kink profile as $T=ux+
\tilde{T}$, the fluctuation spectrum of $\tilde{T}$ on the kink world-volume 
shows that the mass spacing between two neighboring states is $\Delta m^2=
1/\ap$, which is twice the absolute value of the tachyon mass squared. 
\end{itemize}
However, this model does not give the perfect description of the system:
\begin{itemize}
\item The ratio of the tension is wrong, 
\[ \frac{\cT_8}{\tilde{\tau}_9}=2\sqrt{\pi\ap}\neq 2\pi\sqrt{\ap}\frac{1}
{\sqrt{2}}. \]
\item When one examines the gauge fluctuations about the kink background by 
expanding the square root of determinant to second order in $F_{MN}$, he 
finds that the mass spacing is given by $\Delta m^2=1/2\ap$ in the limit 
$u\to\infty$, which is not the expected value $(1/\ap)$, though finite. 
\end{itemize}
From these results, it may be that although the proposed 
form~(\ref{eq:NR}) of the action is simple and interesting, the full boundary 
string field theory action for the tachyon field ought to have more 
complicated structure.

\section{Large Marginal Deformations}\label{sec:marginal}

\begin{figure}[htbp]
\begin{center}
	\includegraphics{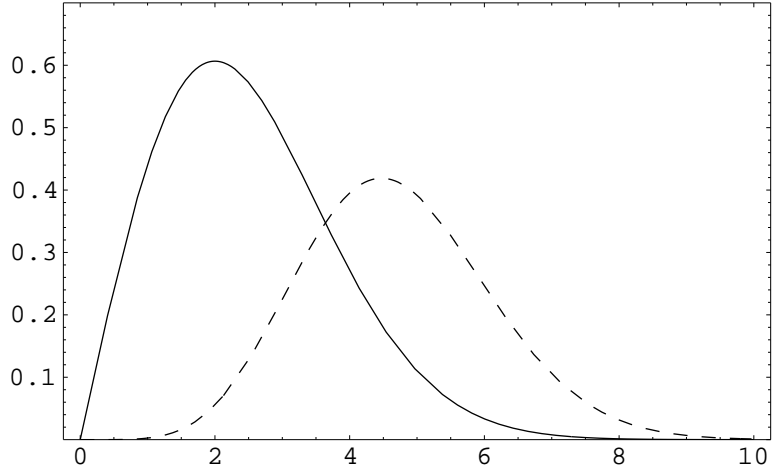}
	\caption{Plot of $\eta_1$ (solid line) and $\eta_5$ (dashed line) as 
	functions of the deformation parameter $a$.}
	\label{fig:EA}
\end{center}
\end{figure}

\begin{figure}[htbp]
\begin{center}
	\includegraphics{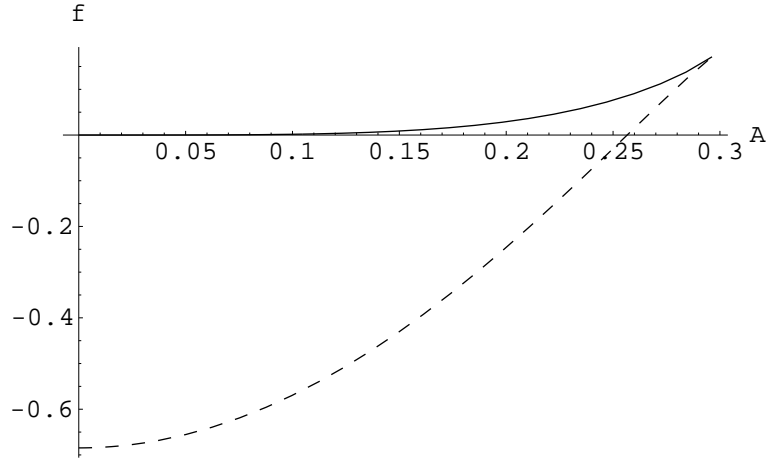}
	\caption{The effective potential for $A_{25}$ at level (1,2). The solid 
	line represents the $M$-branch, while the dashed line the $V$-branch.}
	\label{fig:EB}
\end{center}
\end{figure}

In subsection~\ref{sub:ell}, we saw that the expectation value of the 
massless scalar field $\eta_1(y)$ living on the tachyonic lump, associated 
with the wavefunction $\xi_1(x)$ which is the derivative of the lump profile 
$\overline{\phi}(x)$, represents the infinitesimal translation of the lump 
(eq. (\ref{eq:MD})). In order to implement finite translations, however, not 
only the massless transverse scalar mode but also other massive fields must 
be given nonzero expectation values in string field theory. This fact can 
easily be seen in the $\ell\to\infty$ field theory model~\cite{8231}. When 
we consider displacing the lump $\overline{\phi}(x)=e^{-\frac{x^2}{4\ap}}$ 
centered on $x=0$ by a distance $a$, we write the shifted lump profile 
$\overline{\phi}(x-a)$ as a superposition of the wavefunctions centered at 
$x=0$, 
\begin{eqnarray}
\overline{\phi}(x-a)&=&e^{-\frac{(x-a)^2}{4\ap}}=\sum_{n=0}^{\infty}
\frac{(-a)^n}{n!}\frac{d^n}{dx^n}\overline{\phi}(x) \nonumber \\
&=&\overline{\phi}(x)+\sum_{n=0}^{\infty}\eta_n\xi_n(x), \label{eq:MDA}
\end{eqnarray}
where the wavefunctions 
\[ \xi_n(x)=\frac{1}{2^{n/2}\sqrt{n!}}H_n\left(\frac{x}{\sqrt{2\ap}}\right)
e^{-\frac{x^2}{4\ap}} \qquad \Bigl(\overline{\phi}(x)=\xi_0(x)\Bigr) \]
are normalized as 
\[ \int^{\infty}_{-\infty}dx\ \xi_m(x)\xi_n(x)=\sqrt{2\pi\ap}\delta_{nm}, \]
as mentioned earlier. Due to this orthogonality, we can determine the 
coefficients $\eta_n$'s (constant fields on the lump) as 
\begin{eqnarray*}
1+\eta_0&=&e^{-\frac{a^2}{8\ap}}, \\ \eta_n&=&\frac{1}{2^n\sqrt{n!}}\left(
\frac{a}{\sqrt{\ap}}\right)^ne^{-\frac{a^2}{8\ap}} \quad \mathrm{for} \ \ 
n\ge 1.
\end{eqnarray*}
These results clarify that all $\eta_n$'s must acquire nonvanishing values 
to shift the lump by a finite distance $a$. In particular, we should note 
that the zero mode $\eta_1$, which is responsible for an infinitesimal 
displacement, has a finte range of values: the `string field theory marginal 
deformation parameter' $\eta_1$ is plotted in Figure~\ref{fig:EA} as a 
function of the `CFT marginal deformation parameter' $a$. When we increase 
$a$ beyond the critical value $2\sqrt{\ap}$, $\eta_1$ begins to decrease 
and higher level fields are turned on. For example, we also plotted $\eta_5$ 
in Figure~\ref{fig:EA}, which becomes large around $a\simeq\sqrt{20\ap}$. 
Such a behavior of the marginal deformation parameter 
(\textit{i.e.} massless field) as being restricted to the finite range of 
values, was in fact observed in bosonic cubic string field 
theory~\cite{7153}. To see this, we consider the following string field 
\begin{equation}
|\Phi\rangle=(\phi+A_{25}\alpha_{-1}^{25})c_1|0\rangle, \label{eq:MDB}
\end{equation}
where $A_{25}$ is the constant mode of the 25-th gauge field component, 
namely the Wilson line. But if we take the $T$-dual in the 
$x^{25}$-direction, $A_{25}$ can be regarded as a transverse scalar 
representing the translation of the lower dimensional D-brane. Substituting 
the string field~(\ref{eq:MDB}) into the string field theory action, 
we get the potential 
\begin{equation}
f(\Phi)=-\frac{S(\Phi)}{\tau_pV_{p+1}}=2\pi^2\ap{}^3\left(-\frac{1}{2\ap}
\phi^2+\frac{1}{3}\left(\frac{3\sqrt{3}}{4}\right)^3\phi^3+\frac{3\sqrt{3}}{4}
\phi A_{25}^2\right). \label{eq:MDC}
\end{equation}
We now integrate out the tachyon field $\phi$ by its equation of motion to 
obtain the effective potential for the marginal parameter $A_{25}$. The 
quadratic equation of motion 
\[ K^3\phi^2-\frac{1}{\ap}\phi+KA_{25}^2=0 \qquad \mathrm{with}\ \ 
K=\frac{3\sqrt{3}}{4} \]
is solved by 
\begin{equation}
\phi^M=\frac{32}{81\sqrt{3}\ap}\left(1-\sqrt{1-\frac{729}{64}\ap{}^2A_{25}^2}
\right) \quad \mathrm{or} \quad \phi^V=\frac{32}{81\sqrt{3}\ap}\left(1+
\sqrt{1-\frac{729}{64}\ap{}^2A_{25}^2}\right), \label{eq:MDD}
\end{equation}
where the superscript $M$ indicates that the solution is on the Marginal 
($M$-)branch in the sense that this branch contains the unstable perturbative 
vacuum $\phi^M=A_{25}=0$ and $A_{25}$ is massless around it, while the 
superscript $V$ stands for the `closed string Vacuum' ($V$-)branch. 
It is clear from these expressions that real solutions exist only if the 
value of the marginal parameter $A_{25}$ lies 
within the following finite range 
\[ |A_{25}|\le\frac{8}{27\ap}. \]
Furthermore, the fact that higher level fields must be switched on to 
implement a large translation implies that the effective potential for 
the massless field $A_{25}$ does not have exact flat directions to any given 
finite order in the level truncation scheme. In fact, the effective potential 
for $A_{25}$ on the $M$-branch at level (1,2) can be obtained by substituting 
$\phi^M$ into~(\ref{eq:MDC}) as 
\begin{eqnarray}
f_{(1,2)}^M(A_{25})&=&-2\pi^2\frac{2^{10}}{3^{10}}\left[1-\frac{3^7}{2^7}
\ap{}^2A_{25}^2-\left(1-\frac{729}{64}\ap{}^2A_{25}^2\right)^{3/2}\right] 
\nonumber \\ &\simeq& \frac{3^3\pi^2}{2^4}(\ap A_{25})^4+\frac{3^8\pi^2}{
2^{11}}(\ap A_{25})^6+\ldots , \label{eq:MDE}
\end{eqnarray}
which is not exactly zero, though $A_{25}^2$ term is absent because of the 
masslessness of $A_{25}$. This $M$-branch as well as the $V$-branch is 
illustrated in Figure~\ref{fig:EB}. 
Our expectation is that the effective potential for $A_{25}$ becomes flatter 
as we increase the level of approximation, because 
large translations need the help of an infinite number of higher level 
fields. This expectation was in fact verified in~\cite{7153} up to 
level (4,8), and this result in turn suggests that the effective potential 
for the string field marginal parameter $A_{25}$ would be exactly flat if 
we could deal with the full string field theory without level truncation 
approximation. 

Since the finiteness of the range of definition of the effective potential 
for $A_{25}$ and the branch structure of it originate in the square root 
branches of the quadratic equations of motion just as in the case of the 
effective tachyon potential we saw in subsection~\ref{sub:effpot}, the 
$M$-branch meets the $V$-branch at the critical value of $A_{25}$, 
\textit{e.g.} $A_{25}=8/27\ap$ at level (1,2), at which the discriminant 
vanishes. And the effective potential cannot be extended beyond that value. 
From the observation (Figure~\ref{fig:EA}) of the expectation values of the 
fields $\eta_n$ as functions of the displacement $a$, we can propose the 
following picture~\cite{8227}: For small values of $a$, mainly the massless 
mode $A_{25}$ contributes to the displacement of the brane so that $A_{25}$ 
grows along the $M$-branch as $a$ increases. When $a$ reaches some critical 
value, however, $A_{25}$ takes a maximum value and after that $A_{25}$ 
begins to decrease toward zero through some missing branch on which various 
higher level fields acquire nonzero expectation values to realize the large 
translation. That is to say, there is two-to-one correspondence between the 
CFT marginal parameter $a$ and the string field theory marginal parameter 
$A_{25}$, although the full set of string field deformation parameters, 
of course, corresponds to the displacement $a$ in a one-to-one manner. 
Given that the above explanation is really correct, the effective potential 
for the string field marginal parameter $A_{25}$ should have the finite range 
of definition even in the full string field theory, though we cannot conclude 
it to our present knowledge. 
\medskip

We now turn to the superstring case. Similar calculations were carried 
out~\cite{08127} using Berkovits' open superstring field theory, and we 
will quote the results from it. For a non-BPS D-brane, we first consider 
the string field truncated to level $\frac{1}{2}$, 
\begin{equation}
\widehat{\Phi}=t\cdot\xi ce^{-\phi}\otimes\sigma_1+A_9\cdot\psi^9\xi 
ce^{-\phi}\otimes I, \label{eq:MDF}
\end{equation}
where $A_9$ is the constant mode of the 9-th gauge field component. 
Substituting it into the action~(\ref{eq:Berko}) for a non-BPS D-brane, 
we get the level $(\frac{1}{2},2)$ potential 
\begin{eqnarray}
f_{(\frac{1}{2},2)}(\widehat{\Phi})&=&-\frac{2\pi^2g_o^2}{V_{p+1}}
S_{(\frac{1}{2},2)}(\widehat{\Phi}) \nonumber \\
&=&-2\pi^2\left(\frac{1}{4}t^2-\frac{1}{2}t^4+\left(\frac{1}{2}-\sqrt{2}
\right)t^2A_9^2-\frac{3}{8}A_9^4\right). \label{eq:MDG}
\end{eqnarray}
To integrate out the tachyon field $t$, we derive the equation of motion 
for $t$, 
\[ -2t\left(t^2-\frac{1}{4}-\left(\frac{1}{2}-\sqrt{2}\right)A_9^2\right)
=0, \]
which has two (up to the sign of $t$) solutions 
\begin{equation}
t^M=0 \quad \mathrm{and} \quad t^V=\pm\frac{1}{2}\sqrt{1-2(2\sqrt{2}-1)A_9^2}.
\label{eq:MDH}
\end{equation}
The $V$-branch, as ever, terminates at the critical value 
\[ |A_9|=\sqrt{\frac{2\sqrt{2}+1}{14}}\simeq 0.523, \]
at which the $V$-branch is merged with the $M$-branch, whereas the $M$-branch 
extends over the whole value of $A_9$. Plugging the solution $t^M=0$ back 
into the potential~(\ref{eq:MDG}), we trivially obtain the effective 
potential for $A_9$ on the $M$-branch at level $(\frac{1}{2},2)$ as 
\begin{equation}
f^M_{(\frac{1}{2},2)}(A_9)=\frac{3\pi^2}{4}A_9^4, \qquad -\infty <A_9<\infty. 
\label{eq:MDI}
\end{equation}
The $M$-branch solution $t^M=0$ always exists because the multiplicative 
conservation of $e^{\pi iF}$ or equivalently the trace over the internal 
Chan-Paton matrices forces the fields in the GSO($-$) sector to enter the 
action in pairs. In other words, the GSO($-$) fields can consistently be 
set to zero altogether, provided that we need not give a nonzero expectation 
value to any of them (especially the tachyon). Hence, if we are not 
interested in the $V$-branch which is continuously connected to the closed 
string vacuum, \textit{e.g.} $t=\pm 1/2\neq 0$ at level $(\frac{1}{2},2)$, 
then the effective potential for $A_9$ on the $M$-branch in the non-BPS 
D-brane case can be regarded as identical to that of the BPS D-brane case. 

When we increase the level to $(\frac{3}{2},3)$, we find the $M$-branch 
effective potential to be 
\begin{equation}
f^M_{(\frac{3}{2},3)}(A_9)=\frac{17\pi^2}{108}A_9^4, \qquad -\infty <A_9<
\infty, \label{eq:MDJ}
\end{equation}
which has become flatter than (\ref{eq:MDI}), as expected. Also at this level 
of approximation, the domain of definition of the effective potential for 
$A_9$ is \textit{not} 
restricted to a finite region. This statement is, however, not 
so meaningful in the following sense: Focusing on the $M$-branch, the fields 
which must be given expectation values are all in the GSO($+$) sector and 
have half-integer levels. At level $(\frac{3}{2},3)$ approximation, every 
field other than $A_9$ has level $\frac{3}{2}$, so that the higher level 
fields enter the action only quadratically at level $(\frac{3}{2},3)$. 
As a result, their equations of motion are linear and can be solved uniquely, 
giving rise to no branch structure. Therefore we have to extend the 
calculations to still higher levels in order to see whether or not the string 
field marginal parameter $A_9$ is restricted to some finite range.

\section{Comments on $p$-adic String Theory}
It was pointed out that the $p$-adic string theory (see~\cite{3278} and 
references therein) can be used to check the conjectures on tachyon 
condensation. In this theory, one can compute tree-level $N$-tachyon 
amplitudes for arbitrary $N\ge 3$. This enables us to write the effective 
spacetime (world-volume of a space-filling D($d-1$)-brane) action 
consistently truncated to the tachyon sector as 
\begin{equation}
S(\phi)=2\frac{p+1}{p-1}\tau_{d-1}\int d^dx\left(-\frac{1}{2}\phi 
p^{-\frac{1}{2}\partial^2}\phi+\frac{1}{p+1}\phi^{p+1}\right), 
\label{eq:padA}
\end{equation}
where $p$ is a prime number and $\tau_{d-1}$ denotes the tension of a 
D($d-1$)-brane. The tachyon potential 
\begin{equation}
V(\phi)=-\frac{1}{2}\phi^2+\frac{1}{p+1}\phi^{p+1} \label{eq:padB}
\end{equation}
shows the same qualitative behavior as $V_{\ell+1}(\phi)$ defined 
in~(\ref{eq:LK}) which is plotted in Figure~\ref{fig:BD}. It has a 
maximum (the original D($d-1$)-brane) at $\phi=1$ and a local minimum (the 
closed string vacuum) at $\phi=0$. In fact, the latter has no perturbative 
physical excitation because the quadratic form $\displaystyle \exp\left(
\frac{\ap}{2}k^2\ln p\right)$ has no zero for finite real value of 
$k^2=-m^2$. We determined the normalization of the action~(\ref{eq:padA}) 
such that $S(\phi=0)=0$ and $S(\phi=1)=-\tau_{d-1}V_d$ hold. The equation 
of motion derived from the action~(\ref{eq:padA}) is given by 
\begin{equation}
p^{-\frac{1}{2}\partial^2}\phi=\phi^p. \label{eq:padC}
\end{equation}
This is solved by the following Gaussian profile 
\begin{eqnarray}
\overline{\phi}(x)&=&\prod_{i=q+1}^{d-1}f(x^i), \label{eq:padD} \\
f(x)&=&p^{\frac{1}{2(p-1)}}\exp\left(-\frac{p-1}{2p\ln p}x^2\right), 
\label{eq:padE}
\end{eqnarray}
because the function $f(x)$ satisfies 
\begin{equation}
\exp\left(-\frac{1}{2}\ln p\frac{\partial^2}{\partial x^2}\right)f(x)
=[f(x)]^p,
\end{equation}
which can be verified by moving to the Fourier space. Since $\overline{\phi}
(x)$~(\ref{eq:padD}) is localized in the ($d-q-1$)-dimensional transverse 
space, it represents a solitonic $q$-brane solution which is to be identified 
with a D$q$-brane in $p$-adic string theory. Its tension $\cT_q$ 
is calculated as 
\begin{eqnarray*}
S(\overline{\phi})&=&-\tau_{d-1}\int d^{q+1}\!x_{{}_{/\!\!/}}\int 
d^{d-q-1}\!x_{\perp}\prod_{i=q+1}^{d-1}\left[p^{\frac{p+1}{2(p-1)}}\exp\left(
-\frac{p^2-1}{2p\ln p}x_i^2\right)\right] \\ &=&-\left(\frac{p^{p/(p-1)}}{
\sqrt{p^2-1}}2\pi\sqrt{\ap{}_p}\right)^{d-q-1}\tau_{d-1}\cdot V_{q+1}
\equiv -\cT_qV_{q+1},
\end{eqnarray*}
where we used the relation 
\[ \frac{1}{2\pi\ap{}_p}=\frac{1}{\ln p} \]
for the tension of the $p$-adic string. It is almost obvious from the 
form~(\ref{eq:padD}) of the lump solution that the effective action for the 
(tachyonic) fluctuation around the lump solutions of various codimension has 
self-similarity (descent relation), as is the case for background independent 
open string field theory and the $\ell\to\infty$ field theory 
model~\cite{11002}. As a result, the ground state wavefunction in the 
transverse space which is associated with the tachyon field living on the 
lump world-volume coincides with the lump profile itself so that only the 
tachyon field condenses when the unstable lump decays, meaning that all 
the higher modes can consistently be set to zero. A comparison between the 
world-volume theory of the solitonic $q$-brane and that of the Dirichlet 
$q$-brane was made in~\cite{3278}, and the fluctuations around the lump 
solutions were investigated in~\cite{2071}. Even if the tachyon condensation 
phenomenon in the $p$-adic string theory had nothing to do with that in the 
ordinary bosonic string theory, it would serve as a simple model which 
reproduces the results expected in string theory.

\chapter{Concluding Remarks}

In this paper, we have reviewed the various aspects of the study of tachyon 
condensation, laying special emphasis on the open string field theories. 
From the point of view of the D-brane phenomenology, we have succeeded in 
obtaining direct evidence for the conjectured dynamics of the D-brane, such 
as the decay of the unstable D-brane, pair-annihilation of the 
brane-antibrane system and the formation of the lower dimensional D-branes as 
lumps or topological defects, although many pieces of indirect evidence had 
already been obtained from the arguments in the framework of the first 
quantized string theory. But it is true that almost everyone has believed 
these conjectures without any rigorous proof because these phenomena are 
such simple and have plain analogies in familiar particle physics. In this 
sense, `examinations of tachyon condensation using string field theory' 
might be `confirmations of the correctness of open string field theories 
in the light of tachyon condensation.' In fact, the original proposal of 
cubic superstring field theory seems to be rejected from this standpoint, 
as we saw in section~\ref{sec:cubicsuper}.
\medskip

Although the general framework of string field theory was constructed 
a decade ago, its development has been rather slow as compared with 
other progress represented by string duality for example, since there 
have been no subjects to which the string field theory could be applied. 
Recently, the study of the off-shell tachyon potential in the context of 
the (non-BPS) D-brane physics at last required making use of open string 
field theory. In the course of the research in this direction, the 
understanding of the string field theory itself, \textit{e.g.} the 
usefulness of the level truncation scheme and the reliability of the 
superstring field theory as well, has made great advances. We hope that 
further developments, especially including even the closed string field 
theory, will be driven by deep insight into physics, hopefully in the 
near future. 

\section*{Acknowledgement}
I would like to thank T. Eguchi for his hospitable guidance. 
And I am very grateful to T. Takayanagi for many discussions and useful 
comments. I would also like to thank Y. Matsuo, T. Kawano, S. Terashima, 
T. Uesugi, M. Fujii, K. Ichikawa and K. Sakai for discussions.

%\bibliographystyle{plain}
%\bibliography{ref}

\begin{thebibliography}{999}
 \bibitem{Halp}K. Bardakci, ``Dual Models and Spontaneous Symmetry Breaking," 
  \textit{Nucl. Phys.} \textbf{B68}(1974)331; \vspace{1.5mm} \\
  K. Bardakci and M. B. Halpern, ``Explicit Spontaneous Breakdown in a Dual 
  Model," \textit{Phys. Rev.} \textbf{D10}(1974)4230; \vspace{1.5mm} \\
  K. Bardakci and M. B. Halpern, ``Explicit Spontaneous Breakdown in a Dual 
  Model II: N Point Functions," \textit{Nucl. Phys.} \textbf{B96}(1975)285; \vspace{1.5mm} \\
  K. Bardakci, ``Spontaneous Symmetry Breakdown in the Standard Dual String 
  Model," \textit{Nucl. Phys.} \textbf{B133}(1978)297.
 \bibitem{Descent}A. Sen, ``Descent Relations Among Bosonic D-branes," 
  \textit{Int. J. Mod. Phys.} \textbf{A14}(1999)4061, hep-th/9902105.
 \bibitem{Reck}A. Recknagel and V. Schomerus, ``Boundary Deformation 
  Theory and Moduli Spaces of D-Branes," \textit{Nucl. Phys.} 
  \textbf{B545}(1999)233, hep-th/9811237; \vspace{2mm} \\
  C. G. Callan, I. R. Klebanov, A. W. Ludwig and J. M. Maldacena, ``Exact 
  Solution of a Boundary Conformal Field Theory," \textit{Nucl. Phys.} 
  \textbf{B422}(1994)417, hep-th/9402113; \vspace{2mm} \\
  J. Polchinski and L. Thorlacius, ``Free Fermion Representation of a 
  Boundary Conformal Field Theory," \textit{Phys. Rev.} \textbf{D50}(1994)622, 
  hep-th/9404008.
 \bibitem{cycle}A. Sen, ``BPS D-branes on Non-supersymmetric Cycles," 
  \textit{JHEP} \textbf{9812}(1998)021, hep-th/9812031.
 \bibitem{Sen}A. Sen, ``Tachyon Condensation on the Brane Antibrane System," 
  \textit{JHEP} \textbf{9808}(1998)012, hep-th/9805170; \vspace{2mm} \\ A. Sen, 
  ``$SO$(32) Spinors of Type I and Other Solitons 
  on Brane-Antibrane Pair," \textit{JHEP} \textbf{9809}(1998)023, 
  hep-th/9808141; \vspace{2mm} \\ A. Sen, ``Non-BPS States and 
  Branes in String Theory," hep-th/9904207.
 \bibitem{DK}E. Witten, ``D-Branes and K-theory," \textit{JHEP} 
  \textbf{9812}(1998)019, hep-th/9810188.
 \bibitem{Hora}P. Ho\v{r}ava, ``Type IIA D-Branes, K-Theory and Matrix 
  Theory," \textit{Adv. Theor. Math. Phys.} \textbf{2}(1999)1373-1404, 
  hep-th/9812135.
 \bibitem{Pol}J. Polchinski, ``String Theory I,II," Cambridge University 
  Press.

 \bibitem{WiSFT1}E. Witten, ``Non-commutative Geometry and String Field 
  Theory," \textit{Nucl. Phys.} \textbf{B268}(1986)253.
 \bibitem{GJ}D. Gross and A. Jevicki, ``Operator Formulation of 
  Interacting String Field Theory(I),(II)," \textit{Nucl. Phys.} 
  \textbf{B283}(1987)1, \textbf{B287}(1987)225.
 \bibitem{CST}E. Cremmer, A. Schwimmer and C. Thorn, ``The Vertex 
  Function in Witten's Formulation of String Field Theory," 
  \textit{Phys. Lett.} \textbf{B179}(1986)57.
 \bibitem{Sam86}S. Samuel, ``The Physical and Ghost Vertices in 
  Witten's String Field Theory," \textit{Phys. Lett.} 
  \textbf{B181}(1986)255.
 \bibitem{LPP}A. LeClair, M.E. Peskin and C.R. Preitschopf, 
  ``String Field Theory on the Conformal Plane (I). Kinematical 
  Principles," \textit{Nucl. Phys.} \textbf{B317}(1989)411; \vspace{2mm} \\ 
  ``String Field Theory on the Conformal Plane (II). Generalized 
  Gluing," \textit{Nucl. Phys.} \textbf{B317}(1989)464.
 \bibitem{Univ}A. Sen, ``Universality of the Tachyon Potential," 
  \textit{JHEP} \textbf{9912}(1999)027, hep-th/9911116.
 \bibitem{RasZw}L. Rastelli and B. Zwiebach, ``Tachyon Potentials, 
  Star Products and Universality," hep-th/0006240.
 \bibitem{8252}V. A. Kosteleck\'{y} and R. Potting, ``Analytical 
  construction of a nonperturbative vacuum for the open 
  bosonic string," hep-th/0008252.
 \bibitem{Hata}H. Hata and S. Shinohara, ``BRST Invariance of the 
  Non-Perturbative Vacuum in Bosonic Open String Field Theory," 
  \textit{JHEP} \textbf{0009}(2000)035, hep-th/0009105.
 \bibitem{Trim}B. Zwiebach, ``Trimming the Tachyon String Field with 
  $SU(1,1)$," hep-th/0010190.
 \bibitem{11238}M. Schnabl, ``Constraints on the tachyon condensate 
  from anomalous symmetries," hep-th/0011238.
 \bibitem{0101014}P. Mukhopadhyay and A. Sen, ``Test of Siegel Gauge for 
  the Lump Solution," hep-th/0101014.
 \bibitem{SenZw}A. Sen and B. Zwiebach, ``Tachyon Condensation 
  in String Field Theory," \textit{JHEP} \textbf{0003}(2000)002, 
  hep-th/9912249.
 \bibitem{Tay}W. Taylor, ``D-brane Effective Field Theory from String 
  Field Theory," \textit{Nucl. Phys.} \textbf{B585}(2000)171-192, 
  hep-th/0001201.
 \bibitem{MoeTay}N. Moeller and W. Taylor, ``Level truncation and the 
  tachyon in open bosonic string field theory," 
  \textit{Nucl. Phys.} \textbf{B583}(2000)105-144, hep-th/002237.
 \bibitem{GabZw}M. R. Gaberdiel and B. Zwiebach, ``Tensor 
  Constructions of Open String Theories I: Foundations," 
  \textit{Nucl. Phys.} \textbf{B505}(1997)569-624, hep-th/9705038.
 \bibitem{KS}V.A. Kosteleck\'{y} and S. Samuel, 
  ``On a Nonperturbative Vacuum for the Open Bosonic
  String," \textit{Nucl. Phys.} \textbf{B336}(1990)263.
 \bibitem{Hor}G. T. Horowitz, J. Morrow-Jones, S. P. Martin and R. P. 
  Woodard, ``New Exact Solutions for the Purely Cubic Bosonic String Field 
  Theory," \textit{Phys. Rev. Lett. } \textbf{60}(1988)261.
 \bibitem{IIB}M. Srednicki, ``IIB or not IIB," \textit{JHEP} 
  \textbf{9808}(1998)005, hep-th/9807138.
 \bibitem{Yi}P. Yi, ``Membranes from Five-Branes and Fundamental Strings 
  from D$p$-Branes," \textit{Nucl. Phys.} \textbf{B550}(1999)214-224, 
  hep-th/9901159.
 \bibitem{Conf}O. Bergman, K. Hori and P. Yi, ``Confinement on the Brane," 
  \textit{Nucl. Phys.} \textbf{B580}(2000)289-310, hep-th/0002223.
 \bibitem{9061}G. Gibbons, K. Hori and P. Yi, ``String Fluid from Unstable 
  D-branes," hep-th/0009061.
 \bibitem{10240}A. Sen, ``Fundamental Strings in Open String Theory at 
  the Tachyonic Vacuum," hep-th/0010240.
 \bibitem{12081}M. Kleban, A. Lawrence and S. Shenker, ``Closed strings 
  from nothing," hep-th/0012081.
 \bibitem{0101213}U. Lindstr\"{o}m and M. Zabzine, ``Strings at the Tachyonic 
  Vacuum," hep-th/0101213.
 \bibitem{NonBPSD}A. Sen, ``Supersymmetric World-volume Action for 
  Non-BPS D-branes," \textit{JHEP} \textbf{9910}(1999)008, hep-th/9909062.
 \bibitem{3122}M.R. Garousi, ``Tachyon couplings on non-BPS D-branes and 
  Dirac-Born-Infeld action," \textit{Nucl. Phys.} 
  \textbf{B584}(2000)284-299, hep-th/0003122.
 \bibitem{pyo}E.A. Bergshoeff, M. de Roo, T.C. de Wit, E. Eyras and 
  S. Panda, ``$T$-duality and Actions for Non-BPS D-branes," 
  \textit{JHEP} \textbf{0005}(2000)009, hep-th/0003221.
 \bibitem{klu}J. Kluso\v{n}, ``Proposal for non-BPS D-brane action," 
  \textit{Phys. Rev.} \textbf{D62}(2000)126003, hep-th/0004106.
 \bibitem{8033}W. Taylor, ``Mass generation from tachyon condensation for 
  vector fields on D-branes," \textit{JHEP} 
  \textbf{0008}(2000)038, hep-th/0008033.
 \bibitem{5085}J. R. David, ``$U(1)$ gauge invariance from open string field 
  theory," \textit{JHEP} \textbf{0010}(2000)017, hep-th/0005085.
 \bibitem{0101162}H. Hata and S. Teraguchi, ``Test of the Absence of Kinetic 
  Terms around the Tachyon Vacuum in Cubic String Field Theory," 
  hep-th/0101162.
 \bibitem{TTUpriv}T. Takayanagi, S. Terashima and T. Uesugi, private 
  communication. 
 \bibitem{RSZ}L. Rastelli, A. Sen and B. Zwiebach, ``String Field Theory 
  Around the Tachyon Vacuum," hep-th/0012251.
 
 \bibitem{FMS}D. Friedan, E. Martinec and S. Shenker, ``Conformal Invariance, 
  Supersymmetry and String Theory," \textit{Nucl. Phys. } 
  \textbf{B271}(1986)93. 
 \bibitem{WiSFT2}E. Witten, ``Interacting Field Theory of Open Superstrings," 
  \textit{Nucl. Phys. } \textbf{B276}(1986)291.
 \bibitem{Wendt}C. Wendt, ``Scattering Amplitudes and Contact 
  Interactions in Witten's Superstring Field Theory," 
  \textit{Nucl. Phys.} \textbf{B314}(1989)209.
 \bibitem{4112}P-J. De Smet and J. Raeymaekers, ``The Tachyon Potential 
  in Witten's Superstring Field Theory," \textit{JHEP} 
  \textbf{0008}(2000)020, hep-th/0004112.
 \bibitem{PTY}C. R. Preitschopf, C. B. Thorn and S. A. Yost, ``Superstring 
  Field Theory," \textit{Nucl. Phys. } \textbf{B337}(1990)363.
 \bibitem{Med}I. Ya. Aref'eva, P. B. Medvedev and A. P. Zubarev, 
  ``Background Formalism for Superstring Field Theory," 
  \textit{Phys. Lett. } \textbf{240B}(1990)356.
 \bibitem{Zub}I. Ya. Aref'eva, P. B. Medvedev and A. P. Zubarev, ``New 
  Representation for String Field Solves the Consistency 
  Problem for Open Superstring Field Theory," 
  \textit{Nucl, Phys. } \textbf{B341}(1990)464.
 \bibitem{Uro}B. V. Urosevic and A. P. Zubarev, ``On the component 
  analysis of modified superstring field theory actions," 
  \textit{Phys. Lett.} \textbf{246B}(1990)391.
 \bibitem{Aref}I. Ya. Aref'eva, P. B. Medvedev and A. P. Zubarev, 
  ``Non-Perturbative Vacuum for Superstring Field Theory 
  and Supersymmetry Breaking," 
  \textit{Mod. Phys. Lett.} \textbf{A6}(1991)949.
 \bibitem{11117}I. Ya. Aref'eva, A. S. Koshelev, D. M. Belov 
  and P. B. Medvedev, ``Tachyon Condensation in Cubic 
  Superstring Field Theory," hep-th/0011117.
 \bibitem{OV}H. Ooguri and C. Vafa, ``$N=2$ heterotic strings,"
  \textit{Nucl. Phys. }\textbf{B367}(1991)83.
 \bibitem{twisted}N. Berkovits, ``The Ten-Dimensional Green-Schwarz 
  Superstring is a Twisted Neveu-Schwarz-Ramond String," 
  \textit{Nucl. Phys. } \textbf{B420}(1994)332, hep-th/9308129.
 \bibitem{Uniqueness}N. Berkovits and C. Vafa, ``On the Uniqueness 
  of String Theory," \textit{Mod. Phys. Lett. } 
  \textbf{A9}(1994)653, hep-th/9310170.
 \bibitem{GS}N. Berkovits, ``Covariant Quantization of the Green-Schwarz 
  Superstring in a Calabi-Yau Background," \textit{Nucl. Phys. } 
  \textbf{B431}(1994)258, hep-th/9404162.
 \bibitem{Top}N. Berkovits and C. Vafa, ``$N=4$ Topological Strings," 
  \textit{Nucl. Phys. } \textbf{B433}(1995)123, hep-th/9407190.
 \bibitem{sP}N. Berkovits, ``Super-Poincar\'{e} Invariant 
  Superstring Field Theory," \textit{Nucl. Phys.} 
  \textbf{B459}(1996)439, hep-th/9503099.
 \bibitem{Eche}N. Berkovits and C.T. Echevarria, ``Four-Point Amplitude from 
  Open Superstring Field Theory," \textit{Phys. Lett.} 
  \textbf{B478}(2000)343-350, hep-th/9912120.
 \bibitem{New}N. Berkovits, ``A New Approach to Superstring Field Theory," 
  \textit{Fortsch. Phys.} \textbf{48}(2000)31-36, hep-th/9912121.
 \bibitem{NS}N. Berkovits, ``The Tachyon Potential in Open Neveu-Schwarz 
  String Field Theory," \textit{JHEP} \textbf{0004}(2000)022, hep-th/0001084.
 \bibitem{BSZ}N. Berkovits, A. Sen and B. Zwiebach, ``Tachyon Condensation 
  in Superstring Field Theory," \textit{Nucl. Phys.} 
  \textbf{B587}(2000)147-178, hep-th/0002211.
 \bibitem{3220}P-J. De Smet and J. Raeymaekers, ``Level Four Approximation to 
  the Tachyon Potential in Superstring Field Theory," 
  \textit{JHEP} \textbf{0005}(2000)051, hep-th/0003220.
 \bibitem{4015}A. Iqbal and A. Naqvi, ``Tachyon Condensation on a Non-BPS 
  D-Brane," hep-th/0004015.
 \bibitem{7235}J. R. David, ``Tachyon condensation in the D0/D4 system," 
  hep-th/0007235.
 
 \bibitem{2117}J.A. Harvey, and P. Kraus, ``D-Branes as Unstable Lumps in 
  Bosonic Open String Field Theory," \textit{JHEP} 
  \textbf{0004}(2000)012, hep-th/0002117.
 \bibitem{3031}R. de Mello Koch, A. Jevicki, M. Mihailescu and R. Tatar, 
  ``Lumps and $p$-Branes in Open String Field Theory," \textit{Phys. Lett.} 
  \textbf{B482}(2000)249-254, hep-th/0003031.
 \bibitem{MSZ}N. Moeller, A. Sen and B. Zwiebach, ``D-branes as Tachyon 
  Lumps in String Field Theory," \textit{JHEP} 
  \textbf{0008}(2000)039, hep-th/0005036.
 \bibitem{8053}R. de Mello Koch and J.P. Rodrigues, ``Lumps in level 
  truncated open string field theory," \textit{Phys. Lett.} 
  \textbf{B495}(2000)237-244, hep-th/0008053.
 \bibitem{8101}N. Moeller, ``Codimension two lump solutions in string field 
  theory and tachyonic theories," hep-th/0008101.
 
 \bibitem{WiBI1}E. Witten, ``On Background Independent Open String 
  Field Theory," \textit{Phys. Rev.} \textbf{D46}(1992)5467, hep-th/9208027.
 \bibitem{WiBI2}E. Witten, ``Some Computations in Background Independent 
  Off-shell String Theory," \textit{Phys. Rev.} 
 \textbf{D47}(1993)3405, hep-th/9210065.
 \bibitem{Sh1}S. Shatashvili, ``Comment on the Background Independent Open 
 String Theory," \textit{Phys. Lett.} \textbf{B311}(1993)83, hep-th/9303143.
 \bibitem{Sh2}S. Shatashvili, ``On the Problems with Background Independence 
 in String Theory," hep-th/9311177.
 \bibitem{LiWi}K. Li and E. Witten, ``Role of Short Distance Behavior in 
 Off-shell Open String Field Theory," \textit{Phys. Rev.} 
 \textbf{D48}(1993)853, hep-th/9303067.
 \bibitem{GerSh}A.A. Gerasimov and S.L. Shatashvili, ``On Exact 
  Tachyon Potential in Open String Field Theory," \textit{JHEP} 
  \textbf{0010}(2000)034, hep-th/0009103.
 \bibitem{KMM1}D. Kutasov, M. Mari\~{n}o and G. Moore, ``Some Exact 
  Results on Tachyon Condensation in String Field Theory," 
  \textit{JHEP} \textbf{0010}(2000)045, hep-th/0009148.
 \bibitem{KMM2}D. Kutasov, M. Mari\~{n}o and G. Moore, 
  ``Remarks on Tachyon Condensation in Superstring Field 
  Theory," hep-th/0010108.
 \bibitem{GhS}D. Ghoshal and A. Sen, ``Normalisation of the Background 
  Independent Open String Field Theory Action," \textit{JHEP} 
  \textbf{0011}(2000)021, hep-th/0009191.
 \bibitem{HKM}J.A. Harvey, D. Kutasov and E.J. Martinec, 
  ``On the relevance of tachyons," hep-th/0003101.
 \bibitem{10247}S. Dasgupta and T. Dasgupta, ``Renormalization 
  Group Analysis of Tachyon Condensation," hep-th/0010247.
 \bibitem{Affleck}I. Affleck and A. W. Ludwig, ``Universal 
  noninteger `ground-state degeneracy' in critical quantum systems," 
  \textit{Phys. Rev. Lett. } \textbf{67}(1991)161.
 \bibitem{Ludwig}I. Affleck and A. W. Ludwig, ``Exact conformal 
  field theory results on the multichannel Kondo effect: 
  single fermion Green's function, self-energy, and resistivity," 
  \textit{Phys. Rev. } \textbf{B48}(1993)7297.
 \bibitem{12150}A. Fujii and H. Itoyama, ``Behavior of Boundary 
  String Field Theory Associated with Integrable Massless Flow," 
  hep-th/0012150.
 \bibitem{FT}E.S. Fradkin and A.A. Tseytlin, ``Non-linear Electrodynamics 
  from Quantized Strings," \textit{Phys. Lett.} \textbf{163B}(1985)123.
 \bibitem{11009}A. A. Gerasimov and S. L. Shatashvili, ``Stringy Higgs 
  Mechanism and the Fate of Open Strings," hep-th/0011009.
 \bibitem{11002}S. Moriyama and S. Nakamura, ``Descent Relation of 
  Tachyon Condensation from Boundary String Field Theory," 
  hep-th/0011002.
 \bibitem{10218}O. Andreev, ``Some Computations of Partition 
  Functions and Tachyon Potentials in Background Independent Off-Shell 
  String Theory," hep-th/0010218.
 \bibitem{Tsey}A.A. Tseytlin, ``Sigma model approach to string theory 
  effective actions with tachyons," hep-th/0011033.
 \bibitem{DBIDBI}G. Arutyunov, S. Frolov, S. Theisen and A.A. Tseytlin, 
  ``Tachyon condensation and universality of DBI action," hep-th/0012080.
 \bibitem{12198}P. Kraus and F. Larsen, ``Boundary String Field 
  Theory of the $D\overline{D}$ System," hep-th/0012198.
 \bibitem{TTU}T. Takayanagi, S. Terashima and T. Uesugi, 
  ``Brane-Antibrane Action from Boundary String Field Theory," 
  hep-th/0012210.
 \bibitem{mahoh}M. Alishahiha, H. Ita and Y. Oz, ``On Superconnections and 
  the Tachyon Effective Action," hep-th/0012222.
 \bibitem{Billo}M. Bill\'o, B. Craps and F. Roose, ``Ramond-Ramond couplings of non-BPS D-branes," 
  \textit{JHEP} \textbf{9906}(1999)033, hep-th/9905157.
 \bibitem{Wilk}C. Kennedy and A. Wilkins, ``Ramond-Ramond Coupling on Brane-AntiBrane systems," 
  \textit{Phys. Lett.} \textbf{B464}(1999)206-212, hep-th/9905195.
 \bibitem{SW}N. Seiberg and E. Witten, ``String Theory and 
  Noncommutative Geometry," \textit{JHEP} \textbf{9909}(1999)032, 
  hep-th/9908142.
 \bibitem{10021}L. Cornalba, ``Tachyon Condensation in Large 
  Magnetic Fields with Background Independent String Field Theory," 
  hep-th/0010021.
 \bibitem{10028}K. Okuyama, ``Noncommutative Tachyon from Background 
  Independent Open String Field Theory," hep-th/0010028.
 \bibitem{11108}D. Nemeschansky and V. Yasnov, ``Background 
  Independent Open String Field Theory and Constant $B$-Field," 
  hep-th/0011108.
 \bibitem{12089}J. R. David, ``Tachyon condensation using the disc partition 
  function," hep-th/0012089.
 
 \bibitem{GMS}R. Gopakumar, S. Minwalla and A. Strominger, ``Noncommutative 
  Solitons," \textit{JHEP} 
  \textbf{0005}(2000)020, hep-th/0003160.
 \bibitem{Zhou}C-G. Zhou, ``Noncommutative Scalar Solitons at Finite 
  $\theta$," hep-th/0007255.
 \bibitem{Nest}B. Durhuus, T. Jonsson and R. Nest, ``Noncommutative Scalar 
  Solitons: Existence and Nonexistence," hep-th/0011139.
 \bibitem{DMR}K. Dasgupta, S. Mukhi and G. Rajesh, ``Noncommutative 
  Tachyons," \textit{JHEP} \textbf{0006}(2000)022, hep-th/0005006.
 \bibitem{HKLM}J. A. Harvey, P. Kraus, F. Larsen and E. J. Martinec, 
  ``D-branes and Strings as Non-commutative Solitons," \textit{JHEP} 
  \textbf{0007}(2000)042, hep-th/0005031.
 \bibitem{8064}J.A. Harvey, P. Kraus and F. Larsen, ``Tensionless Branes 
  and Discrete Gauge Symmetry," \textit{Phys. Rev.} 
  \textbf{D63}(2001)026002, hep-th/0008064.
 \bibitem{Rey}G. Mandal and S-J. Rey, ``A Note on D-Branes of Odd 
  Codimensions from Noncommutative Tachyons," \textit{Phys. Lett.} 
  \textbf{B495}(2000)193-200, hep-th/0008214.
 \bibitem{7217}C. Sochichiu, ``Noncommutative Tachyonic Solitons. 
  Interaction with Gauge Field," \textit{JHEP} 
  \textbf{0008}(2000)026, hep-th/0007217.
 \bibitem{7226}R. Gopakumar, S. Minwalla and A. Strominger, ``Symmetry 
  Restoration and Tachyon Condensation in Open String Theory," 
  hep-th/0007226.
 \bibitem{Seib}N. Seiberg, ``A Note on Background Independence in 
  Noncommutative Gauge Theories, Matrix Model, and Tachyon Condensation," 
  \textit{JHEP} \textbf{0009}(2000)003, hep-th/0008013.
 \bibitem{9038}A. Sen, ``Some Issues in Non-commutative Tachyon 
  Condensation," \textit{JHEP} \textbf{0011}(2000)035, hep-th/0009038.
 \bibitem{10060}J. A. Harvey, P. Kraus and F. Larsen, ``Exact Noncommutative 
  Solitons," \textit{JHEP} \textbf{0012}(2000)024, hep-th/0010060.
 \bibitem{0101125}L-S. Tseng, ``Noncommutative Solitons and Intersecting 
  D-Branes," hep-th/0101125.
 \bibitem{Finn}F. Larsen, ``Fundamental Strings as Noncommutative Solitons," 
  hep-th/0010181.
 \bibitem{6071}E. Witten, ``Noncommutative Tachyons And String Field Theory," 
  hep-th/0006071.
 \bibitem{10034}M. Schnabl, ``String field theory at large $B$-field and 
  noncommutative geometry," \textit{JHEP} \textbf{0011}(2000)031, 
  hep-th/0010034. 
 
 \bibitem{8227}B. Zwiebach, ``A Solvable Toy Model for Tachyon Condensation 
  in String Field Theory," \textit{JHEP} 
  \textbf{0009}(2000)028, hep-th/0008227.
 \bibitem{8231}J.A. Minahan and B. Zwiebach, ``Field theory models for 
  tachyon and gauge field string dynamics," \textit{JHEP} 
  \textbf{0009}(2000)029, hep-th/0008231.
 \bibitem{9246}J. A. Minahan and B. Zwiebach, ``Effective Tachyon Dynamics 
  in Superstring Theory," hep-th/0009246.
 \bibitem{11226}J. A. Minahan and B. Zwiebach, ``Gauge Fields and Fermions 
  in Tachyon Effective Field Theories," hep-th/0011226.
  
 \bibitem{7153}A. Sen and B. Zwiebach, ``Large Marginal Deformations in 
  String Field Theory," \textit{JHEP} \textbf{0010}(2000)009, hep-th/0007153.
 \bibitem{08127}A. Iqbal and A. Naqvi, ``On Marginal Deformations in 
  Superstring Field Theory," hep-th/0008127.
 \bibitem{3278}D. Ghoshal and A. Sen, ``Tachyon Condensation and Brane 
  Descent Relations in $p$-adic String Theory," \textit{Nucl. Phys.}
  \textbf{B584}(2000)300-312, hep-th/0003278.
 \bibitem{2071}J. A. Minahan, ``Mode Interactions of the Tachyon Condensate 
  in $p$-adic String Theory," hep-th/0102071. 
 
\end{thebibliography}

\end{document}